\documentclass[12pt,letterpaper]{report}

\newcommand{\R}{\mathbb{R}}

\newcommand{\Z}{\mathbb{Z}}
\newcommand{\p}{\partial}
\newcommand{\eps}{\epsilon}
\newcommand{\dt}{\Delta t}
\newcommand{\dx}{\Delta x}
\newcommand{\du}{\Delta u}
\newcommand{\dv}{\Delta v}
\newcommand{\ds}{\Delta s}

\newcommand{\cross}{\times}
\newcommand{\bb}{\mathbf}

\newcommand{\grad}{\nabla}

\usepackage{pdflscape}

\usepackage{amssymb,amsmath,amsthm} 
\usepackage{graphicx,ctable,booktabs}
\usepackage{verbatim}
\usepackage{multirow}

\usepackage{algorithm, algorithmic}

\usepackage{epstopdf}

\newcommand{\thesistitle}{Modeling the Mitral Valve}
\newcommand{\thesisauthor}{Alexander D. Kaiser}
\newcommand{\thesisadvisor}{Professor Charles S. Peskin}
\newcommand{\graddate}{September, 2017}

\usepackage{setspace}

\usepackage[hidelinks]{hyperref}
\usepackage{hyperref}
\hypersetup{
    allcolors=black,
    linktoc=all 
}

\begin{document}

\pagenumbering{roman}
%
\thispagestyle{empty}
\begin{center}
  {\Large\textbf{\thesistitle}}
  \vspace{.7in}

  by
  \vspace{.7in}

  \thesisauthor
  \vfill

\begin{doublespace}
  A dissertation submitted in partial fulfillment\\
  of the requirements for the degree of\\
  Doctor of Philosophy \\
  Department of Mathematics\\
  New York University\\
  \graddate
\end{doublespace}
\end{center}
\vfill

\noindent\makebox[\textwidth]{\hfill\makebox[2.5in]{\hrulefill}}\\
\makebox[\textwidth]{\hfill\makebox[2.5in]{\hfill\thesisadvisor\hfill}}
\newpage

\begin{center}

\thispagestyle{empty}

$\;$
 
\vspace{1in}

\copyright $\,$ Alexander D. Kaiser 

\vspace{.3in}

All Rights Reserved, 2017

\vfill 

\end{center}
\newpage 



\addcontentsline{toc}{section}{Acknowledgements}
\section*{Acknowledgements}

First, thank you to my advisor, Charles S. Peskin. 
It was through a stroke of good timing and good luck that I took his thrilling course on differential equations in biology when I was seeking an advisor. 
He put questions of physiology into a familiar mathematical framework, yet the answers they returned were wild and different. 
I was particularly struck by his work on the aortic valve, which influenced me to investigate the mitral valve. 
He introduced me to many new methods and ways of thinking. 
From this perspective, boundaries between fields were permeable, and writing my thesis in mathematics on heart valves became a plausible and exciting thing to do. 
Charlie's creativity is inspiring; his enthusiasm is contagious. 
Working under his guidance has been a great joy.

Thank you to David McQueen, who provided assistance on so many aspects of this project. 
He helped interpret and process medical image data and supplied beautiful flow visualization programs. 
I am grateful for his incredible volume of knowledge on running fluid-structure interaction simulations, and his willingness to share this wisdom on this and many other things with me.

Thank you to Boyce Griffith for his consistent guidance and commentary. 
His fluid solvers were indispensable to this project. 
I appreciate his advice and assistance on running immersed boundary computations and valve modeling.

Thank you also to Esteban Tabak and Antoine Cerfon for serving on my committee.

Thank you to Thomas Bringley, whom I wish I could have met, and his parents Jim and Maureen, for establishing a memorial fellowship in his honor. 
The fellowship was of much help on laboratory expenses, and it was an honor to receive it.

Many friends and colleagues across many fields helped me on this project, and I am happy to acknowledge them here. 
Thank you to Mark Alu and Cynthia Loomis, of the Experimental Pathology Research Laboratory at the New York University Langone Medical Center. They graciously agreed to start a collaboration to examine mitral valve fiber structure and anatomy, which resulted in most of the work described in Charter \ref{anatomy}. 
Thank you to Jason Wexler, for his consistent wisdom on fluid and solid mechanics and over a decade of great friendship; 
to Stephanie Chang, for teaching me to dissect hearts in my kitchen; 
to Michael Lewis, for many helpful discussions on numerical methods and all things calculus; 
to Shenglong Wang, for being the most responsive and helpful manager of a high performance computing center possible; 
and 
to Sarah Leone, for arranging for permission to take me to the anatomy laboratory at NYU medical center and showing me human hearts and mitral valves.

Further, 
thank you to Jonathan Goodman, for advising me on my master's thesis, 
and 
to my early mentors David Bailey, Michael Cheng and Emily Peters. 
Thank you to Marina Ratner, who taught my real analysis and recently passed away; I wish I could have told her that I finished my doctorate. 
Thank you to musical mentor C.K. Ladzekpo, whose lessons about rhythms are (not so secretly) also lessons about life. 
And thank you to my many friends, for listening to me talk about heart valves and always being so great.

Finally, I would like to acknowledge parents my Nancy and David, sister Anna and grandmother Joy for their consistent, unwavering support.

\newpage

\section*{Abstract}\addcontentsline{toc}{section}{Abstract}

This thesis is concerned with modeling and simulation of the mitral valve, one of the four valves in the human heart. 
The valve is composed of leaflets attached to a ring, the free edges of which are supported by a system of chordae, which themselves are anchored to muscles inside the heart.  
First, we examine valve anatomy and show the results of original dissections. 
These display the gross anatomy and information on fiber structure of the mitral valve. 
Next, we build a model valve following a design-based approach to elasticity. 
We incorporate information from the dissections to specify the fiber topology of this model. 
We assume the valve achieves mechanical equilibrium while supporting a static pressure load. 
The solution to the resulting differential equations determines the pressurized configuration of the valve model. 
To complete the model we then specify a constitutive law based on experimental stress-strain relations from the literature. 
Finally, using the immersed boundary method, we simulate the model valve in fluid in a computer test chamber. 
The aim of this work is to determine the basic principles and mechanisms underlying the anatomy and function of the mitral valve.


\newpage

\tableofcontents
\newpage

\addcontentsline{toc}{section}{List of Figures}
\listoffigures
\newpage

\addcontentsline{toc}{section}{List of Tables}
\listoftables
\newpage

\pagenumbering{arabic} 

\chapter{Introduction}

The mitral valve is one of the four valves in the human heart. 
It lies between the left atrium, which serves as a staging chamber for oxygenated blood coming from the lungs, and the left ventricle, which is the main muscular pumping chamber that sends blood to all of the tissues and organs of the body. 
The valve is composed of leaflets, thin membranous flaps of tissue attached to a ring. 
The free edges of the leaflets are supported like a parachute by a system of strings called chordae tendineae, which themselves are anchored to muscles called papillary muscles  that protrude from the ventricular wall.

This thesis concerns modeling and simulation of the mitral valve.

First, we examine valve anatomy. 
We review classical textbooks and literature. 
I performed a number of dissections, the results of which will be presented herein. 
The specimens were obtained fresh from a local butcher as meat and dissected in my kitchen. 
Later, I began a collaboration with the histopathology core facility at New York University Langone Medical Center. 
We stained and imaged a number of valves in an effort to obtain more information about valve anatomy, specifically the fiber structure of the valve leaflets. 
Observations from my studies and literature are described in Chapter \ref{anatomy}.

Next, we build a model of the valve following a design-based approach to elasticity. 
We assume the valve achieves mechanical equilibrium while supporting a static pressure load. 
The solution of the resulting partial differential equations specifies the pressurized configuration of the valve model. 
This provides information about the tension throughout the model valve. 
Combining this with the experimental information about the type of constitutive laws that such tissue produces, we generate a constitutive law and reference configuration for the model. 
This creates a general model suitable for simulations. 

Finally, using the immersed boundary method, we simulate the model valve in fluid. 
The starting geometry and constitutive law are determined by the previous step. 
The valve is placed in a model test chamber, and simulations are driven by prescribed waveforms of the pressures upstream and downstream of the valve.

My goal is to provide evidence of what makes a valve function effectively and reliably, and what types of problems occur when valves fail. 
All of these steps work together to form a comprehensive strategy for modeling and simulation of the mitral valve. 
They can be summarized as follows: 

\begin{enumerate}

\item Specify and solve a problem of mechanical equilibrium for the pressurized configuration of the valve. 

\item Use the pressurized configuration to assign a geometry, reference configuration and constitutive laws to a model valve. 

\item Simulate this valve in fluid by the immersed boundary method. Run computational experiments to answer questions about valve mechanics. 

\end{enumerate}

Most of the valve geometry is derived from mathematics and general anatomical observations, rather than measured. 
Although we make use of measurements of the elastic properties of valve tissue, the geometry is not made from a scan of a dissected valve, nor is it a simple analytic shape. 
This is the essential difference of this line of research from many other modeling studies of the mitral valve.

I would like to add one note to the reader before we begin.  
This work contains many images, as well as videos that are online.\footnote{ 
\noindent 
See \url{https://cims.nyu.edu/~kaiser} and  

\noindent \url{https://www.youtube.com/channel/UCWsLmHAiQFnYILsgI8WwJXg}

\noindent 
The institutional URL is subject to change but new ones may be found by searching.
The youtube link should remain valid indefinitely.}
Movies of simulations convey the details of flows in ways that still images cannot. 
If you are interested in the details of this work, please seek out and view these animations.

\newpage

\chapter{Mitral Valve Anatomy}
\label{anatomy}

A standard, general medical reference on cardiac anatomy is Frank Netter's heart atlas \cite{netter}. 
He describes the mitral and tricuspid valves, together called the atrioventricular valves, as follows: 

\begin{quotation}
\singlespacing 
``Each \emph{atrioventicular-valve apparatus} consists of a number of \emph{cusps, chordae tendineae} and \emph{papillary muscles}. 

The cusps are thin, yellowish-white, glistening membranes, roughly trapezoidal shaped with fine irregular edges. 
They originate from the \emph{annulus fibrosus}, which is a somewhat-ill-defined and rather unimpressive fibrous ring around each atrioventricular orifice...  

The atrial surface of the atrioventricular valve is rather smooth... The ventricular surface is very irregular, owing to the insertion of the chordae tendineae... 

The extreme edges of the cusps are thin and delicate and have a sawtoothlike appearance because of the insertion of equally fine chordae. 

The [chordae tendineae] originate from or near the apices of the papillary muscles. 
They form a few strong, tendinous chords which subdivide into several thinner strands as they approach the valve edges... 

The depth of the \emph{commissures} between the cusps is variable, but the commissures never reach the \emph{annulus}, so the cusps are only incompletely separated from each other. 

The \emph{mitral} (bicuspid) \emph{valve} actually is made up of four cusps. 
These are two large ones -- the \emph{anterior} (aortic) and \emph{posterior} (mural) cusps -- and two small \emph{commissural cusps}. 
Here, as in the case of the tricuspid valve, the commissures are never complete.'' 
\end{quotation}

These gross features of mitral valve anatomy will all be present in our model. 
Some are specified; others emerge from the solutions to the problems of mechanical equilibrium described in Chapter \ref{static}. 

To build models, we wish to understand further details of mitral valve anatomy. 
Mitral valve tissue is fibrous, which means that there are one or more local directions in which the leaflet exerts more force. 
We will perform a number of dissections and stains to attempt to answer a fundamental question about mitral valve anatomy: what is the fiber structure of the mitral valve? 
More specifically, how are collagen fibers oriented to give the valve its large scale mechanical properties?

\section{Dissections}
\label{Dissections}
For this project, I performed a number of dissections to study mitral valve anatomy. 
Additionally, I visited the anatomy laboratory at NYU Langone Medical Center and viewed the hearts and mitral valves of numerous human cadavers. 
Many, if not most, of my opinions about mitral valve anatomy details were formed through dissections and inspections, rather than literature, and I will summarize them here. 
All dissections shown in this chapter are original work. 
Photographs were taken with a stock iPhone 7 and a Canon EOS Rebel T5i, and lit with a mixture of natural light from windows and artificial ambient light unless specified otherwise.
The camera and photographer (either the author or Mark Alu from the Experimental Pathology Research Laboratory) are cited in the captions. 

Fresh porcine (pig) hearts were purchased as meat from a local butcher (The Meat Hook, Brooklyn, New York). 
The ventricles, mitral and tricuspid valves were generally free from excessive damage. 
The atria were largely removed from the specimens before we received them.

\begin{figure}[H]
\centering
\includegraphics[scale=0.13]{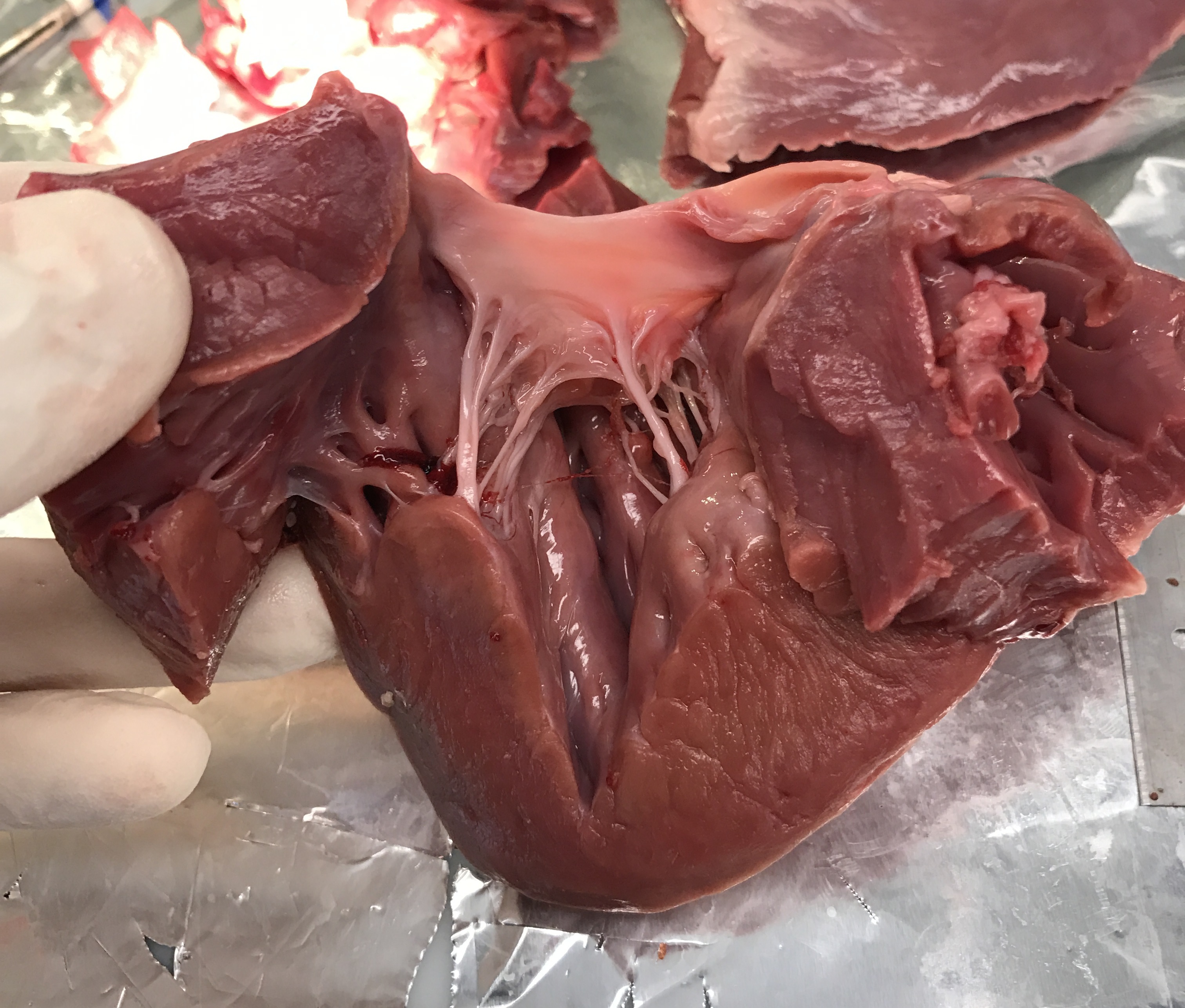}
\caption{Porcine mitral valve in left ventricle, mid dissection. Camera, iPhone 7; photographer, the author.}
\label{valve_in_heart}
\end{figure}

We dissected the ventricles using the following procedure to expose the mitral valve. 
First, the right ventricle is removed by cutting along the boundary between the ventricular wall and septum. 
Next, a cut is made in the aortic valve ring at the point most distant from the center of the anterior mitral valve leaflet. 
This cut is continued down the left ventricle, parallel to the long axis of the heart, to the apex. 
This allows the left ventricle to be opened. 
With the muscle spread apart, the anterior leaflet, papillary muscles and chordae tendineae are visible.  
From this position, the ventricular tissue can be gradually removed, starting with the apex and septum. 
This is shown in figure \ref{valve_in_heart}, which I took with the hand that is not visible in the frame. 
The anterior leaflet and its connection to the papillary muscles are visible in the center of frame. 
The posterior leaflet is behind the anterior and is not visible in this view. 
Portions of the septal wall and apex are removed. 
The free wall and entire mitral valve structure remain intact. 

Next, we remove additional ventricular muscle tissue to the extent that this is possible without damaging the valve. 
We excise the papillary muscles from the their attachments to the ventricle wall, trying to to maintain the natural attachment of the chordae to the papillary tips. 
We slice coarsely around the valve ring and fully free the mitral valve structure. 
Finally, we cut the center of the posterior leaflet, allowing the valve to sit flat. 

\begin{figure}[H]
\centering 
\includegraphics[width=\textwidth]{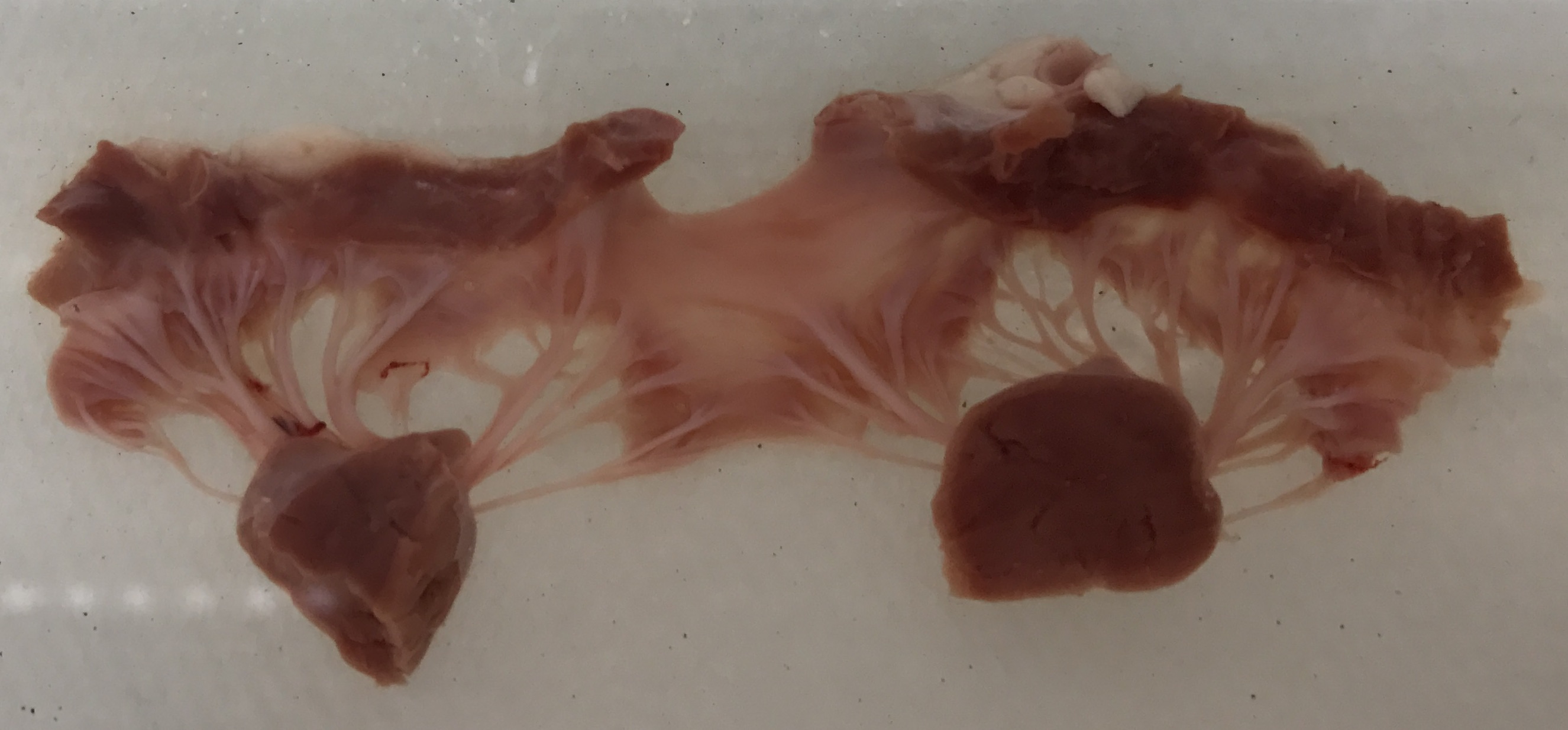}
\caption{Porcine mitral valve resting in ethanol. Camera, iPhone 7; photographer, the author.}
\label{valve_basic}
\end{figure}

\begin{figure}[H]
\centering 
\includegraphics[width=\textwidth]{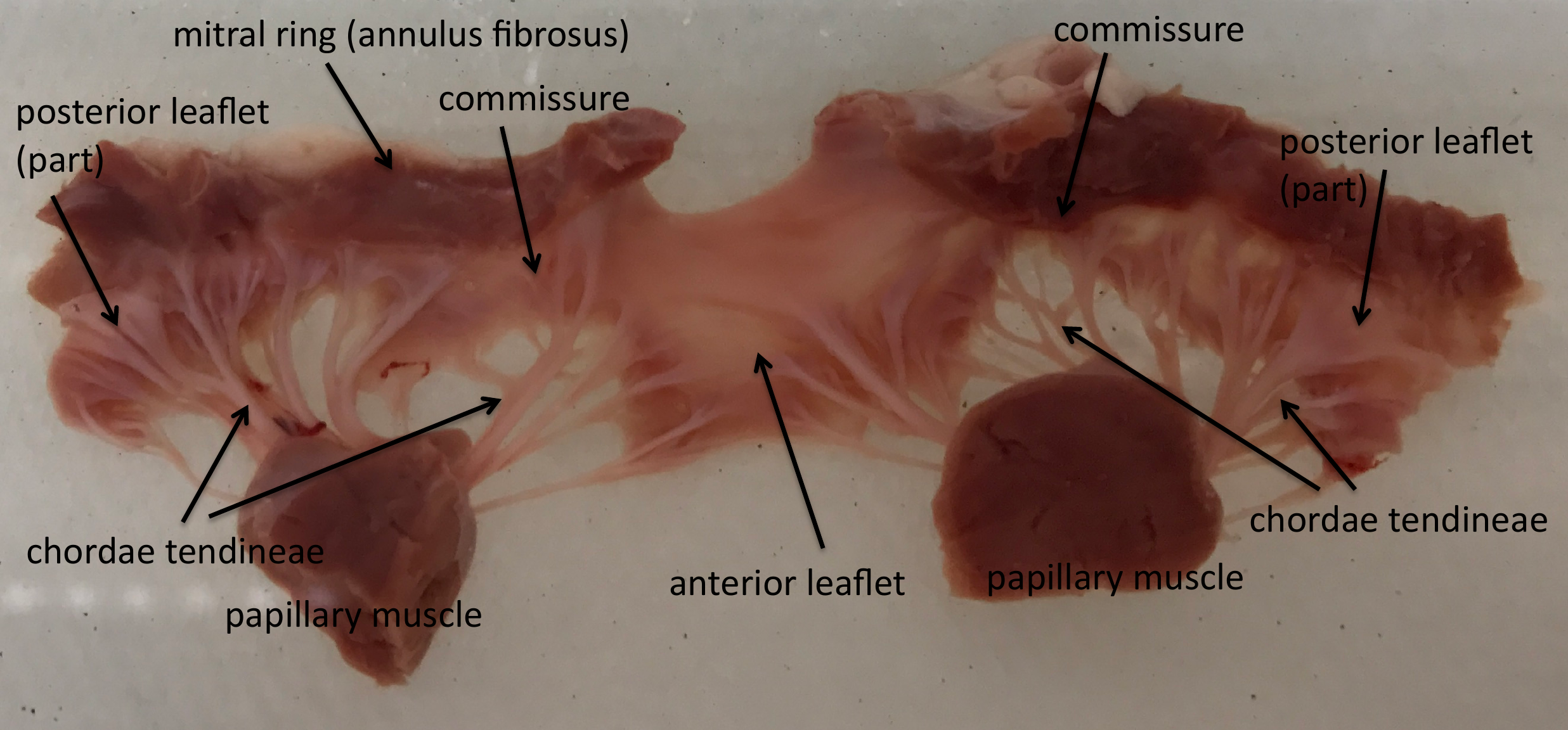}
\caption{Porcine mitral valve with basic structures labeled. Camera, iPhone 7; photographer, the author.}
\label{valve_labeled}
\end{figure}

A fully dissected valve is shown in figure \ref{valve_basic}, and repeated with labels in figure \ref{valve_labeled}. 
This image shows all the main structures of the mitral valve, with the ventricular side face up in the frame.
The mitral ring runs on the top of the frame. 
The anterior leaflet is centered; the posterior is sectioned in two and lies on the left and right of the frame. 
The commissural leaflet on the left is fairly well-separated from the posterior leaflet in this specimen, but the boundary between the commissural leaflet on the right and the posterior leaflet is ill-defined.  

\section{Literature}

Having shown the gross anatomy of the mitral valve, we will here discuss relevant findings from literature on valve anatomy before continuing with our own studies.

McCarthy et al. describe the histology of the valve \cite{mccarthy2010anatomy}. 
They note that the mitral valve has four distinct histological layers. 
The closest to the atrial side is the \emph{atrialis}, which they claim is composed of ``aligned elastic and collagen fibers.''
Next is the spongiosa, which does not have a well defined fiber structure. 
We hypothesize that this layer is present so that some degree of movement is possible between the more fibrous, load-bearing layers. 
The next layer is the fibrosa, in which ``collagen fibres are compact and aligned providing strength and stiffness to the leaflet.''
Finally, comes the ventricularis, also with fibrous collagen. 
Other authors claim that it is difficult to separate the mechanical effects of the fibrosa and ventricularis layers when building constitutive laws for mitral valve models, suggesting that these two layers are tightly related \cite{sacks_collagen_constitutive}.

The existence of such layers says that the atrial-side fiber structure (in the atrialis layer) is well-separated from the ventricular-side fiber structure (in the fibrosa and ventricularis layers.)
Experiments or views that suggest one orientation on the ventricular side do not imply that the same structure exists on the atrial side. 
We do not incorporate these layers in detail, since our model views the leaflets as two-dimensional surfaces.  
Yet we do take away this important point: 
The atrialis may have very different fiber structure from that of the fibrosa and ventricularis.

One group partially mapped the fiber structure of the anterior leaflet using small angle light scattering \cite{PMID:1751231}. 
Their summary diagram suggests that the anterior leaflet has a heavy circumferential fiber orientation, at least on the venticularis/fibrosa side. 
This includes arching circumferential fibers that connect one side of the free edge of the anterior leaflet to the other side of the free edge of the anterior leaflet.

In a study on canine (dog) mitral valves which largely uses optical and electron microscopy, the authors suggest points that we use repeatedly throughout this project \cite{fenoglio1972canine}. 
They say: 
\begin{quotation}
\singlespace 
``The chordae continue intact to their insertion into the ventricular aspect of the valve leaflet. At this point...the dense collagen core divides. The bulk of the collagen is continuous with the fibrosa and courses towards the annulus. A portion of the central core, however, turns and courses toward the free edge of the leaflet, forming the fibrosa of the smooth areas (Fig. 4A) and the free edge. As the collagen fibers course toward the annulus they fan out to form a curtain of fibrous tissue. The collagen bundles crisscross and layer as they move toward the annulus. This arrangement is best seen in the midportion of the leaflet, where the collagen bundles appear to lie at right angles to one another.''
\end{quotation}

That is, the chordae tendineae attach to the leaflet in a continuous manner; 
they branch on attachment, with some of the branches moving toward the valve ring, radially, and others moving circumferentially, towards the opposing papillary muscle; 
and finally, there is a large region of leaflet in which there appear to be two distinct fiber families.

\section{Mitral valve staining}
\label{staining_section}

Hoping to find additional details and reveal more about the fiber structure of the mitral valve, I started a collaboration with the Experimental Pathology Research Laboratory at the New York University Langone Medical Center. 

First, we stained an entire mitral valve with van Gieson's picrofuchsin, which selectively stains collagen. 
This stain was chosen following the methodology of a study on the histology and mechanics of the aortic valve \cite{SAUREN198097}. 
The protocol was modified from the laboratory's internal protocol to stain a whole mount (entire piece of tissue) specimen, rather than paraffin section. 
The valve was stained in Verhoeff's solution for 1 hour. 
It was then differentiated in ferric chloride for just over 2 mins, a process that removes excess dye from the tissue, with the stopping time determined by visual inspection. 
It was then counterstained for 3 minutes, followed by further differentiation in 70\% ethanol. 
The stains were supplied by Polysciences Incorporated.

The valve tissue is highly reactive to the stain; most of the tissue takes on a dramatic purple color. 
This suggests that there is collagen throughout the leaflets and chordae, as expected. 
It is possible that another stain, or the same stain used at lower concentration or for less time, would give a more complete picture of the fiber architecture of the valve, but even the present images clearly reveal certain aspects of the fiber structure.
Exposures in this section were set high to reveal contrast and details in the dark colored valve.

\begin{figure}[H]
\centering 
\includegraphics[width=\textwidth]{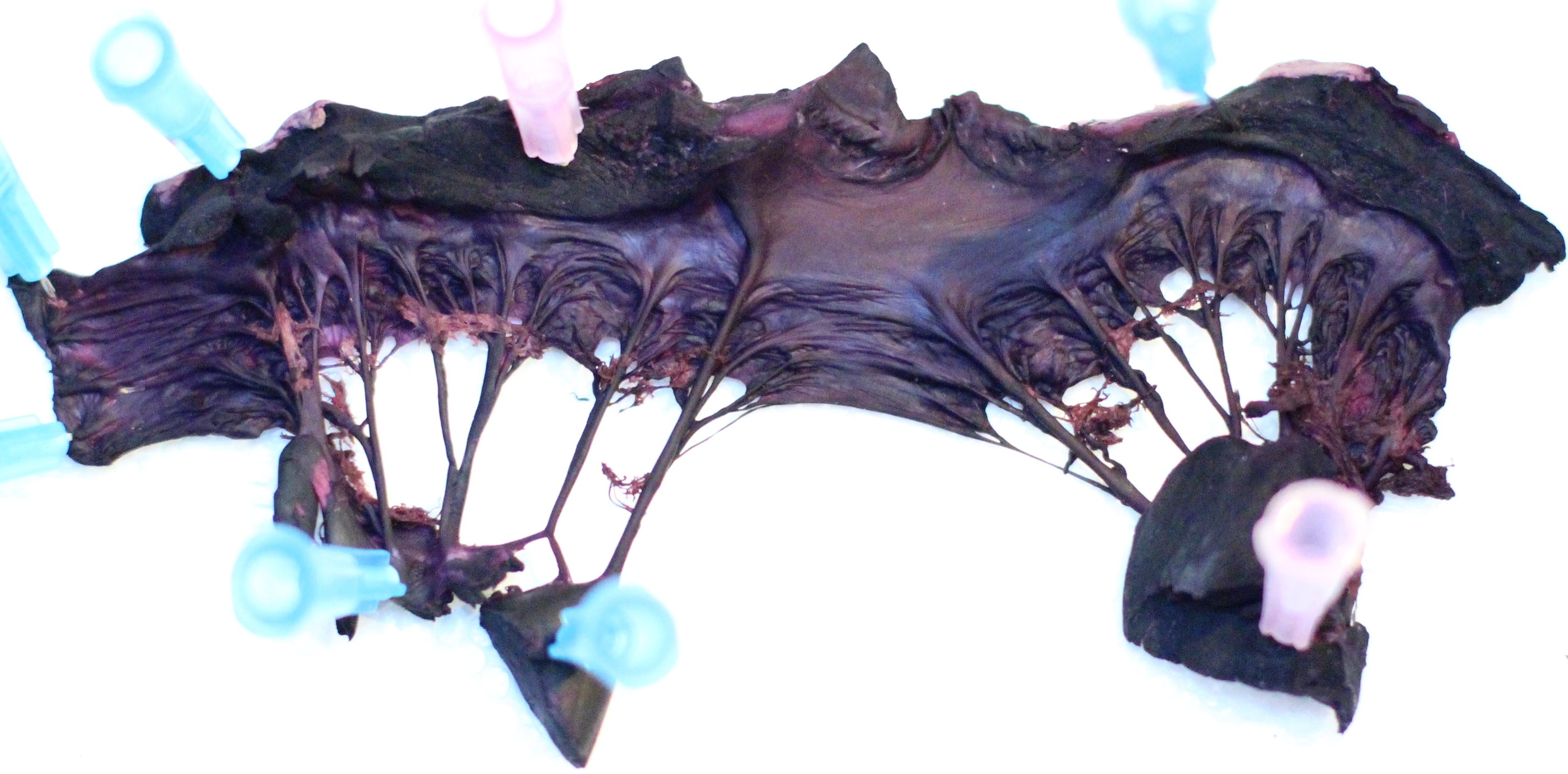}
\caption{Stained porcine valve pinned, ventricular side. Camera, Canon; photographer, Alu.}
\label{pinned_ventricular_zoom_out}
\end{figure}

Figure \ref{pinned_ventricular_zoom_out} shows the full valve from the ventricular side, and includes all the same major objects present in the unstained version. 
The ring and papillary muscles are pinned into styrofoam. 
The slight tension that this induces spreads out the tissue and makes details in the leaflet and free edge more visible.

\begin{figure}[H]
\centering 
\includegraphics[scale=0.12]{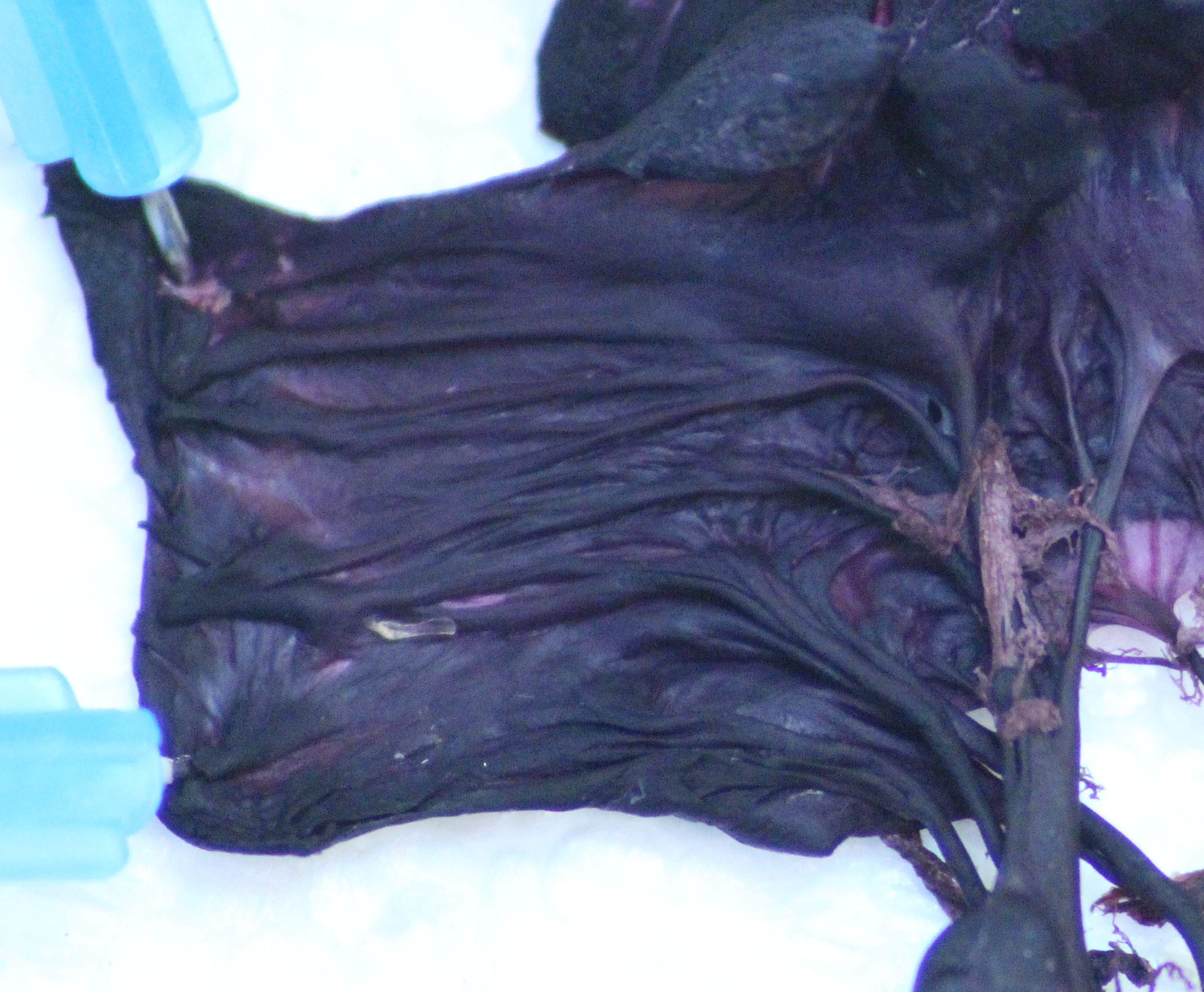} 
\caption{Detail of a portion of the posterior leaflet showing circumferential fibers and branching structure. The valve ring is at the top of the frame; circumferential orientation is horizontal. Camera, Canon; photographer, Alu.}
\label{pinned_posterior_detail}
\end{figure}

At locations where the chordae insert into the leaflet, there is frequently the appearance of a split or fanning out of the chordae. 
At times, this appears to go two ways. 
One large, stringy fiber turns towards the opposing papillary muscle, taking on a circumferential orientation. 
The other turns up, taking on a radial orientation in the direction of the valve ring. 
This branch is frequently less conspicuous when viewed from the ventricular side. 
It sometimes appears to sink into the leaflet, perhaps going under other fibers. 
This is possibly because it lies further from the ventricular surface of the valve. 
An example is shown in figure \ref{pinned_posterior_detail}.

We hypothesize that there are radially oriented fibers on the atrial side of the posterior and anterior leaflets, and quite likely in the commissures as well. 
Since, at their insertions, some of the branches appear to point in the radial direction, it seems likely that these fibers continue towards the valve ring. 
If there was not fibrous tissue in this direction, there would be nothing to support these branches. 
This is less visible, perhaps because the fibers may duck ``under'' the circumferential fibers that are visible on the ventricular side towards the atrial side. 
Later, I will use this observation in building models of the mitral valve in Chapter \ref{static}.

The foregoing has been challenging to confirm, since the atrial side has a smoother, more uniform appearance than the ventricular side of the valve. 
Perhaps these fibers simply have a finer or more uniform structure than the more visible circumferential fibers of the ventricular side. 
Literature suggests that the atrial side of the valve has more radially oriented fiber structure than the ventricular side, see figure 10.7c, right panel in \cite{Lee2015}. 
This is taken at high magnification in a small portion of the leaflet, and does not show collagen arrangement as clearly as it shows elastin. 
Thus, this supports our hypothesis but does not prove it.

\begin{figure}[H]
\centering 
\includegraphics[width=\textwidth]{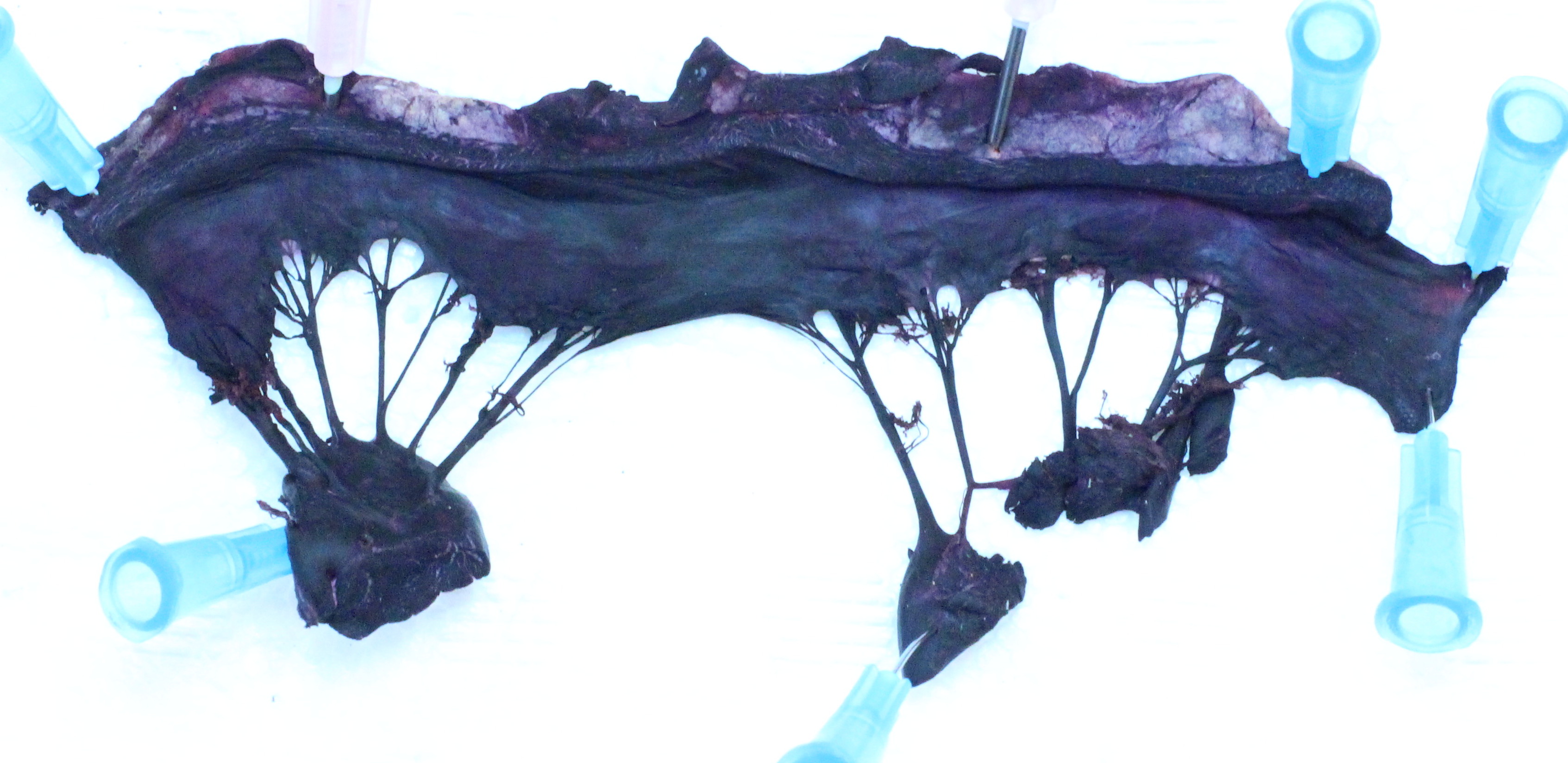}
\caption{Stained porcine valve pinned, atrial side. Camera, Canon; photographer, Alu.}
\label{pinned_atrial_zoom_out}
\end{figure}

Figure \ref{pinned_atrial_zoom_out} shows the same valve re-pinned from the atrial side. 
From this angle, it is clear that there is a significant region below the mitral ring in which there is contiguous tissue. 
There is a tube of tissue that looks and feels like leaflet, but lacks any separation between the anterior, commissural and posterior leaflets. 
The leaflets, then, split from each other only a certain distance away from the annulus fibrosus. 
This is consistent with Netter's observation that the commissures do not reach fully to the valve ring.

The texture of the atrial side is smoother than that of the ventricular side, and this is clearly visible by comparing figures \ref{pinned_atrial_zoom_out} and \ref{pinned_ventricular_zoom_out}. 
All of the complexity and roughness of the chordae insertion takes place on the ventricular side. 
At some locations, especially around the commissures, there are small, thin pieces of membranous tissue that lie below some of the insertions of the chordae.

\begin{figure}[H]
\centering 
\includegraphics[width=\textwidth]{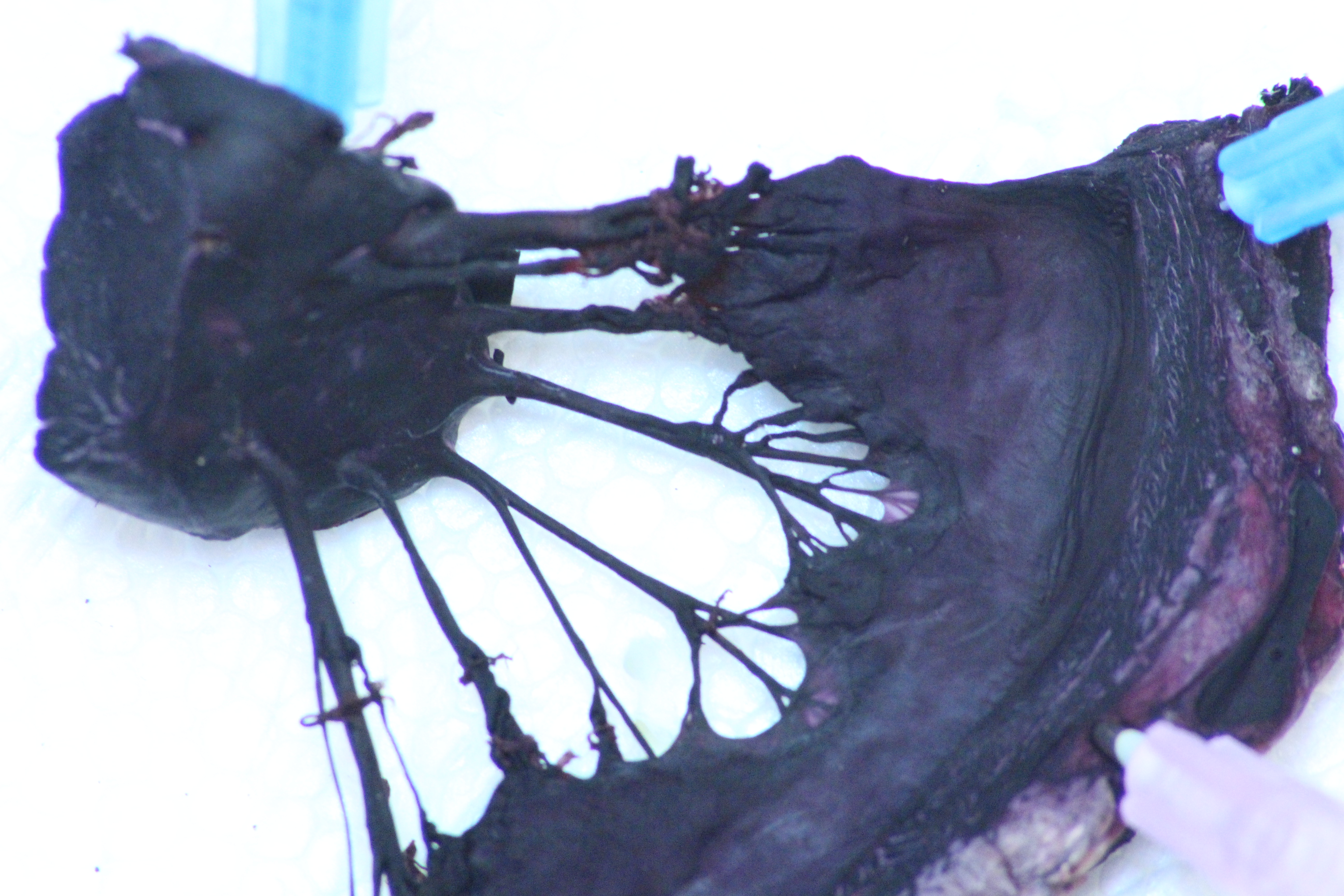}
\caption{Detail of papillary muscle and chordae attachments. Camera, Canon; photographer, Alu.}
\label{papillary_detail}
\end{figure}

Figure \ref{papillary_detail} shows details of the attachment of the chordae tendineae to one papillary muscle. 
The attachments are arranged along an approximately circular arc; they occur over about three fourths of the circle. 
The remaining fourth, with no attachment, is roughly pointed at the opposing papillary muscle. 
The opposing papillary muscle was less intact after the dissection, so we do not discuss its attachments here.

\begin{figure}[H]
\centering 
\includegraphics[width=\textwidth]{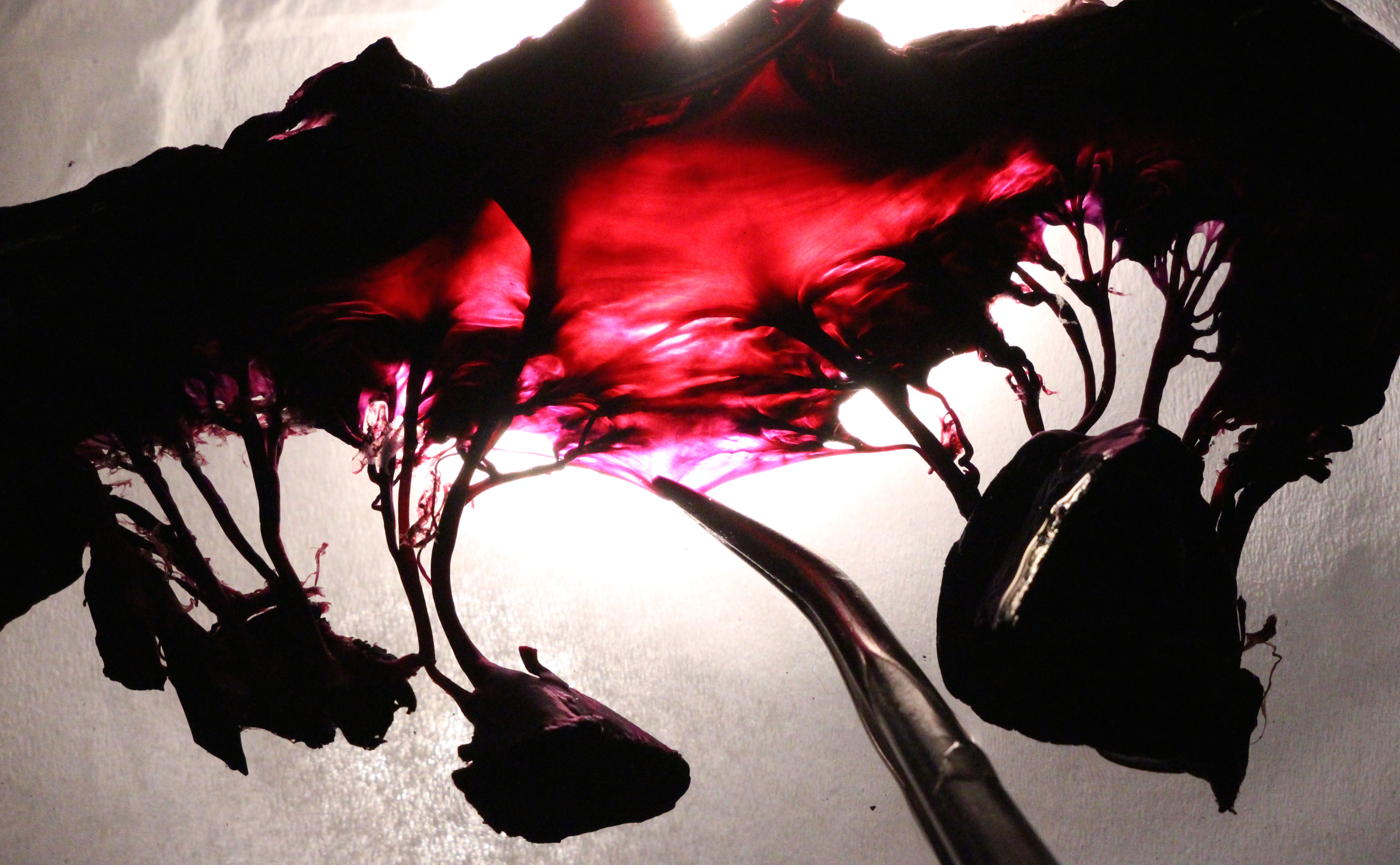}
\caption{Valve in alcohol solution lit from behind. The anterior leaflet is manually pulled apart slightly with forceps. Camera, Canon; Photographer, Alu.}
\label{light_box}
\end{figure}

We then unmounted the valve and took a number of pictures with the valve lit from behind, shown in figure \ref{light_box}. 
Having the valve in liquid sometimes revealed fine structure tissue around the free edge, so we placed the valve into diluted ethanol for this image. 
Pulling the valve tissue apart with forceps again reveals some structures. 
In the center of the anterior leaflet, approaching the free edge there are large, circumferentially oriented, fibrous pieces of tissue. 
These originate from the chordae tendineae, proceed circumferentially across the leaflet, then connect to chordae from the other papillary muscle at approximately the same distance from the valve ring. 
These are visible to the eye in about half of the anterior leaflet; the same qualitative phenomenon may occur at finer scales closer to the valve ring.

\begin{figure}[H]
\centering 
\includegraphics[width=\textwidth]{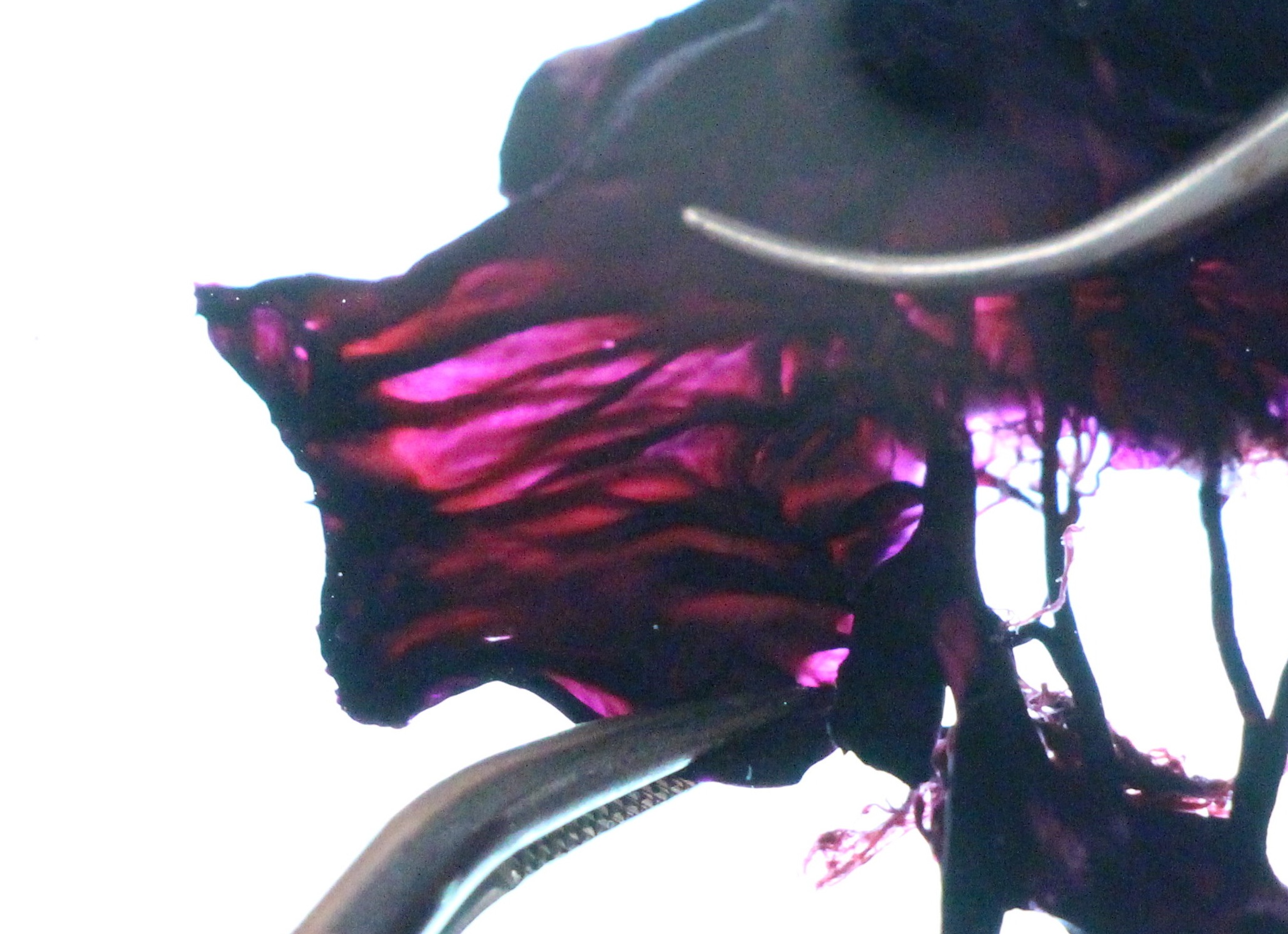}
\caption{Detail of posterior leaflet lit from behind. Camera, Canon; photographer, Alu.}
\label{light_box_posterior}
\end{figure}

Figure \ref{light_box_posterior} shows half of the the posterior leaflet, pulled by forceps and backlit. 
There are macroscopic, circumferentially oriented fibers visible to the naked eye. 

\section{Mitral valve tissue clearing}
\label{clearing}

We also processed a valve using tissue clearing, a process that makes the tissue translucent or even transparent, which allows for different optical imaging and subsequent staining \cite{treweek2015whole}. 
This is the same valve shown being dissected in Section \ref{Dissections}. 
This is typically done on much smaller pieces of tissue than a mitral valve. 
As a result, the valve did not become transparent quickly, and we left it in the clearing solution for approximately seven months. 
We stopped the process at this point, even though the tissue was not fully cleared. 

\begin{figure}[H]
\centering 
\includegraphics[width=\textwidth]{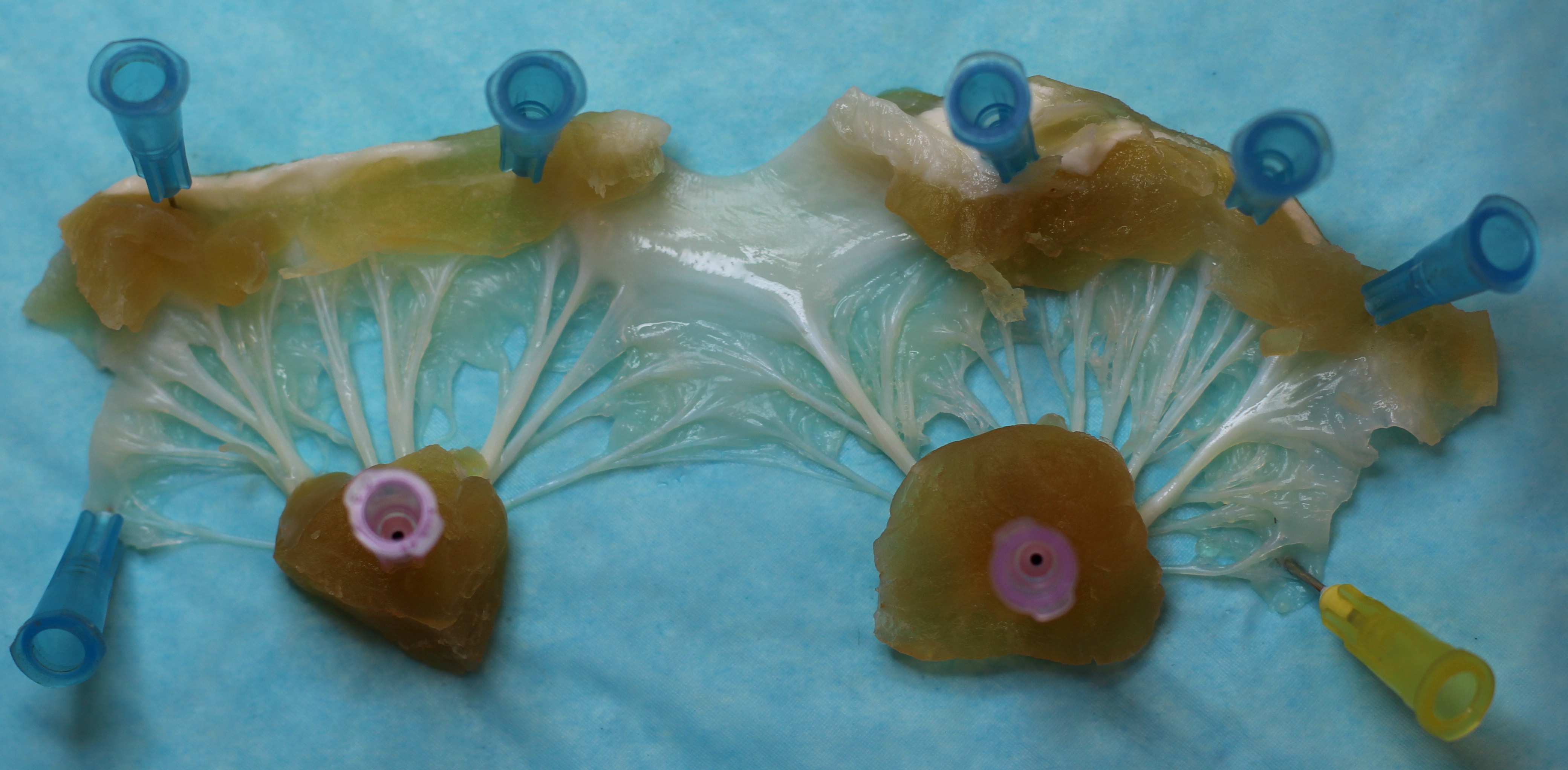}
\caption{Cleared porcine valve pinned, ventricular side. Camera, Canon; photographer, Alu.}
\label{pinned_ventricular_zoom_out_cleared}
\end{figure}

The valve achieves a translucent color, with some of the thicker portions remaining opaque. 
A pinned view of the valve viewed from the ventricular side is shown in figure \ref{pinned_ventricular_zoom_out_cleared}. 
Blue paper is placed behind the valve to differentiate it in color from the white mounting surface.  

\begin{figure}[H]
\centering 
\includegraphics[width=\textwidth]{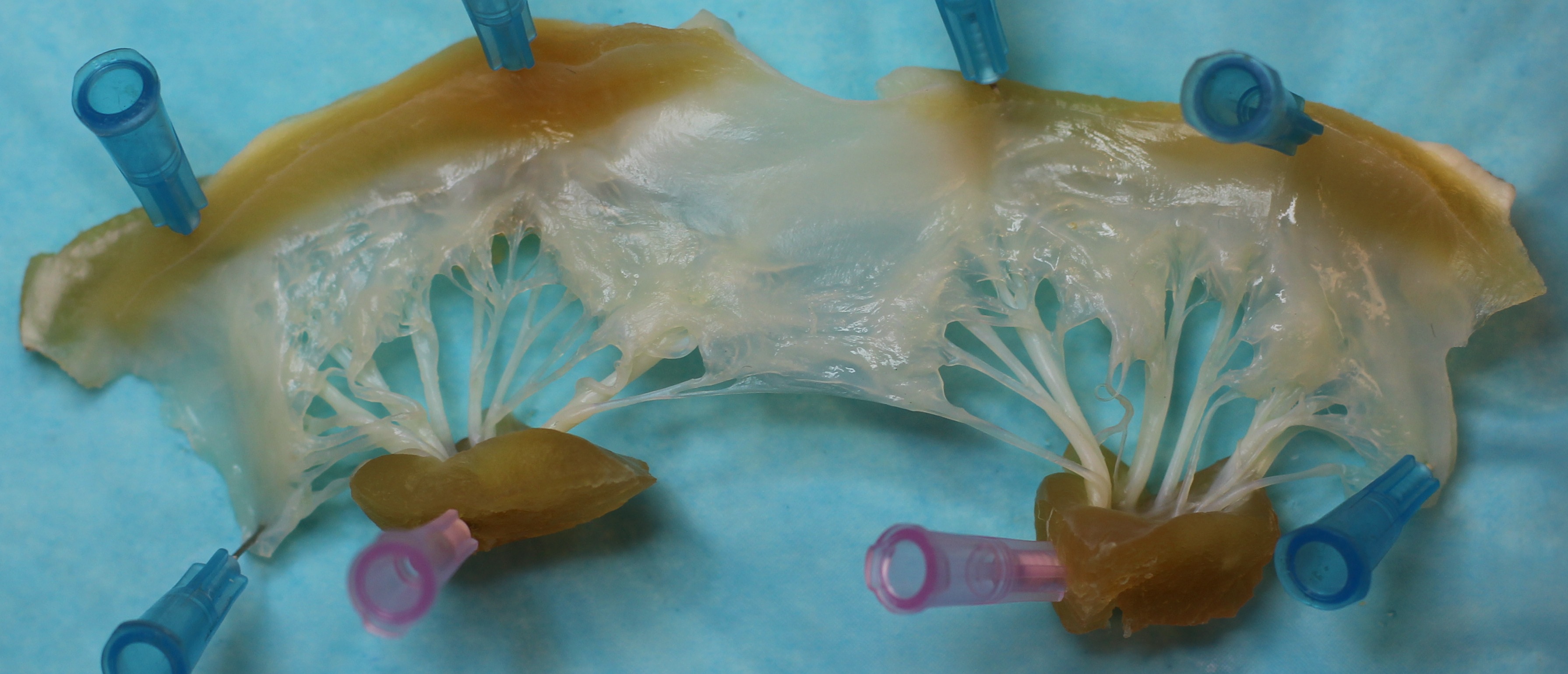}
\caption{Cleared porcine valve pinned, atrial side. Camera, Canon; photographer, Alu.}
\label{pinned_atrial_zoom_out_cleared}
\end{figure}

Figure \ref{pinned_atrial_zoom_out_cleared} shows the same valve from the atrial side. 
This specimen possesses a smooth surface on this side of the valve; we believe this feature to be universal across non-pathological mitral valves. 

\begin{figure}[H]
\centering 
\includegraphics[width=\textwidth]{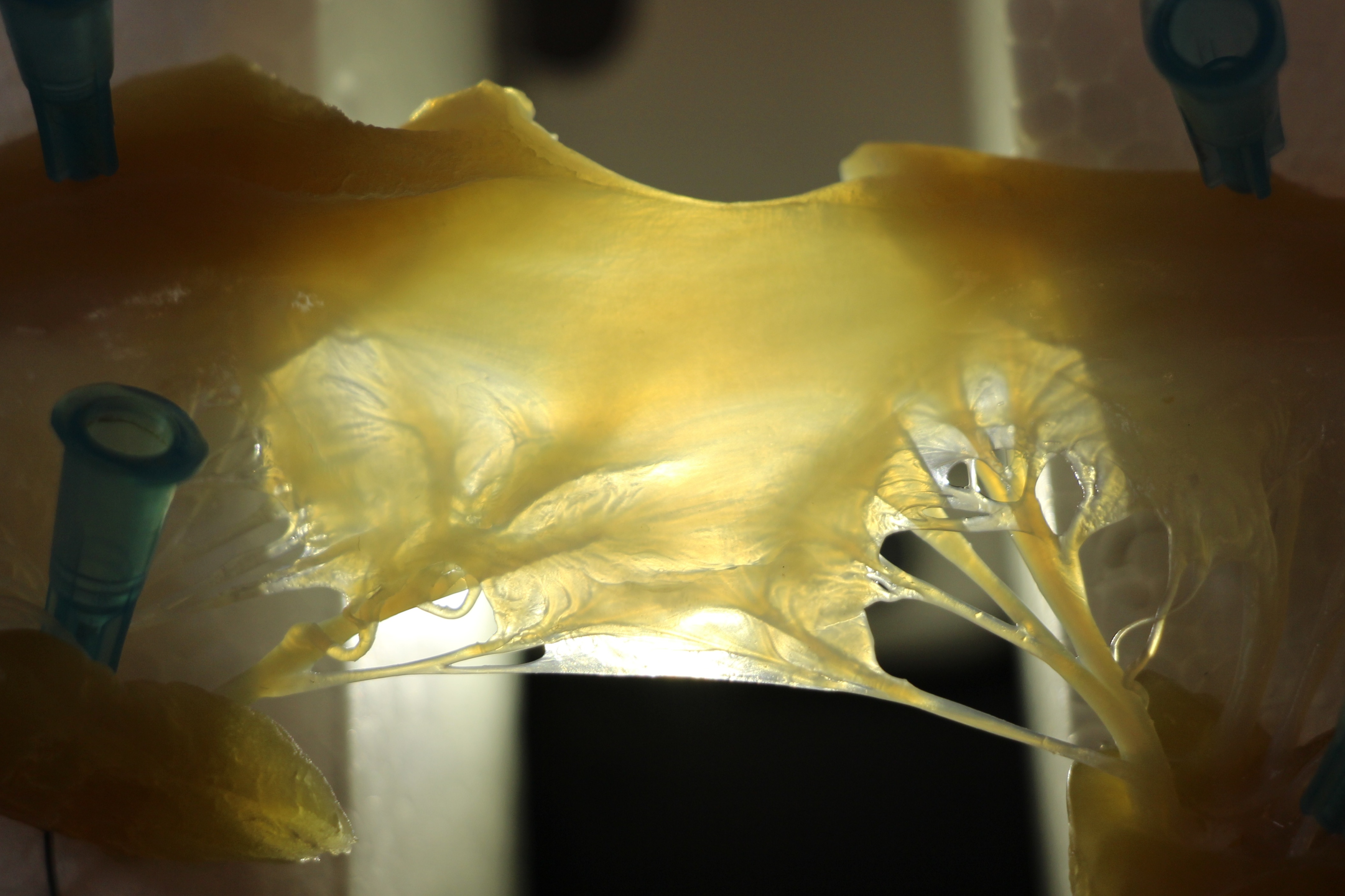}
\caption{Cleared porcine valve pinned, atrial side, pinned and lit from below. Camera, Canon; photographer, Alu.}
\label{pinned_atrial_zoom_out_cleared_lit_below}
\end{figure}

Figure \ref{pinned_atrial_zoom_out_cleared_lit_below} shows the cleared valve, pinned and lit from behind to take advantage of the partially cleared leaflets translucence. 
The atrial side of the valve faces the camera in this frame, though with this lighting the two sides look similar to each other. 
The chordae tendineae branch and spread out around their insertions to the leaflet. 
At each point of insertion there is no single, well-defined angle at which the strand attaches; rather, the strand inserts over a wide range of angles.
This view also gives some visual clues about the thickness of different locations of the valve.

The Appendix of this thesis describes additional preliminary work on this specimen, in which the valve was stained and imaged using polarized-light microscopy. 

\section{Valve skeleton from MRI data}
\label{MRI}

When we later construct models of the mitral valve, we need the radius of the valve ring and the relative locations of the papillary muscles. 
To obtain this information, we use images from a cardiac magnetic resonance imaging (MRI) scan. 
The subject of the scans was a healthy male in his thirties, but little other information is available. 
The images are taken in many times through the cardiac cycle, and show four standard MRI  views: the vertical long axis view, the three chamber view, the four chamber or horizontal long axis view, and the short axis view \cite{mri_basics}. 
The short axis view refers to an angle, rather than a specific slice, and the scans contain many such slices. 

\begin{figure}[H]
\centering
$ \begin{array}{cc}
\includegraphics[width=.475\textwidth]{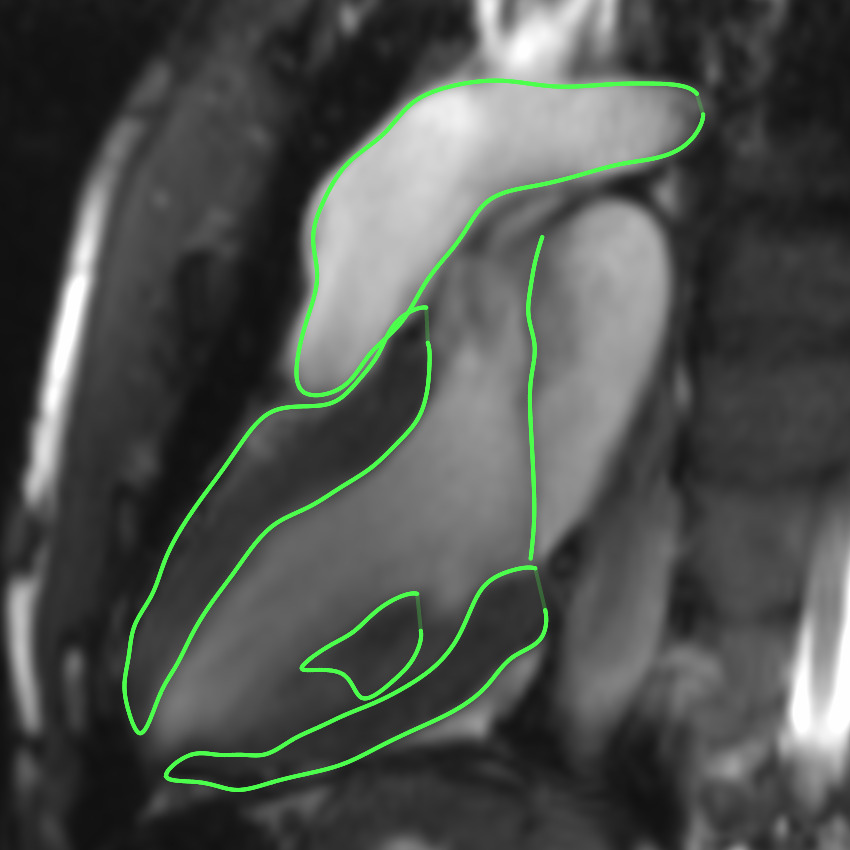}  &
\includegraphics[width=.475\textwidth]{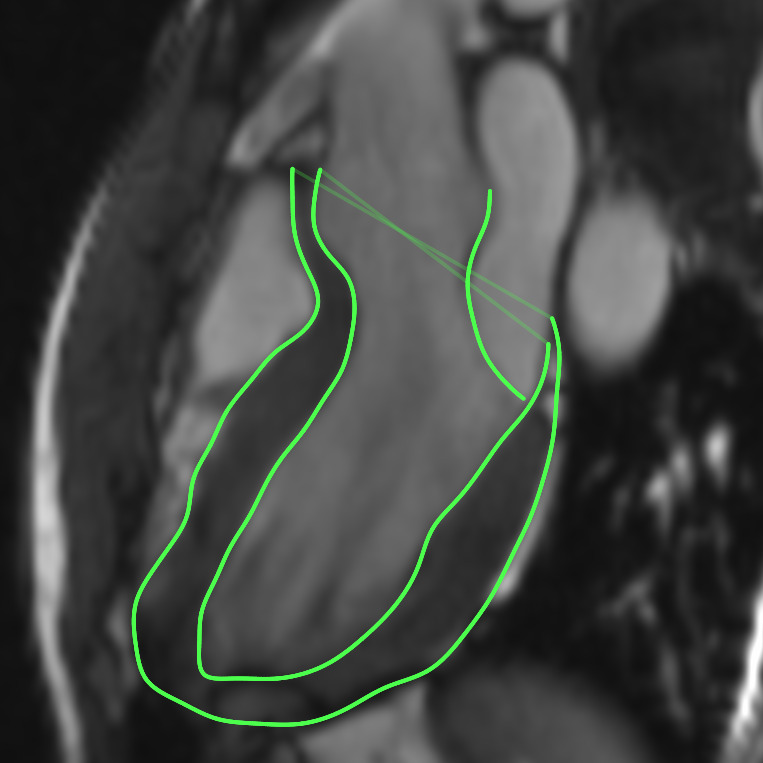}  \\ 
\includegraphics[width=.475\textwidth]{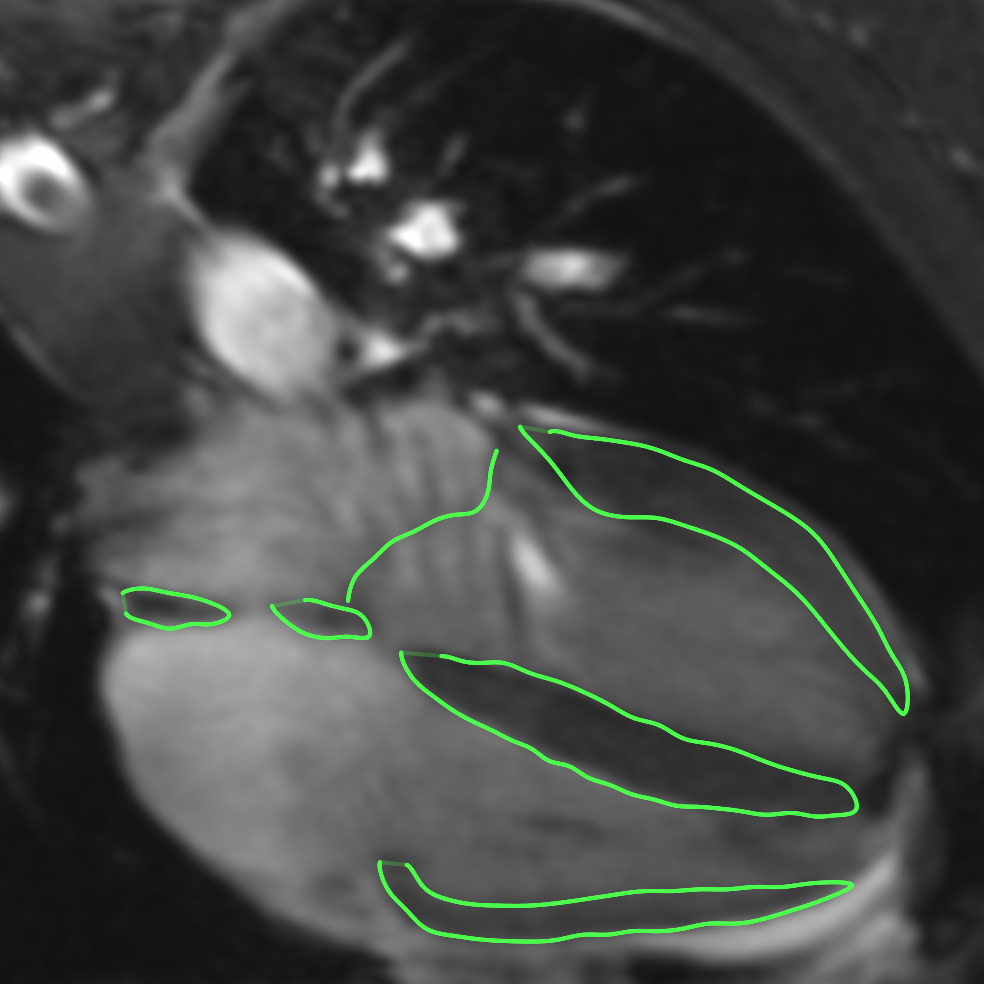}  
& \includegraphics[width=.475\textwidth]{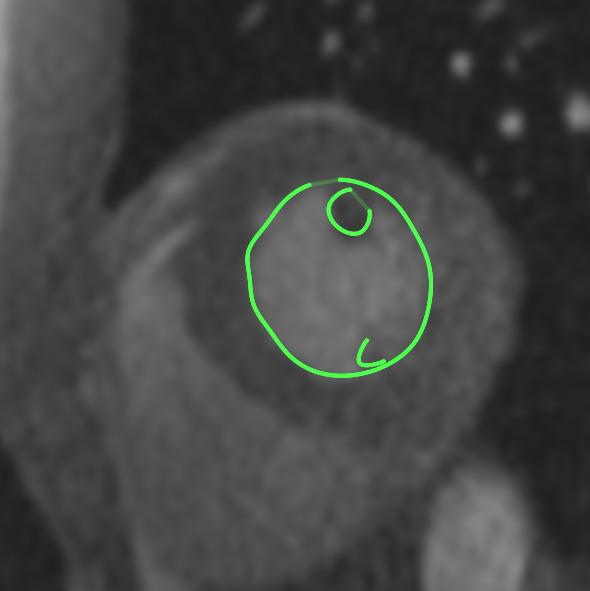}  \\ 
\end{array} $
\caption{Four MRI views with segmentation contours overlaid. From top to bottom, left to right: vertical long axis view, three chamber view, four chamber view, short axis view.}
\label{mri_segmentation}
\end{figure}

Figure \ref{mri_segmentation} shows these four views, with contours overlaid in green. 
This segmentation was performed manually by David McQueen, with input from the author, using the Osirix medical image viewer \cite{Rosset2004}. 
This software allows users to draw contours, and ensures that they are output in coordinates that are consistent across all views.

The vertical long axis view shows the left ventricle in the center bottom of the frame, the walls of which are segmented. 
Inside the left ventricle, a papillary muscle is visible and segmented. 
When visualizing the data in three dimensions, this will help confirm that the papillary muscles are aligned across all views. 
The location of the mitral valve is shown with a thin, nearly linear contour on the right of the frame. 
This segment is likely longer than the mitral valve leaflet, and we believe it includes some of the adjacent tissue.

The three chamber view also includes the walls of the left ventricle at a different orientation. 
In this view, the aortic sinus is clearly visible at the top of the left ventricle, center top of the frame. 
The mitral valve is segmented with a curve that appears contiguous with part of aortic sinus. 
This supports our opinion that mitral valve segmentation on the vertical long axis view also includes some tissue that is above the mitral ring. 
The four chamber view also shows the left ventricle, and a single curve denotes the mitral valve.

The papillary muscles are visible on the short axis views. 
They appear as two gray dots, which are segmented here. 
On some slices of the short axis view, one papillary muscle appears to reach closer to the valve ring than the other papillary muscle. 
Despite this, all anatomy descriptions and dissections from Chapter \ref{anatomy} suggest that the papillary muscles should be approximately symmetric with respect to their distance to the ring. 
We believe this appearance of asymmetry is the result of chordae or even leaflet that looks in the image like a continuation of the papillary muscle. 
Thus, we chose the slice closest to the ring on which both papillary muscles are visible to give the location of the tips.

These segmentations give a number of points that we believe to be on the mitral valve ring. We compute best fit circle to these points in the least squares sense.  
Some points on the circle are likely on the part of the aorta that is contiguous with the mitral valve.  
We assume that the ``D-shaped'' nature of the ring, shown in figure \ref{valve_ring_2d}, accounts for this.

\begin{figure}[H]
\centering
\includegraphics[width=\textwidth]{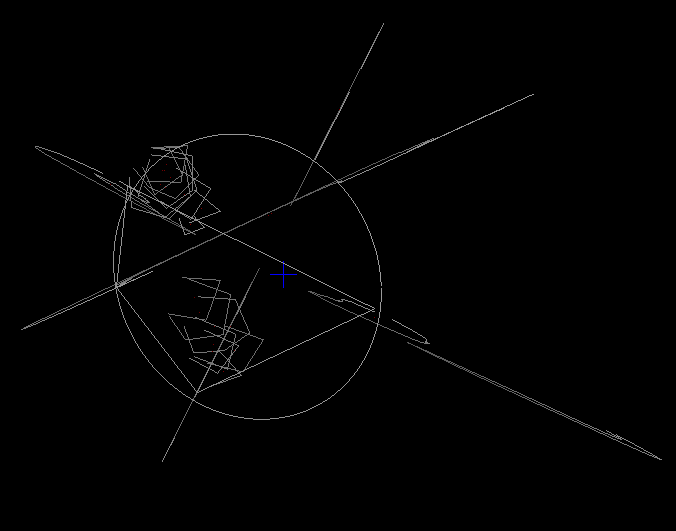}  
\caption{Contours visualized in three dimensions simultaneously.  }
\label{ring_avove_lines3D.png}
\end{figure}

The segmented contours are visualized in three dimensions in figure \ref{ring_avove_lines3D.png}. 
The best fit circle from which we construct the D-shaped mitral ring is shown to the left of the center of the frame. 
The radius of the circle is 2.19 cm. 
The point of view is approximately aligned with planes from the vertical long axis view, three chamber view and four chamber view. 
The point of view is not normal to the circle, which is why the circle appears as an ellipse in the image. 
Contours from those three views appear nearly as line segments. 
Numerous smaller contours outline the papillary muscles in the short axis views, which from this view are mostly on top of each other.

To use this information in building the model valve, the valve ring and tips are rigidly rotated and translated such that the valve ring is in the plane $z = 0$ and is centered at the origin.
Then, we rotate around the $z$ axis so that the plane $x=0$ corresponds to the approximate symmetry plane of the valve.  
To determine this angle, we require that the $y$ coordinates of the papillary tips are equal in magnitude and opposite in sign, which makes the papillary tips approximately symmetric with respect to the plane $x = 0$.  
The symmetry is approximate because the $x$ and $z$ components of the papillary tips are not precisely equal after performing this sequence of transformations.
The location of the papillary tips in the MRI data reflects the approximate symmetry of the real valve. 
The tips are roughly the same distance from the center of the valve ring in two directions, the direction orthogonal to the approximate symmetry plane of the valve and the direction normal to the ring. 
Under this transformation, these directions are aligned with the $y$ and $z$ coordinate axes, so the $x$ and $z$ components of the papillary tips become approximately equal.

\newpage 

\chapter{Construction of the Model Mitral Valve}
\label{static}

We aim to build a model of the mitral valve for use in in fluid-structure interaction simulations. 
Generic information about anatomical features is known. 
It is possible to access information about the location of the papillary tips and valve ring from magnetic resonance imaging (MRI) data. 
However, many pieces of information about the valve are not immediately available.  
Specifically, we do not have detailed geometry of the leaflets or chordae tendineae, nor a constitutive law or reference configuration for the valve.

To combat these challenges, I proposed to build a model by solving a static mechanics problem for the configuration of the loaded valve. 
We assume that the valve supports a static pressure, and in doing so achieves a state of mechanical equilibrium. 
That is, we specify how the valve has to function -- what forces it must support -- and determine its configuration by solving the associated differential equations.  
This is sometimes referred to as a \emph{design based} approach to elasticity, or colloquially \emph{reverse engineering} the configuration of the valve. 

This modeling philosophy has been applied previously to the aortic valve \cite{PeskinH319}. 
However, the mitral valve has a much more complicated structure.
In particular, the aortic valve lacks chordae, and the fibers within its leaflets run primarily in the circumferential direction.

\section{Literature}
\label{literature}

Before we discuss the construction of the model, we will briefly review some related literature on modeling the mitral valve. 

Longterm leaders in this field are Karyn Kunzelman and her collaborators. 
In one recent paper, they present their new method for building a computational model of the mitral valve \cite{Toma2016}. 
Their fluid-structure interaction technique is described more fully in their subsequent paper \cite{CNM:CNM2815}. 
We believe this paper to be an excellent recent contribution to mitral valve modeling, as well as the most similar in spirit to the current work. 
Thus, we discuss their techniques, as well as differences with our approach, in some detail here. 

First an ovine (sheep's) mitral valve is scanned using micro-computed tomography. 
This gives a highly anatomical mitral valve geometry. 
Their model valve is placed in a box-like computer test chamber, as is ours, and simulated with a finite element numerical method. 
Simulations in their work are driven by velocity boundary conditions, in which the boundary itself moves to create pressures, modeling a piston at one edge of the test chamber. 
They consider the fiber structure of the model valve in detail, and moreover use certain mathematical models to fill in information that is not readily available. 
They use their earlier work \cite{PMID:1751231}, which involves small angle light scattering (SALS), to estimate the valve fiber direction at certain locations. 
They consider there to be a single mean fiber direction. 
This rules out the possibility of having two local fiber directions, like a woven cloth, as we later specify. 
However, they then construct a sophisticated constitutive law in which the mean fiber direction plays a role but is not the only direction in which the material can exert stress.

On building a valve fiber structure, they comment that despite efforts to use in-vitro experiments in building models, a 
\begin{quotation}
\singlespacing 
``
challenge is how to determine fiber orientations non-invasively. 
\\

\noindent 
...we have some certainty with respect to some areas of alignment. These are our boundary conditions. For example, we know that the fiber orientation is mostly aligned with the axis of the chordae and the free edges of the leaflets... we know that the fibers will be oriented vertically [radially] at the fibrous trigones, and approximately circumferentially along the annulus at the midpoints of the anterior and posterior leaflets. Lastly, we assume that to be mechanically optimal the fiber field must be slowly varying between these fixed directions. Thus we solve a modified Laplace problem to determine the unspecified fiber directions in between.'' \cite{Toma2016}
\end{quotation}

This statement is the most aligned with our approach that we believe exists in mitral valve modeling literature. 
It acknowledges a fundamental difficulty: it is challenging to know the fiber structure of the mitral valve, and even more so to know the fiber structure in a non-invasive manner. 
Doing so non-invasively is essential to making patient-specific models, because if one is trying to model and ultimately advise treatments for a living patient, of course, dissections of their heart are not available. 
They propose to resolve this by solving an appropriate partial differential equation for the orientation of the fibers. 
Their choice of the Laplace equation is interesting. 
On one hand, smooth solutions to the Laplace equation exist on any reasonable geometry. 
The equation is theoretically well understood, and robust numerical solvers are available.  
Their model produces fiber directions that are continuous with the directions in the chordae tendineae. 
On the other hand, the Laplace equation does not fully capture the physics of the structure.

In comparison to the work described above, in this thesis we use a different partial differential equation to derive the fiber structure of the mitral valve.
Our partial differential equation is directly related to the physical function of the valve, namely that it supports a pressure when it is closed. 
We use different numerical methods throughout, and ultimately achieve agreement with experiments in flux through the mitral valve, whereas they largely evaluate their model in terms of matching forces at the papillary muscles. 
It is possible, further, that these information from these models can be combined, which we leave as interesting future work.

These same authors have a long history in the field, and produced a review and retrospective of modeling the mitral valve and left heart in 2010 \cite{CNM:CNM1280}. 
The review discusses a number of studies on the mitral valve. 
Some of these look at the valve as a solid object and study its dynamics by applying pressure forces to the surfaces of the valve. 
They also discuss some fluid-structure interaction models and related fully Lagrangian methods for fluid-structure interaction simulations. 
The anatomy of the valve models is not as realistic as that of their newer work, and so we believe the specifics of the models described in the review have been largely superseded.

Many authors contributed to an extensive four-part review of heart valve engineering in 2015, including our collaborator Boyce Griffith \cite{kheradvar2015emergingIV}.
This review is recent, broad and, we believe, fairly comprehensive. 
The fourth and final part of this review discusses the role of computational modeling in heart valve engineering. 
This reviews the immersed boundary method for fluid-structure interaction and related techniques in solid mechanics for building the ``structure'' part of the model. 
This addresses, how to put structural models that are expressed using solid-mechanics vocabulary in terms of tensors into the immersed boundary framework. 
It presents simulations of a model natural aortic valve driven by pressures, which achieve realistic flow rates across multiple beats. 
It also presents work on a model of a mechanical prosthetic aortic valve that similarly functions under realistic pressures. 

Notably, there is no mitral valve model included that achieves this type of success. 
They comment that 
\begin{quotation}
\singlespace 
``There appear to be relatively few FSI [fluid-structure interaction] valve models that can perform multiple cardiac cycles and also simulate the closed, loaded configuration of the valve... Closure seems to be especially challenging to simulate because it fundamentally involves a delicate balance between the fluid dynamics and elasticity of the valve's leaflets.''
\end{quotation}

In this thesis, we will achieve exactly what this comment says most models cannot perform. 
We build a model that performs robustly under multiple cardiac cycles when driven by physiological pressures. 
Moreover, we will show this model to be robust under pressures that are much higher or much lower than normal physiological pressures.

This general review also discusses some sophisticated multi-scale modeling techniques applied to the aortic valve. 
We regard building multi-scale models of the mitral valve that are effective under physiological pressures to be an open question.

Efforts are underway to resolve this; a review of multi-scale methods as applied to the mitral valve is given by Michael Sacks and collaborators in \cite{Lee2015}.
One difficulty in creating such models is that there are even more parameters that are both unknown and physically important than in a phenomenological, macro-scale valve model. 
The same group takes a novel approach to this problem by using Bayesian inference and Monte Carlo sampling methods to fit parameters to experimental data \cite{sacks_collagen_constitutive}. 
In other work, they review methods to incorporate experimental techniques into computational models \cite{bloodworth2017ex}. 
Their entire workflow, which is pieced together from many shorter papers, is reviewed in \cite{CNM:CNM2921}.
Their work only focuses on solid mechanics and modeling fully loaded or unloaded states. 
They do not include fluid-structure interaction in these papers.

Boyce Griffith (who is also a collaborator of ours) and colleagues previously studied a model prosthetic mitral valve \cite{griffith2009simulating}. 
This paper shares much in common with this thesis in numerical techniques. 
The valve model, however, is specifically meant to model a polyurethane prosthetic valve. 
This prosthetic (atypically) includes chordae tendineae, which continue from attachment at the free edge into the leaflets.  
The polyurethane is isotropic and not fibrous, but the leaflets are not isotropic because the chordae continue through them to the mitral valve ring. 
It is notable that the authors artificially lower the Young's modulus in the model chordae tendineae by a factor of ten.  
In this thesis, our simulations are very slow because of timestep restrictions caused by explicit treatment of the Lagrangian structure. 
Writing nearly a decade later, with significant improvements in computing power, we do not take this route, but still have to wait quite a while for simulations to complete.

Another study by the same group looked to build a model of the natural mitral valve from MRI data \cite{MA2013417}. 
Their leaflet geometry matches many gross features of the anatomy. 
They view the leaflets as made of a linear isotropic material (before they discretize the material), plus a fixed number of radially-oriented anisotropic fibers that reinforce the the leaflet.
However, they use a fiber based discretization of the isotropic material, and fibers in the discretization are roughly radial and circumferential, which we believe is anatomical. 
The chordae are modeled as cables without branching; there are a fixed number of such chordae. 
When the chordae connect with the leaflet at the free edge, they subsequently become radial fibers.
These continue through the leaflet as fiber reinforcement of the otherwise isotropic material in the leaflet. 
We hypothesize that this is anatomical, as discussed in Chapter \ref{anatomy}, though believe the chordae are contiguous with the leaflet in a much more complicated manner including branching and attaching to circumferential fibers as well. 
They present results over one beat, and the flux has reasonable agreement with the physiology.

Mark Ratcliffe and collaborators have also built MRI-based model valve geometry, but use a finite element approach \cite{wenk2010first}. 
They gave a number of living sheep myocardial infarctions (heart attacks) then built an animal-specific left ventricle and mitral valve model. 
They look only at valve and ventricular mechanics, without fluid, and force the models with scalar pressures. 
They claim that the mitral valve fibers are aligned radially and circumferentially in the valve leaflets (although they do not discuss their motivation or evidence) which agrees with our assumption that this is so. 
They model the muscular damage from the infarct, and using this pathological model are able to see holes appear in the loaded mitral valve. 
This provides numerical confirmation of functional mitral regurgitation due to ischemia (loss of blood to the heart), which is known to be a common clinical finding.
This work is the most surgery-focused of all studies we are familiar with involving the mitral valve. 
It is also the most focused on a specific pathology, although that pathology is functional, rather than primary, with respect to the valve.

This work uses a form for their collagen constitutive laws that is taken from a materials study on knee ligaments \cite{quapp1998material}. 
Ligaments are also rich in collagen; they propose a constitutive model that incorporates stress from collagen fibers, as well as a background matrix and interactions between the two. 
Remarkably, their reported strain energy density for collagen fibers (equation 7 in their paper) has the exact same functional form as the constitutive law we select, shown in equation \ref{constitutive_summary}. 
We were not aware of this model before finalizing this review of literature after all simulations shown in this thesis were complete. 
This group, working on different part of the body, but one with similar materials, invented constitutive laws that are essentially the same as ours. 
We believe this supports our choice of constitutive laws for model fibers. 

Ratcliffe and collaborators also wrote a review on finite element modeling of the mitral valve that is focused on surgical applications \cite{morgan2016finite}. 
This brought many of the papers cited here to our attention.

To conclude, there is much literature on modeling the mitral valve. 
Some literature focuses on deriving information from laboratory experiments and does not include fluid-structure interaction. 
Some groups have focused on applications to surgery. 
Others include fluid-structure interaction and various strategies involving scans and MRI data. 
One group, Kunzelman and collaborators, solve a partial differential equation to learn about valve fiber structure, but this differential equation is chosen to ensure smooth solutions, instead of being physically motivated.

Our approach derives the layout of valve fiber structure by solving a partial differential equation. 
This equation is derived from the function of the valve, namely that the valve supports a pressure load and should be in mechanical equilibrium when doing so. 
We believe this approach to be completely different from any other mitral valve model previously published. 
Further, our model produces realistic flow rates when simulated under physiological pressures over multiple beats.

\section{Assumptions}

The model geometry is built according to the following principles, which form a summary of the idealized anatomy and function of the mitral valve.

\begin{enumerate}

\item The valve is composed of two leaflets, which are made up of fibers. 
These fibers exert tension only in the fiber directions. 
They do not exert any forces besides tensile forces. 

\item At any point internal to the leaflet, there are two families of fibers under tension. 
The first family of fibers is oriented radially. 
It connects the free edge to the valve ring. 
The second is circumferential. 
It runs approximately parallel to the valve ring. 
Each circumferential fiber connects free edge to free edge or forms a closed ring. 

\item The leaflets are supported by a system of chordae tendineae, which anchor into two papillary muscles. 
Like the fibers in the leaflets, the chordae exert tensile forces only. 

\item Tension in the leaflets supports a static, uniform pressure load.
This is possible because the leaflets are curved.  
There is no pressure load acting directly on the chordae (since they are idealized as being one-dimensional), but the tension in the chordae indirectly supports the pressure load on the leaflets.
The whole structure, composed of leaflets and chordae, achieves a mechanical equilibrium in which all of these forces balance.

\end{enumerate}

\section{Problem formulation}

First, we derive the continuous formulation of the equations of equilibrium in the leaflet. 
We represent the leaflet as an unknown parametric surface in $\bb R^{3}$, 
\begin{align}  
\bb X(u,v) : \Omega \subset \R^{2} \to \R^{3} . 
\end{align}
In this formulation, there are two families of fibers, one running along the curves $v$ = constant, and the other along the curves $u$ = constant.  
The fibers $v$ = constant, on which $u$ varies, will be called $u$-type fibers, and the fibers $u$ = constant, on which $v$ varies, will be called $v$-type fibers.  
The assumption that the fibers are arranged in this manner implies that the parametrization conforms to the fibers.  
Also by convention, we take $u$ and $v$ to have units of length. 
Let subscripts denote partial derivatives and let single bars, $| \cdot |$, denote the Euclidean norm. 
The unit tangents to these two fiber families are 
\begin{align}
\frac{\bb X_{u}}{ | \bb X_{u} |} \quad \text{ and } \quad \frac{\bb X_{v}}{ | \bb X_{v} |}, 
\end{align}
respectively. 
Let $S(u,v)dv$ be the tension transmitted by the $u$-type fibers with v in the interval $(v,v+dv)$, and similarly let $T(u,v)du$ be the tension transmitted by the $v$-type fibers with $u$ in the interval $(u,u+du)$.  
Note that $S$ and $T$ have units of force/length (since we assumed $u$ and $v$ have units of length), but they are best described as ``force per unit $v$'' and ``force per unit $u$'' respectively.  
In particular, the value of $S$ changes if $v$ is replaced by some function of $v$, and the value of $T$ changes if $u$ is replaced by some function of $u$.  
These are the most general changes of parameters that are allowed, since the parameterization is assumed to conform to the fibers.

Consider the static mechanical equilibrium of an arbitrary patch of leaflet $[u_{1}, u_{2}] \cross [ v_{1} , v_{2} ]$. 
Let $p$ denote the pressure, which acts in the normal direction to the patch. 
We assume that $p$ is constant; that is, the pressure load is spatially uniform. 
The total pressure force on the patch is given by 
\begin{align} 
\int_{v_{1}}^{v_{2}}  \int_{u_{1}}^{u_{2}}   p  (  \bb X_{u}(u,v) \cross \bb X_{v}(u,v) ) \;  du dv . 
\end{align}
The tension force due to $u$-type fibers acts on the edges of constant $u$, so the total force transmitted across the arc $u=u_{2}$, $v \in (v_{1}, v_{2} )$ is given by  
\begin{align} 
\int_{v_{1}}^{v_{2}}   S(u_{2}, v) \frac{ \bb X_{u} (u_{2}, v) }{|  \bb X_{u} (u_{2}, v) |}  \;  dv  . 
\end{align}
Then, the total force due to $u$-type fibers on the patch is then given 
\begin{align} 
\int_{v_{1}}^{v_{2}}   S(u_{2}, v) \frac{ \bb X_{u} (u_{2}, v) }{|  \bb X_{u} (u_{2}, v) |} - S(u_{1},v)   \frac{ \bb X_{u} (u_{1}, v) }{|  \bb X_{u} (u_{1}, v) |}     \;  dv  . 
\end{align}
Similarly, the total force to to $v$-type fibers is given 
\begin{align} 
\int_{u_{1}}^{u_{2}}   T(u, v_{2})  \frac{\bb X_{v} (u, v_{2})}{|\bb X_{v} (u, v_{2})|} - T(u,v_{1})  \frac{\bb X_{v} (u, v_{1}) }{|\bb X_{v} (u, v_{1})|} \;  du . 
\end{align}

\begin{figure}[H]
\centering
\includegraphics[width=\textwidth]{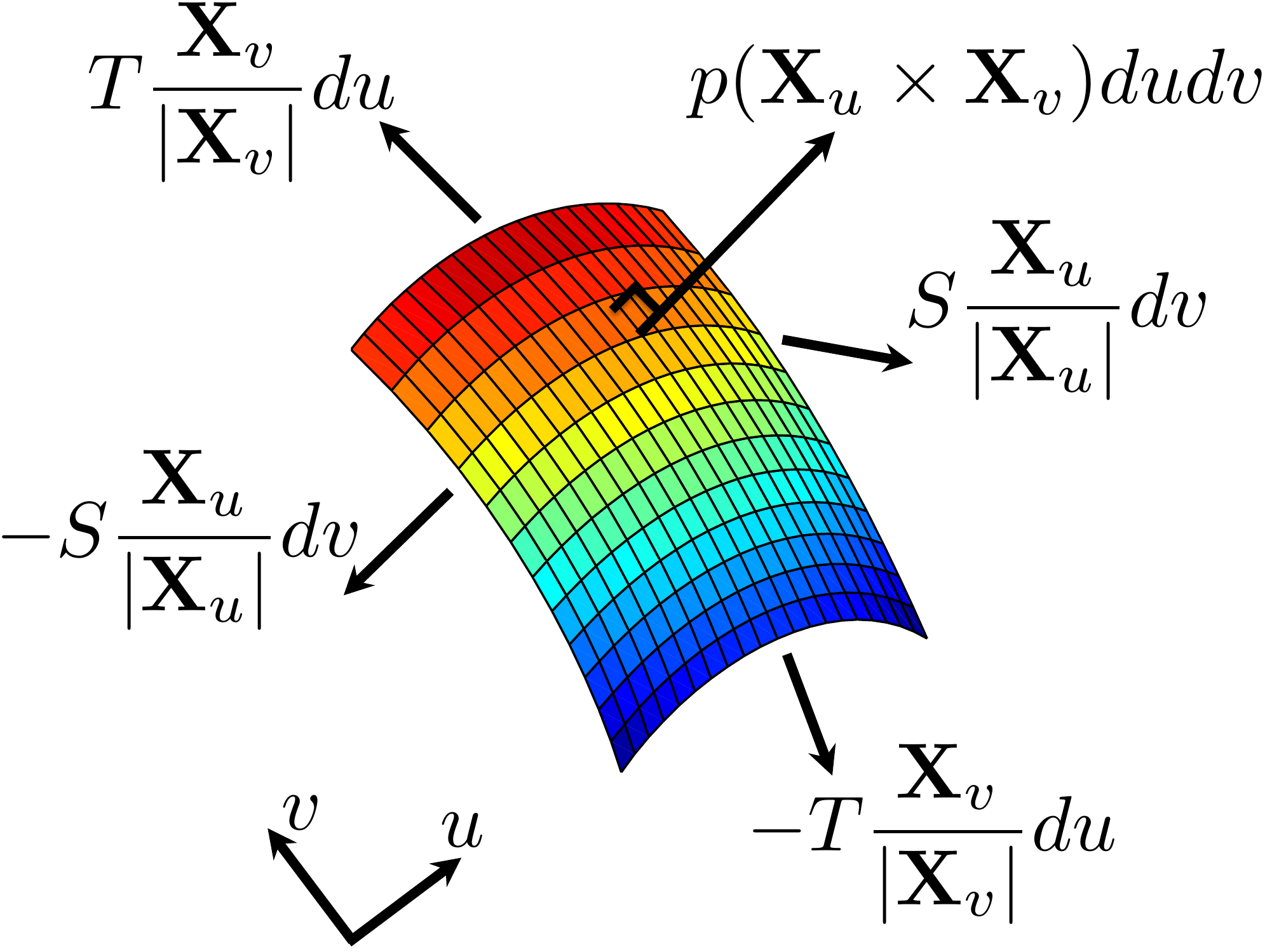}
\caption{Free body diagram for mechanical equilibrium on the leaflet. 
Note that the tension variables are evaluated on opposing boundaries of the patch, and thus do not cancel, but we drop arguments to all functions for visual clarity.}
\label{free_body_diagram_leaflet.pdf}
\end{figure}

A free body diagram of these forces applied to a patch of leaflet is shown in figure \ref{free_body_diagram_leaflet.pdf}. 

The condition of static mechanical equilibrium dictates that the forces on the patch $[u_{1}, u_{2}] \cross [v_{1}, v_{2}]$ must sum to zero, so the integral form of the equations of equilibrium is
\begin{align} 
0 &= \int_{v_{1}}^{v_{2}}  \int_{u_{1}}^{u_{2}}   p  (  \bb X_{u}(u,v) \cross \bb X_{v}(u,v) ) \;  du dv  \\ 
     &+ \int_{v_{1}}^{v_{2}}   S(u_{2}, v) \frac{ \bb X_{u} (u_{2}, v) }{|  \bb X_{u} (u_{2}, v) |} - S(u_{1},v)   \frac{ \bb X_{u} (u_{1}, v) }{|  \bb X_{u} (u_{1}, v) |}     \;  dv \nonumber  \\ 
     &+ \int_{u_{1}}^{u_{2}}   T(u, v_{2})  \frac{\bb X_{v} (u, v_{2})}{|\bb X_{v} (u, v_{2})|} - T(u,v_{1})  \frac{\bb X_{v} (u, v_{1}) }{|\bb X_{v} (u, v_{1})|} \;  du . \nonumber
\end{align}
Apply the fundamental theorem of calculus to convert all the integrals into double integrals over the patch. 
Swap the order of integration formally as needed. 
This gives 
\begin{align} 
0 = \int_{v_{1}}^{v_{2}}  \int_{u_{1}}^{u_{2}} \left(   p  (  \bb X_{u} \cross \bb X_{v} )  +   \frac{\p}{\p u}  \left( S \frac{ \bb X_{u} }{ |\bb X_{u}| } \right)  +  \frac{\p}{\p v}  \left( T \frac{ \bb X_{v} }{|\bb X_{v}|} \right)    \right)  \;  du dv , 
\end{align}
where all unknowns are evaluated at $u,v$. 
Since the patch is arbitrary, the integrand must be zero. 
This gives the partial differential equation form of the equations of equilibrium as 
\begin{align} 
0 = p  (  \bb X_{u} \cross \bb X_{v} )  +   \frac{\p}{\p u}  \left( S \frac{ \bb X_{u} }{ |\bb X_{u}| } \right)  +  \frac{\p}{\p v}  \left( T \frac{ \bb X_{v} }{|\bb X_{v}|} \right) .    
\label{eq_eqns}
\end{align}
As mentioned previously, we let the parameters $u$ and $v$ have units of length, and therefore $S$ and $T$ have units of force/length.
It follows that all terms in the above equation have the same units as pressure, that is, force/area.

\section{Closing the equilibrium equations}

This form of the equilibrium equations shown in equation \ref{eq_eqns} is is not closed. 
It is three equations, one for each component of the net force on the patch, and five unknowns, three for $\bb X$ and two for the tensions $S,T$. 
To close it, we must specify a law that allows us to compute tension from known quantities. 
Here, we hope to find a tension law that does not require a reference configuration, since we do not have access to such a configuration.

The simplest example of a tension law without a reference configuration would a prescribed constant tension for each fiber family, 
\begin{align}
S(u,v) = \alpha, \quad T(u,v) = \beta . 
\label{const_tension}
\end{align} 
This has elegant consequences for the geometry of the surface, but presents significant practical and numerical challenges.  
Since tension does not depend on lengths of fibers, this model allows to fibers to slip against opposing families when solving the equations of equilibrium. 
We show below that \ref{const_tension} implies that both fiber families are geodesics on the surface $\bb X$. 
While this sounds desirable, the fact that the fibers can slip past each other means that they sometimes collide (meaning that points on the discretized fibers that are required to be distinct coincide), so we will need to modify this formulation later.

Suppose, temporarily, that tensions are constant and the equilibrium equation is given 
\begin{align} 
0 = p  (  \bb X_{u} \cross \bb X_{v} )  +  \alpha \frac{\p}{\p u}  \left(  \frac{ \bb X_{u} }{ |\bb X_{u}| } \right)  +  \beta \frac{\p}{\p v}  \left( \frac{ \bb X_{v} }{|\bb X_{v}|} \right) . 
\label{eq_eqn_const_tension}
\end{align}
Note that $\bb X_{u} / |\bb X_{u}|$ and $\bb X_{v}/ |\bb X_{v}| $ span the tangent plane of the surface, and $\bb n =  \bb X_{u} \cross \bb X_{v} $ is normal to the tangent plane. 
We are assuming here that the fiber system is non-degenerate; that is, the two families are not parallel at any point on the leaflet.

The vector $ \frac{\p}{\p u}  \left(   \bb X_{u} / |\bb X_{u}|  \right)$ is a scalar multiple of the principal normal of a $u$-type fiber, and similarly the vector $\frac{\p}{\p v}  \left( \bb X_{v} / |\bb X_{v}|  \right)$ is a scalar multiple of the principal normal of a v-type fiber. 
Take the inner product of each side of equation \ref{eq_eqn_const_tension} with $\bb X_{u} / |\bb X_{u}|$. 
The pressure term is zero, since 
\begin{align} 
\frac{ \bb X_{u} }{ |\bb X_{u}| }  \cdot p  (  \bb X_{u} \cross \bb X_{v} ) 
= \frac{p}{|\bb X_u|} (\bb X_{u} \cross \bb X_{u}) \cdot \bb X_{v} 
= 0 .  
\end{align} 
The second term is also zero, since 
\begin{align} 
\frac{ \bb X_{u} }{ |\bb X_{u}| }  \cdot \frac{\p}{\p u}  \left(  \frac{ \bb X_{u} }{ |\bb X_{u}| } \right) 
=  \frac{\p}{\p u}  \left( \frac{1}{2} \frac{ \bb X_{u} }{ |\bb X_{u}| } \cdot \frac{ \bb X_{u} }{ |\bb X_{u}| } \right) 
=  \frac{\p}{\p u}  \left(  \frac{1}{2} \right) 
= 0 
\label{unit_vector_derivative}
\end{align} 
because $ \bb X_{u} / |\bb X_{u}| $ is a unit vector. 
This implies that 
\begin{align} 
\frac{ \bb X_{u} }{ |\bb X_{u}| }  \cdot \frac{\p}{\p v}  \left(  \frac{ \bb X_{v} }{ |\bb X_{v}| } \right) 
= 0 
\end{align} 
Thus, the principal normal of a $v$-type fiber is orthogonal to $\bb X_{u}$. 
But it is also orthogonal to $\bb X_{v}$, by the same argument as in \ref{unit_vector_derivative} with $v$ replacing $u$. 
Since $\bb X_{u}$ and $\bb X_{v}$ span the tangent plane to the surface, the principal normal to a $v$-type fiber must be orthogonal to the surface. 
This implies that every $v$-type fiber is a geodesic \cite{millman1977elements}.
The same reasoning with $u$ and $v$ interchanged shows that every $u$-type fiber is a geodesic as well.

A geodesic locally minimizes length on a surface. 
When we solve the equilibrium equations, we observe these surfaces becoming degenerate. 
Lengths tend to zero, and regions that are supposed to represent a full leaflet shrink to points or thin rings that do not resemble the real mitral valve.

We propose the following tension law as an alternative, which we call the \emph{decreasing tension} model.  
Suppose that the maximum tension in $u$-type fibers is limited by $\alpha$, but goes smoothly to zero as $|\bb X_{u}|$ goes to zero. 
Take 
\begin{align}
S(u,v)  = \alpha \left( 1 - \frac{1}{1 + |\bb X_{u}|^{2} / c_{\alpha}^{2} } \right)   , 
\label{dec_tension}
\end{align}
where $c_{\alpha}$ is a tunable parameter. 
Similarly, let 
\begin{align}
T(u,v)  = \beta \left( 1 - \frac{1}{1 + |\bb X_{v}|^{2} / c_{\beta}^{2} } \right)   . 
\end{align}
This tension law drastically improves the quality of geometries which we were able to achieve compared to the constant tension formulation of equation \ref{const_tension}. 

Note that arc length $ds$ along a $u$-type fiber is given by $ds = |\bb X_{u}| du$, and similarly arc length $ds$ along a $v$-type fiber is given by $ds = |\bb X_{v}|dv$. 
Informally, as $|\bb X_{u}|$ goes to zero, $u$-type fibers become ``closer together.'' 
In Section \ref{Discretization} we will introduce the discretization of this model, and in that version, this corresponds to the lengths of individual links of the fiber mesh going to zero.

\begin{figure}[H]
\centering
\includegraphics[scale=.7]{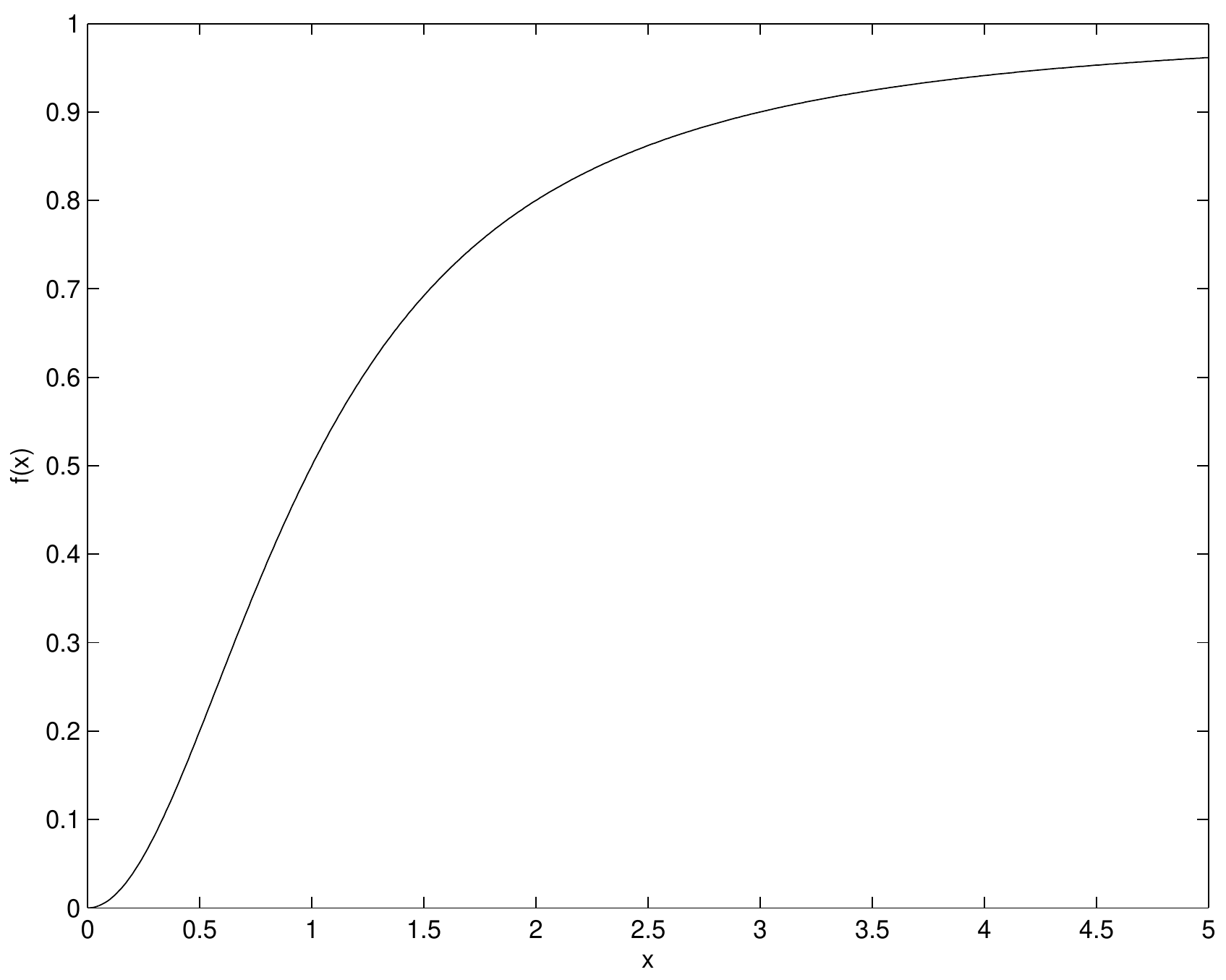}
\caption{Form of length-tension curve with $c=1$.}
\label{dec_tension_curve.pdf}
\end{figure}

To understand this tension law, consider the function 
\begin{align} 
f(x) = 1 -  \frac{1}{1 + x^{2}/c^{2}} , 
\end{align}
shown in figure \ref{dec_tension_curve.pdf} with $c = 1$. 
The function $f \to 0$ as $x \to 0$, and $f \to 1$ as $x \to \infty$. 

\noindent 
Taking its derivative, we get 
\begin{align} 
f'(x) = \frac{2x/c^{2}}{ (1 + x^{2}/c^{2})^{2} } . 
\end{align}
This says that says that $f'(x) > 0$ for positive $x$. 
Thus, using this tension law gives a tension that goes smoothly to zero as $|\bb X_{u}| \to 0$, and asymptotes to $\alpha$ as $|\bb X_{u}| \to \infty$. 
The parameter $c$ controls how quickly this transition occurs. 
In the $c \to 0$ limit, the tension assumes the constant value $\alpha$. 
Also, $\lim_{x \to 0^{+}} f'(x) = 0$, so the derivative of tension also goes smoothly to zero as $|\bb X_{u}| \to 0$. 
We chose the power two because it is the smallest integer power such that both the tension and its derivative go smoothly to zero in this manner.

We can use this parameter in numerical solutions to influence how spread out the fibers are in a given configuration of the valve. 
Informally, if a fiber approaches degeneracy, the condition that $|\bb X_{u}| \to 0$, then the fiber no longer exerts much or any force. 
Since the fibers must support the specified pressure, this causes them to stay apart and so resists collisions and equilibrium configurations can be found effectively. 

Note that this law is not designed to be a physical constitutive law, and does not represent the tensions that would occur in the range of deformations that the mitral valve experiences through the cardiac cycle. 
This tension law is in place as design principle, which is selected to be a compromise on the design principle of uniform tension on the leaflet.

Using this tension law, the form of the equilibrium equation in the leaflets is given as
\begin{align} 
0 &= p  (  \bb X_{u} \cross \bb X_{v} )  \label{eq_eqn_dec_tension}  \\
&+  \frac{\p}{\p u}  \left( \alpha \left( 1  -  \frac{1}{1 +  |\bb X_{u}|^{2} / c_{\alpha}^{2}  } \right) \frac{ \bb X_{u} }{ |\bb X_{u}| } \right) \nonumber \\ 
&+  \frac{\p}{\p v}  \left(  \beta \left( 1  -  \frac{1}{1 +  |\bb X_{v}|^{2} / c_{\beta}^{2} }     \right) \frac{ \bb X_{v} }{|\bb X_{v}|} \right) .  \nonumber 
\label{equilibrium_dec_tension}
\end{align}

This is a boundary value problem, the solution of which represents the loaded mitral valve leaflets. 
Note that the coefficients $\alpha, \beta$, as well as the decreasing tension parameters $c_{\alpha}, c_{\beta}$ need not be constants. 
Indeed, tuning the values of these for specific regions of the valve has proven important to getting good results.

\section{Fiber and mesh specification}
\label{fiber_spec}

We will discretize and solve the equations of equilibrium numerically using a finite difference approach. 
Before discussing the discretization of the equations, we will specify the domain on which they are solved in terms of the parameters $u$ and $v$. 
The leaflets and chordae tendineae have a complex layout composed of many parts. 
We chose the finite difference mesh for the leaflets and the topology of the model chordae to mimic this structure. 
The details are as follows.

\begin{figure}[H]
\centering
\includegraphics[width=\textwidth]{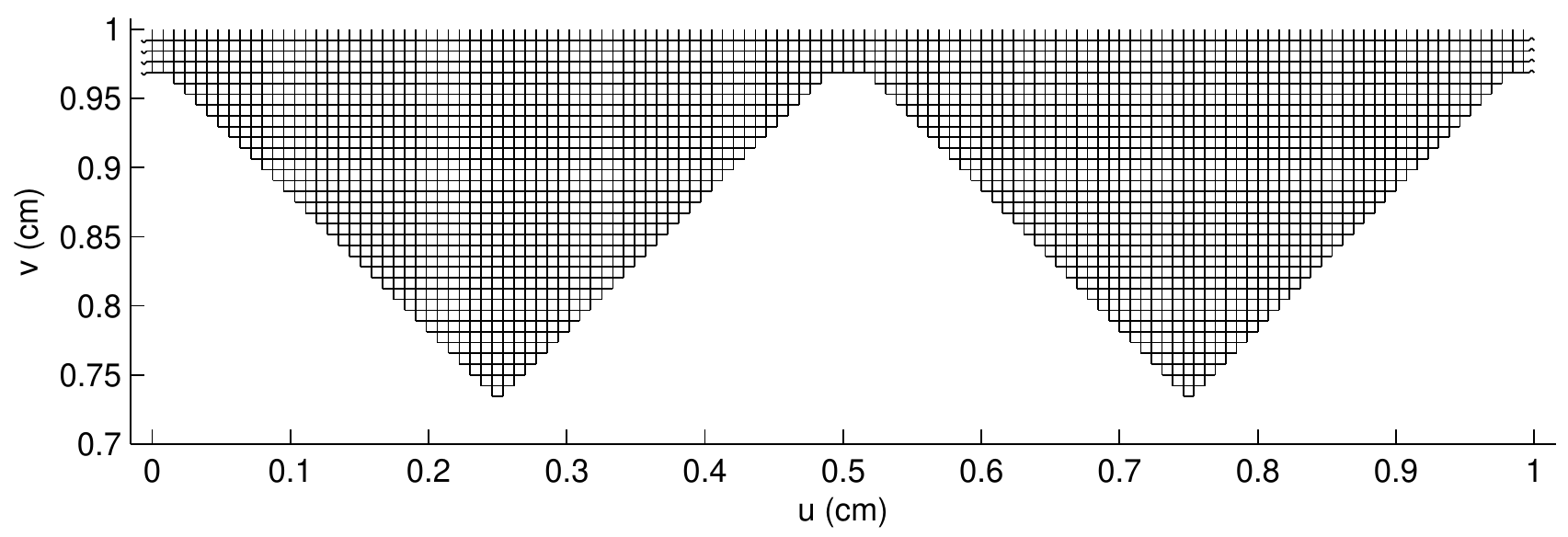}
\caption{Schematic of leaflet mesh. No fiber is drawn at the top of the diagram because a fiber is not placed at the ring when solving the equations of equilibrium; the points are boundary conditions and not connected to each other in this phrasing. $N = 128$.}
\label{mesh_schematic_leaflet.pdf}
\end{figure}

The domain on which the parameters $u,v$ are defined is a subset of a 1 cm by 1 cm square, shown in figure \ref{mesh_schematic_leaflet.pdf}.  
The valve ring, which is treated as a boundary condition, corresponds to $v = 1$. 
This boundary condition specifies the configuration of the valve ring, and determines how the leaflets attach to it. 
The ring is discretized with $N$ points, and the mesh spacing in $u$ is defined to be $\du = 1/N$ cm. 
The domain is taken to be periodic in $u$ with period 1 cm.
For any given $v$, the set of $u$ that are included in the domain may be a proper subset of $[0,1] \cross v$. 

The curve $v_{min}(u)$ is a location dependent minimum for $v$. 
This is selected to be piecewise linear, making triangles in parameter space that correspond to the anterior and posterior leaflets.
The slope is generally $\pm 1$, with the exception of small, flat sections corresponding to the commissures in which the slope is zero. 
This is so that outline of the finite difference mesh resembles that of real valve leaflets. 
The ``point'' of the triangles in parameter space will spread out when the equilibrium equations are solved. 
Both of these triangular regions have width $1/2$ cm and height $1/4$ cm. 
The curve $v_{min}(u)$ is selected such that the mesh spacings are equal in both directions; that is, $\du = \dv$. 
We only use discretizations for which $v_{min}(u) > 0$, but this is not required.

The curve $v_{min}(u)$ is treated as a free boundary when solving the equations of equilibrium. 
The location of this free boundary is determined by the discrete equations of equilibrium for the chordae tendineae simultaneously with the equations of equilibrium in the leaflets.
(The fiber specification for the chordae tendineae is discussed shortly below.)

The $u$-type fibers, on which $u$ varies and $v$ is constant, are referred to as \emph{circumferential}. 
Their preimages in parameter space are parallel to the valve ring. 
Upon solving the equations of equilibrium, the $u$-type fibers are no longer parallel to the valve ring (or each other), but they will still be roughly parallel to the valve ring. 
The top few fibers are taken to be \emph{rings}, which are topologically circles. 
This represents the fact in the real valve there is a region of leaflet tissue near the valve ring that is topologically an annulus, as can be seen in dissections in figure \ref{pinned_atrial_zoom_out}. 
The $N_{rings}$ rings are placed below the boundary condition at $v = 1$, and occupy the region 
\begin{align}
(u,v) \in [0,1] \cross [ 1 - 1/64 ,1)
\end{align}
in parameter space, where the factor of $1/64$ is selected such that there are eight rings when $N = 512$. 
Other $u$-type fibers begin and end at the free edge of the domain, which corresponds to the free edge of the leaflet.

The $v$-type fibers, on which $v$ varies and $u$ is constant are \emph{radial}. 
They are orthogonal to the valve ring in parameter space.  
Every $v$-type fiber begins at the free edge of the domain, and ends at the valve ring, which corresponds to $v = v_{max}$. 
Note that while the two fiber families are orthogonal in parameter space, they will generally not be in solutions to the equilibrium equations.

Regions of the mesh that correspond to different parts of the valve have different properties. 
We view the anterior leaflet as $u \in [0,1/2]$, and the posterior as $u \in [1/2,1]$. 
The anterior and posterior regions have location dependent values of the coefficients $\alpha, c_{\alpha}, \beta, c_{\beta}$. 
Circumferential fibers with $v > 1 - 1/64$ are seen as rings, and are topologically circles. 
They too have their own $\alpha, c_{\alpha}$. The parameter $\beta$ does not have a distinct value in this region, since radial fibers in that region are continuous with radial fibers in the leaflets.
However, $c_{\beta}$ does have a distinct value, so that we may use it to influence fiber spacing in the rings.

Finally, we add a small section of ``edge connectors'' at the commissures. 
We found that many models were prone to leaking in this region, and adding a patch of extra tissue helped this. 
It too has its own constants. 
Radial fibers in this region simply end at the free edge, rather than connect to chordae tendineae. 
Whatever tension they support must be supported by curvature in the circumferential fibers in that region. 
Adding these fibers makes more fibers into topological circles, but we still think of the regions as being separate because their parameters are different from those of the rings and leaflets. 

\begin{figure}[H]
\centering
\includegraphics[width=\textwidth]{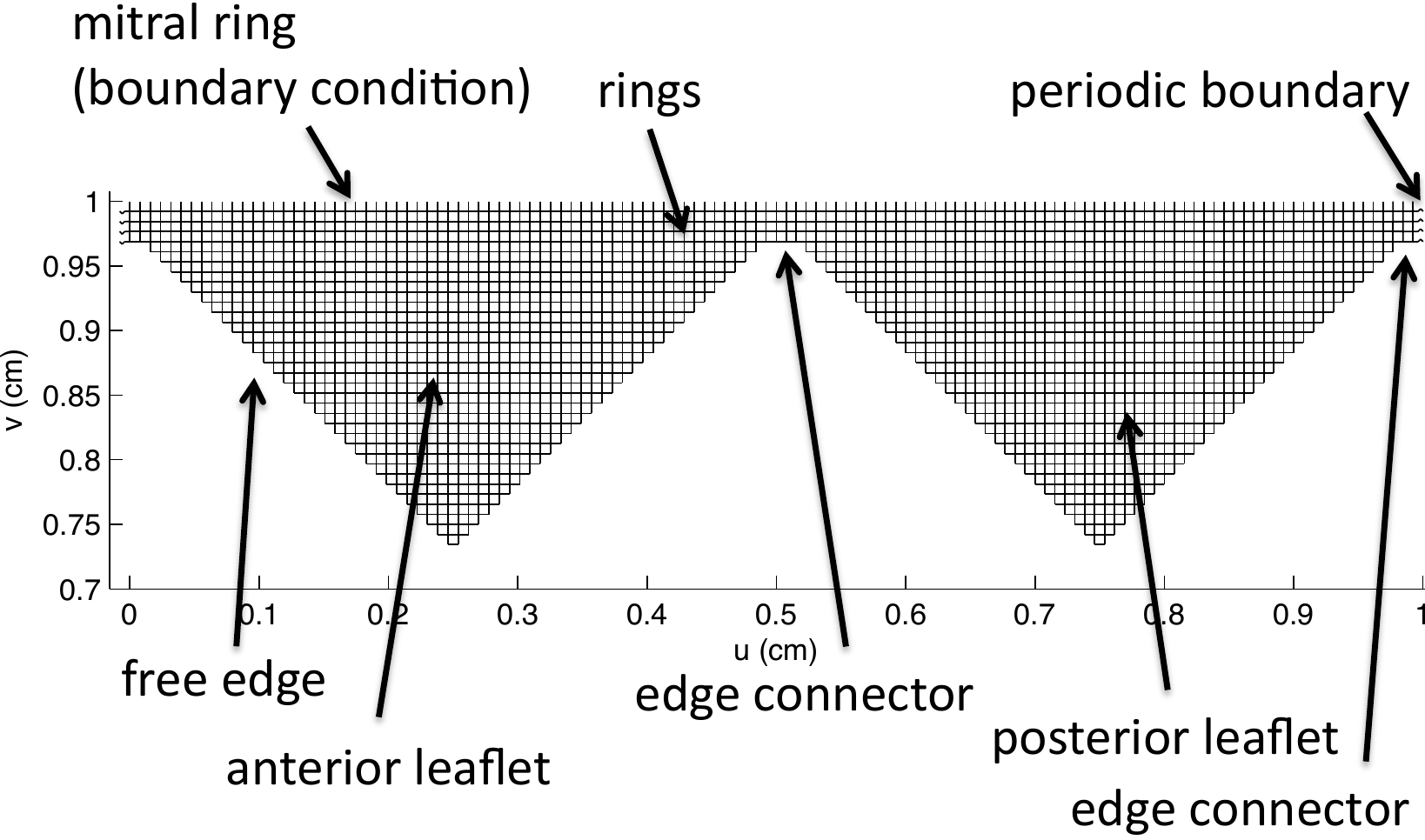}
\caption{Schematic of leaflet mesh with anatomical structures labeled. $N = 128$.}
\label{mesh_schematic_labeled.pdf}
\end{figure}

Having discussed these regions in more detail, we show the same schematic with anatomical and mesh features labeled in figure \ref{mesh_schematic_labeled.pdf}.

To specify boundary conditions at $v = v_{max}$, we first comment on the anatomy of the mitral ring. 
It is observed that the valve ring takes on a ``rounded-D'' or ``lima-bean'' shape, especially in systole.  
The anterior leaflet is centered on this slight depression in the ring, as shown in an anatomy study, figure 1 in \cite{DEGANDT20071250}. 
One review of valve anatomy describes this as a ``mild concave form, because it is directly related to the circular aspect of the aortic orifice'' \cite{Misfeld1421}. 
That is, the center of the anterior leaflet corresponds with the location at which the valve ring is adjacent to the aorta.
The slight concavity may be caused by the valve ring being squeezed against the aorta during ventricular contraction. 
It is also observed that the anterior leaflet takes up less than half of the valve ring in terms of angle \cite{mccarthy2010anatomy}.

\begin{figure}[H]
\centering
\includegraphics[scale=.6]{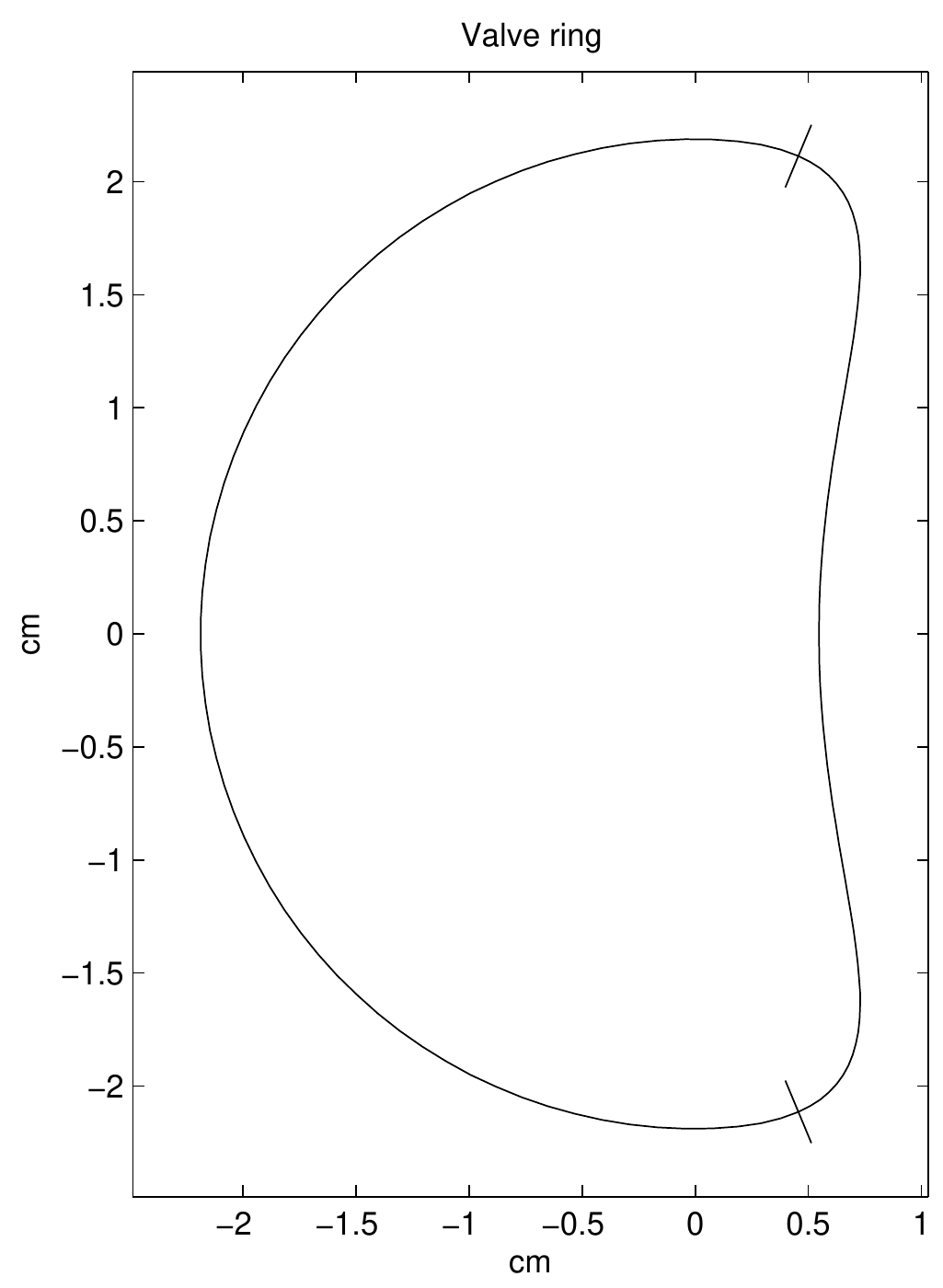}
\caption{Valve ring. The anterior leaflet attaches on the right of the frame, between the two ticks, and the posterior on the remainder of the ring.}
\label{valve_ring_2d}
\end{figure}

Thus, we specify that the shape of the ring is a semicircle on the posterior side, and is modified from a semicircle to be slightly concave on the anterior side. 
We specify that each leaflet takes up a prescribed angle of this lima-bean shaped ring, the anterior taking up under half of the total angle. 
The ring is shown in figure \ref{valve_ring_2d}. 
To implement this, we define a number of parameters and functions.
Let $\theta$ denote the angle around the valve ring, and arrange the coordinates such that $\theta = 0$ corresponds to the center of the anterior leaflet.
This implies that $\theta = \pi$ corresponds to the center of the posterior leaflet.   
Let  
\begin{align}
\varphi(\theta) = 
\begin{cases}
\cos^{2}(\theta)   & : \; | \theta | < \pi/2 \\ 
0                         & : \; | \theta | \geq \pi/2 , \\ 
\end{cases}
\end{align}
where $\theta$ is restricted to $\theta \in [-\pi,\pi]$.
Let $a_{ant} = 5\pi/6$ and $a_{post} = 2\pi - a_{ant} = 7\pi/6$ denote the angles that the anterior and posterior leaflet respectively occupy on the ring. 
The anterior leaflet is fixed to the ring for $\theta \in (-a_{ant}/2, a_{ant}/2)$, and the posterior leaflet is fixed to the ring for $\theta \in (\pi -a_{post}/2, \pi + a_{post}/2)$. 
Let $\rho_{dip}$ be a tunable parameter that controls the slight concavity of the valve ring around the anterior leaflet, which we select as $\rho_{dip} = 0.75$. 
The anterior leaflet attaches to the ring on $u \in [0,1/2], v = v_{max}$, and the posterior attaches to the leaflet on $u \in [1/2,1], v = v_{max}$. 
The value of $\theta$ is expressed as a linear function of $u$, which is given as 
\begin{align}
\theta(u) = 
\begin{cases}
a_{ant}2u - a_{ant}/2   & : \;  u \in [0,1/2]  \\ 
a_{post}2(u - 1/2) - a_{post}/2 + \pi                        & : \;   u \in (1/2,1)  \\ 
\end{cases}
\end{align}
Let $r = 2.19$ cm denote the valve radius read from MRI data, as discussed in Section \ref{MRI}. 
We define the valve ring boundary condition as 
\begin{align}
\bb X(u,1) = r ( \cos(\theta(u)) -  \rho_{dip} \varphi(\theta(u)) , \sin(\theta(u)), 0)
\end{align}
Note that if $\rho_{dip} = 0$ the valve ring would be a circle, and for $\rho_{dip} > 0.5$ the ring has the desired concavity. 

Commentary from medical device manufacturers suggests a 3:4 ratio of the anterior-posterior diameter to the transverse (commissure to commissure) diameter is desirable \cite{Annuloplasty_ring}, which in our framework implies $\rho_{dip} = 0.5$ and no concavity in the center of the anterior leaflet. 
However, this device is used for patients with annular dilation and it may be that a lower ratio creates other problems such as reducing forward flow during diastole. 
Using $\rho_{dip} = 0.75$ gives a ratio of 5:8, which is slightly narrower in the anterior-posterior diameter than specifications from this manufacturer suggest, but we believe that the slight convexity given by this value is more effective and anatomical for modeling a normal valve.

The chordae tendineae are modeled as trees that exert tensile forces.  
They support the leaflets, which in turn support pressure. 
However, since they are modeled as one-dimensional strings, there is no pressure applied directly to the chordae. 
The chordae must maintain mechanical equilibrium of tensile forces only. 

The branching structure of the true trees is complex and varies heavily with individuals. 
We idealize this by assuming that every tree is binary and that each internal node has precisely two descendants. 
The root has a single descendent, since we think of each of the chordae as it emerges from the papillary muscle as a single strand. 
The branching occurs some distance from the papillary muscles as the chordae approach the free edge, as is visible in figure \ref{papillary_detail}. 
The center of the papillary tip is read from MRI data. 
The roots of the trees are spread across a portion of a small circle surrounding this point.
Their locations are treated as boundary conditions.

This creates an obvious way to change the trees as the mesh is refined. 
If a mesh is twice as fine and thus has twice as many points, add an additional generation to the tree, which doubles the number of leaves. 
This also allows us to form consistent scaling laws for the included constants. 
The tension law in the chordae still takes the form shown in equation \ref{dec_tension}. 
Experiments with constant tension chordae had similar consequences to prescribing constant tension in the leaflet. 
Trees attempted to collapse into points or otherwise become degenerate. 

Note that we specify the topology of the trees, but do not specify the lengths of any links. 
Like in the leaflets, the physical configuration emerges from solving the equilibrium equations.

\begin{figure}[H]
\centering
\includegraphics[width=\textwidth]{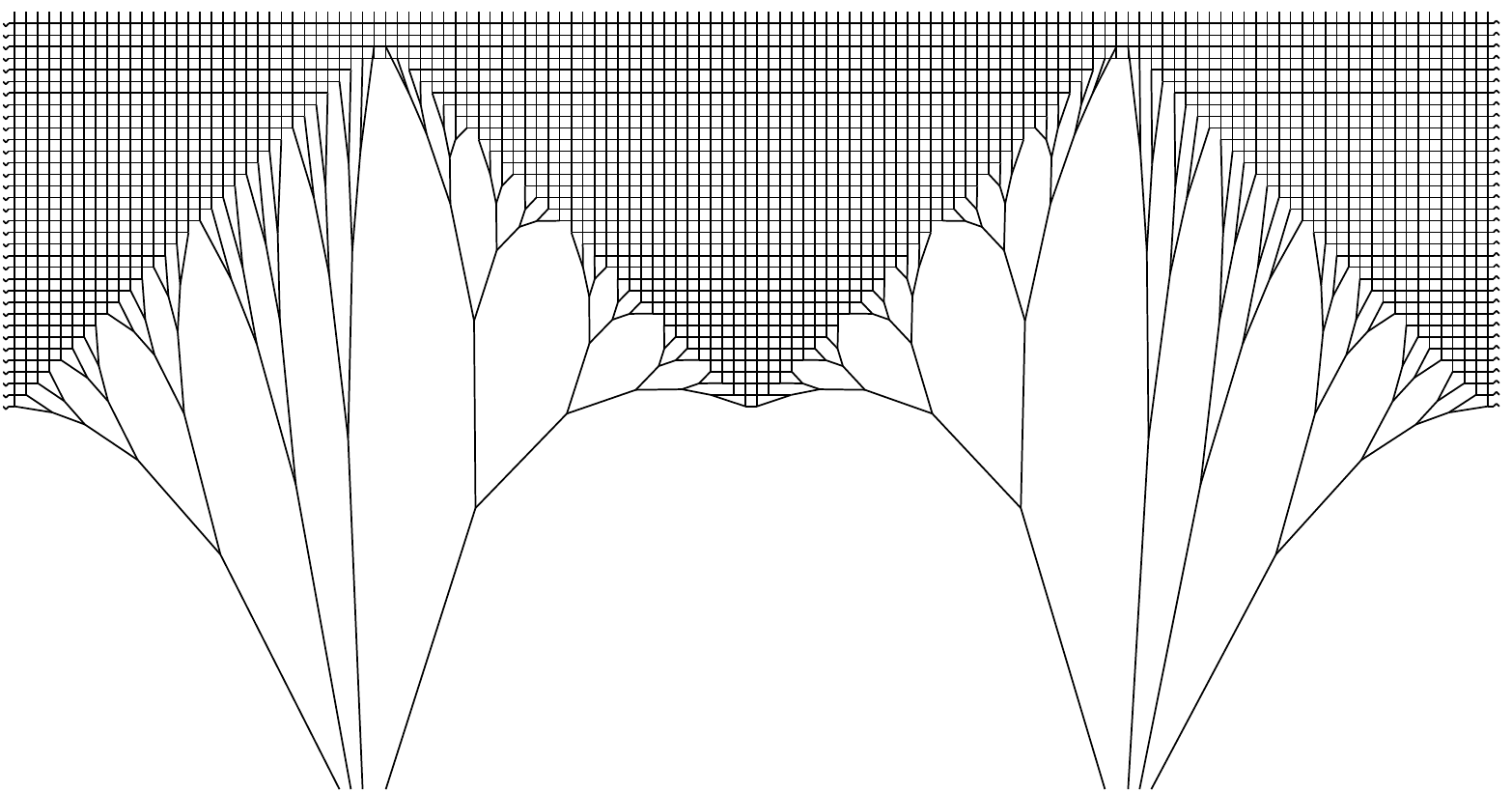}
\caption{Schematic of valve mesh, including trees of chordae. The anterior leaflet is centered, and the posterior leaflet is split in two, as in dissections in figure \ref{valve_basic}.
For visual clarity, the mesh shown is four times more coarse than the final mesh that is later used in fluid. 
}
\label{mesh_schematic.pdf}
\end{figure}

Figure \ref{mesh_schematic.pdf} shows a schematic of the mesh, including the trees. 
The lengths in this diagram are not used; it is provided for illustrative purposes only. 
As such, no axes are drawn because they have no meaning in relation to the chordae. 
Also, this version has been rearranged to match the dissection photographs shown in figures \ref{valve_basic} and \ref{pinned_ventricular_zoom_out}; the anterior leaflet is centered and the posterior is split in two.

The spatial configuration of this entire system -- both leaflets, the small region of rings and the chordae tendineae -- is solved for simultaneously. 
This implies that free edge is an unknown in this formulation. 
It's configuration is solved for simultaneously with the equilibrium configuration of the leaflets and chordae tendineae.  
Formally, this makes the solution to the leaflet partial differential equations a free boundary problem, in which the free boundary is determined by this system of coupled equations.

\section{Discretization}
\label{Discretization}

We discretize equation \ref{eq_eqn_dec_tension} with a centered finite difference scheme. 
Let $j,k$ denote the index of a particular point, and suppose that all four of its neighbors lie on the leaflet. 
The pressure term is discretized with two point wide centered differences. 
\begin{align}
0 &=   p \left(  \frac{(\bb X^{j+1,k} - \bb X^{j-1,k} )}{2\du}  \cross \frac{ ( \bb X^{j,k+1} - \bb X^{j,k-1} ) }{2\dv} \right)  
\end{align}
The tangent terms are differenced with one point wide differences  
\begin{align}
\frac{ \bb X^{j+1,k}  -  \bb X^{j,k}  }{ | \bb X^{j+1,k}  -  \bb X^{j,k} |  } , 
\end{align}
as are the tension terms 
\begin{align}
\alpha \left( 1 - \frac{1}{1 + \dfrac{ | \bb X^{j+1,k}  -  \bb X^{j,k} |^{2}}{  c_{\alpha}^{2} (\du)^{2}  } }  \right)  . 
\end{align}
The second difference is then centered, and computed as 
\begin{align}
            &
             \frac{\alpha}{\du} 
	     \left( 1 - \frac{1}{1 + \dfrac{ | \bb X^{j+1,k}  -  \bb X^{j,k} |^{2}}{  c_{\alpha}^{2} (\du)^{2}  } }  \right)            
             \frac{ \bb X^{j+1,k}  -  \bb X^{j,k}  }{ | \bb X^{j+1,k}  -  \bb X^{j,k} |  }  
             \label{dec_tension_discretized} \\ 
           - &     
             \frac{\alpha}{\du} 
	     \left( 1 - \frac{1}{1 + \dfrac{ | \bb X^{j-1,k}  -  \bb X^{j,k} |^{2}}{  c_{\alpha}^{2} (\du)^{2}  } }  \right)            
              \frac{ \bb X^{j,k}  -  \bb X^{j-1,k}  }{ | \bb X^{j,k}  -  \bb X^{j-1,k} |  }  . 
		\nonumber 
\end{align}
This gives the discretization of the full system at this point as 
\begin{align}
0 &=   p \left(  \frac{(\bb X^{j+1,k} - \bb X^{j-1,k} )}{2\du}  \cross \frac{ ( \bb X^{j,k+1} - \bb X^{j,k-1} ) }{2\dv} \right)  \label{eq_eqn_version_one}  \\ 
            & \quad +  
             \frac{\alpha}{\du} 
	     \left( 1 - \frac{1}{1 + \dfrac{ | \bb X^{j+1,k}  -  \bb X^{j,k} |^{2}}{  c_{\alpha}^{2} (\du)^{2}  } }  \right)            
             \frac{ \bb X^{j+1,k}  -  \bb X^{j,k}  }{ | \bb X^{j+1,k}  -  \bb X^{j,k} |  }  \nonumber  \\ 
            & \quad -    
             \frac{\alpha}{\du} 
	     \left( 1 - \frac{1}{1 + \dfrac{ | \bb X^{j,k}  -  \bb X^{j-1,k} |^{2}}{  c_{\alpha}^{2} (\du)^{2}  } }  \right)            
              \frac{ \bb X^{j,k}  -  \bb X^{j-1,k}  }{ | \bb X^{j,k}  -  \bb X^{j-1,k} |  }     \nonumber   \\   
             & \quad + 
             \frac{\beta}{\dv} 
	     \left( 1 - \frac{1}{1 + \dfrac{ | \bb X^{j,k+1}  -  \bb X^{j,k} |^{2}}{  c_{\beta}^{2} (\dv)^{2}  } }  \right)            
             \frac{ \bb X^{j,k+1}  -  \bb X^{j,k}  }{ | \bb X^{j,k+1}  -  \bb X^{j,k} |  } \nonumber  \\ 
             & \quad -               
              \frac{\beta}{\dv} 
	     \left( 1 - \frac{1}{1 + \dfrac{ | \bb X^{j,k}  -  \bb X^{j,k-1} |^{2}}{  c_{\beta}^{2} (\dv)^{2}  } }  \right)            
             \frac{ \bb X^{j,k}  -  \bb X^{j,k-1}  }{ | \bb X^{j,k}  -  \bb X^{j,k-1} |  }    .  \nonumber 
\end{align}

The terms in this expression have units of pressure or stress, or force per unit area. 
Let square brackets around a quantity denote the units of the enclosed quantity. 
We assume that $\du = \dv = 1\text{ cm}/N$, where $N$ is the number points on the valve ring, and that these parameters have units of length, $[u] = [v] = $ length. 
This is only a convention; they could also be assumed to be dimensionless and analysis could proceed similarly. 
This implies that the unit tangent terms are dimensionless. 
We assume further that the ``decreasing tension'' term, introduced in equation \ref{dec_tension} and taking the form $1 - 1/(1 + x^{2}/c^{2})$, is also dimensionless. 
The decreasing tension coefficients must satisfy 
\begin{align}
\left[  | \bb X_{u} | \right] = [c_{\alpha}], 
\end{align}
so such constants are dimensionless. 
Dimensional consistency implies that 
\begin{align}
[\alpha] = [\beta] = \frac{\text{force}}{\text{length}},
\end{align}
so the coefficients have units of force per unit length. 
Note that these coefficients do not change as meshes are refined, rather length elements scale to change total force. 
This is consistent with the continuous formulation in equation \ref{equilibrium_dec_tension}, in which these coefficients do not change. 

Equation \ref{eq_eqn_version_one} is invariant under the following scaling. 
Suppose that
\begin{align}
u' = \lambda u  
\end{align}
so that 
\begin{align}
\du' = \lambda / N = \lambda \du . 
\end{align}
Then if the coefficients are scaled as 
\begin{align}
\alpha' = \lambda^{-1} \alpha, \quad 
c_{\alpha}' =  \lambda^{-1} c_{\alpha}
\end{align}
the equations are identical. 
A similar scaling is valid in $v$. 
This implies that the somewhat arbitrary choices of domain in Section \ref{fiber_spec} ultimately have no effect on the solution, so long as coefficients are scaled appropriately as the domain of $u,v$ is altered.

In the chordae, we wish to use equations in which the terms have units of force. 
Thus for consistency, we multiply by an area element $\du \dv = (\du)^{2}$ to convert the leaflet equation from having units of force per unit area to units of force. 
 
Also as a convention on signs in this equation, we will write all tension-tangent terms with a positive sign in front. 
This has the effect that the tangent term always points from the current node to the neighbor. 
Some points also have tension force from the chordae, and these structures and forces do not have a continuum limit. 
If a connection to one of the chordae exists at a leaflet point, the \hspace{.1pt} chordal force is added to the force from neighboring points \hspace{.1pt} within the leaflet following this convention.

Applying these changes gives the form that we solve numerically, which is 
\begin{align}
0 &=  \frac{p}{4} \left(  (\bb X^{j+1,k} - \bb X^{j-1,k} )  \cross  ( \bb X^{j,k+1} - \bb X^{j,k-1} ) \right)  \label{equilbrium_eqn_discrete}\\ 
            & \quad +  
             \alpha \du 
	     \left( 1 - \frac{1}{1 + \dfrac{ | \bb X^{j+1,k}  -  \bb X^{j,k} |^{2}}{  c_{\alpha}^{2} (\du)^{2}  } }  \right)            
             \frac{ \bb X^{j+1,k}  -  \bb X^{j,k}  }{ | \bb X^{j+1,k}  -  \bb X^{j,k} |  }  \nonumber  \\ 
            & \quad +    
             \alpha \du  
	     \left( 1 - \frac{1}{1 + \dfrac{ | \bb X^{j-1,k}  -  \bb X^{j,k} |^{2}}{  c_{\alpha}^{2} (\du)^{2}  } }  \right)            
              \frac{ \bb X^{j-1,k}  -  \bb X^{j,k}  }{ | \bb X^{j-1,k}  -  \bb X^{j,k} |  }   \nonumber  \\   
             & \quad + 
             \beta \du  
	     \left( 1 - \frac{1}{1 + \dfrac{ | \bb X^{j,k+1}  -  \bb X^{j,k} |^{2}}{  c_{\beta}^{2} (\du)^{2}  } }  \right)            
             \frac{ \bb X^{j,k+1}  -  \bb X^{j,k}  }{ | \bb X^{j,k+1}  -  \bb X^{j,k} |  }  \nonumber  \\ 
             & \quad +                
              \beta \du  
	     \left( 1 - \frac{1}{1 + \dfrac{ | \bb X^{j,k-1}  -  \bb X^{j,k} |^{2}}{  c_{\beta}^{2} (\du)^{2}  } }  \right)            
             \frac{ \bb X^{j,k-1}  -  \bb X^{j,k}  }{ | \bb X^{j,k-1}  -  \bb X^{j,k} | } . \nonumber 
\end{align}

Some points on the leaflet lie on the free edge, and thus may not have four neighbors. 
For these points, the neighbors that attach to the current points are included in the tension terms, and those that do not exist are simply omitted. 
For computing the normal, if centered differences are not available because the current point is on the free edge, we take a one sided difference instead. 
We construct the mesh such that all points in the leaflet have at least one neighbor in each direction.
Thus, there are always two neighbors from which to compute the normal at any given point. 
The weights on the pressure term are adjusted accordingly. 

Internal to the chordae, there is no pressure directly applied and the tensions applied to a given junction between chordae must then sum to zero. 
In the trees of chordae, there is no notion of a continuum limit in the tension equations and we use the discrete equations alone. 
To make the finest level in the trees blend seamlessly into the valve mesh, we specify that there are $N$ total leaves in the trees.
For each point on the discretization of the valve ring, there is exactly one point on or near the free edge at which the trees attach. 
These points have the same index $j$ in the discretization and have equal values of $u$ in parameter space. 
Note that if $N$ doubles, then the number of leaves in the trees must double to maintain this relationship. 
Thus we double the number of leaves in each tree, adding another generation, when $N$ is doubled. 
This also doubles the total number of leaves across all trees.

The tension in the trees takes the form shown in equation $\ref{dec_tension_discretized}$, but without a continuum limit we remove the finite difference mesh parameters. 
Suppose that $\bb C$ denotes the current point in the tree, and $\bb C^{n}$ denotes the current neighbor. 
The force due to tension at $\bb C$ due to its connection with $\bb C^{n}$ is defined to be  
\begin{align}
k \left( 1 - \frac{1}{1 + |\bb C^{n} - \bb C|^{2} / c_{k}^{2} } \right)    \frac{ \bb C^{n}  -  \bb C }{ | \bb C^{n}  -  \bb C| } . 
\end{align}
The coefficients $k$ have units of force, because there is no continuum limit available to make it a force density of some kind. 
This implies that we require a rule to determine how $k$ and $c_{k}$ scale in the trees of chordae, the leaves of which scale with $N$ but the roots of which do not scale at all.

For each tree in the model, we wish the total force at the attachment to the leaflet to be approximately constant as the mesh is refined.  
Let $k_{0,L}$ be the tension coefficient at level zero, the finest level, of the tree that contains $L$ leaves, where $L = 2^{l}$ is a power of two. 
The coefficient for each generation in the tree is $k_{g,L}$, numbering from the leaves.  
Under a refinement by a factor of two, there are twice as many leaves. 
This means that each leaf takes half the force of the leaves in the coarse resolution tree, so the coefficient $k_{0,M}$ scales as 
\begin{align}
k_{0,2L} = \frac{k_{0,L}}{2} . 
\end{align}
Applying this relationship repeatedly, we find the relationship of $k_{0,L}$ to $k_{0,1}$ the coefficient in the degenerate case of a tree with one leaf. 
\begin{align}
k_{0,L} = \frac{k_{0,1}}{2^{l}} = \frac{k_{0,1}}{L} 
\end{align}
The constant $k_{0,1}$ is a tunable free parameter, representing the total of the coefficients in the leaf generation. 

Each tree has a single link that is connected to its root, and the tension coefficient in this link, $k_{root,L}$ should remain constant as the mesh is refined. 
This specifies the (degenerate, constant) scaling relationship 
\begin{align} 
k_{root, 2L} = k_{root,L} = k_{root} .
\end{align}
where $k_{root}$ is a tunable free parameter. 

Finally, to determine the coefficients throughout, we assume that the tension coefficient at each generation in the tree is a scalar multiple of the next finer generation. 
That is, 
\begin{align}
k_{g+1,L} = k_{m} k_{g,L}
\end{align}
for any generation index $g$ in the tree. 

The factor $k_{m}$, for k multiplier, scales these coefficients and is selected to preserve the prescribed values of $k_{0,1}$ and $k_{root}$. 
Since there are $l$ leaves in the tree, there are $l + 1$ total levels in the tree. 
\begin{align} 
k_{root} = k_{m}^{l} k_{0,L} = k_{m}^{l} \frac{ k_{0,1} }{2^{l}}  . 
\end{align}
This implies 
\begin{align} 
k_{m} = \left(  \frac{k_{root}}{ k_{0,1} / 2^{l} } \right)^{1/ l} = 2 \left(  \frac{k_{root}}{k_{0,1}} \right)^{1/ l }   . 
\end{align}

Similarly, the decreasing tension coefficient on the trees needs to have a scaling. 
As the mesh is refined, we expect the length of the links in the finest level in the tree to go down by half. 
Thus the the decreasing tension coefficient in the finest level should also scale with $M$. 
The root decreasing tension coefficient should not scale. 
In intermediate generations, we change the coefficients following a multiplicative rule. 
Note that this is the same relationship as the scaling of the tension coefficients.  

The level zero coefficient $c_{0,L}$ scales as 
\begin{align} 
c_{0,2L} = \frac{c_{0,L}}{2}
\end{align}
Let $c_{g,L}$ denote the level $g$ decreasing tension coefficient for a size $L$ tree, and $c_{m}$ denote the constant multiplicative scaling factor. 
Then 
\begin{align}
c_{g+1,L} = c_{m} c_{g,L} . 
\end{align}
Call the decreasing tension coefficient at the root as $c_{root}$, which as above is a constant.  
Then 
\begin{align} 
c_{root} = c_{m}^{l} c_{0,N} = c_{m}^{l} \frac{c_{0,1}}{2^{l}} 
\end{align}
so 
\begin{align}  
c_{mult} = \left(\frac{c_{root}}{c_{0,1} / 2^{l}} \right)^{1/l} = 2  \left(\frac{c_{root}}{c_{0,1} } \right)^{1/l} . 
\end{align} 

Speaking informally, turning $k_{0,1}$ up or turning $c_{0,1}$ down causes the leaves of the tree to clump together and become less spread out. 
Turning $k_{root}$ up pulls the entire structure towards the papillary muscles.

This gives a full set of scalings for all coefficients in the model, both tree and leaflet.

\section{Numerical solutions}

This nonlinear system is solved with Newton's method, with the addition of a number of modifications to increase robustness. 
Let $\Phi_{m}$ denote the vector of all unknowns on iteration $m$, $F(\Phi_{m})$ denote the entire system of nonlinear difference equations and $J_{m}$ the Jacobian of $F(\Phi_{m})$. 
The iteration for Newton's method is given by 
\begin{align} 
\Phi_{m + 1} = \Phi_{m} - J(\Phi_{m})^{-1} F(\Phi_{m}) 
\end{align} 
where we solve the linear system rather than explicitly inverting $J$.

This implies that we need the Jacobian of the entire nonlinear system of difference equations. 
Any experiments involving various algorithms to approximate the Jacobian and use various optimization algorithms failed completely. 
Thus, we compute the entire Jacobian analytically and implement a program to evaluate the analytic form numerically on a given configuration. 

\begin{figure}[H]
\centering 
\includegraphics[scale=.95]{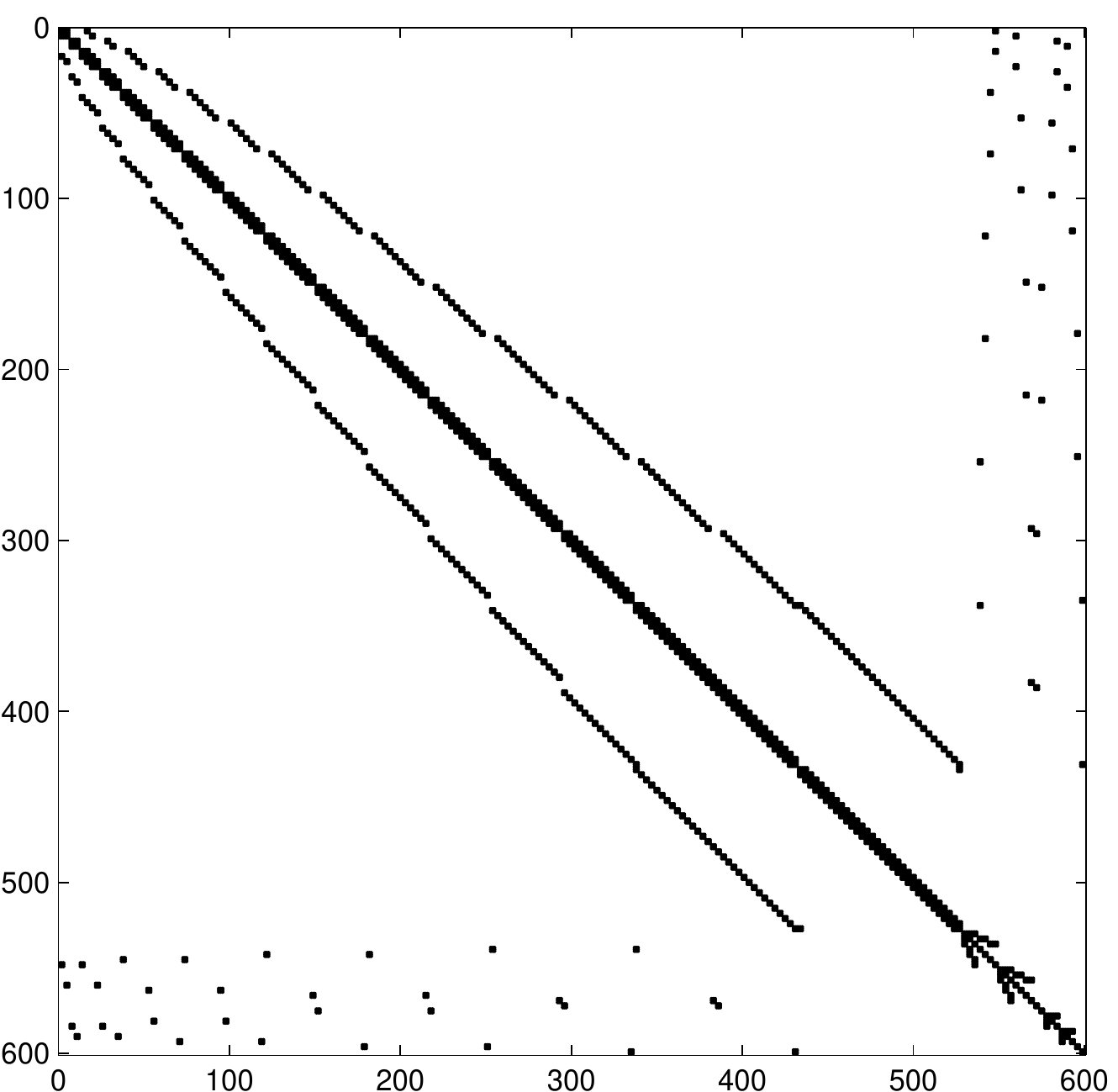}
\caption{Nonzero pattern of Jacobian. Low resolution, $N = 32$, is used to better show structure.}
\label{jacobian_nnz}
\end{figure}

The Jacobian, shown in figure \ref{jacobian_nnz}, has a large block in the upper left that corresponds to the leaflets. 
This has nonzero structure similar to the two-dimensional Laplacian matrix, but scalars are now three by three blocks. 
The band away from the diagonal is now a variable distance from the diagonal, owing to the variable number of points in each row of the mesh. 
The lower right contains blocks internal to the various trees of chordae, and the two very sparse remaining blocks correspond to insertions of the tree into the leaflet. 
The nonzero pattern of this matrix is symmetric, but the matrix is not, owing to blocks corresponding to pressure that are skew symmetric.

The Jacobian is built by taking Jacobians of individual vector equations. 
All types of derivatives, gradients and Jacobians necessary are computed by hand. 
Data structures determine which nodes are internal to the leaflet and which are boundary conditions, and manage constants and coefficients through all included structures. 
They specify which nodes are internal to the chordae, and how they connect to the leaflet. 
Additional information specifies indices, numbered globally, including or excluding boundary conditions. 
The Jacobian program reads these data to find which types of derivatives need to be computed for each term, calculates the associated three by three blocks, and places them into the appropriate location in the overall Jacobian.

Pressure blocks take the form of matrices which apply the cross product up to signs. 
Let 
\begin{align} 
C(\bb X) = 
\begin{bmatrix}
 0 &  -X^{(3)} &  X^{(2)} \\  
   X^{(3)}  &   0 & -X^{(1)}  \\
  -X^{(2)}  & X^{(1)}  &  0
 \end{bmatrix}
\end{align} 
Note that $C(\bb X) \bb Y = \bb X \cross \bb Y$, so $C(\bb X)$ is the matrix representation of the linear operator whose action is taking the cross with a fixed vector $\bb X$. 
The pressure Jacobian with respect to $\bb X^{j+1,k}$ is 
\begin{align}
J_{\bb X^{j+1,k}} &( (\bb X^{j+1,k} - \bb X^{j-1,k} )  \cross  ( \bb X^{j,k+1} - \bb X^{j,k-1} ) )  
&= 
- C(\bb X^{j,k+1} - \bb X^{j,k-1})
\end{align}
Other pressure terms take the same form up to signs.

Tension blocks are computed as follows. 
We refer to the current term as $\bb X$ and its neighbor as $\bb X^{n}$ and omit indices for clarity. 
For functions
\begin{align}
f:\R^{3} \to \R , \quad G : \R^{3} \to \R^{3} , 
\end{align}
we have the product rule 
\begin{align}
J_{\bb X}(  f(\bb X) \bb G (\bb X)  ) =    \grad_{\bb X} f  \bb G^{T} + f J_{\bb X}(\bb G) . 
\end{align}
The Jacobian for a tangent term is 
\begin{align}
J_{\bb X}\left(  \frac{\bb X^{n} - \bb X}{|  \bb X^{n} - \bb X | } \right) 
=  \frac{ (\bb X^{n} - \bb X) (\bb X^{n} - \bb X)^{T} }{ |  \bb X^{n} - \bb X |^{3} }  -  \frac{1}{ |  \bb X^{n} - \bb X |} I   , 
\end{align}
where $I$ is the three by three identity matrix. 
The Jacobian for a tension multiplied by tangent term is then 
\begin{align} 
J_{X} &\left ( \left( 1 - \frac{1}{1 + \dfrac{ |\bb X^{n} - \bb X|^{2} }{ c^{2} (\du)^{2} }} \right)  \frac{\bb X^{n} - \bb X}{ |  \bb X^{n} - \bb X |} \right) \\ 
= &\left( \frac{   \dfrac{ 2 |\bb X^{n} - \bb X| }{ c^{2} (\du)^{2} } }{ \left( 1 + \dfrac{ |\bb X^{n} - \bb X|^{2} }{ c^{2} (\du)^{2} } \right)^{2} } \right) \frac{(\bb X^{n} - \bb X)(\bb X^{n} - \bb X)^{T}}{ |  \bb X^{n} - \bb X |^{2}} \nonumber  \\ 
+ &\left( 1 - \frac{1}{1 + \dfrac{ |\bb X^{n} - \bb X|^{2} }{ c^{2} (\du)^{2} }} \right)  \left(  \frac{ (\bb X^{n} - \bb X) (\bb X^{n} - \bb X)^{T} }{ |  \bb X^{n} - \bb X |^{3} }  -  \frac{1}{ |  \bb X^{n} - \bb X |} I  \right) \nonumber 
\end{align}
This gives all the forms of Jacobians that we need to build the global Jacobian. 

\begin{figure}[H]
\centering 
\includegraphics[scale=.75]{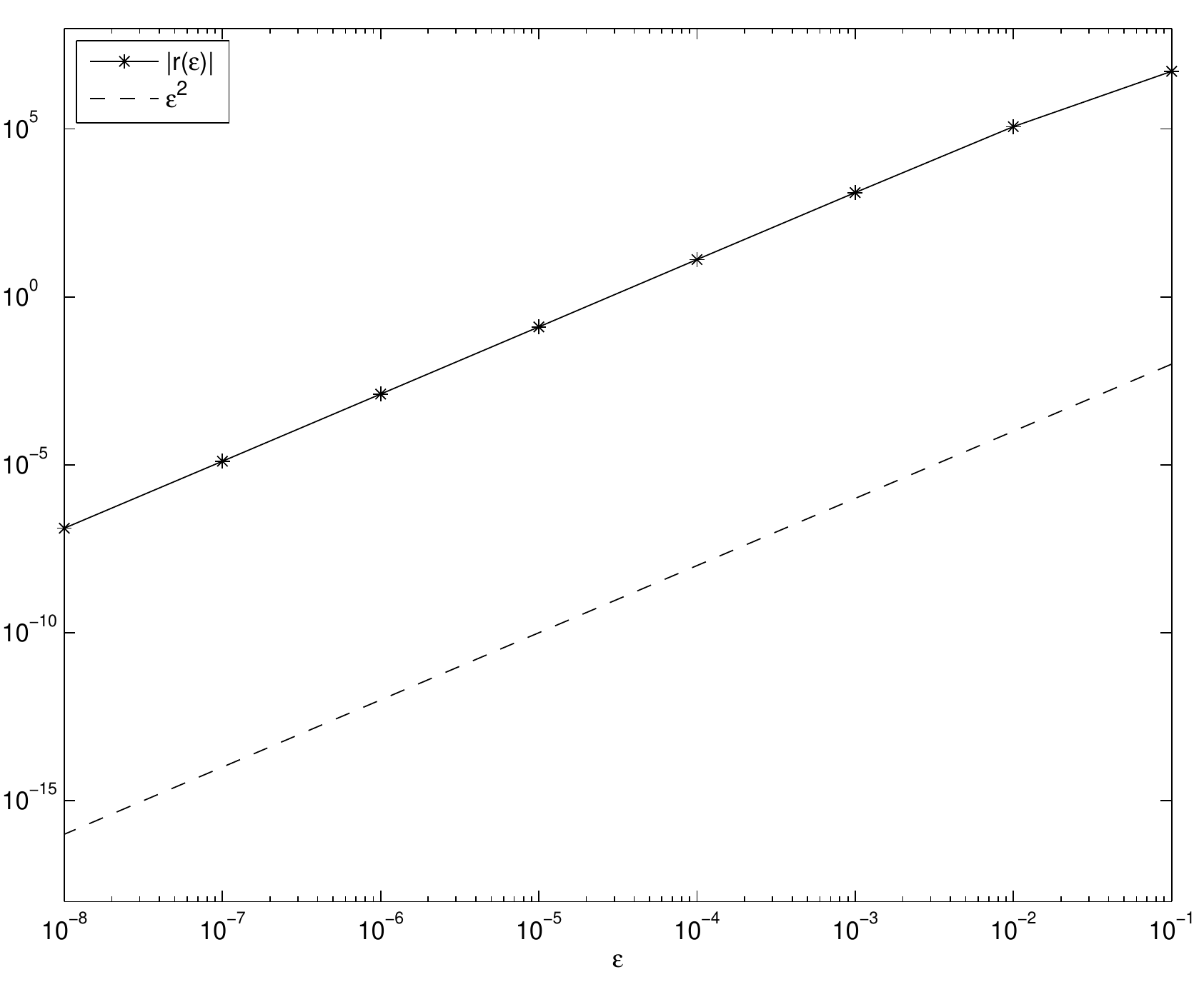}
\caption{Second order decrease in Jacobian comparison.} 
\label{jacobian_conv}
\end{figure}

To check the Jacobian, we use a test based on a Taylor expansion to show that the computed Jacobian is indeed the derivative of the system of difference equations. 
The Jacobian appears in the Taylor expansion of $F$ as 
\begin{align} 
F(\Phi + \eps \Psi) = F(\Phi) + \eps J(\Phi) \Psi  + O(\eps^{2}) 
\end{align} 
To check, let 
\begin{align} 
r(\eps) = F(\Phi + \eps \Psi) - F(\Phi) - \eps J(\Phi) \Psi 
\end{align} 
and compute $ | r(\eps) |$ for a variety of $\eps$. 
When plotted on a log-log plot, this should form a line parallel to $\eps^{2}$, and when computed in a table, the exponent of $|r(\eps)|$ should decrease by $10^{-2}$ for every $10^{-1}$ decrease in $\eps$. 
Note that this tests the relationship between $F$ and $J$, that $J$ is the Jacobian of $F$, rather than anything about $F$ itself. 
Results are shown in table \ref{jac_test} and figure \ref{jacobian_conv}. 

\begin{table}[H]
\centering 
\begin{tabular}{ c  |  c |  }
$\eps$	 & $| F(\Phi + \eps \Psi) - F(\Phi) - \eps J(\Phi) \Psi |$ \\ 
\hline 
1.0e-01	 & 5.194941e+06 \\ 
 \hline 
1.0e-02	 & 1.167860e+05 \\ 
 \hline 
1.0e-03	 & 1.283970e+03 \\ 
 \hline 
1.0e-04	 & 1.286958e+01 \\ 
 \hline 
1.0e-05	 & 1.287329e-01 \\ 
 \hline 
1.0e-06	 & 1.287367e-03 \\ 
 \hline 
1.0e-07	 & 1.287377e-05 \\ 
 \hline 
1.0e-08	 & 1.294217e-07 \\ 
 \hline 
\end{tabular}
\caption{Second order decrease in Jacobian comparison.}
\label{jac_test}
\end{table}

To solve the system, we use a direct method provided by Matlab's backslash operator. 
The matrix is stored in a sparse data structure, and the operator checks the matrix for sparsity and symmetry, and selects an algorithm automatically. 
Here, Matlab selects a solver provided by UMFPACK, which computes an LU factorization of the matrix using pivoting strategies to maintain sparsity and numerical stability \cite{UMFPACK}. 
At full resolution, profiling reveals that solving the linear systems takes under three percent of the total iteration time. 
This is excellent considering that it is the most numerically intensive step of the entire solution. 

Alternatives such as Krylov methods were ineffective on this problem. 
In the event that some fibers become very close together condition numbers can skyrocket and convergence is slow or fails. 
Given the speed of the direct solves, we abandoned these quickly and take no more effort to improve on this aspect.  

Simple optimizations to control flow improved performance significantly. 
Matlab uses a \emph{lazy copy} or \emph{copy on write} memory management strategy \cite{matlab_memory}. 
This means that when an array is modified within a function, the array is copied to maintain the contents of the original. 
As a result, na\"ive attempts to build the Jacobian were slow. 
Functions that computed blocks and placed three by three blocks led to many unnecessary copies. 
Profiling revealed that assignment lines were a dominant cost in the entire solve, much slower than even solving the linear systems.  
Thus, some utility functions were arranged to be nested, which allows them to access the workspace of the calling function and modify the single copy of an array. 
This is generally dangerous practice, because it follows strange and specific patterns of variable scoping that may be confusing to the user. 
The speedup was significant, so this slight complication is worthwhile here.

Following these changes, large portions of time go into utility routines that manage the leaflet and tree data structures and their connections. 
These are spread across simple functions that manage tree and leaflet connectivity and indexing. 
These are not numerically intensive, which suggests there is not too much optimization left.
Perhaps more could be done by placing repetitive code and being more clever with nested functions. 
However, this would make the code that much more complex. 
Additionally, the solves are essentially a preprocessing step to the later solves with fluid, and run in minutes of wallclock time at the maximum resolution we have used here.

There are two enhancements to the standard Newton's method algorithm that proved essential. 
The first is Newton's method with line search. 
We introduce a scalar parameter $s \in (0,1]$ and replace the standard Newton's update with 
\begin{align} 
\Phi_{m + 1} = \Phi_{m} - s_{m} J(\Phi_{m})^{-1} F(\Phi_{m}) 
\label{newtons_line_search}
\end{align} 
There are cases in which versions of this algorithm are guaranteed to be globally convergent if solved in exact arithmetic. 

At each step the value of $s$ is initialized to one. 
If the norm of $F$ decreases with this value of $s$, then the guess is accepted. 
Otherwise the value of $s$ is decreased by half repeatedly until the norm decreases. 
This gives extreme improvements in the overall robustness and convergence of the Newton solves. 
Pseudocode is shown in algorithm \ref{newton_line_search}. 
In the implementation, the solution of the linear system, $J^{-1}(\Phi)F(\Phi)$, is stored at line 5 and does not need to be recomputed at line 14 in the inner loop. 
The difference equations, however, must be reevaluated at line 15 every inner loop iteration.

\begin{algorithm}[H]                                          
\caption{Newton's method with line search}     
\label{newton_line_search}
\begin{algorithmic}[1] 

	\STATE err = $ | F(\Phi) | $
			
	\WHILE{err $>$ tol}
			
		\STATE err\_prev = err 
			
		\STATE s = 1 		
		
		\STATE $\Phi_{tmp} = \Phi - s J^{-1}(\Phi)F(\Phi)$
		
		\STATE err = $ | F(\Phi) | $
		
		\WHILE{true}
		
			\IF {err $<$ err\_prev}
				\STATE // Decrease has occurred. Continue to next Newton step.
				\STATE $\Phi = \Phi_{tmp}$
				\STATE {\bf break} 
			\ENDIF 
				
			\STATE s = s/2
			
			\STATE $\Phi_{tmp} = \Phi - s J^{-1}(\Phi)F(\Phi)$

			\STATE err = $ | F(\Phi) | $
				
		\ENDWHILE
		 
	\ENDWHILE
		
\end{algorithmic}
\end{algorithm}

An alternative is to consider the one-dimensional function 
\begin{align}
s \mapsto | \Phi_{m} - s J(\Phi_{m})^{-1} F(\Phi_{m}) |
\end{align} 
and call a black box optimization routine on the function to minimize with respect to $s$. 
This was less robust, perhaps because the built-in, one-dimensional optimization routines do not use exact derivatives, or some subtle interaction with the convergence analysis on regular Newton's method. 
We do not use this technique as a result.

The second is the method of continuation. 
If there are values of a parameter for which the convergence is more effective or reliable, solve with that parameter set more ``nicely'' first. 
Then, use the converged solution as an initial guess and gradually turn the parameter towards the goal. 
Here, that parameter is pressure. 
With the decreasing tension model, lower pressures converge more robustly. 
We discovered this by playing with the code and trying zero pressure solves, which are generally the easiest of all.

We take the following adaptive algorithm to turn up the pressure. 
First, try to solve at a specified goal pressure.
If that fails, try to solve at a specified easiest pressure. 
Create an increment, add to the pressure and try to solve. 
Every failure, reduce the size of the increment by four (which is empirically better than two) and continue the process. 
Repeat until we have a successful solve at the goal pressure. 
This is shown in pseudocode in algorithm \ref{continuation}.

\begin{algorithm}[H]                         
\caption{Adaptive method of continuation on pressure} 
\label{continuation}
\begin{algorithmic}[1] 

	\STATE Input $p_{goal}$, $p_{easy}$. 
		
	\STATE Run Newton's method with line search with $p = p_{goal}$. 
	
	\IF {solve passed}
		\STATE Return configuration. 
	\ENDIF
	
	\STATE Run Newton's method with line search with $p = p_{easy}$. 
	
	\IF {solve failed}
		\STATE Return error. No continuation possible. Easiest parameter failed. 
	\ENDIF
	
	\STATE Set $p_{current} = p_{goal}$, $p_{increment} = p_{goal} - p_{easy}$, $p_{passed} = p_{easy}$. 
	
	\WHILE{\TRUE}
		\STATE Run Newton's method with line search with $p = p_{current}$.
		
		\IF{solve passed}
		
			\IF{$p_{current} == p_{goal}$}
				\STATE Return configuration. 
			\ENDIF
		
			\STATE Set $p_{current} = p_{current} + p_{increment}$. 
		
		\ELSE
		
			\STATE Set $p_{increment} = p_{increment} / 4$, $ p_{current} = p_{passed} + p_{increment}$. 
		
		\ENDIF
		 
	\ENDWHILE
		
\end{algorithmic}
\end{algorithm}


\section{Results}

\begin{figure}[H]
\centering
\includegraphics[scale=.74]{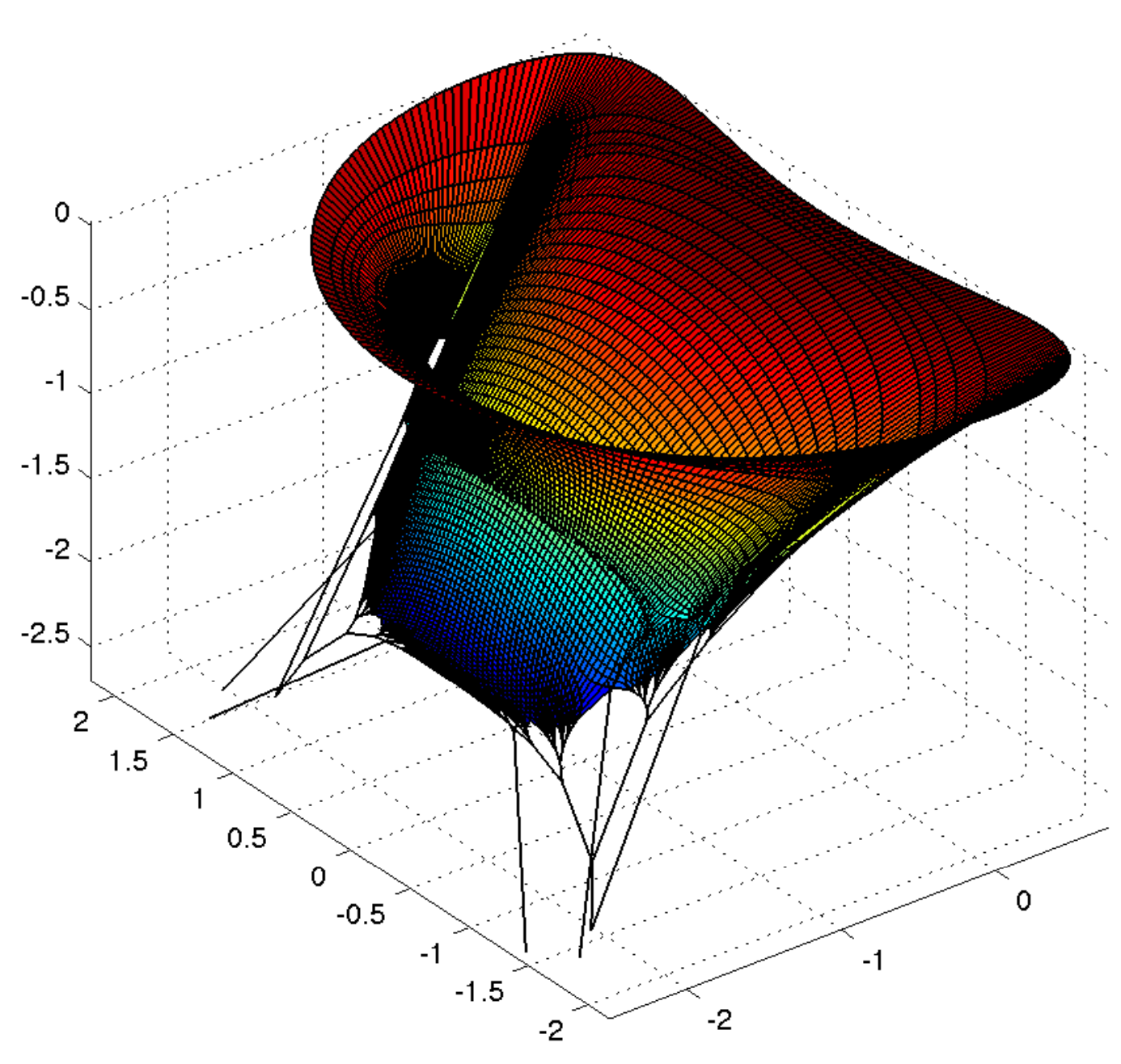} 
\includegraphics[scale=.74]{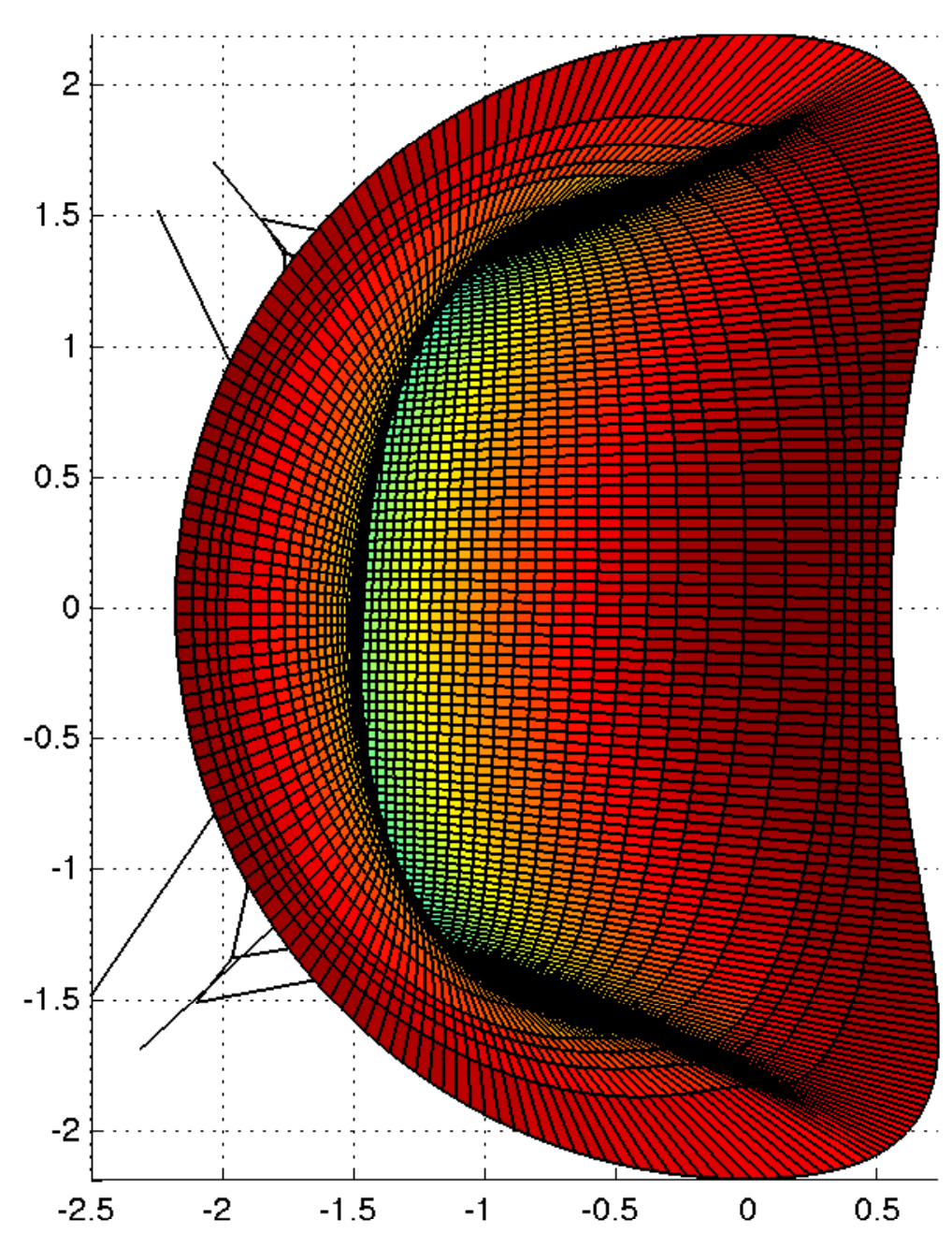}
\caption{Closed geometry of model valve. $N=256$. }
\label{static_valves}
\end{figure}

Finally, we have results. 
The valve effectively forms a closed shape. 
Parameters are selected such that the configuration achieves contact, or some degree of self-intersection, near the free edge. 
This is so that when the valve is placed in fluid, there is a little bit of extra model tissue to let the leaflets coapt, or colloquially ``mash together,'' on an large area. 
Figure \ref{static_valves} shows two views on a closed valve. 
The final resolution we use in fluid uses $N = 512$, but here we present results with $N = 256$. 
This is for visual clarity, as at the finest resolution there are so many fibers they are difficult to distinguish and the valves appear solid black.

Note that the equilibrium equations, \ref{equilbrium_eqn_discrete}, do not include any notion of contact. 
If the solution is such that that the leaflets pass through each other, then they simply do. 
We found that tuning for a solution with some contact in a region of the free edge created more robust valve models.
Suppose that we specify a target pressure, and the leaflets are spaced slightly apart. 
Then when placed in Chapter \ref{Fluid-Structure_Interaction}, if the pressure is even slightly less or the configuration is slightly different, the valve may not achieve good coaptation. 
However, a small amount of contact in the solution acts to reduce the precision necessary in lining up all parts of the free edge. 
When placed in fluid, the extra model tissue will be pressed together, and hopefully fill any small holes that may arise. 
Thus, solving for a closed configuration that has some contact between the leaflets gives a more forgiving structure when placed in fluid. 
Before placing in fluid, this geometry will be relaxed so there is no contact, as the fluid-structure interaction techniques require a structure with no self-intersection for the valve to open correctly. 
This process is described in Section \ref{general_model}.

Many details emerge from the solves. 
Here we look at a detail at the free edge where chordae insert into the leaflet. 
The free edge, at which in the mesh has a staircase shape, forms what is approximately a smooth curve. 
The chordae form nested arches, inheriting a recursive structure from the trees.

\begin{figure}[H]
\centering
\includegraphics[scale=.65]{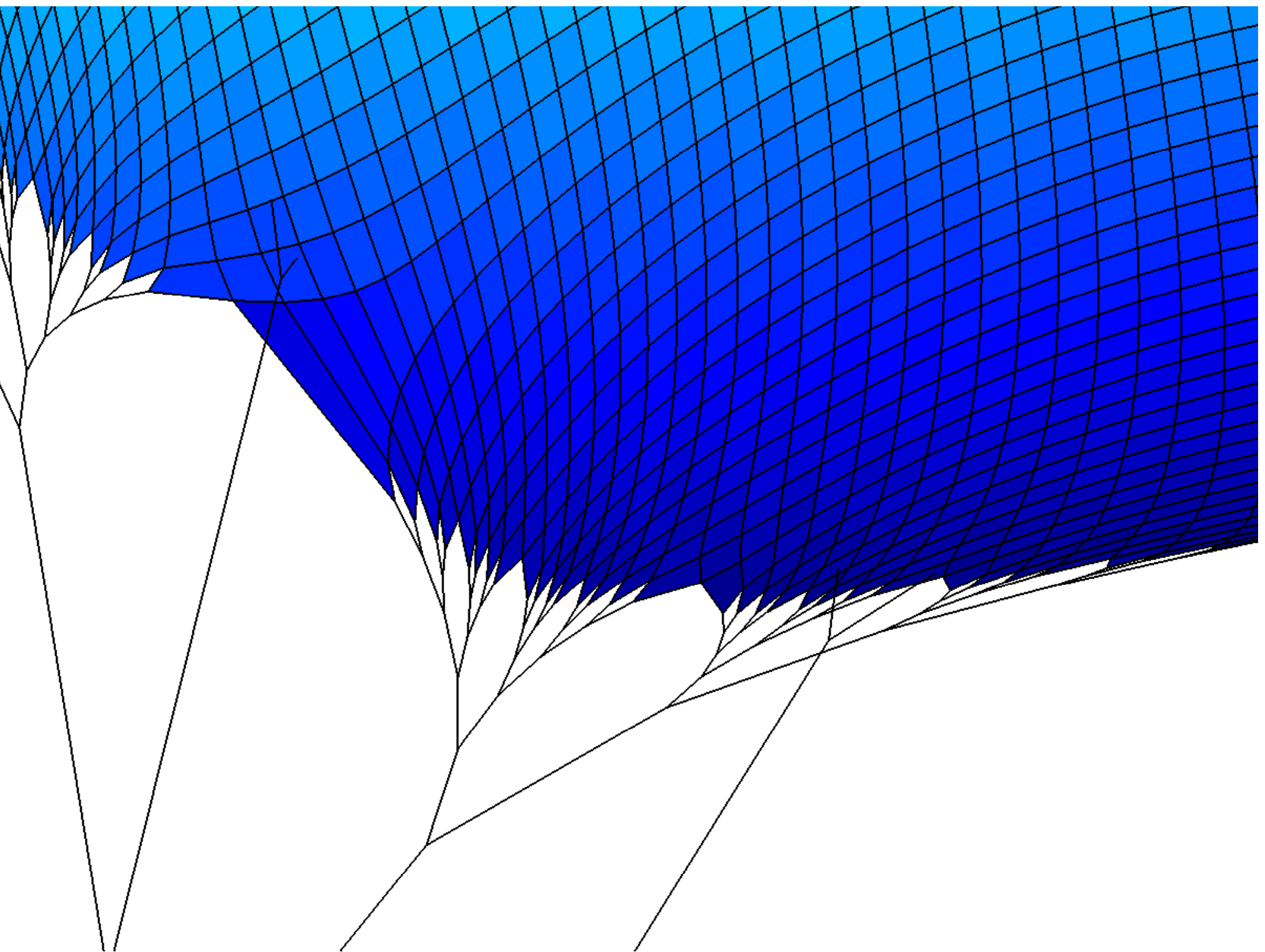} 
\caption{Detail on trees in closed configuration. $N=256$.}
\label{tree_detail}
\end{figure}

\begin{figure}[H]
\centering
\includegraphics[scale=.73]{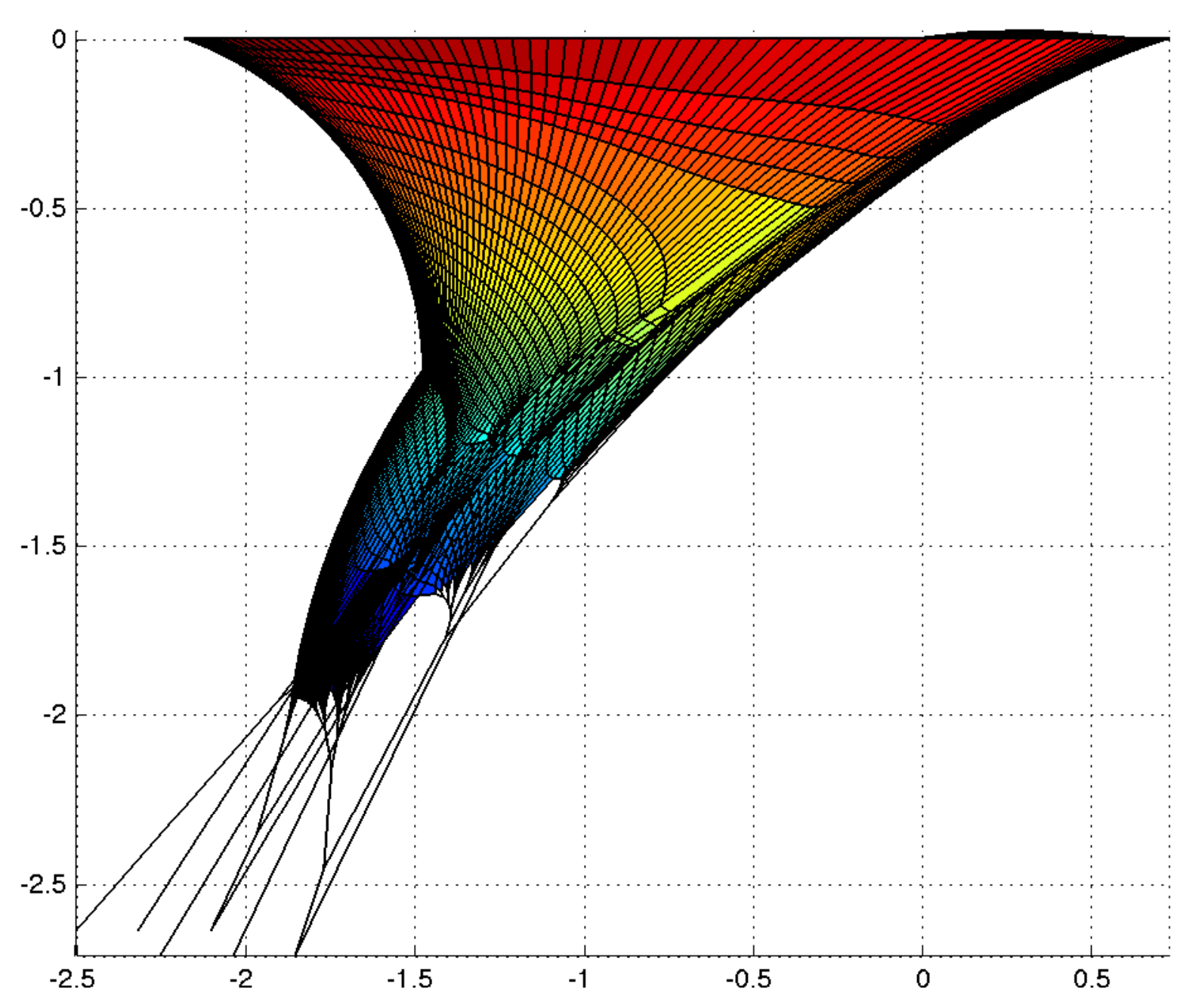} 
\includegraphics[scale=.73]{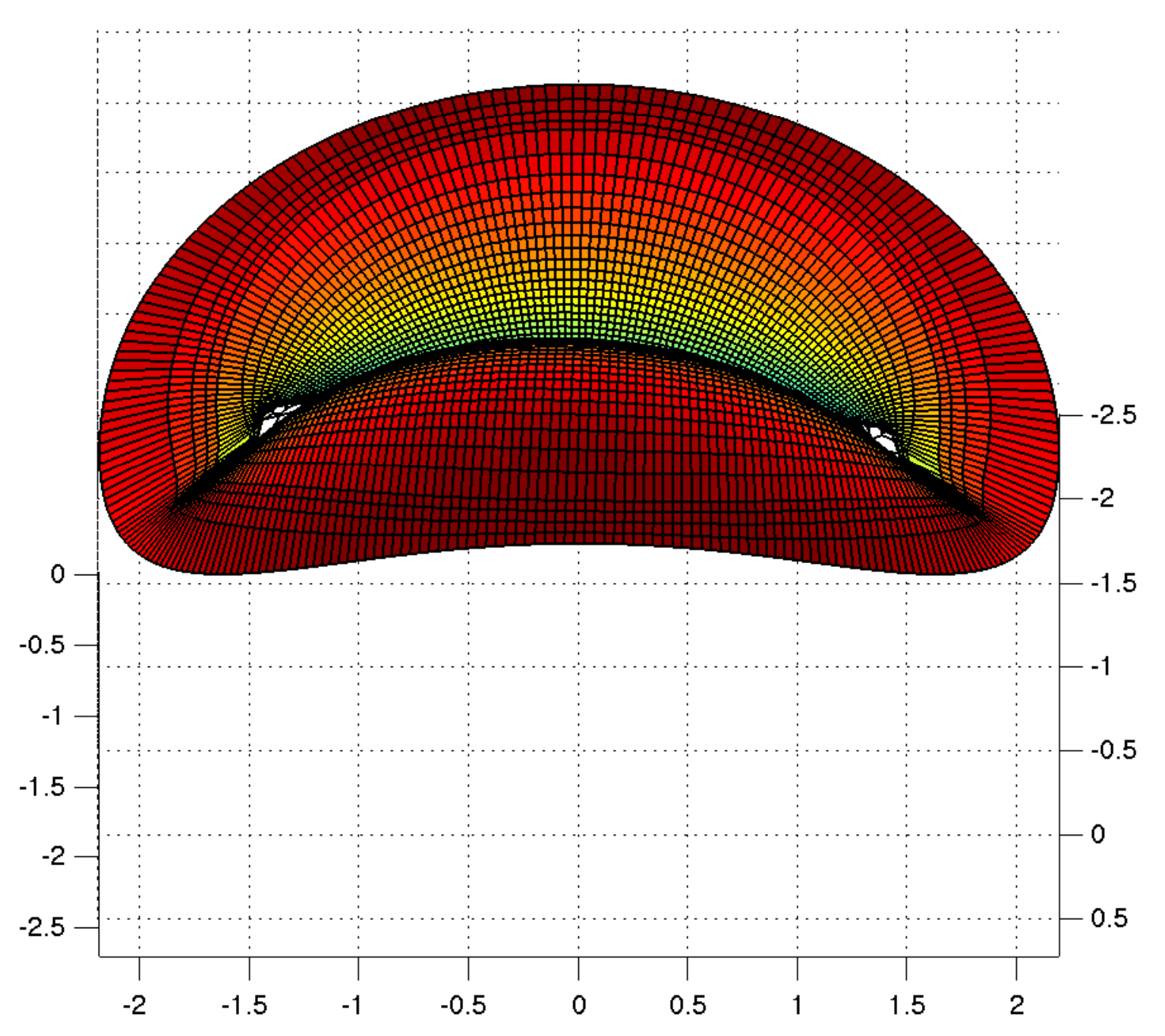}
\caption{Closed geometry of model valve, views to emphasize closure. $N=256$.}
\label{static_valves_closure}
\end{figure}

Viewing the valve from the side and top at a slightly different angle emphasizes the closure of the configuration, as shown in figure \ref{static_valves_closure}.

To verify that the solution converges as the mesh is refined, we check that the differences between solutions go to zero as $N$ is refined, and check the empirical order of accuracy of the solution. 
There is a slight quirk in doing this, since the parameterization has both a physical and a numerical interpretation. 
We do not wish there to be a fiber centered on either leaflet, so select an even number of points on the ring. 
That is, there is no radial fiber in the center of the valve. 
This implies that on each leaflet there are an even number of points at the valve ring, and thus an odd number of intervals. 
Points are placed corresponding to $0 ... (N-2)/(N-1)$, and as $N$ is refined by factors of two, these points do not correspond exactly. 
Thus, when comparing valves of varying resolution, we use linear interpolation to evaluate the finer mesh at corresponding points on the coarse valve. 
Linear interpolation introduces second order errors, see theorem 4.3.3 in \cite{dahlquist2003numerical}. 
We expect less than second order convergence here because of the free boundary, so this is not a problem. 

We also experimented with a piecewise constant interpolant. 
This corresponds to the value at one point of the fine solution to compare to the coarse solution, even though they do not precisely align in parameter space. 
This gave universally higher differences between solutions of various resolutions, as expected, and we do not use this further. 

There is no such problem in $v$. 
One row is used for the boundary condition at the valve ring. 
The number of rings below the boundary condition is a power of two, as is the number of points on the leaflet below the rings. 
This implies that the total number of points in the $v$ direction is odd, so there are an even number of intervals, and the points correspond exactly. 

Let $\bb X_{N}$ denote the solution for the valve with $N$ mesh points on the valve ring, and $\bb X_{N, interp}$ denote the solution interpolated onto the $N/2$ mesh. 
To check that the solution is converging, we compute 
\begin{align}
| \bb X_{N} - \bb X_{2N, interp} |  
\end{align}
in the discretized $L^{1}, L^{2}$ and supremum norms. 
Results from are shown in table 
\begin{table}[H]
\centering 
\begin{tabular}{ c  |  c |  c  |  c |  }
$N,2N$	 & $L^{1}$ & $L^{2}$ & $L^{\infty}$   \\ 
\hline
64,128 & 5.65e-03 & 1.26e-02  &  1.06e-01    \\  
 \hline 
 128,256 & 4.38e-03 & 9.74e-03  &  7.83e-02    \\  
 \hline 
 256,512 & 3.51e-03 & 7.98e-03  &  6.34e-02    \\  
 \hline 
 512,1024 & 3.27e-03 & 8.07e-03  &  6.28e-02    \\  
 \hline 
 \end{tabular}
\caption{Norms of differences in solutions for various $N$. All norms decrease monotonically with $N$ (except for the 2 norm in the final step) which suggests the solution is converging.}
\label{static_differences_check}
\end{table}
To check the order of accuracy, we compute 
\begin{align}
\frac{ | \bb X_{N} - \bb X_{2N, interp} |  }{ | \bb X_{2N} - \bb X_{4N, interp} | } = 2^{p},
\end{align}
which defines $p$, the empirical order of accuracy in the solution. 
To check the convergence tester program, we also evaluate the convergence at the valve ring. 
This should be second order, and testing with Richardson extrapolation confirms that it is so. 
Results from on the valve for a variety of $N$ are shown in table \ref{static_order_check}. 
\begin{table}[H]
\centering 
\begin{tabular}{ c  |  c |  c  |  c |  }
$N,2N,4N$	 & $2^{p},\;  L^{1}  \text{ norm}$ & $2^{p},\;  L^{2}  \text{ norm}$ & $2^{p},\;  L^{\infty} \text{ norm}$   \\ 
\hline 
64,128,256 & 1.29 & 1.30  &  1.35    \\  
 \hline 
 128,256,512 & 1.25 & 1.22  &  1.24    \\  
 \hline 
 256,512,1024 & 1.08 & 0.99  &  1.01    \\ 
 \hline
 \end{tabular}
\caption{Empirical check on order of accuracy. Values of $2^{p}$ are shown.}
\label{static_order_check}
\end{table}
The estimated order of accuracy is approximately $\log_{2}(1.25) \approx 0.3$ for $4N=256$ and $512$ in all three norms. 
This is a very limited type of convergence, not even making first order. 
We believe this is because of differences at the free boundary. 
For $4N=1024$, the order of accuracy is even lower on the $L^{1}, L^{\infty}$ norms, and no convergence at all on the $L^{2}$ norm.

\begin{figure}[H]
\centering
\includegraphics[width=\textwidth]{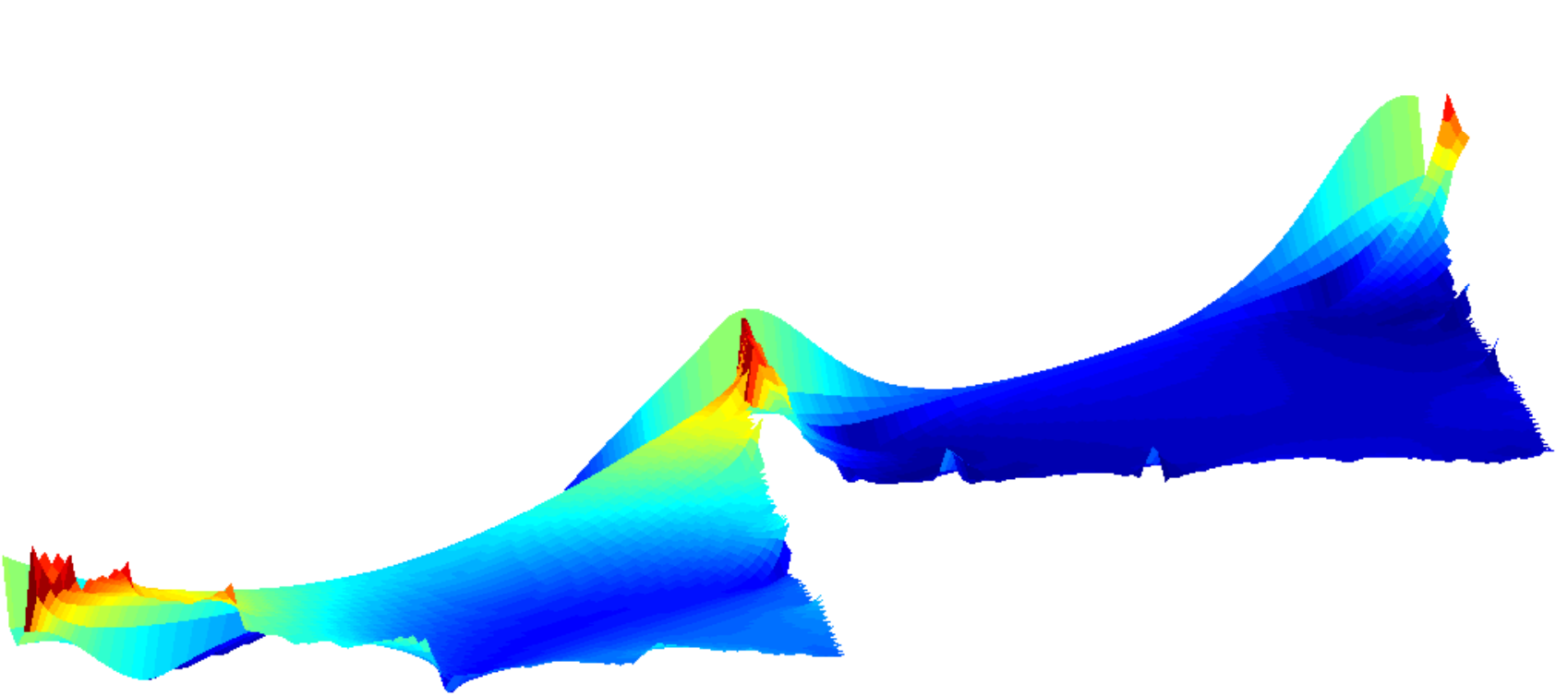} 
\includegraphics[width=\textwidth]{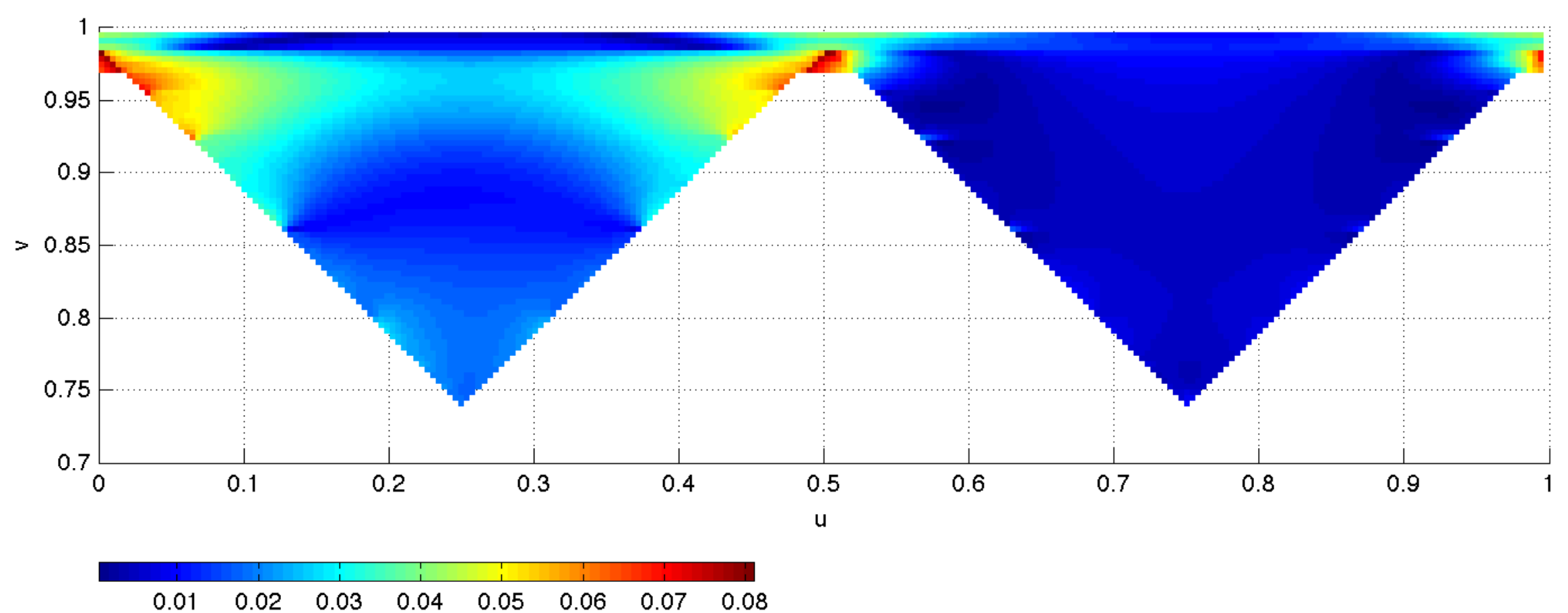} 
\caption{Component-wise two norm difference of $N=256$ and $N=512$ leaflet mesh. The first panel shows this as a surface, in which height and color correspond to magnitude of the difference. The second shows a pseudocolor plot of the same image, using the same color scale.}
\label{static_convergence}
\end{figure}

Figure \ref{static_convergence} shows the component-wise $L^{2}$ difference between the leaflet solution at $N=256$ and $N=512$. 
While the solutions do appear to converge, differences in the solution at the free boundary appear to propagate into the surrounding regions. 
Certain locations, such as the edge connector regions at the commissures, create localized errors. 
This is notable, since this is the only location at which the free edge and free boundary are not directly attached to the chordae.

\begin{figure}[H]
\centering
\includegraphics[width=.64\textwidth]{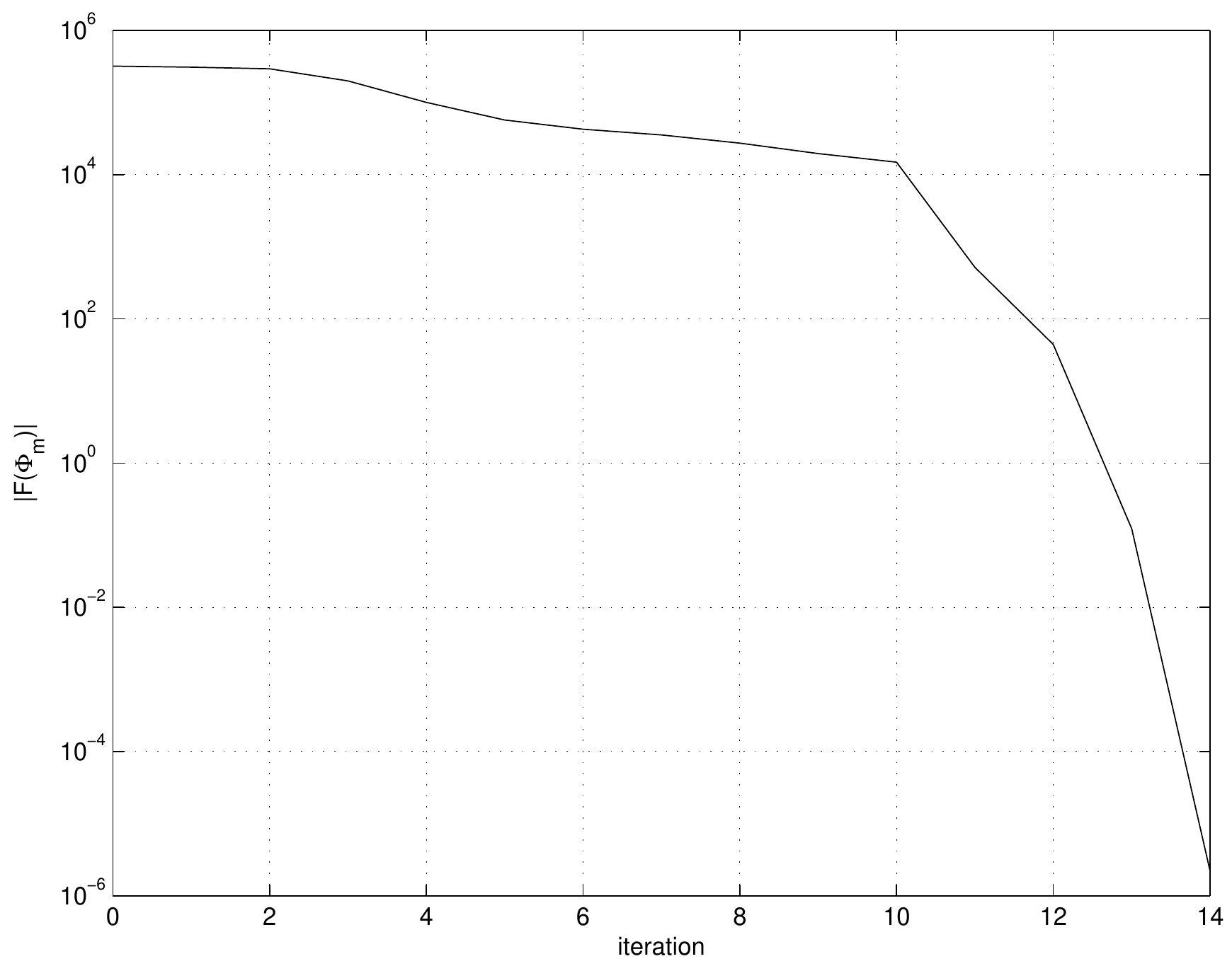} 
\caption{Error in Newton's method.}
\label{newton_error}
\end{figure}

We also check the convergence of Newton's method to the desired solution. 
In the first few iterations, the line search parameter $s_{m}$ in equation \ref{newtons_line_search}, is some value less than one. 
The algorithm avoids a bad guess and ensures monotonic decrease of the error. 
After a number of iterations, the solution appears to be in the region surrounding a zero of the difference equations. 
The algorithm finds that the norm of the error decreases with $s_{m} = 1$. 
Suddenly, the convergence takes on the squaring of error, that is, approximate doubling of the number of correct digits, or ``quadratic convergence'' that characterizes Newton's method in the neighborhood of a zero. 
This is shown in table \ref{newton_converge_table}, which also shows $s_{m}$ and the time taken for each iteration. 
Iterations with $s_{m} = 1$ are generally faster, since the difference equations are evaluated fewer times. 
Over the course of fourteen iterations, the error improves eleven orders of magnitude, five orders of magnitude on the final iteration alone. 
The log-scale error is shown in figure \ref{newton_error}, which shows the characteristic ``waterfall'' of convergence that is more rapid than geometric.

\begin{table}[H]
\centering 
\begin{tabular}{ c  |  c |  c  |  c |  }
Iteration	 &  $ |F(\Phi_{m})| $  & $s_{m}$ & Time elapsed (s)   \\ 
\hline 
initial & 3.200829e+05 & - & - \\ 
\hline 
1&   3.090390e+05& 1/32 &   33.65    \\
\hline  
2&   2.940966e+05&  1/8 &   38.02  \\ 
\hline 
3&   1.995802e+05&  1/4 &  31.84  \\ 
\hline 
4&   1.007015e+05& 1/2  &  25.79 \\ 
\hline 
5&   5.718715e+04&  1/2 &  26.97  \\ 
\hline 
6&   4.270672e+04&  1/2 &  21.32  \\ 
\hline 
7&   3.540817e+04&  1/2  & 28.38  \\ 
\hline 
8&   2.740952e+04&  1/2  &  20.23    \\ 
\hline 
9&   1.953890e+04&  1/2  &  20.29  \\ 
\hline 
10&   1.488641e+04& 1  &  17.21   \\  
\hline 
11&   5.159897e+02&  1 &  23.56    \\ 
\hline 
12&   4.422052e+01&  1 &  15.76    \\ 
\hline 
13&   1.229451e-01&  1  & 15.75 	     \\ 
\hline 
14&   2.166187e-06&  1 &  18.60 	    \\ 
\hline 
\end{tabular}
\caption{Convergence on Newton's method. $N = 256$.}
\label{newton_converge_table}
\end{table}

\begin{figure}[H]
\centering
$ \begin{array}{ccc}
\includegraphics[width=.3\textwidth]{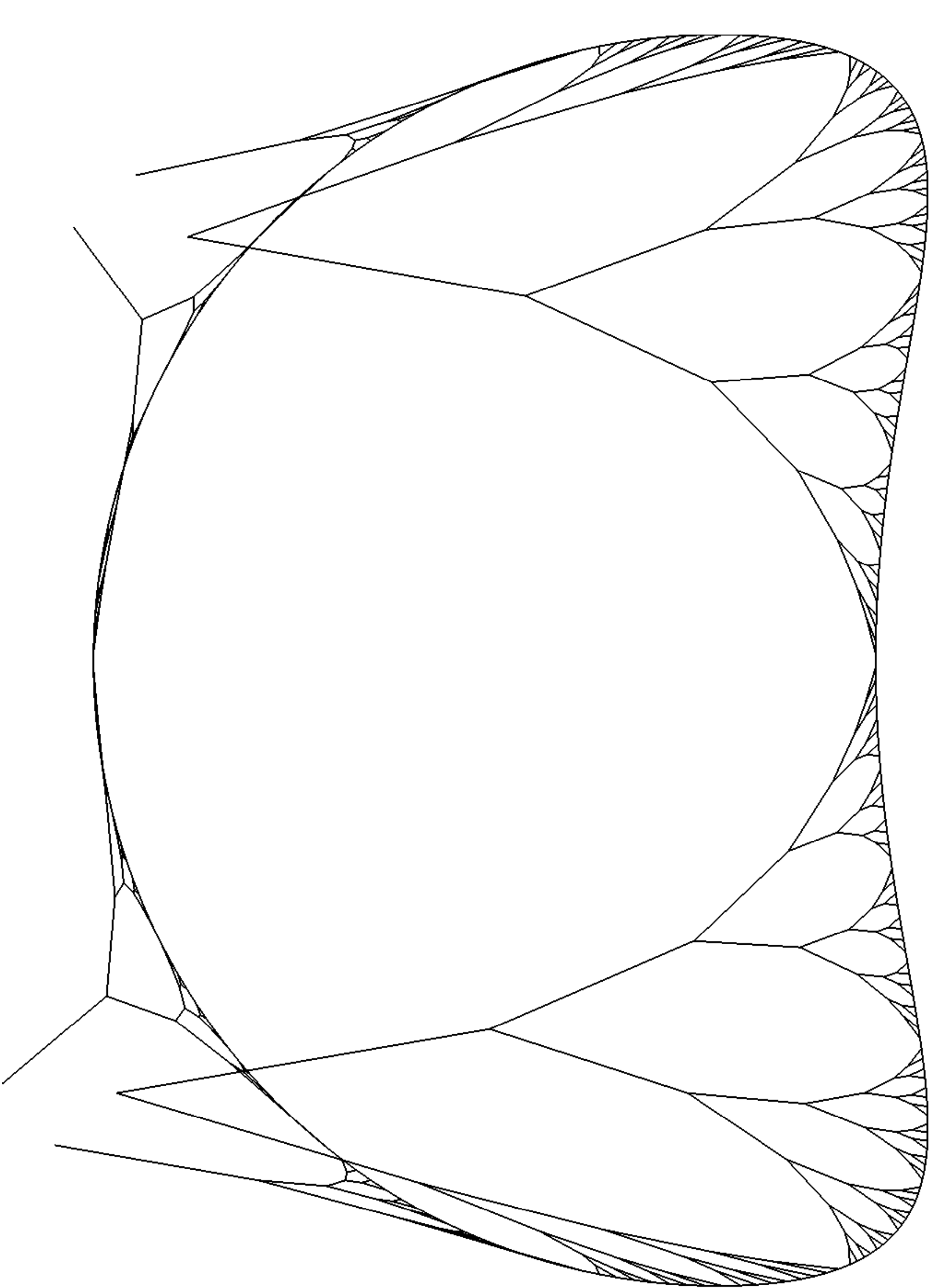} & 
\includegraphics[width=.3\textwidth]{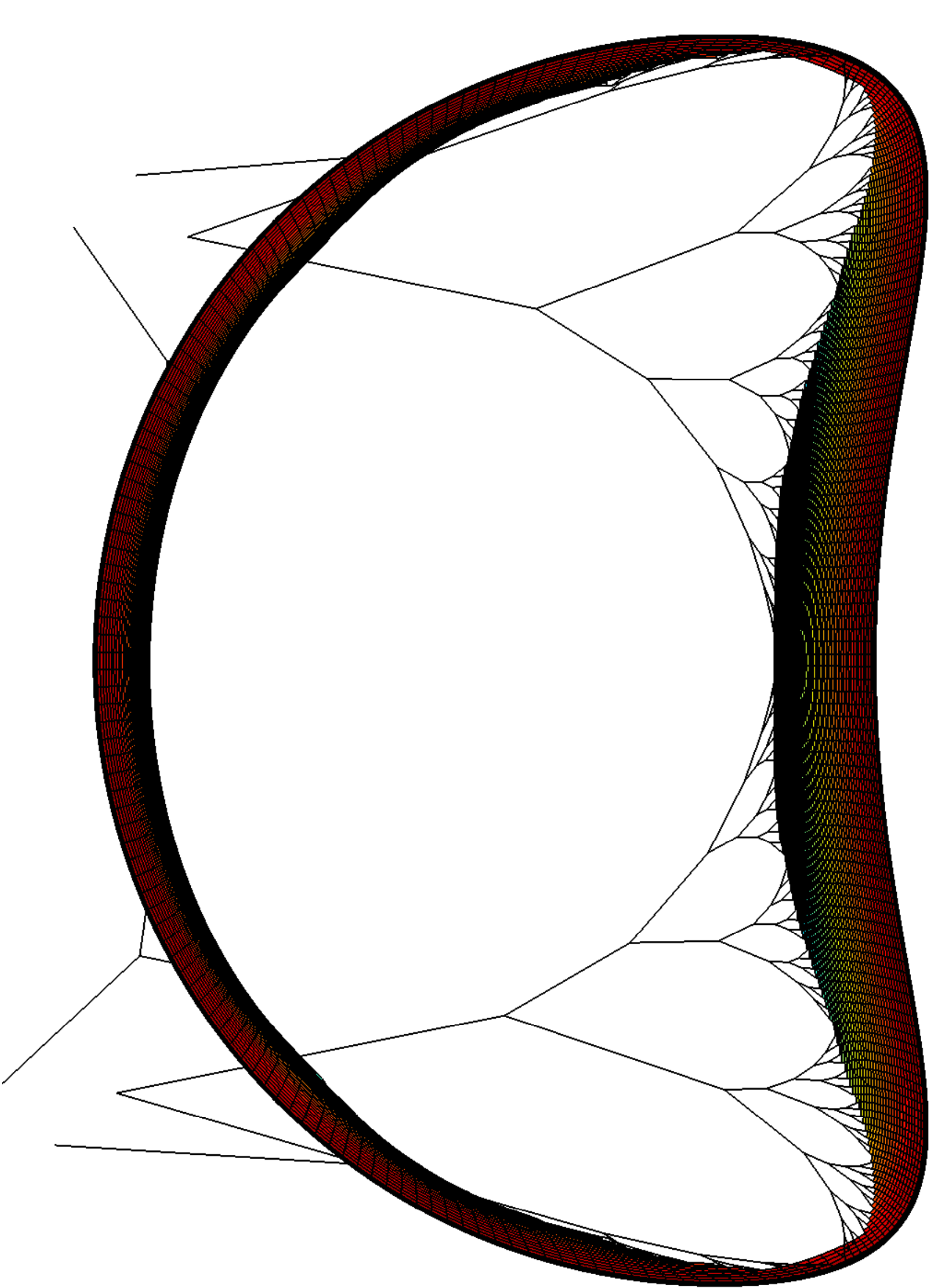}  &  
\includegraphics[width=.3\textwidth]{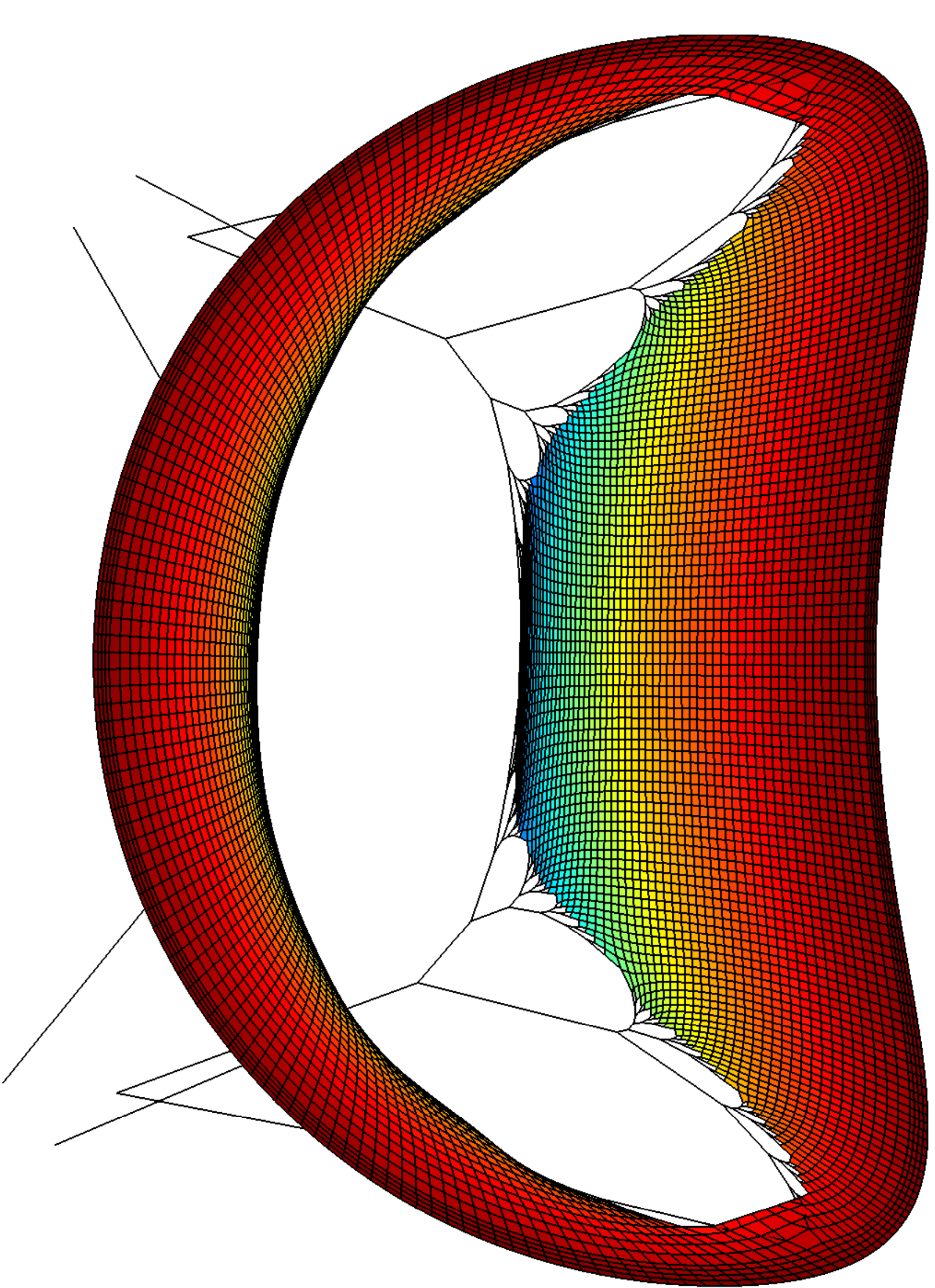} \\
\includegraphics[width=.3\textwidth]{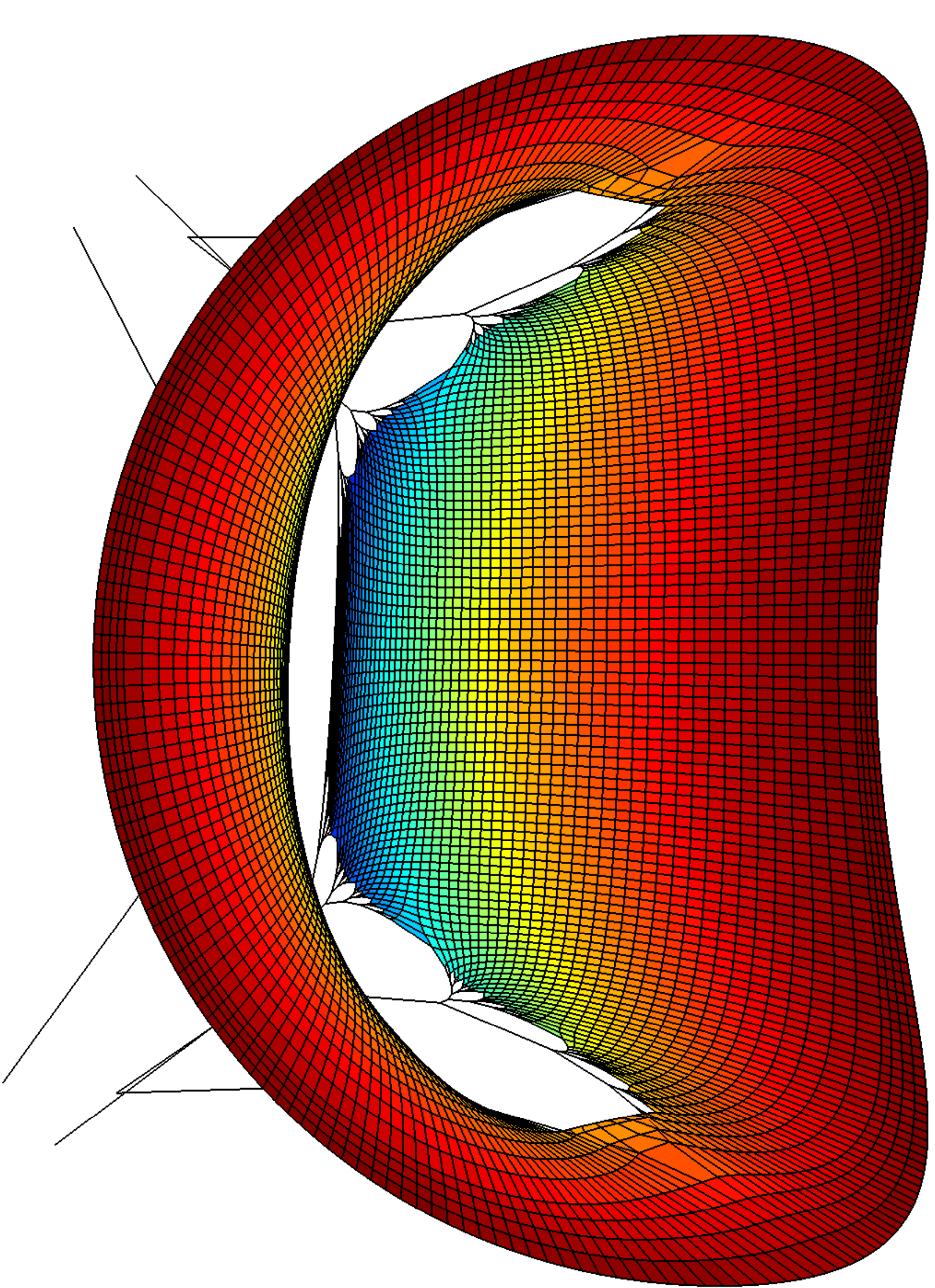}  &
\includegraphics[width=.3\textwidth]{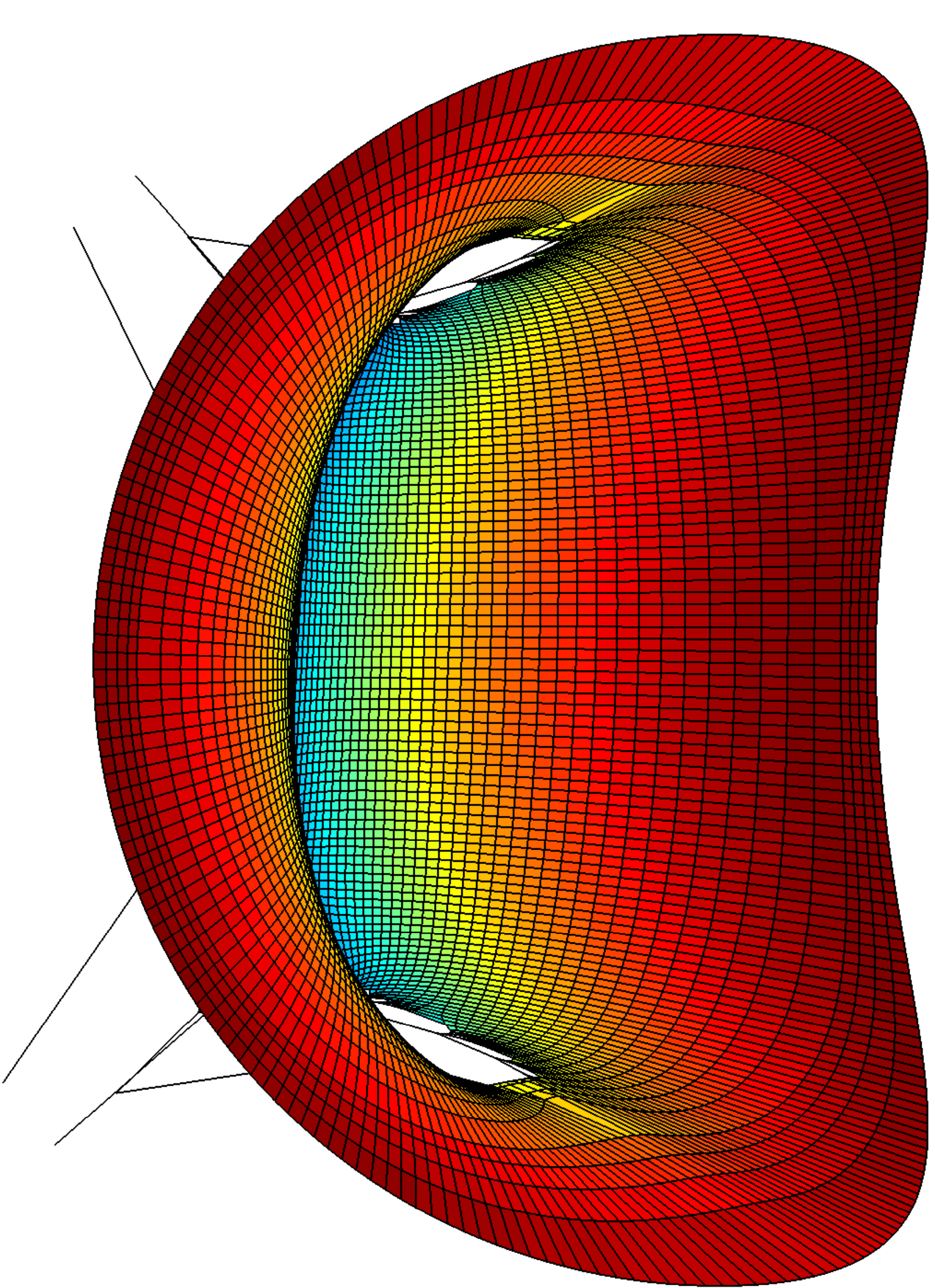} & 
\includegraphics[width=.3\textwidth]{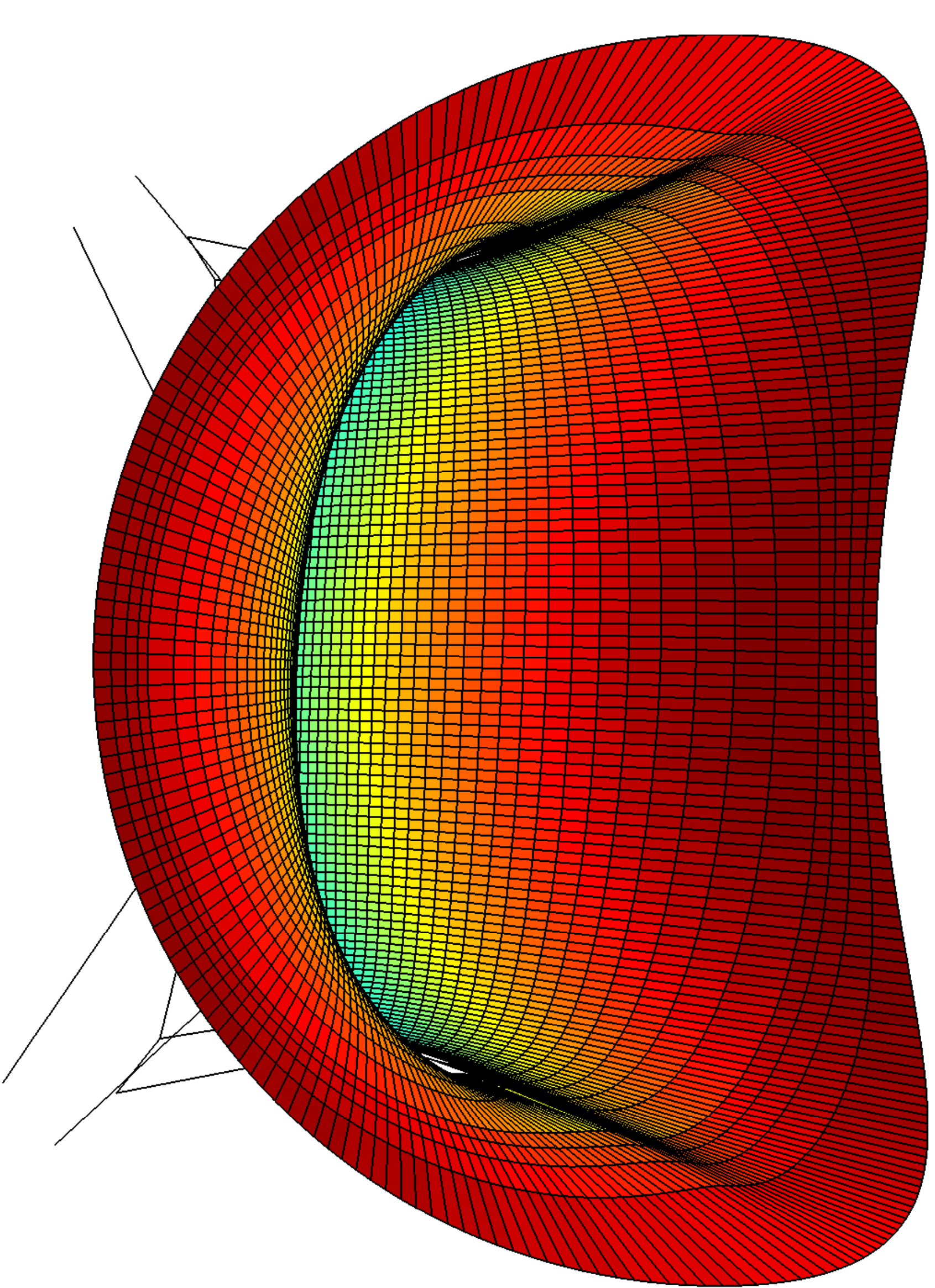}  \\ 
\includegraphics[width=.3\textwidth]{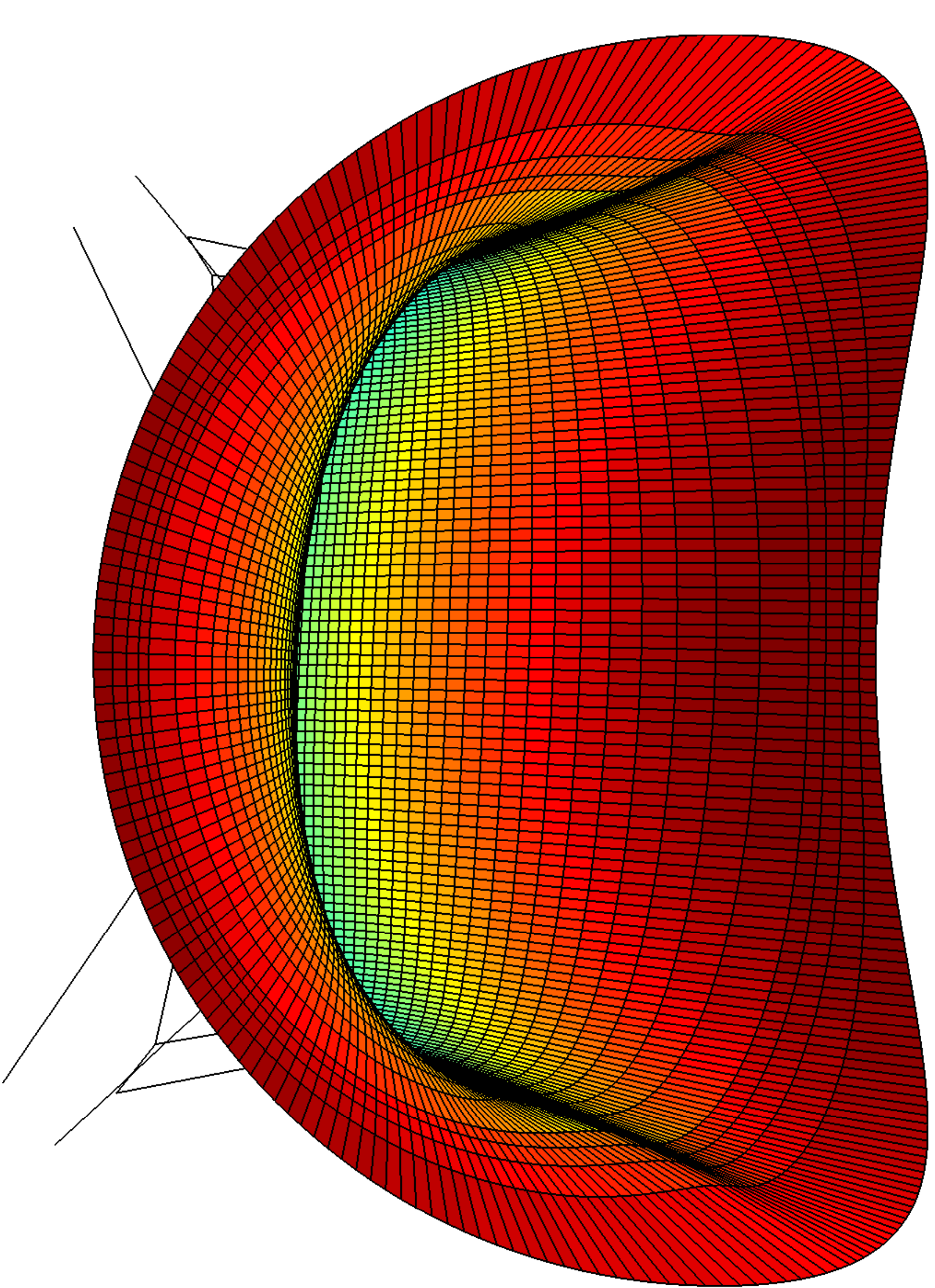} & 
\includegraphics[width=.3\textwidth]{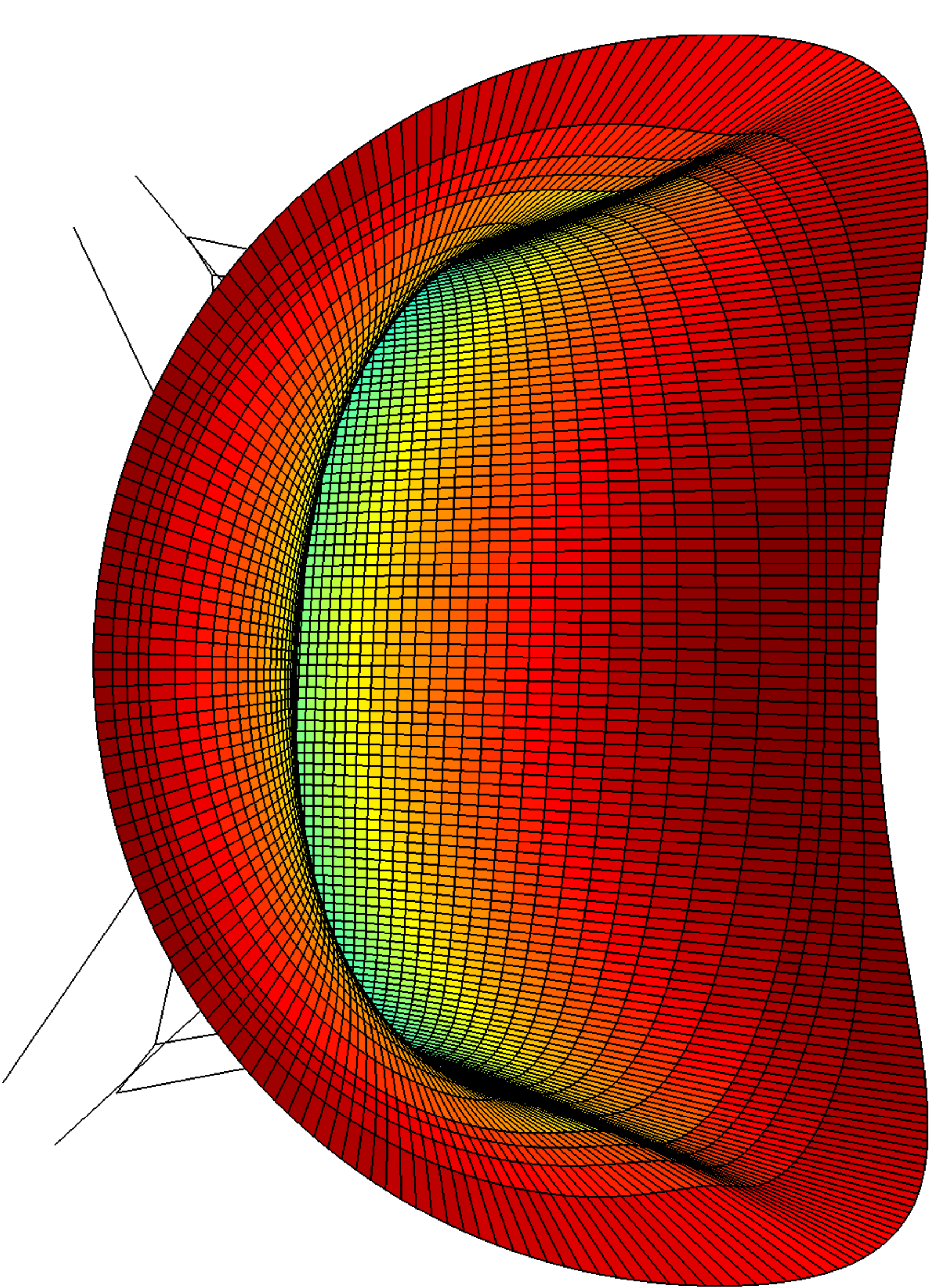}  &  
\end{array} $
\caption{Convergence of Newton's method on the model valve viewed from above. Every other iteration is shown. $N = 256$.}
\label{newton_converge_top}
\end{figure}

\begin{figure}[H]
\centering
$ \begin{array}{ccc}
\includegraphics[width=.3\textwidth]{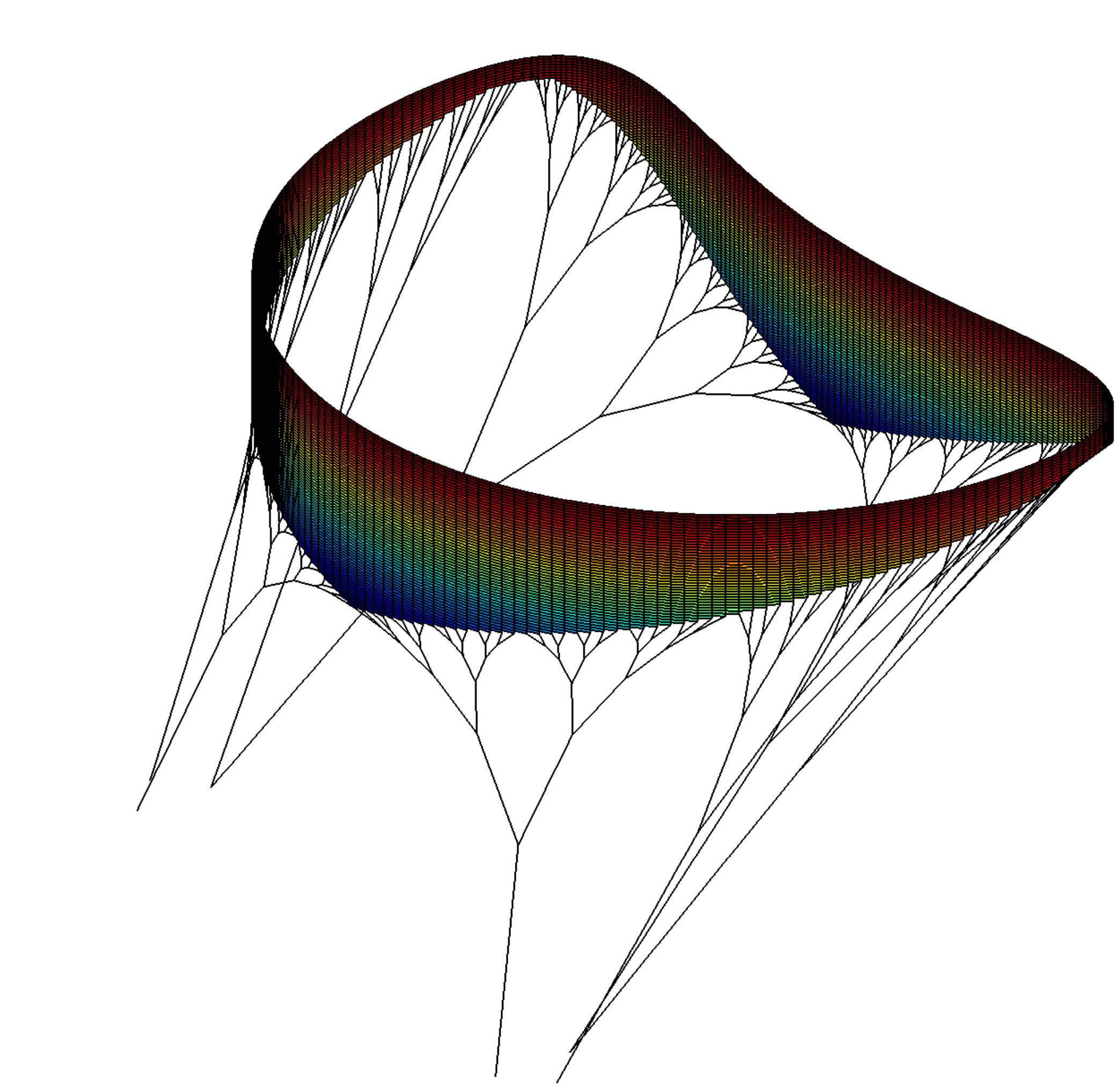} & 
\includegraphics[width=.3\textwidth]{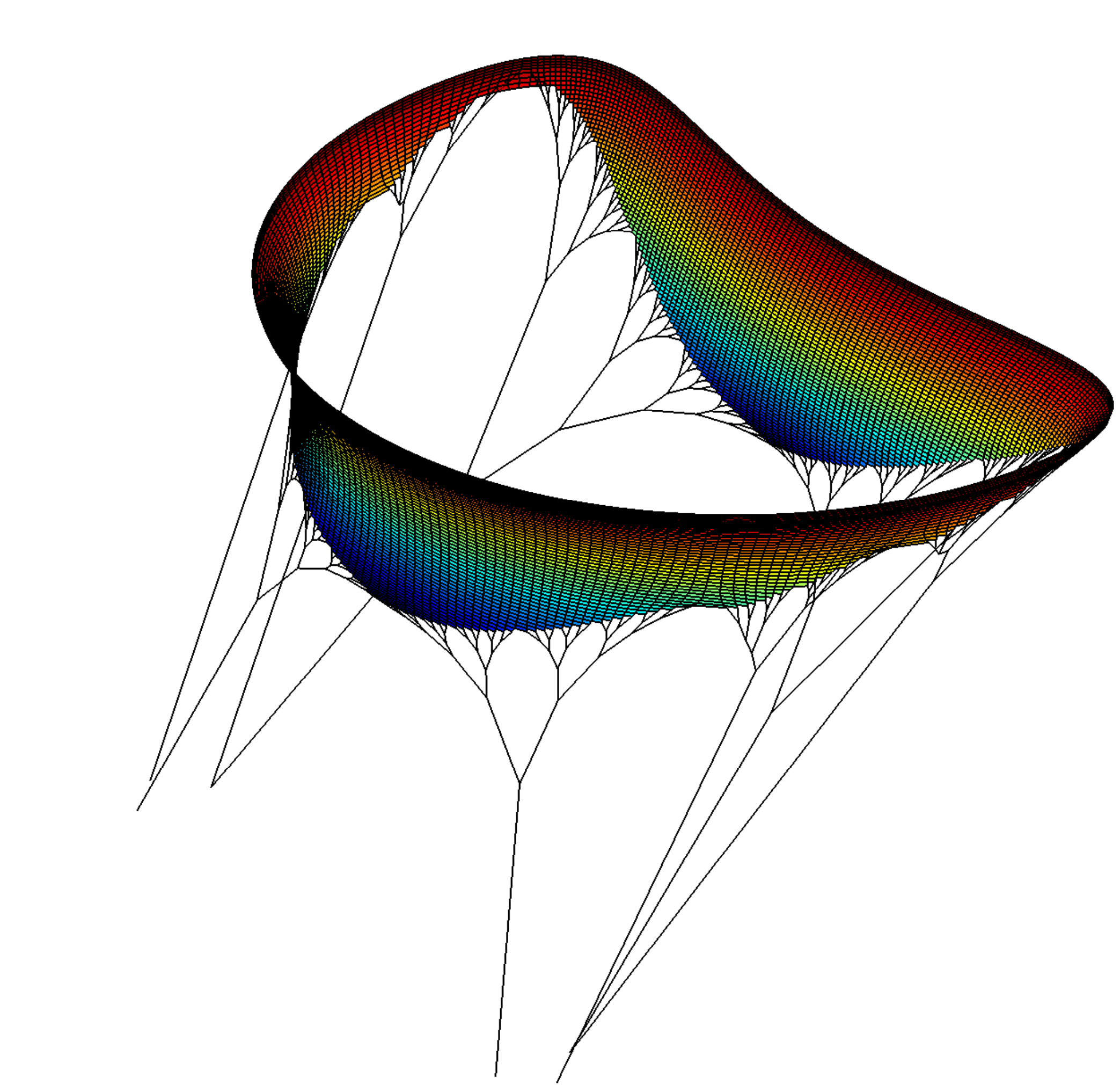}  &  
\includegraphics[width=.3\textwidth]{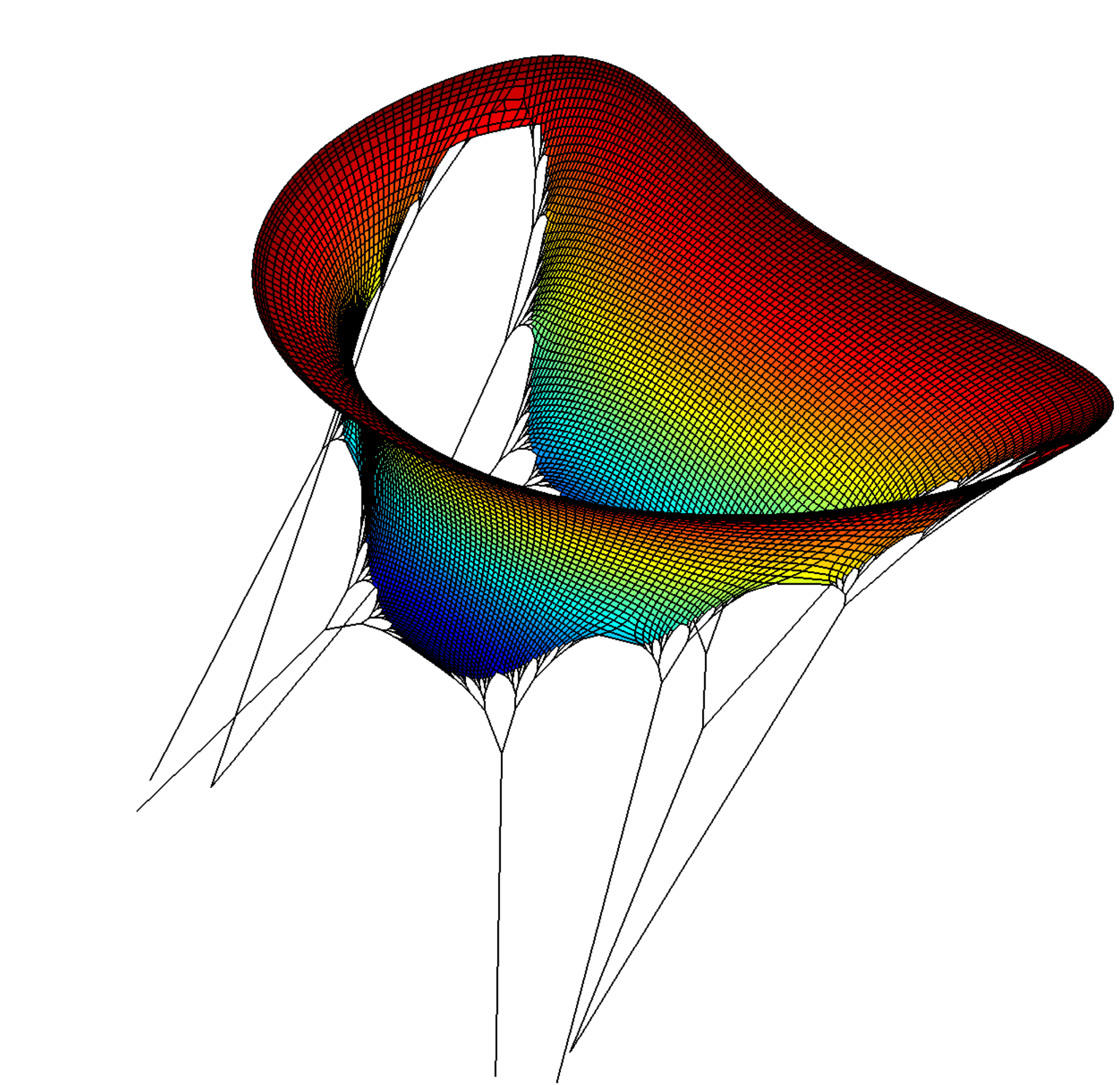} \\
\includegraphics[width=.3\textwidth]{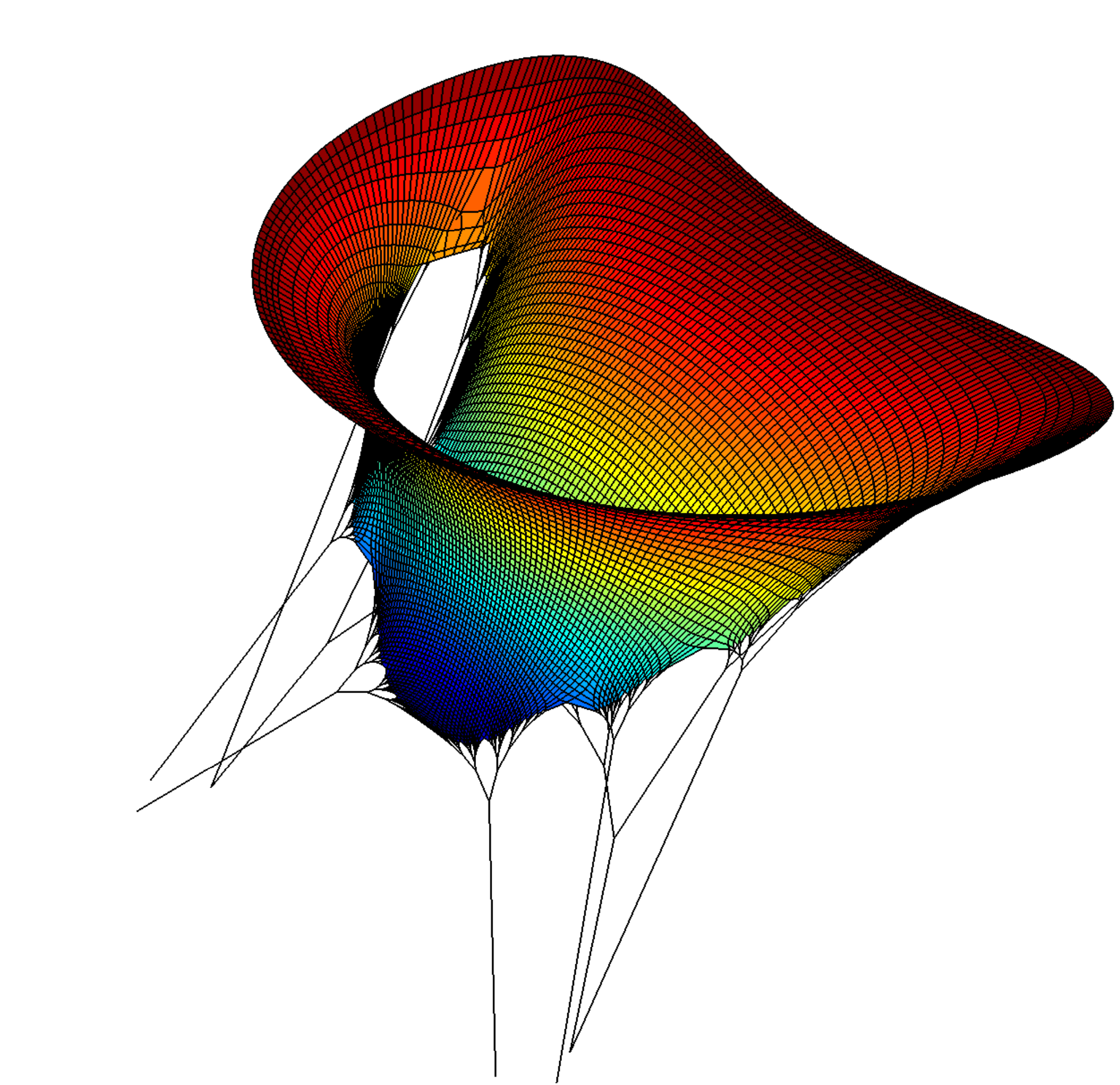}  &
\includegraphics[width=.3\textwidth]{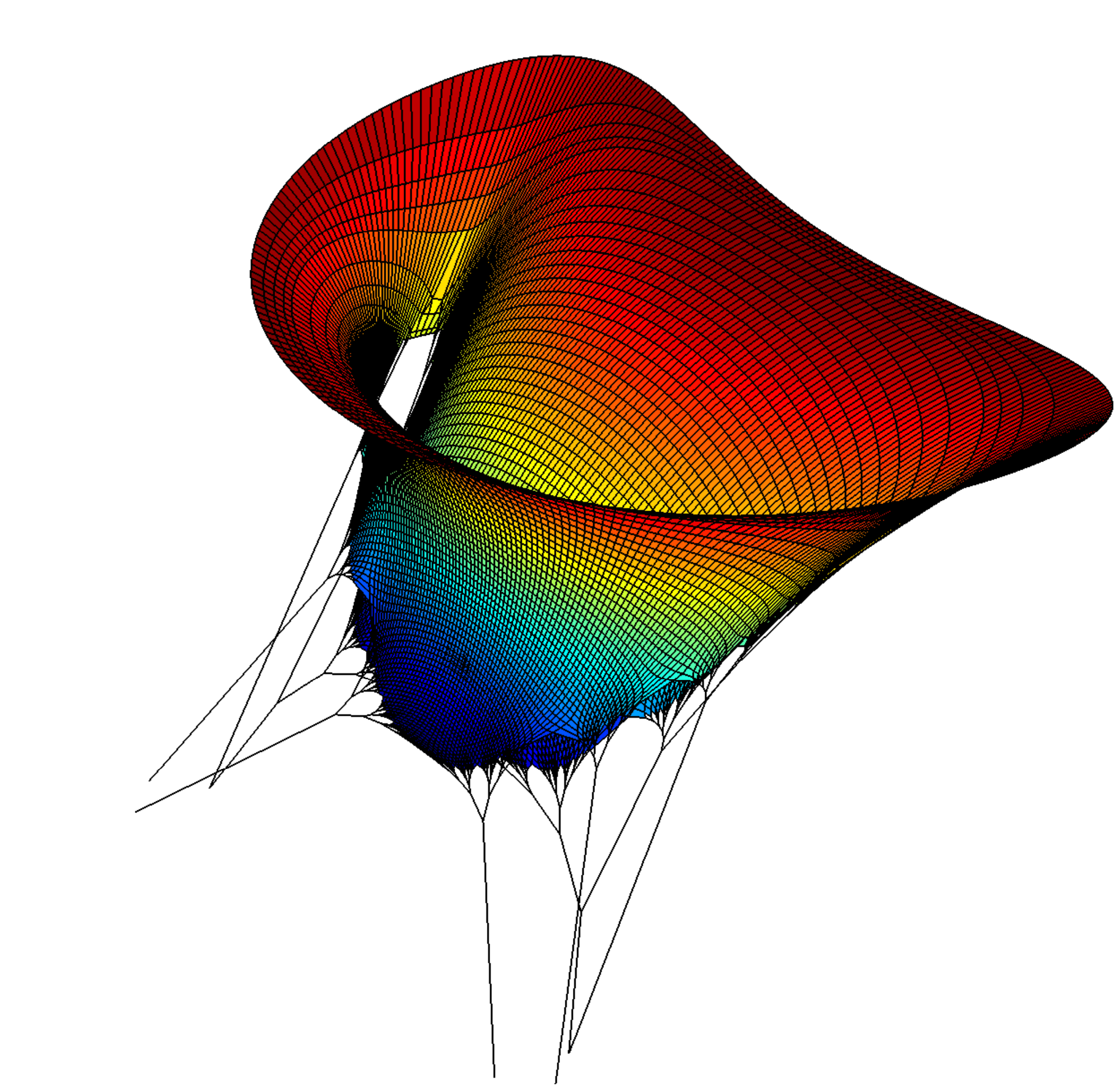} & 
\includegraphics[width=.3\textwidth]{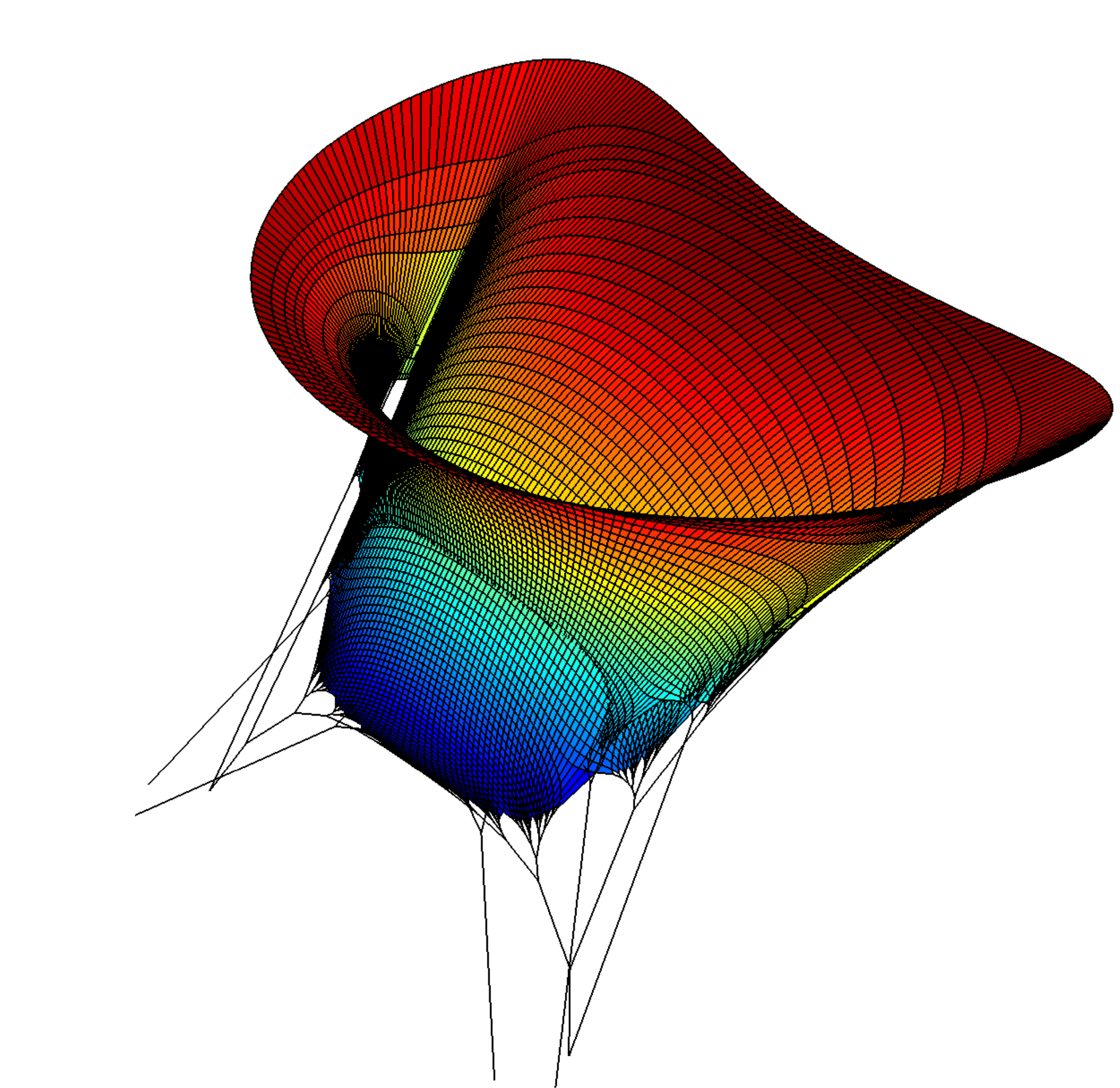}  \\ 
\includegraphics[width=.3\textwidth]{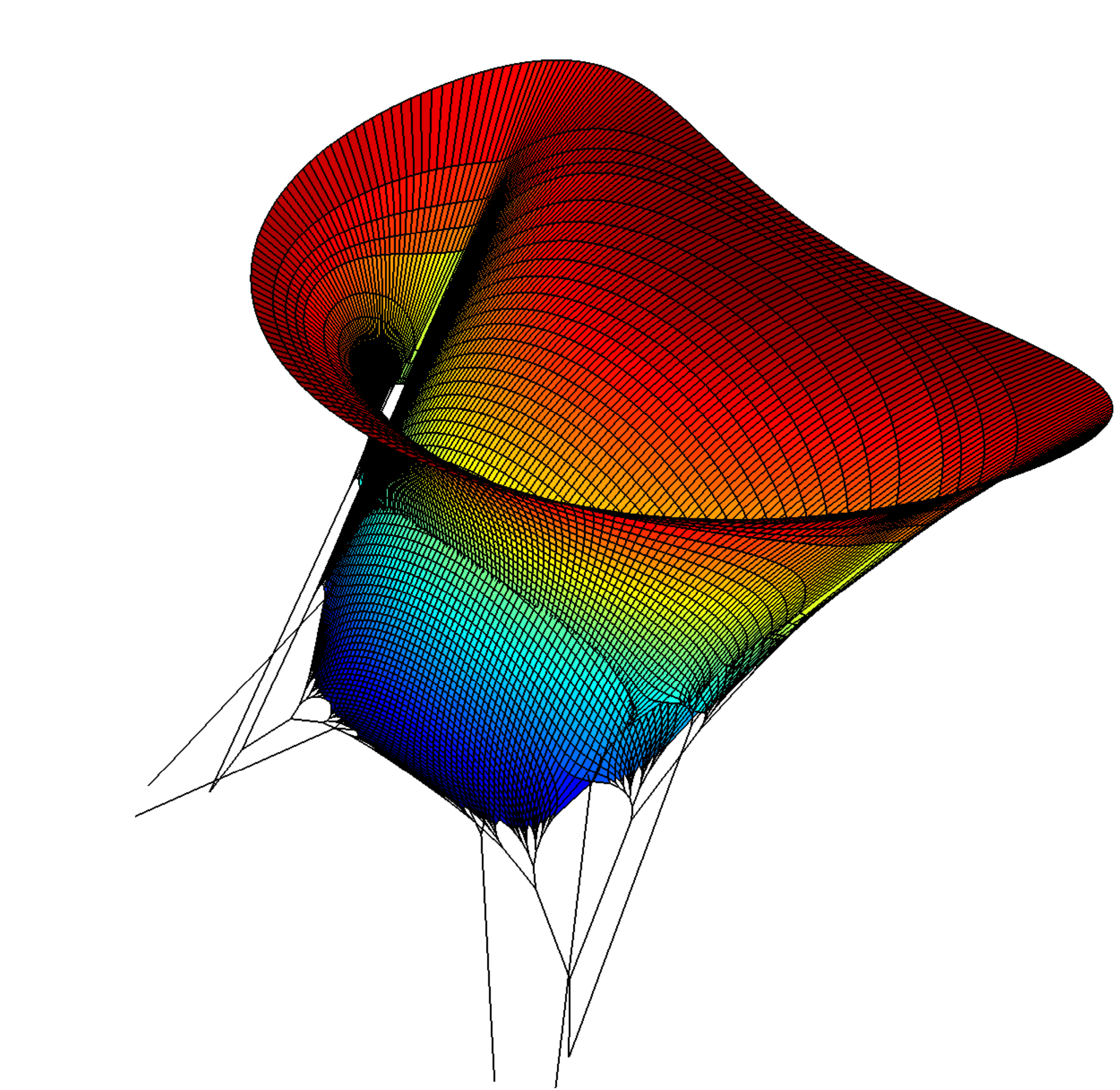} & 
\includegraphics[width=.3\textwidth]{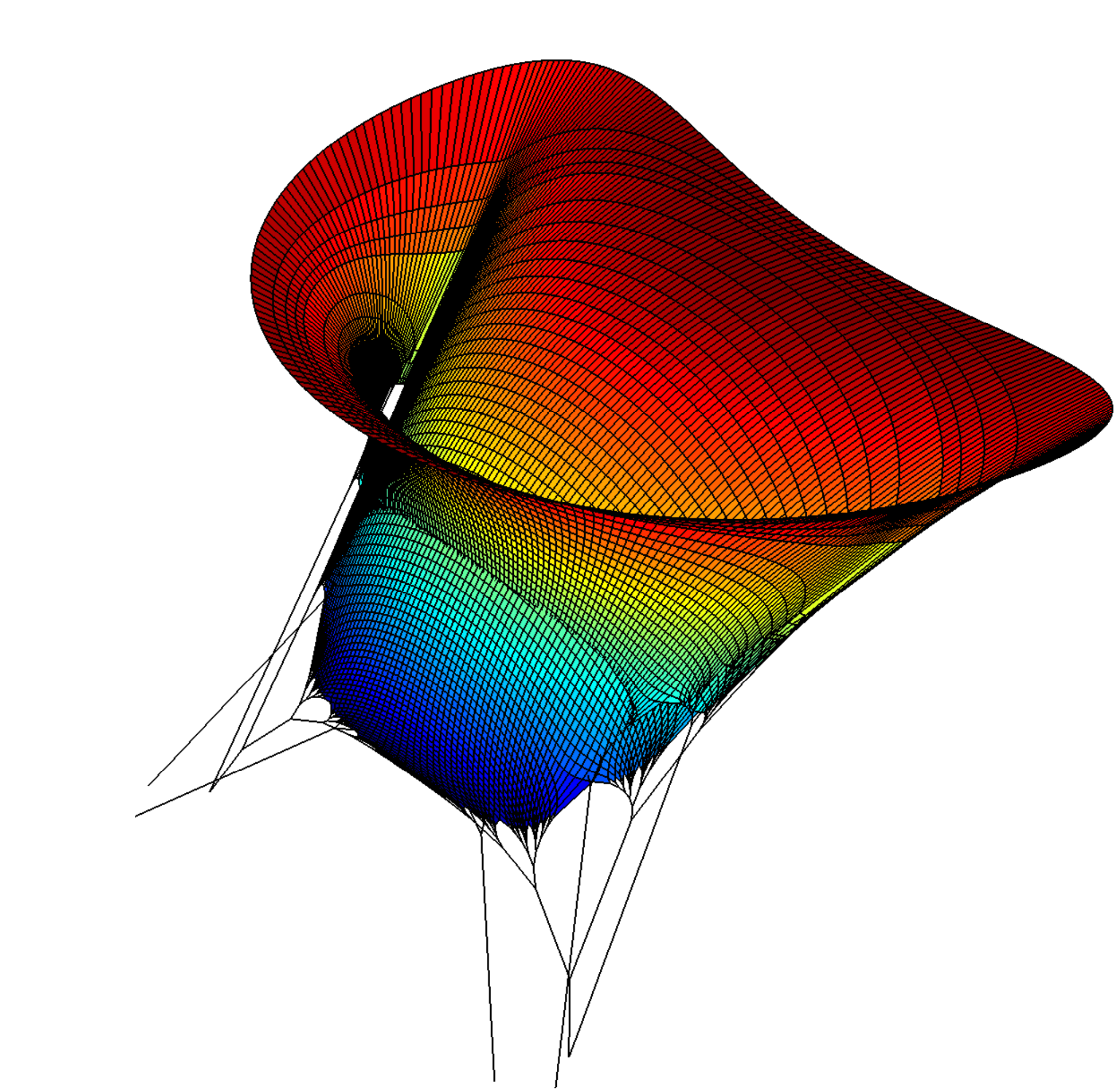}  &  
\end{array} $
\caption{Convergence of Newton's method on the model valve viewed from above. Every other iteration is shown. $N = 256$.}
\label{newton_converge_diagonal}
\end{figure}

It is instructive to view the steps of the Newton's iteration visually. They are shown in figures \ref{newton_converge_top} and \ref{newton_converge_diagonal}. 
These show, as expected, slower changes in the first few iterations when $s_{m}$ is smaller, then very rapid changes. 
The final few iterations do not move much to the eye, but greatly improve on the norm of the error.

\section{A general model for use in fluid}
\label{general_model}

This process gives us a loaded geometry for the valve. 
We wish to use this to generate a model that is effective when simulated with fluid. 

The ``decreasing tension'' model is effective at building closed geometries. 
However, there is no reason why such a constitutive law would be effective or match the real stresses through the full range of strains that the valve undergoes through the cardiac cycle. 
Thus, we use the tensions in the loaded model to assign a physical constitutive law to the configuration of the valve.

\begin{figure}[H]
\centering 
\includegraphics[scale=.8]{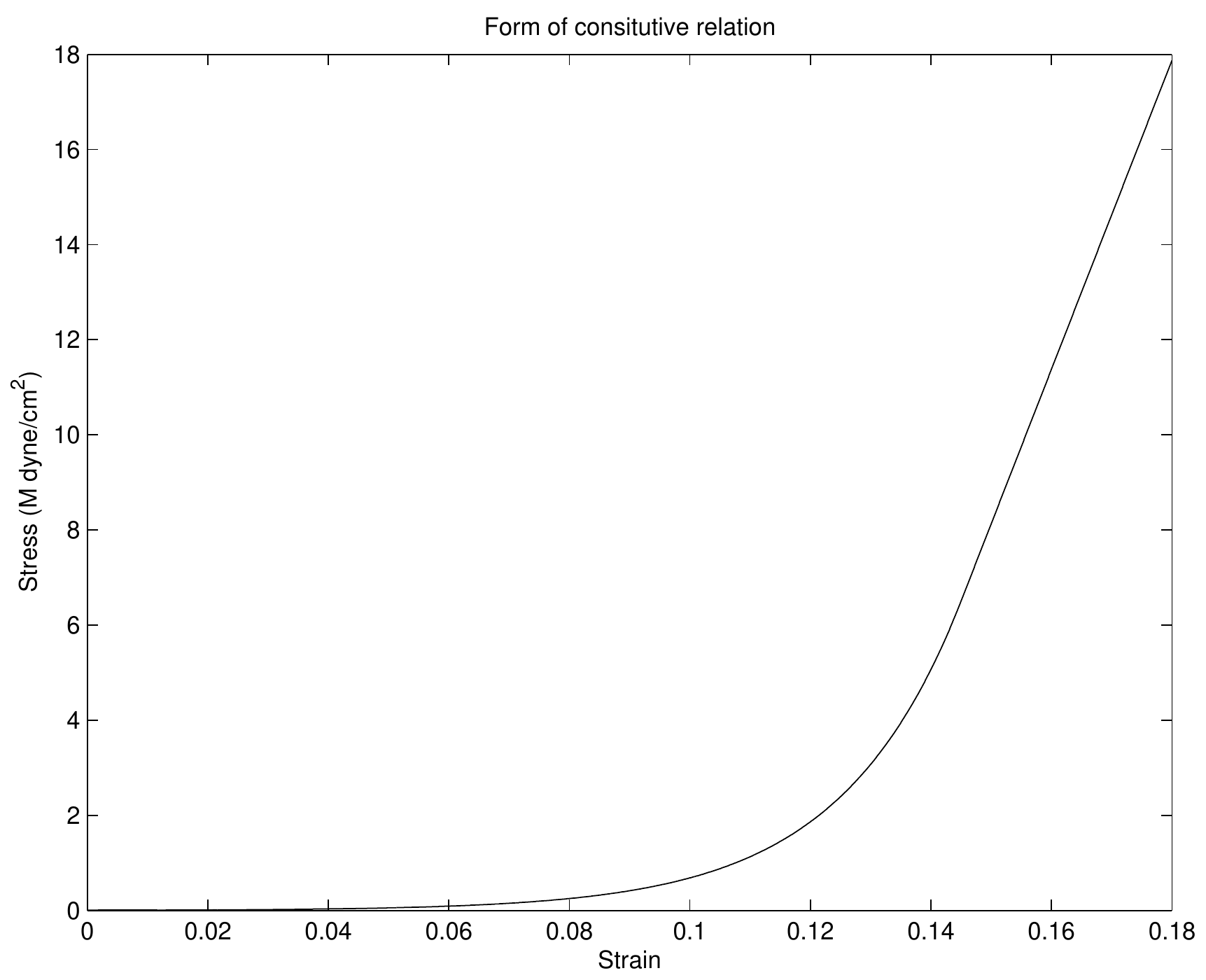}
\caption{Experimental stress-strain relation, based on \cite{sacks_collagen_constitutive}, figure 3a.}
\label{collagen_curve.pdf}
\end{figure}

The constitutive law on each link takes a form from experiments. 
A strip of valve tissue is extended, and a stress-strain curve is reported. 
This is shown in figure \ref{collagen_curve.pdf}, which is derived from \cite{sacks_collagen_constitutive}, figure 3a. 
This paper also discusses a number of more sophisticated constitutive laws that take into account histologic layers of the valve. 
These are not suitable to the pure fiber-based model that I have developed. 
Additionally, they are challenging to interpret as the many constants in the models are found through an elaborate Monte Carlo parameter estimation process.

The curve is somewhat linear with a lower slope from zero strain up to a certain value. 
This reflects the elastin structures present in the valve, which are very roughly isotropic and lower elastic modulus compared to collagen.  
At a certain point, the curve increases slope rapidly. 
This occurs as collagen fibers begin to straighten and exert force. 
Finally, all the collagen straightens and the curve is approximately linear past this point. 
The final slope is much greater than the initial, reflecting the greater elastic modulus of the material. 
Other earlier experiments report similar functional forms under mechanical testing with only differences in numerical values \cite{kunzelman_experimental}.

Following these experimental results, we assume that the stress-strain relation is affine for $E > 0.145$, taking the functional form 
\begin{align}
\sigma (E)= c E + d 
\end{align}
At this point the collagen fibers have reached ``full recruitment'' and are completely straightened 
The intercept and slope of this affine portion are taken from the experiment in  \cite{sacks_collagen_constitutive}. 
For $E \leq 0.145$, we take the functional form 
\begin{align}
\sigma (E) = a(e^{bE} - 1)
\end{align}
and select $a,b$ to ensure that $\sigma(E)$ and its first derivative are continuous at E = 0.145. 
Note that this stress-strain relation goes through the origin for any constants. 
The fibers do not exert any compressive forces; the stress is zero for negative strains. 
The tension is then given by 
\begin{align}
t(E) = 
\begin{cases}
0                           & : \; E < 0 \\ 
ka(e^{bE} - 1)     & : \; 0 \leq E < 0.145 \\  
k (c E + d)                 & : \; 0.145 \leq E  \\  
\end{cases}
\label{constitutive_summary}
\end{align}
where the coefficient $k$ determines the stiffness of the given link.

This constitutive law is phenomenological, simple and effective. 
It is applied to all links in all fibers in the model valve.

We make the assumption that the every link in the discretized valve is under uniform strain with value $E = 0.16$. 
The current configuration determines lengths, and with this we solve 
\begin{align}
E = \frac{L}{R} - 1 
\end{align}
for the reference length $R$. 

The stressed configuration provides a force in every link in the model. 
Now possessing a rest length and tension $t$, we solve 
\begin{align}
t = k (c E + d)
\end{align}
for the spring constants $k$. 
Note that since the prescribed strain is greater than the full-recruitment strain, we always solve for the coefficient in the affine region of equation \ref{constitutive_summary}.

Note that this is a fully fiber based elasticity model.
The valve has two local directions at any point in the leaflet in which it exerts force. 
This is very different from an isotropic model, even in locations where the fiber angles approach orthogonal, because the leaflet reacts differently to stretches in the fiber directions from how it reacts to stretches in other directions.

We solve this equation of equilibrium without pressure, then use this solution as initial conditions in fluid-structure interaction simulations. 
This results in a new configuration that lacks any self-intersection, which is necessary for simulating in fluid.

\begin{figure}[H]
\centering
\includegraphics[scale=.74]{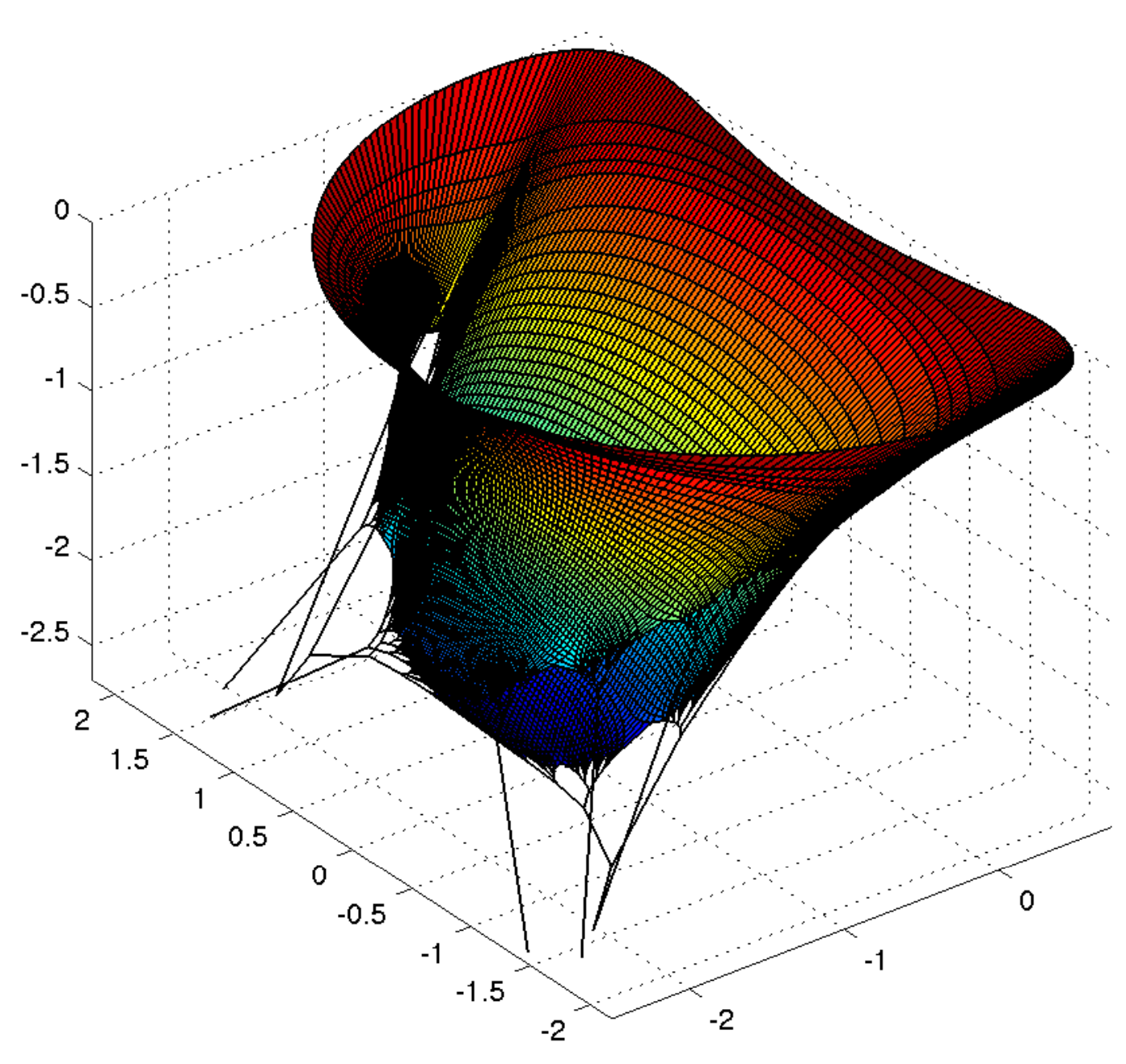} 
\includegraphics[scale=.74]{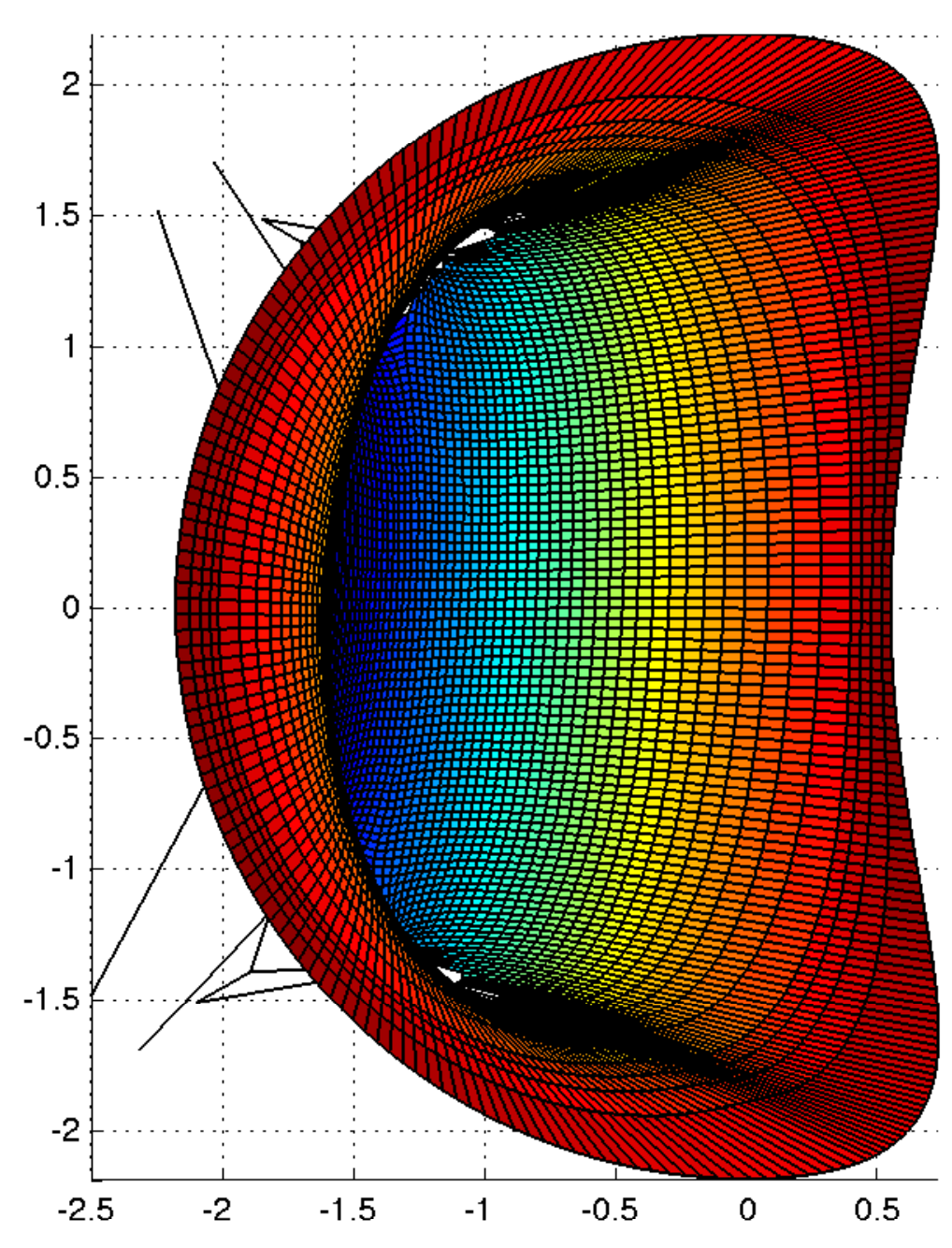}
\caption{Partially open geometry of model valve.}
\label{with_ref}
\end{figure}

\begin{figure}[H]
\centering
\includegraphics[scale=.73]{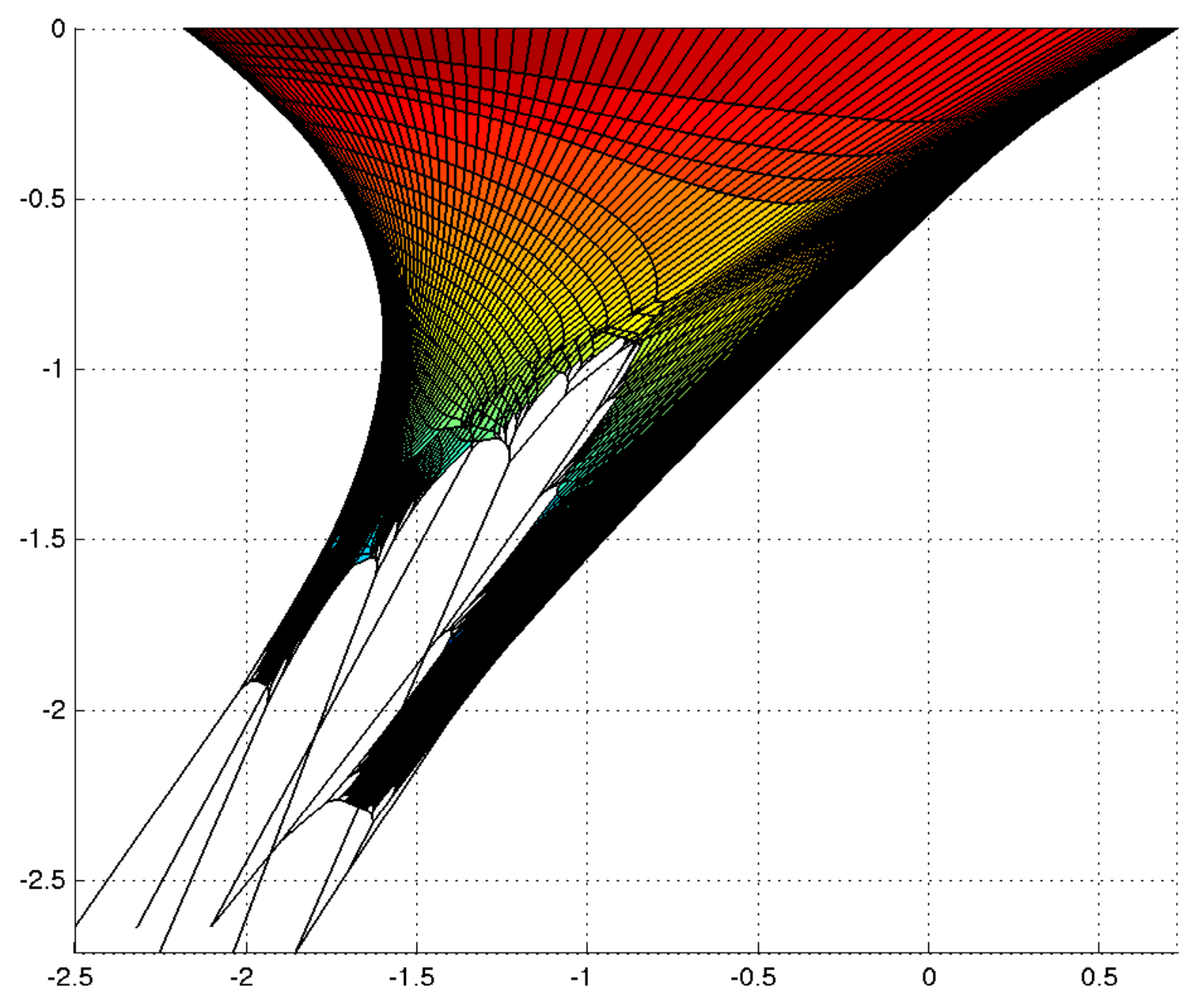} 
\includegraphics[scale=.73]{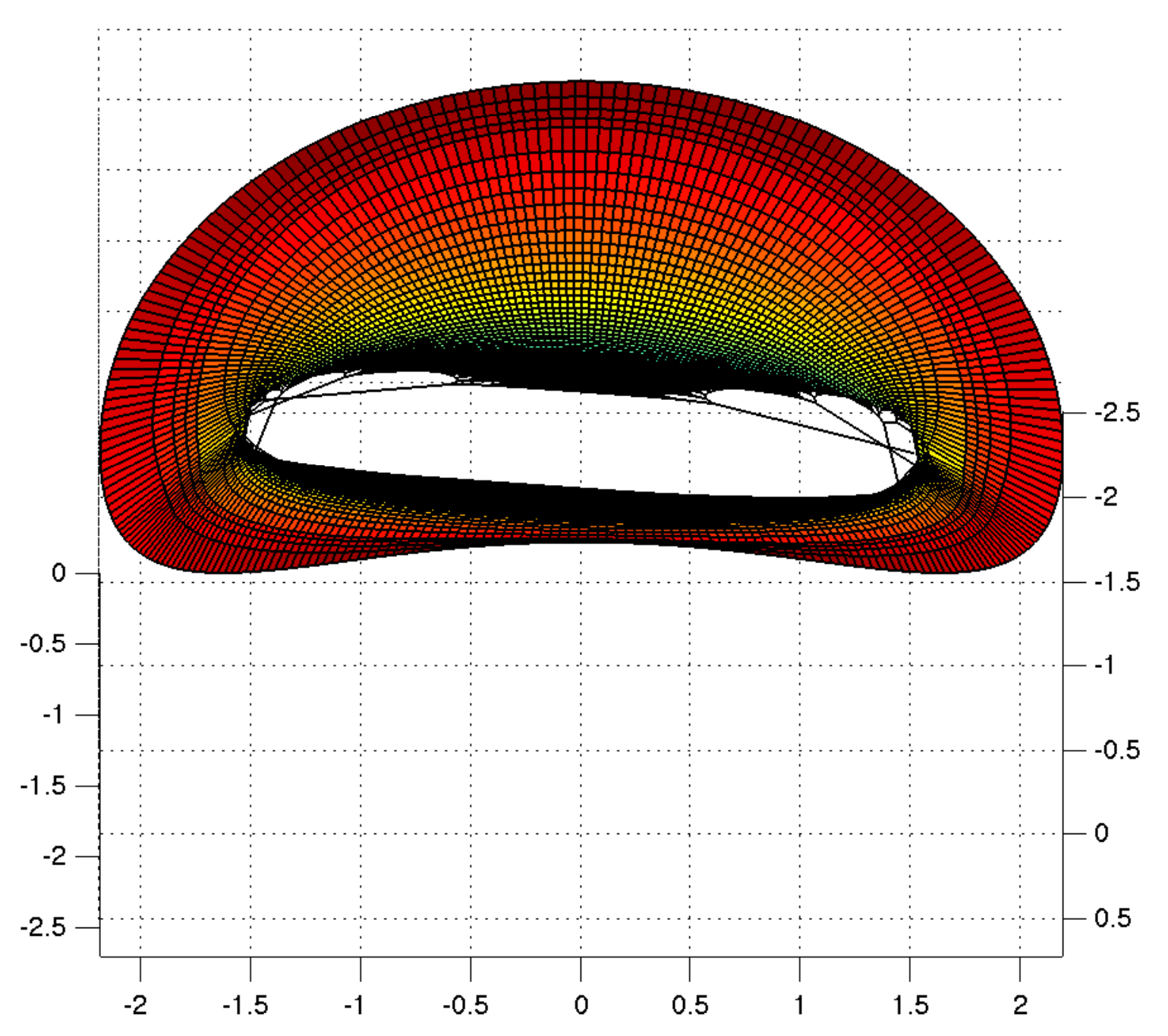}
\caption{Partially open views of model valve, additional views.}
\label{with_ref_2}
\end{figure}

Figures \ref{with_ref} and \ref{with_ref_2} show four views of the general valve model. 
We do not move the papillary locations, as that will be later handled when simulated in fluid. 
Also, the initial conditions of the simulation are taken at end systole, which corresponds to the position of the papillary tips that is shown here. 
No pressure is applied, so the leaflets are no longer pushed to closure. 
The anterior leaflet is generally deeper than the posterior, as is true in the anatomy. 
This was not specified a priori, but rather emerged from the equilibrium equations and geometry of the papillary tips relative to the valve ring.

\begin{figure}[H]
\centering
\includegraphics[width=\textwidth]{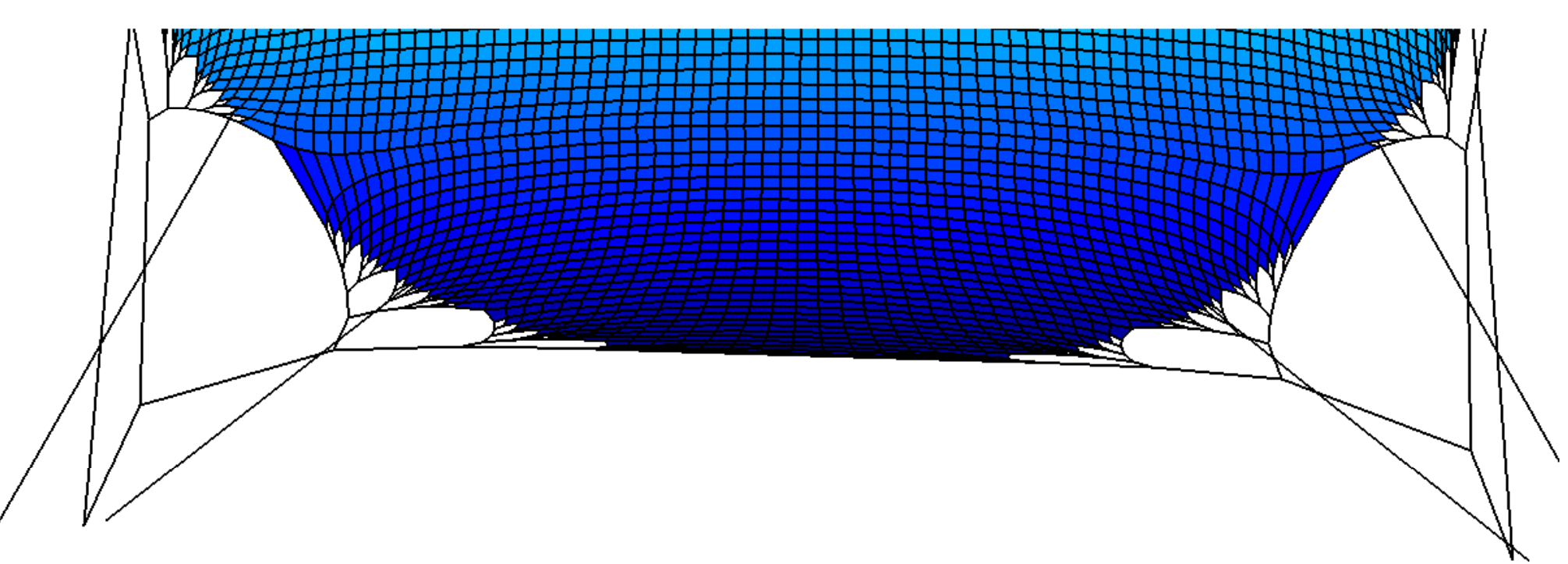} 
\caption{Partially open views of model valve, detail of anterior leaflet free edge.}
\label{valve_with_ref_front.pdf}
\end{figure}

Finally, figure \ref{valve_with_ref_front.pdf} shows a detail of the anterior free edge. 
Note that the point of the leaflet mesh has spread out. 
The center of the anterior leaflet is pulled flat and supported by chordae.  
The free edge inherits fine structure from the chordae insertion, and the envelope of the free edge takes on a blunt, oval like shape. 

To summarize, the model is built from the static version, which provides a length and tension for every spring in the discrete model. 
We solve for rest length and spring constants using this information. 
This provides a general constitutive law that has the same lengths and tensions in the closed configuration as the static solve model. 
This constitutive law (unlike the decreasing tension model) is valid through the large range of strains and tensions experienced through the cardiac cycle. 
Using the new constitutive law, the pressure is reduced as much as possible and equilibrium equations are solved again. 
This configuration is used for fluid-structure interaction simulations.

\newpage 

\chapter{Fluid-Structure Interaction}
\label{Fluid-Structure_Interaction}

From Chapter \ref{static} we now have a configuration for the valve. 
This includes an initial condition, a reference lengths for each link in each fiber, and a physical constitutive law to determine how the model exerts force. 
We now use this to study how the valve behaves in fluid with the immersed boundary method. 

To run such a simulation requires many additional technical details.
We review the immersed boundary method, which is a general technique for simulations of structures immersed in a viscous, incompressible fluid. 
We mount the valve in a model test chamber, which involves building a model partition and papillary muscles. 
We will show a numerical artifact that may occur when a thin membrane supports a high pressure, and present a solution to mitigate its effects. 
We motivate and derive a new discrete delta function for the use in the immersed boundary method. 
We discuss experimentally derived driving pressures for the simulations. 

We then simulate the valve under normal conditions and discuss its behavior. 
We then drive the valve under higher and lower pressures, remove the atrial kick from the pressures, and finally make comparisons of our ``standard'' model with other slightly different versions.

\section{The immersed boundary method}

The continuous form of the immersed boundary (IB) method is a system of partial differential and integral equations \cite{ib_acta_numerica}. 
The equations for fluid motion are written with respect to a laboratory, or Eulerian, frame of reference. 
The equations for structural motion, however, are written with respect to a material, or Lagrangian, frame of reference. 
There is no separation in space of these two frames; the two domains lie on top of each other. 
The fluid domain resides in the entirety of the space, as defined to be some finite region in this problem. 
The structure domain takes whatever subset of the fluid domain it happens to occupy at that time.  
Additionally, the structure is assumed to be massless; that is, it is a pure force generator. 
All momentum is carried by the fluid, and the structure moves with the local fluid velocity. 
Two additional integral equations are added to couple these two frames.

Suppose that $\bb x$, an independent variable, denotes position with respect to the Eulerian frame. 
Let $\bb u$ denote the fluid velocity, $\rho$ denote constant fluid density, $p$ denote pressure, $\mu$ denote dynamic viscosity of the fluid, and $\bb f$ denote a body force. 
We will sometimes write components of $\bb u$ as $\bb u = (u,v,w)$. 
Time is written as $t$.
Further, let $u,v$ be independent variables in the Lagrangian structure, each of which runs along one of two families of fibers in the leaflets as described in Chapter \ref{static}. 
To avoid confusion with fluid-velocity variables, we introduce the shorthand $s = (u,v)$ and $ds = du dv$. 
The parameter $s$ may be supplemented by more independent variables if the fibers form a three-dimensional continuum, or some other form in the case of an arbitrary network of springs or more complex structure model. 
Let $\bb X(s,t)$ denote the position of the structure corresponding to a particular parameter value $s$ at time $t$.
The force density with respect to $ds = du dv$ in the Lagrangian frame is denoted $\bb F$. 
Finally, let $\delta$ denote the Dirac delta function, which is not an ordinary function at all, but a distribution. 

The continuous form of the immersed boundary method equations is given by 
\begin{align}
\rho ( \bb u_{t}  + \bb u \cdot \grad \bb u ) &= - \grad p + \mu \Delta \bb u  + \bb f   \label{momentum} \\
\grad \cdot \bb u  &= 0   		\label{mass}		\\
\bb F(s,t) &=  \mathcal S(\bb X(s,t))  \label{nonlinear_force}  \\ 
\bb X_{t}(s,t)&=  \bb u(\bb X(s,t), t) 		
		= \int \bb u(\bb x, t)   \delta (  \bb x  - \bb X(s,t) )  \;  d  \bb x   \label{interpolate}   \\ 
\bb f   &= \int  \bb F(s,t)  \delta(  \bb x - \bb X(s,t)  )   \; ds          \label{spreading} 
\end{align}  
All functions written without arguments are evaluated at the independent Eulerian variables, $\bb x,t$. 
Equations \ref{momentum} and \ref{mass} are the Navier Stokes equations for conservation of momentum and mass, respectively, of a viscous, incompressible fluid. 
The term $\bb f$ is a body force that represents whatever force the structure generates in the Eulerian frame. 
This is one of the distinctive and important features of the immersed boundary method: the fluid feels the structure as a body force. 
Equation \ref{nonlinear_force} represents force density in the Lagrangian frame. 
The function $\mathcal S$, which is arbitrary here and usually nonlinear, determines how the model exerts force. 
The method is agnostic to the form of this function; any function with a physical meaning can be used here.

Equations \ref{interpolate} and \ref{spreading} are interaction equations. 
They determine how the two frames relate to each other, through convolutions with the Dirac delta function. 
Equation \ref{interpolate} is called the interpolation equation. 
It allows the Eulerian velocity field to be evaluated at an arbitrary point in space. 
(The difference is more apparent in the discrete version, in which $\bb u$ is only defined on a finite difference mesh.)
Note that the Eulerian velocity $\bb u$ is equal to the Lagrangian velocity $\bb X_{t}(s,t) = \bb u(\bb X(s,t),t)$ at corresponding points. 
We write $\bb U$ for the Lagrangian velocity. 
In the continuous equations, these functions are identical, $\bb U(\bb X(s,t),t) = \bb u (\bb X(s,t),t)$. 
In the discrete version, however, $\bb u$ is only defined on a grid, and $\bb U$ is defined for all real arguments through the interpolation formula. 
Thus, we sometimes use the notation $\bb U$ to emphasize this distinction. 
Equation is \ref{spreading} is referred to as the force spreading equations. 
It specifies the Eulerian frame body force $\bb f$, which determines  how the force in the Lagrangian frame affects the fluid. 
Note that $\bb f(\bb x,t)$ is not equal to the Lagrangian force $\bb F(s,t)$ at corresponding points, as a Jacobian determinant appears if the integral is transformed to an integral over $\bb X$. 

All simulations are run with the software library IBAMR, or Immersed Boundary Adaptive Mesh Refinement \cite{IBAMR, GRIFFITH200710, griffith2010parallel}.
The Eulerian variables are discretized with a staggered grid scheme \cite{mac_scheme}.
Details of the implementation of the staggered grid scheme and time-stepping algorithms for the entire immersed boundary system of difference equations in this library is described in \cite{Griffith_aortic}. 
The pressure and velocity are solved for simultaneously, using a sophisticated preconditioning scheme \cite{griffith_projection}. The projection method itself \cite{chorin1968numerical} is applied as a preconditioner for the discretized incompressible Stokes equations. 
The advective term and body force are treated explicitly. 
This is why the solver works on the discretized Stokes equations, rather than full Navier-Stokes equations. 
By solving the full system, this allows for efficient and effective application of general boundary conditions. 
The valve mesh and other Lagrangian structures are treated as an arbitrary network of linear and nonlinear springs. 
The method for computing force from such a network is described in \cite{ib_force_notes}. 

We refer to the discrete version of the interaction equations, \ref{interpolate} and \ref{spreading}, later, so summarize their spatial discretization here. 
In the discrete version, the integral convolutions are replaced by discrete convolutions taken as sums. 
The delta function is replaced by a discrete approximation to the delta function. 
Let $g_{h}$ denote the discrete Eulerian grid, and $G_{h}$ denote the Lagrangian grid. 
\begin{align}
\frac{d\bb X(s,t)}{dt} &= \sum_{\bb x \in g_{h}} \bb u (\bb x,t) \delta_{h} (\bb x - \bb X(s,t))  h^{3}    \label{discrete_interpolate} \\ 
\bb f(\bb x, t) &= \sum_{s \in G_{h}} \bb F (s,t) \delta_{h} (\bb x - \bb X(s,t))          \label{discrete_spread}
\end{align}
The function $\bb F$ in the discrete spreading equation, \ref{discrete_spread}, includes the measure $ds$ that is included in the continuous form, equation \ref{spreading}. 
IBAMR requires user-defined functions for evaluating $\bb F$ to return forces, rather than force densities, so we think of the discrete equations in this form.

\section{Simulation setup}
\label{setup}

\begin{figure}[H]
\centering
\includegraphics[width=\textwidth]{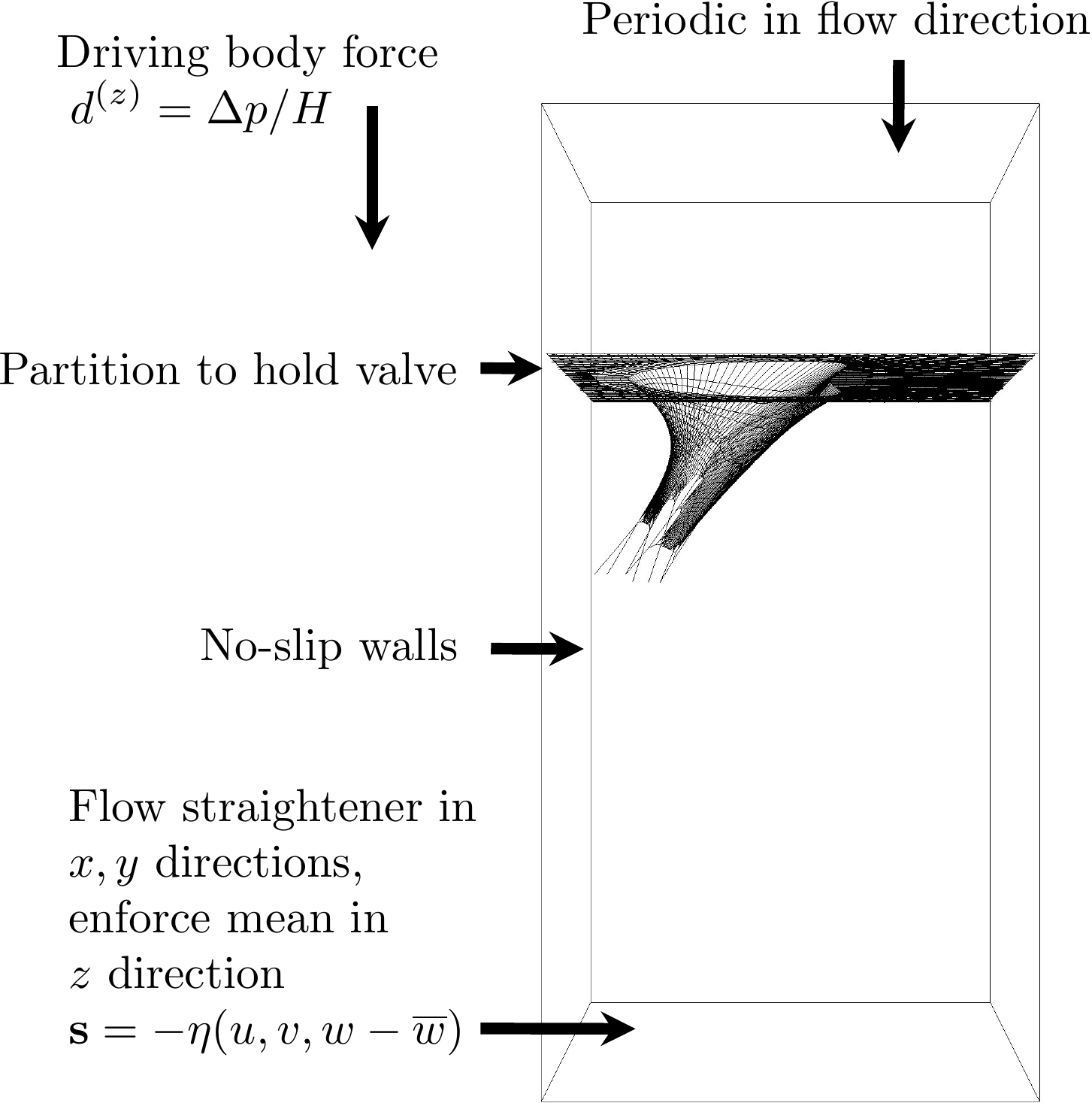}
\caption{Schematic of valve in computer test chamber. }
\label{simulation_schematic.pdf}
\end{figure}

To run simulations, the model valve is placed in a computer test chamber. 
A schematic of this is shown in figure \ref{simulation_schematic.pdf}

In this view, which we will use repeatedly, the anterior leaflet is on the right and posterior is on the left.
If the valve was oriented the same way, but placed in a heart rather than the computer test chamber, the left atrium would be above the valve. 
The aortic valve would be just to the right of the mitral valve. 
The left ventricle would sit below the valve, with left ventricular free wall on the left and the interventricular septum on the right.

The fluid domain is a rectangular box of dimensions $[-L,L] \times [L,L] \times [-3L,L]$, where $L = 3$ cm. 
This domain is discretized with $128 \times 128 \times 256$ points, and has an Eulerian mesh spacing of $\dx = 4.6875 \cdot 10^{-2}$ cm, or just under half a millimeter. 
The mitral ring radius is approximately 2.1 cm, so there are approximately 90 fluid points across the valve ring at its widest cross section. 
The valve is mounted to a model partition in the $x,y$ plane; flow in the negative $z$ direction corresponds to forward flow through the mitral ring. 
On the walls, where $x$ or $y$ is equal to $\pm L$, we impose a no-slip boundary condition. 
The heart works normally regardless of the angle of the body, or without gravity at all, so we ignore gravity throughout.

The time is set to $\dt = 1.5 \cdot 10^{-6}$ s in all simulations shown in the thesis, which is the largest step that we found to be stable. 
This was chosen empirically; even $\dt = 1.65 \cdot 10^{-6}$ s was shown to be (ever so slightly) unstable. 
Individual steps take approximately 0.4 seconds when running on 84 cores; the simulation takes around 4 days to solve one 0.8 second beat. 
Thus, this timestep restriction is burdensome. 
The reason for this restriction is that time discretization is explicit in the Lagrangian structure.
The stiffness of the valve components, as well as the viscoelastic target forces described in Section \ref{papillary_target}, all contribute to this restriction. 
We could easily solve an individual time step with more resolution and more cores, as robust parallel linear solvers are available. 
However, the timestep restriction will be even worse if we do so and the simulations will become even slower. 
A robust method that is implicit in the Lagrangian structure would improve this tremendously.

The domain is taken to be periodic in $z$.
This is to avoid complications from the so called ``open boundary instability.''
We originally used a non-periodic domain, and observed strange artifacts in the velocity field. 
Colleagues (Boyce Griffith) pointed out that such problems are common with pressure boundary conditions, and that removing them reliably is an active research problem.
There are many schemes that attempt to fix it, yet we do not view any as completely proven to be totally reliable in all cases \cite{CNM:CNM2918}. 
The authors claim that there are two situations that cause open boundary instabilities. 
The first is when the flux changes sign across an open boundary. 
The second is when a vortex or other structure causes a small region of inflow on a boundary with net outflow, or vice versa. 
That is, at a boundary with normal $\bb n$, the local velocity $\bb u \cdot \bb n$, at some subset of the domain, has a different sign than the flux $\int \bb u \cdot \bb n \; dS$. 
(Note that with no-slip walls this is a nonissue, as $\bb u$ is identically zero on no-slip walls.)
These conditions may increase kinetic energy in the flow, and strange behavior occurs. 
In our simulations, we expect both of these situations to occur regularly if pressure boundary conditions are applied.
During valve closure, vibrations will cause the signs of the flux across the entire domain to change signs rapidly. 
Vortices will be shed from the leaflets and cause local reversals of flow, and complicated patterns during systole may do so as well. 
We do not believe the mechanisms behind these problems are fully understood, nor fully resolved. 
Thus, we make the simulation periodic in $z$ to remove open boundaries from the setup and avoid them entirely.

The simulation is driven by prescribing a pressure difference between the top and bottom of the domain.
Conventionally, a \emph{modified pressure} is introduced to remove the body force due to gravity from the momentum equations  \cite{batchelor1973introduction}. 
Here, we add an additional body force that has the effect of prescribing a pressure difference, denoted by $P(t)$, across the computer test chamber. 
The force has value 
\begin{align}
\bb d = \left(0, 0, \frac{-P(t) }{H} \right), 
\end{align}
where $H = 4L$ is the total domain height.  
This is equivalent, through a change of variables, to prescribing a pressure difference across the domain. 

Consider a solution $\bb u, p_{mod}$ to the Navier-Stokes momentum equation, where $p_{mod}$ denotes the modified pressure. 
Let the domain for this solution be periodic in $z$, and include the body force $\bb d$.
The Navier-Stokes momentum equation with this pressure and force is then
\begin{align}
\rho ( \bb u_{t}  + \bb u \cdot \grad \bb u ) &= - \grad p_{mod} + \mu \Delta \bb u + \bb d . 
\label{periodic_ns}
\end{align}
The modified pressure is periodic in $z$, which says that 
\begin{align} 
p(x,y,-3L) = p(x,y,L) .  
\end{align}
Recall that the boundaries of the domain are $z = -3L$ and $z = L$. 
Now, define a change of variables on pressure, 
\begin{align}
p = p_{mod} + \frac{P(t)}{H} z . 
\end{align}
This $p$, along with the original $\bb u$, then solves the Navier-Stokes momentum equation,
\begin{align}
\rho ( \bb u_{t}  + \bb u \cdot \grad \bb u ) &= - \grad p + \mu \Delta \bb u , 
\label{nonperiodic_ns}
\end{align}
and satisfies the boundary condition 
\begin{align}
p(x,y,-3L) = p(x,y,L) + P(t) . 
\end{align}
We refer to this pressure $p$ as the physical pressure. 
Thus, applying this body force is identical to prescribing a pressure difference across the domain.

At the downstream end of the chamber, away from the valve, we add a mathematical flow straightener in the $x,y$ directions, and an averaging force in the $z$ direction. 
The straightener serves to reduce the values of $u,v$, which removes vortices to prevent them from appearing at the inflow boundary due to periodicity.  
The averaging force stops localized regions of high velocity from reentering due to periodicity, and ensures a nearly uniform inflow profile. 
Thus, this force serves to hide the periodicity of the domain as much as possible. 

This is a body force of the form 
\begin{align}
\bb s =  -\eta (u, v, w - \overline w) , 
\label{straightener_force}
\end{align}
where 
\begin{align}
\overline w = \frac{1}{A} \int_{z = 0} w \; dx dy 
\end{align}
is the mean of the $z$ component of flow and $A = (2L)^{2}$ be the cross sectional area of the domain. 
Note that because of no-slip condition on the walls, this flux is the same for any plane of constant $z$. 
We use the flux through the mitral ring here, because it is already computed for output purposes. 
The force is applied in a thin slab, $z \in (1/2 - \dx, 1/2 + \dx)$, and is scaled by 
\begin{align}
\frac{1}{2} \left( \cos \left(  \frac{\pi ( z - 1/2) }{ \dx}   \right) + 1 \right)  
\end{align}
to slightly smooth the force. (With the staggered grid discretization, there are two planes in which this force is nonzero in the $x,y$ directions and one in the $z$ direction.)
The value of the coefficient  is selected as  $\eta = \rho / (4 \dt)$. 
As a model, consider an approximate solution to the ordinary differential equation $u_{t} = -\eta  u$ solved with forward Euler. 
The iteration is given 
\begin{align}
u_{n+1} = u_{n} - \frac{\dt \eta}{\rho} u_{n} = \frac{3}{4} u_{n} 
\end{align}
The full solution includes many other terms, but this heuristic suggests we are removing approximately one fourth of the $x$ and $y$ components at each time step. 

The averaging force may be of concern, since one may expect it to change the flow. 
Subtracting the mean reduces the $L^{2}$ norm, and thus kinetic energy of the flow.
This is because  
\begin{align}
\int_{z=z_{0}} (w - \overline w)^{2} \; dx dy  
&=  \int_{z=z_{0}} w^{2} \; dx dy  - 2 \overline w \int_{z=z_{0}} w \; dx dy  +  \int_{z=z_{0}} (\overline w)^{2} \; dx dy  \label{ke_reduction} \\ 
&=  \int_{z=z_{0}} w^{2} \; dx dy  - A (\overline w)^{2} \nonumber  \\ 
&\leq \int_{z=z_{0}} w^{2} \; dx dy ,  \nonumber 
\end{align}
for any planar subset of the domain $z = z_{0}$. 
This results in slightly lower forward flows during diastole, as we show in Section \ref{comparison}, figures \ref{flux_no_average} and \ref{flux_no_average_diff}. 
Flux with and without the averaging force is compared, and shows that this change is small and unimportant. 
The oscillations during closure are slightly damped by the presence of this force. 
We do not model the compliances of structures surrounding the valve, and expect these compliances to be considerable.
Because they are absent, small additional damping is welcome. 
Thus, we believe the benefits of hiding the periodic structure and increasing damping are significant.

The meshes on all Lagrangian structures have a desired spacing $\ds$, which must be set in relation to the fluid mesh spacing $\dx$. 
This value must be taken sufficiently fine, but not too fine. 
If the Lagrangian spacing is too large, due to compact support of the delta functions, then points which should interact may not. 
Interpolation may be inaccurate and Eulerian points may not receive force at locations at which they would in the continuum phrasing of the problem. 
If there are large gaps in the mesh, then there is nothing to stop fluid from going through those gaps. 
This creates numerical ``leaks'' through Lagrangian structures. 
This is a major problem when trying to simulate a valve that closes.  
If the Lagrangian mesh is too fine, this increases computational cost since there are more points on which to evaluate forces. 
This may cause time-step restrictions. 
Further, the spreading operator, equation \ref{discrete_spread}, which is a linear operator, maps forces due to some local subset of Lagrangian points to some local subset of Eulerian points. 
If number of such Lagrangian points is larger than the number of Eulerian points it interacts with, then this operator has a nontrivial null space. 
This in turn implies that certain Lagrangian forces are not felt by the fluid.

Conventional wisdom among collaborators is that one should set the Lagrangian spacing to be approximately one half of the fluid mesh spacing, or $\ds = \dx / 2$ (unpublished). 
The new delta functions introduced in Section \ref{delta_function_section} have wider support than older delta functions, which may influence this wisdom. 
Some literature has suggested that a slightly wider spacing is more effective in terms of drag and torque, but these tests were in Stokes flow and concerned one-dimensional bodies in three-dimensional flow, for which leak is not relevant \cite{BRINGLEY20085397}. 
Systematic studies in a variety of conditions and Reynolds numbers would thus be beneficial, but we leave this as future work.

Thus, we tune the Lagrangian mesh to have local spacings on all discrete links of approximately $\ds_{goal} = \dx / 2$, and no larger than $\dx$. 
The fluid spacing is $\dx = $ 6cm / 128 $\approx  4.69 \cdot 10^{-2}$ cm. 
In Chapter \ref{static}, we discretize the ring with $N$ points, so the spacing of Lagrangian points on the valve ring is then approximately $\ds = 2\pi r / N$. 
We use $N = 512$ for all simulations, and $r = 2.19$, which implies that the ring spacing is $\approx 2.69 \cdot 10^{-2}$ cm.
This is very roughly half the fluid spacing, and we use this rough equivalence to determine what structure resolution corresponds with what fluid resolution.

After solving the equations of equilibrium, the links take lengths that are determined as the solution to the equations of equilibrium. 
Some may be above or below the desired Lagrangian spacing. 
If a link is longer than twice the desired Lagrangian spacing, it is split into additional links. 
That is, if a link has length $L \geq 2\ds_{goal}$, we split it into $\lfloor  L / \ds_{goal} \rfloor$ links. 
This ensures that no link is longer than $2\ds_{goal} = \dx$, the fluid mesh spacing.  
If a link in the Lagrangian mesh is shorter than the estimated spacing, we view that as acceptable and do nothing. 
Points on the partition are not the solution to the equations of equilibrium and thus can be spaced to our preferences.
They are taken with spacing $\ds_{goal} = \dx / 2$.

\begin{figure}[H]
\centering
\includegraphics[width=\textwidth]{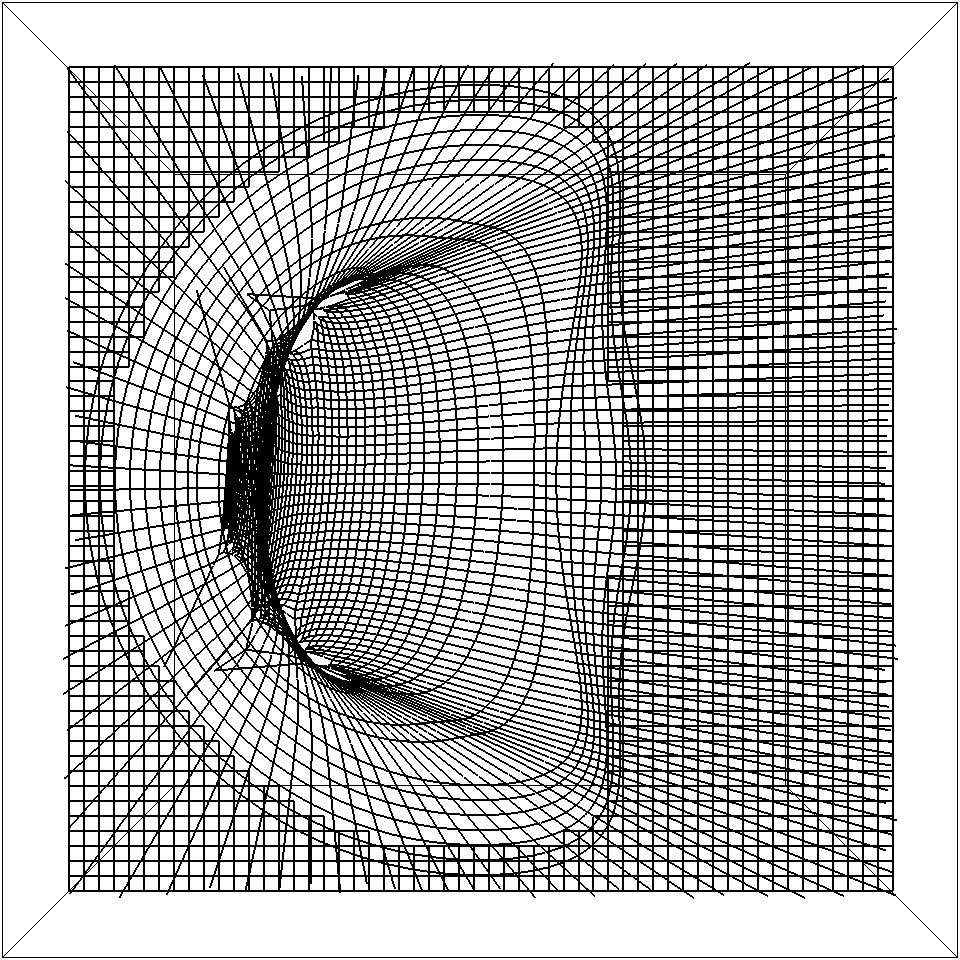}
\caption{Valve in computer test chamber, viewed from above (left atrial side). $N=128$.}
\label{top_view0000}
\end{figure}

Figure \ref{top_view0000} shows the valve in the test chamber from the left atrial side, which we refer to as ``above'' because we think of it as upstream of the valve and it is the maximum of the numerical domain. 

\begin{figure}[H]
\centering
\includegraphics[width=\textwidth]{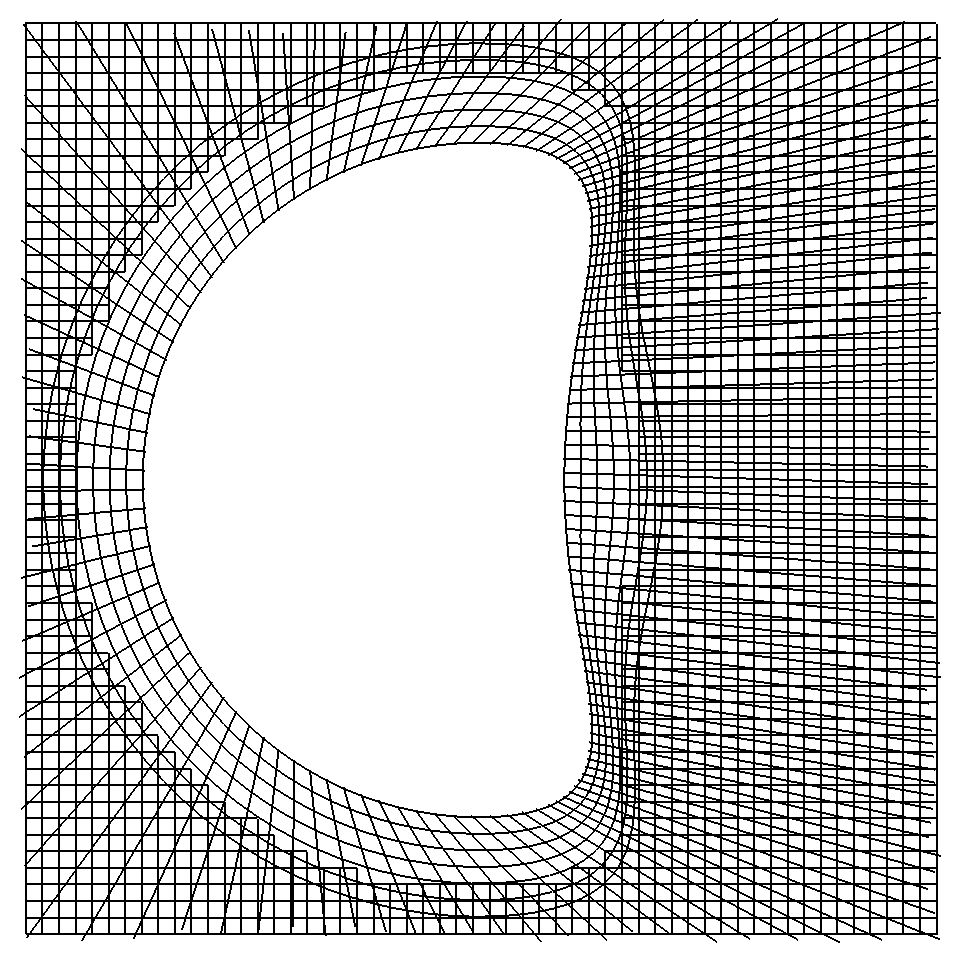}
\caption{Partition without valve, viewed from above (left atrial side). $N=128$.}
\label{top_view_mesh_only0000}
\end{figure}

The partition is shown in figure \ref{top_view_mesh_only0000}, without the valve for clarity. 
The partition also uses a fiber based elastic model. 

There are three distinct types of fibers included. 
The first family is a system of rings, the innermost of which is the mitral valve ring. 
This family of fibers models the physical mitral ring, and ensures that the model mitral ring curve shown in figure \ref{valve_ring_2d} remains approximately stationary throughout the simulation. 
Each point on the subsequent rings is propagated outwards following its current coordinates at a length $\ds$. 
That is, if the current point is $\bb p$, the subsequent point is $\bb p + \ds \bb p / | \bb p |$. 
(This $\bb p$ is not to be confused with pressure.)
More rings are added so long as the entire ring stays in the domain.

The second is a system of rays. 
These are arranged to be continuous with the radial fibers in the valve model. 
These are placed so that the radial fibers in the leaflet do not ``end'' and continue into the partition.  
This allows tension in the radial fibers to be supported by tension and target point forces (described in Section \ref{papillary_target}) in the partition. 
In older models without this family, we observed radial fibers ``pulling away'' from the ring. 
Adding these fibers effectively prevents this problem. 
We compute the angle at which the ray should propagate as follows. 
Let $\bb r$ be the point on the ring, and $\bb x$ be the previous point on the radial fiber that is being continued. 
The point $x$ is rotated around the local normal to the ring at $r$ such that it lies in the plane $z = 0$ and is inside the valve ring. 
Then, this point is reflected across the local normal to be outside the ring. 
The increment that this reflection determines is scaled to have magnitude $\ds$. 
Points are placed with spacing $\ds$ along this ray so long as they remain in the domain.

Finally, a certain distance from the valve ring, we place a two-dimensional Cartesian mesh.
This serves to prevent leaks in the corners of the box, where the previous fiber families have inadequate coverage. 
The mesh begins $\dx/2$ from the no-slip walls on the side of the chamber. 
Links in this mesh are included if they are outside of the fourth hoop placed; that is, if they are outside the polygon defined by 
\begin{align}
\bb p + 4\ds  \frac{ \bb p}{| \bb p |} 
\end{align} 
for all points $\bb p$ on the ring. 

All points in the partition are held approximately still with forces due to \emph{target points}. 
The details of how such forces are applied are described in Section \ref{papillary_target}. 

\section{Papillary muscles and target points}
\label{papillary_target}

Both the partition and the model papillary tips are held in place with \emph{target points}. 
Target points provide a method to specify an approximate location for an immersed boundary point. 
An immersed boundary point is connected to the target point, and a force is applied to keep them close together. 
The location of the target points is prescribed, and for the papillary muscles, time dependent. 
This is an example of a penalty method, which approximately enforces a boundary condition. 
Note that enforcing these locations exactly as boundary condition would  need to be phrased as a constraint, and thus complicate the equations dramatically. 

Suppose that $\bb X_{node}$ denotes the position of an immersed boundary point. 
This point interacts with the fluid through equations \ref{interpolate} and \ref{spreading}. 
Let $\bb X_{target}$ denote the target location.
The target point does not interact with the fluid directly. 
The location of this point is prescribed as a boundary condition, as is its derivative in the time dependent case. 
It appears in the equation for computing Lagrangian force, \ref{nonlinear_force}, but not elsewhere. 

Target points exert force as follows. 
Each target point is connected to an immersed boundary point with a Kelvin-Voigt viscoelastic element \cite{howell}.
This exerts tension as a linear spring, chosen to have zero rest length, in parallel with a dashpot. 
That is, the tension is the sum of a term proportional to length $L$ of the element, plus a term proportional to the derivative of length,
\begin{align}
T = k L + \eta \frac{dL}{dt}.  
\label{target_force}
\end{align}
In the immersed boundary framework, the relevant length for use in equation \ref{target_force} is given by 
\begin{align}
 L = | \bb X_{target} - \bb X_{node} |  . 
\end{align} 
Take the time derivative of the expression 
\begin{align}
 L^{2} = | \bb X_{target} - \bb X_{node} |^{2}  
\end{align}  
to obtain 
\begin{align}
L \frac{dL}{dt} = \sum_{\alpha = 1}^{3}  \left( X^{(\alpha)}_{target} -  X^{(\alpha)}_{node} \right) \left(  \frac{d X^{(\alpha)}_{target} }{dt} - \frac{d X^{(\alpha)}_{node} }{dt} \right),   
\end{align} 
which we use to calculate the time derivative of the spring length. 
Since we prescribe the position of $\bb X_{target}$, we know both its position and derivative. 
The time derivative of the Lagrangian variable $\bb X_{node}$ is available from interpolation of the velocity field, following the interpolation formula \ref{interpolate}; we denote it $\bb U_{node}$. 
Because this force is computed with known Lagrangian velocity, it is explicit in time in the discretized version of the model.

The scalar tension associated with the dashpot is thus computed as 
\begin{align}
\eta \left ( \frac{  \bb X_{target} - \bb X_{node}  }{L} \right)  \cdot  \left(  \frac{d \bb X_{target} }{dt} -  \bb U_{node} \right) 
\end{align} 
The full force, including the spring force, is then given 
\begin{align}
F &= \left( k  L  +  \eta \left( \frac{  \bb X_{target} - \bb X_{node}  }{L}  \cdot  \left( \frac{d \bb X_{target} }{dt} -  \bb U_{node} \right) \right)  \right)  \frac{ \bb X_{target} - \bb X_{node}}{L}   \\ 
   &= k (\bb X_{target} - \bb X_{node})   \\
   &\; \;  +  \eta \left( \frac{  \bb X_{target} - \bb X_{node}  }{L}  \cdot  \left( \frac{d \bb X_{target} }{dt} -  \bb U_{node} \right) \right) \frac{ \bb X_{target} - \bb X_{node}}{L}   \nonumber
\end{align}
Note that the dashpot force depends on the difference in the prescribed target velocity and the fluid velocity at the node. After scaling by the damping coefficient $\eta$, this quantity is then orthogonally projected onto a vector between the target point and node.

The partition is not designed to mimic any particular material. 
Similarly, we are not modeling any details of the papillary muscle mechanics. 
Thus, we set both the spring and damping coefficients empirically in both the partition and the papillary muscle target points.  
Setting the papillary damping coefficient correctly, in particular, proved important to getting good closure and vibrations that match observations.

The papillary location in systole is determined from MRI data, as described in Section \ref{MRI}. 
The location in diastole is chosen by eye to allow the anterior chordae to be relatively slack, and anterior leaflet to move clear of forward flow. 
When moving from systolic to diastolic positions and back, all papillary target points move with the same velocity. 
The motion between the positions is piecewise linear, so the prescribed velocity is piecewise constant.

\section{Thickening the structure}
\label{layers}

\begin{figure}[H]
\centering 
\includegraphics[width=\textwidth]{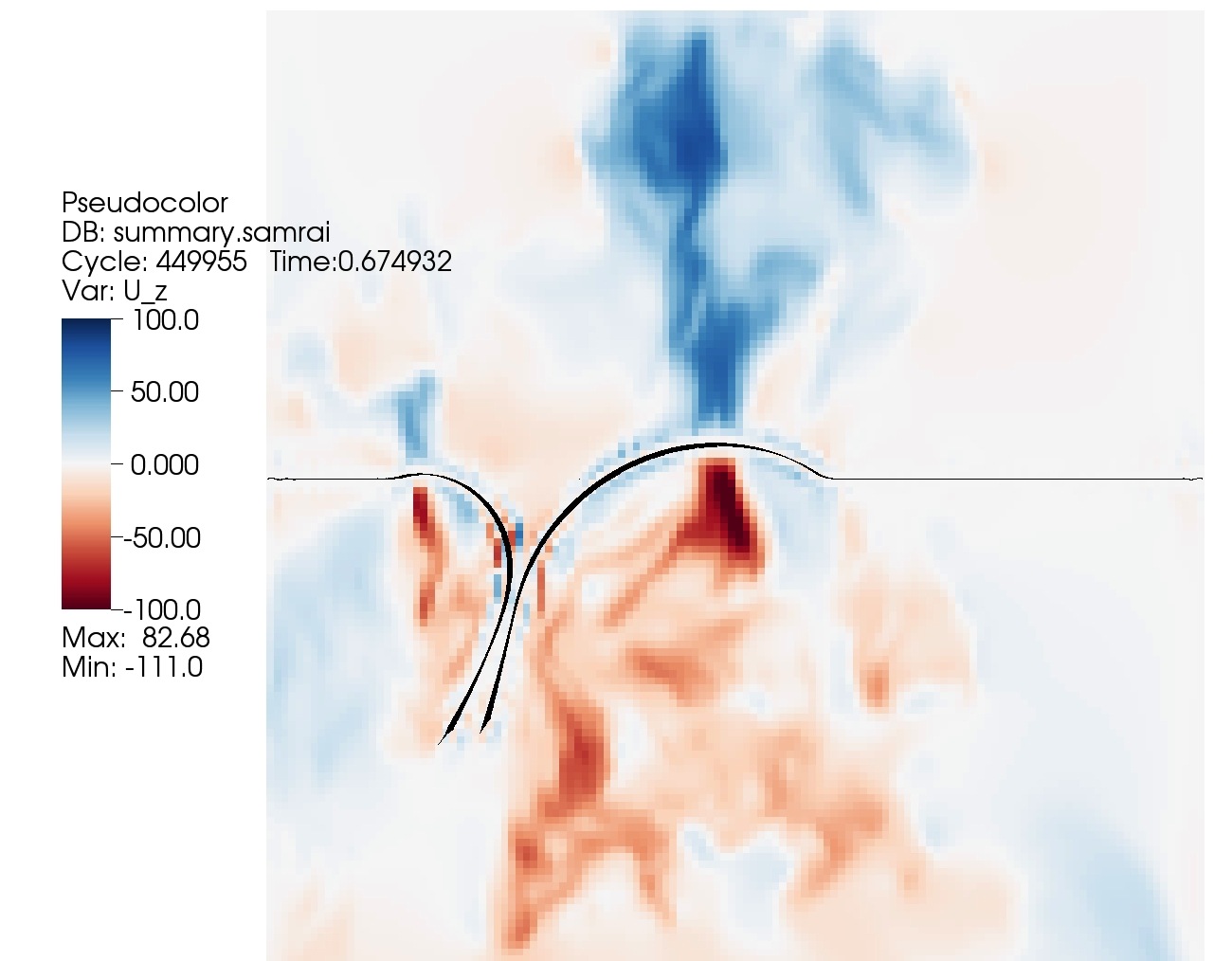} 
\caption{Grid aligned artifacts on a mitral valve simulation with an infinitely thin (two-dimensional surface) valve model. This slice view shows the $z$ component of velocity and the model valve.}
\label{artifact_valve}
\end{figure}

In many simulations, we noticed a strange spurious flow.  
At locations at which the valve is aligned with the coordinate direction, a jet appears going away from the valve in both directions. 
After examining flow fields, we could find no physical explanation for such flows, which suggests that the problem may be numerical. 
An example is shown in figure \ref{artifact_valve}, which shows a slice view of the $z$-component of velocity during systole. 
Approximately centered in the frame, the model valve has a tangent plane that is approximately flat; that is, it is aligned with the $x,y$ plane. 
Near this point, a jet shoots away from the membrane.

\begin{figure}[H]
\centering 
\includegraphics[width=\textwidth]{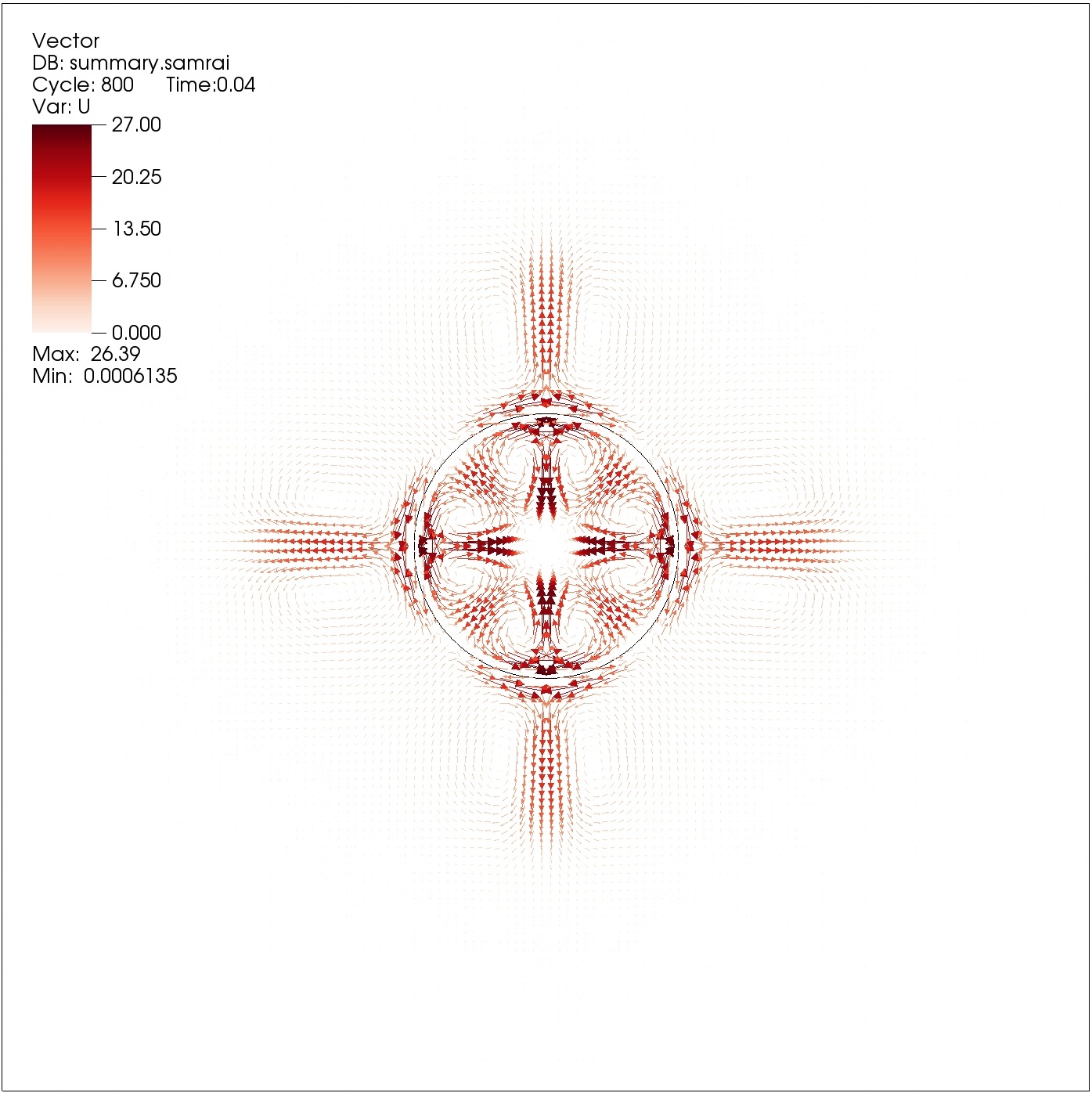} 
\caption{A thin membrane under tension. Large grid aligned artifacts are clearly visible. Note that all flow is numerical error, since the exact solution to this problem is $\bb u = 0$.}
\label{artifact_thin}
\end{figure}

This phenomenon is not particular to our valve model; we found out that it is easy to create such flows. 
Consider a one-dimensional circular membrane of radius $1/4$ cm in a two-dimensional flow under tension. 
The fluid domain is a 1cm by 1cm box with periodic boundary conditions. 
This has an exact solution with $\bb u = 0$ and piecewise constant pressure.
The membrane is discretized as springs with zero rest length and initial spacing $\ds = \dx / 2$. 
Springs with zero rest length exert tension as 
\begin{align}
T = k L, 
\end{align}
where $L$ is the length of the spring in question. 
Spring constants are set to the arbitrary value of $k = 10^{6}$ Dynes/cm, which gives a pressure difference of approximately 23.4 mmHg. 
Precisely the same type of artifact occurs, shown in figure \ref{artifact_thin}. 
It is repeated at locations in which the membrane aligns with the grid, here forming an approximate four-fold symmetry of artifacts. 
Further, we observed similar effects in unrelated physical simulations involving flows in arteries (unpublished coursework).  
In the artery simulations, we moved the structure by various angles, yet the artifact always remains aligned with the grid. 
It seems that any situation involving a discontinuous pressures across an infinitely thin membrane is prone to these issues. 

This effect is severe and needs to be mitigated. 
It persists with changes in delta functions, discretization of the advective term and time step (though the precise velocity field may change slightly).  
One might suppose that the problem is caused by the Cartesian product nature of the discrete delta function, as shown in equation \ref{delta_cartesian_product}, because this is a phenomenon that is aligned with Cartesian grids. 
However, the effect persists with a radially symmetric delta function. 
Increasing the resolution decreases the magnitude of the artifact and makes it more localized, but it remains at all reasonable fluid resolutions for use in valve simulations. 

\begin{figure}[H]
\centering 
$ \begin{array}{cc}
\includegraphics[width=.475\textwidth]{1_shell_bad.jpeg} 
\includegraphics[width=.475\textwidth]{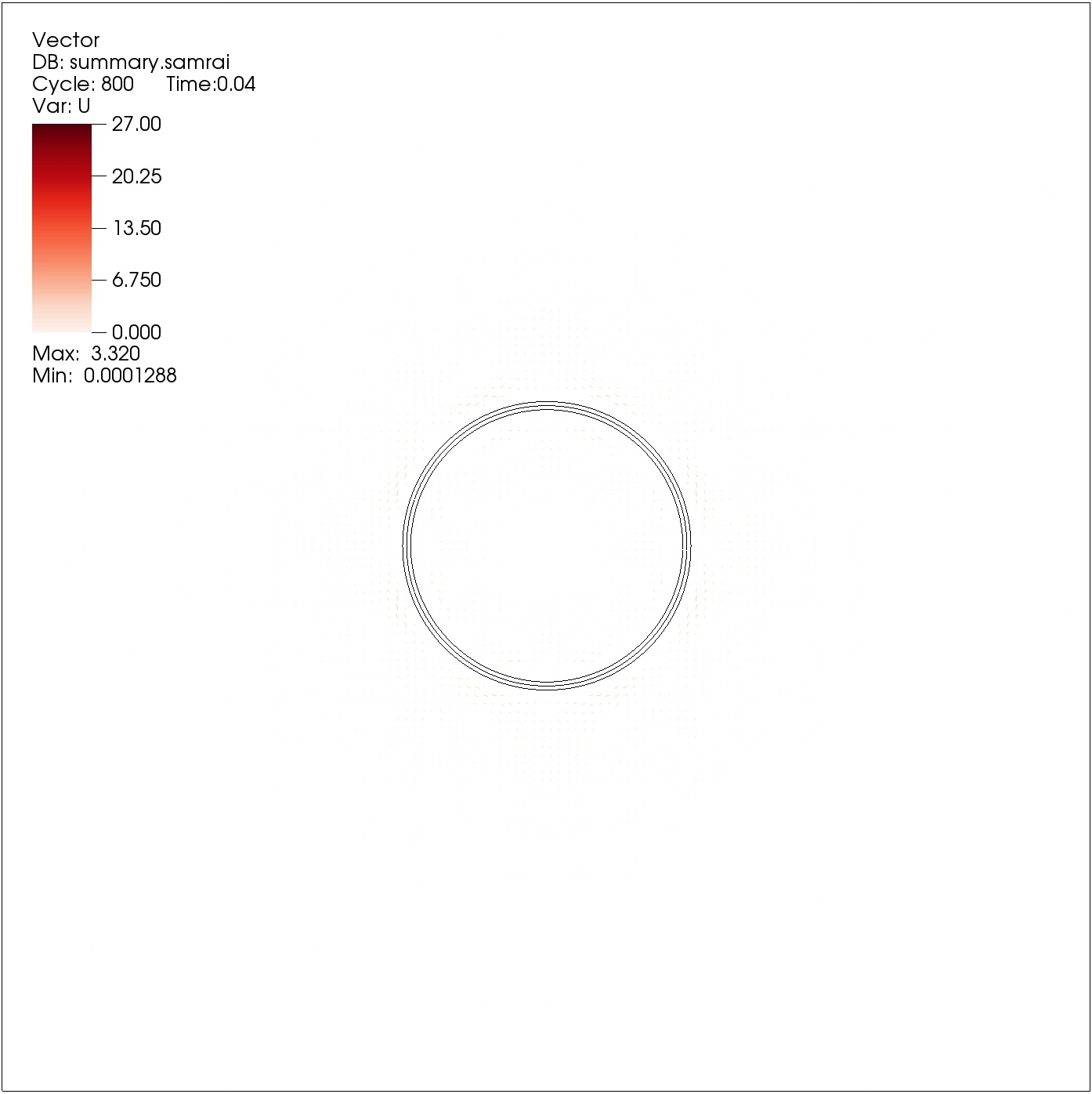} 
\end{array} $
\caption{Comparison of a simulation with a thin structure (left) and a thickened structure (right) under tension. Artifacts with the thickened structure are an order of magnitude smaller and are invisible at this scale.}
\label{artifact_shell}
\end{figure}

We have found, however, that thickening the structure reliably decreases the size of the artifact by an order of magnitude or more. 
Figure \ref{artifact_shell} shows the result shown in figure \ref{artifact_thin} again, along with the same experiment with a thickened shell. 
Here we replace the single ring under tension with three rings with one third the strength. 
With the same scales in both panel no visible artifact appears in the thickened model. 
	
This suggests that all immersed structures that support a pressure should be thickened to avoid this type of problem. 
To quantify this, we run a number of simulations with various numbers layers and spacings in two-dimensional flows. 
This will advise us on how to avoid such numerical artifacts on a real flow. 
Table \ref{shell_test} shows the numerical error in the maximum norm for these experiments. 
The minimum spacing we investigate is one half the Eulerian mesh width, or $\dx/2$, because that is the target Lagrangian mesh width in general, and we do not wish to have this dimension be much thinner than any other mesh spacings. 
This test reveals that three layers spaced $\dx/2$ apart reduces the error the most. 

\begin{table}[H]
\centering 
\begin{tabular}{ c | c | c | c | c | c | c | }
	\multicolumn{2}{c}{ }  & \multicolumn{5}{c}{Number of layers}  \\ 
\cline{3-7}
\multicolumn{2}{c|}{ } & 1 & 2 & 3 & 4 & 5 \\ 
\cline{2-7}
\multirow{3}{*}{Spacing} 
& $ \dx / 2$ & 26.10 & 15.47 & 3.38 & 4.22 &  4.72 \\ 
\cline{2-7}
&  $ \dx $ & 26.10 & 14.80 & 9.54 & 6.79  &  5.83 \\ 
\cline{2-7}
&  $ 2 \dx $ & 26.10  & 13.81  &  9.87 & 7.85 & 6.93 \\ 
\cline{2-7}
\end{tabular}
\caption{Maximum modulus of $\bb u$, which is equal to the maximum norm of the numerical error, for various spacings and number of layers}
\label{shell_test}
\end{table}

Thus, we thicken the entire Lagrangian structure in all mitral valve simulations. 
We add additional copies of the entire valve structure and partition, simply placing the additional copies the desired spacing uniformly offset in the $z$ direction. 
We select three layers spaced $\ds = \dx/2$ apart. 
Conveniently, this corresponds to the target Lagrangian mesh width through the entire valve structure. 
Spring constants are uniformly reduced by the number of copies, three, throughout.  
At full resolution, this models the Lagrangian structure as approximately $0.07$ cm thick.
Also conveniently, this is approximately the thickness of a mitral valve leaflet. 

\begin{figure}[H]
\centering 
\includegraphics[width=\textwidth]{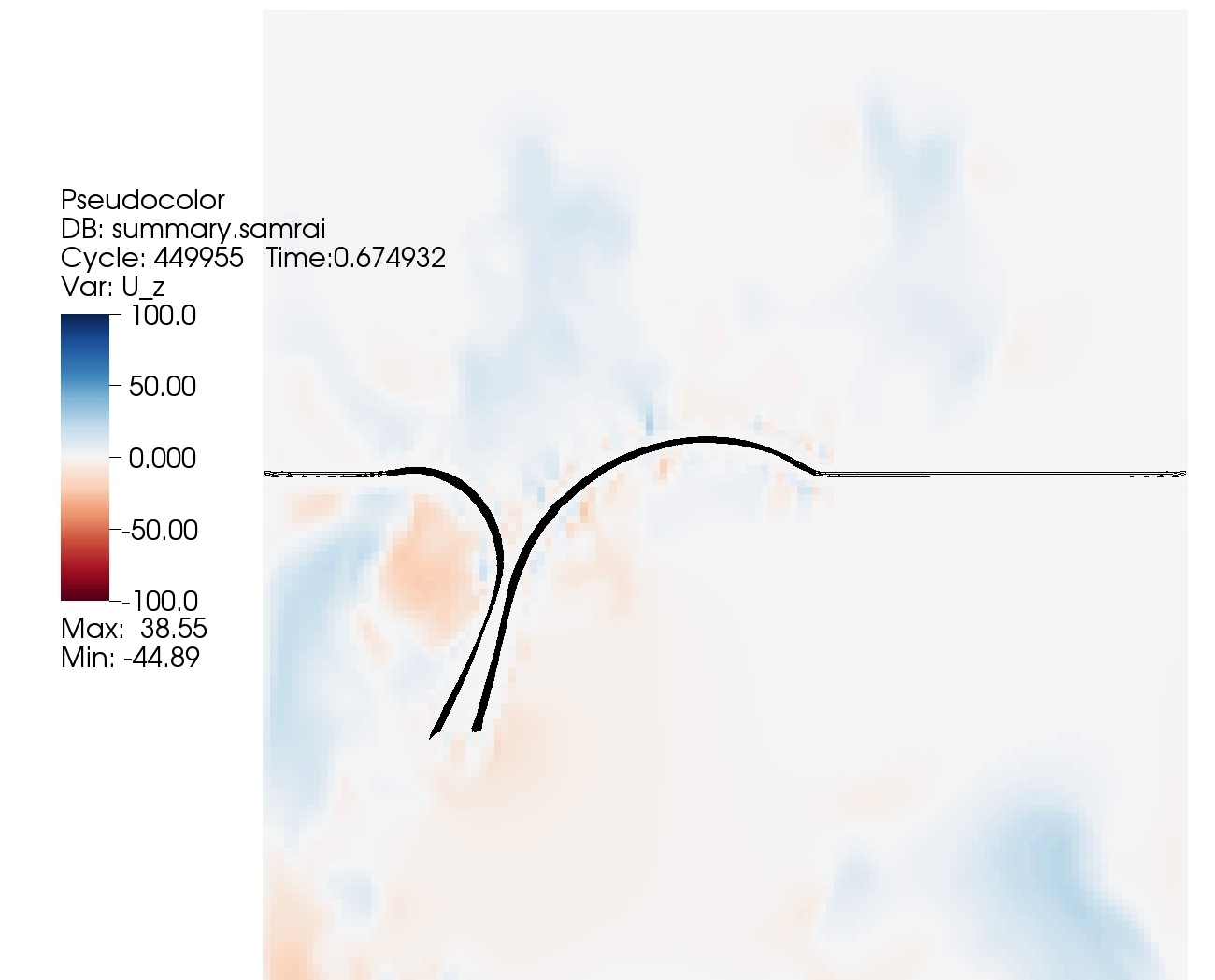} 
\caption{Reduction of artifacts with thickened model valve. 
Note that this simulation is otherwise identical to that of figure \ref{artifact_valve}, and both figures show the same time step.}
\label{artifact_gone_thickened}
\end{figure}

This dramatically reduces the appearance of the artifact, as shown in figure \ref{artifact_gone_thickened}. 
The field is qualitatively different, with much lower magnitude throughout. 

The layers are placed close together, and because they move in a continuous velocity field, they may be expected to stay somewhat close together. 
However, we observed that oscillations during closure grew as the simulation progressed (see figure \ref{flux_no_layers}), and in some locations on the valve mesh it appeared that the layers were not well aligned. 
Thus, we tie the layers together with linear springs of rest length $\ds$. 
We determine the constants, which are uniform, empirically by picking the largest value that does not contribute to time-step restrictions. 
This only serves to keep the layers together, is not intended to represent specific mechanical properties of the anatomy.  
As the mesh is refined, there are four times as many springs in the leaflet mesh. 
We wish to have the same total force with the same strain on each link, so the relative spring constants are reduced by 4 when $N$ is increased by 2. 
The rest lengths also reduce by $2$ when $N$ increases by 2, so the absolute spring constants reduce by 2. 
The chordae are modeled as one-dimensional strings, rather than a surface, but we use the same scaling as on the surface for simplicity. 
Layers on the partition are not connected, as they are held in place by target points.

\section{A discrete delta function with three continuous derivatives}
\label{delta_function_section}

In the discretized version of the interaction equations, \ref{interpolate} and \ref{spreading}, the integral convolutions are replaced by discrete convolutions taken as sums, shown in equations \ref{discrete_interpolate}  and \ref{discrete_spread}. 
The delta function is replaced by a discrete approximation to the delta function. 
We take the approximate delta function to be the Cartesian product of three one-dimensional delta functions, each of which is a scaling of a function $\phi : \R \to \R$ which is nonsingular. 
This is given as 
\begin{align} 
\delta_{h}(\bb x) = \frac{1}{h^{3}} \phi(x/h) \phi(y/h) \phi(z/h)  , 
\label{delta_cartesian_product}
\end{align}
where $h = \dx$ is the Cartesian grid spacing.

Many versions of $\phi$ have been proposed. 
Before we derive a new function $\phi$ in this section, we will discuss the motivation behind various postulates. 
Most functions $\phi$ possess \emph{moment conditions}, which guarantee that interpolation and spreading are exact for constants and linear functions.
This implies they are second order accurate when interpolating smooth functions.
The zeroth and first moment condition are given as 
\begin{align}
\sum_{j} \phi(r - j) &= 1  \label{zero_moment}  \\ 
\sum_{j} (r-j) \phi(r - j) &= 0 
\end{align}
where all sums are over $j \in \Z$ and the conditions hold for any $r \in \R$.

The zeroth moment condition implies that $\int \phi = 1$. 
To see this, suppose that $\phi$ is continuous and supported on the interval $[-L,L]$ where $L \in \Z$. 
Then 
\begin{align}
\int_{\R} \phi(x) \; dx 
&= \sum_{k=-L}^{L-1} \int_{k}^{k+1} \phi(x) \; dx    
= \sum_{k=-L}^{L-1} \int_{0}^{1} \phi(r + k) \; dr \\ 
&= \int_{0}^{1} \sum_{k=-L}^{L-1} \phi(r + k) \; dr 
= \int_{0}^{1} \sum_{j \in \Z} \phi(r - j) \; dr 
= \int_{0}^{1} 1 \; dr 
= 1
\end{align}
This implies that the family of functions $\delta_{h}$ integrates to one for all positive $h$, and that convolution with $\delta_{h}$ is an approximation to the identity \cite{stein2005real}.  
Note that the same argument holds for any nonnegative $\phi$ without compact support, using the monotone convergence theorem to swap the sum and integral, or any integrable $\phi$ and using the dominated convergence theorem. 
However, in this project we only use continuous, compactly supported $\phi$.

Some discrete delta functions used in the immersed boundary method satisfy an even/odd condition. 
This replaces the zeroth moment condition, equation \ref{zero_moment}, with the two conditions 
\begin{align}
\sum_{j \text{ even}} \phi(r - j) &= 1/2  \\ 
\sum_{j \text{ odd}} \phi(r - j) &= 1/2  . 
\end{align}
These equations together imply that the zeroth moment condition is satisfied. 
They were introduced for use with collocated grid methods. 
In such methods, two-point-wide first derivative stencils are used; that is, 
\begin{align}
\frac{\p u(x)}{\p x} \approx \frac{u(x + \dx) - u(x - \dx)}{2\dx} . 
\end{align}
Grid functions that takes constant values on even or odd indexed grid points are in the null space of this operator. 
Any generic function and thus forces from immersed structures generally have a nonzero component of ``checkerboard modes,'' or functions that are constant on even or odd indexed grid points. 
Because they lie in the null space of derivative operators, their contribution to pressure is not applied correctly. 
Pressures and subsequently velocities then acquire spurious checkerboard oscillations, and the overall solutions become meaningless. 
We do not use collocated grid methods in this project. 

\begin{figure}[H]
\centering
\includegraphics[width=\textwidth]{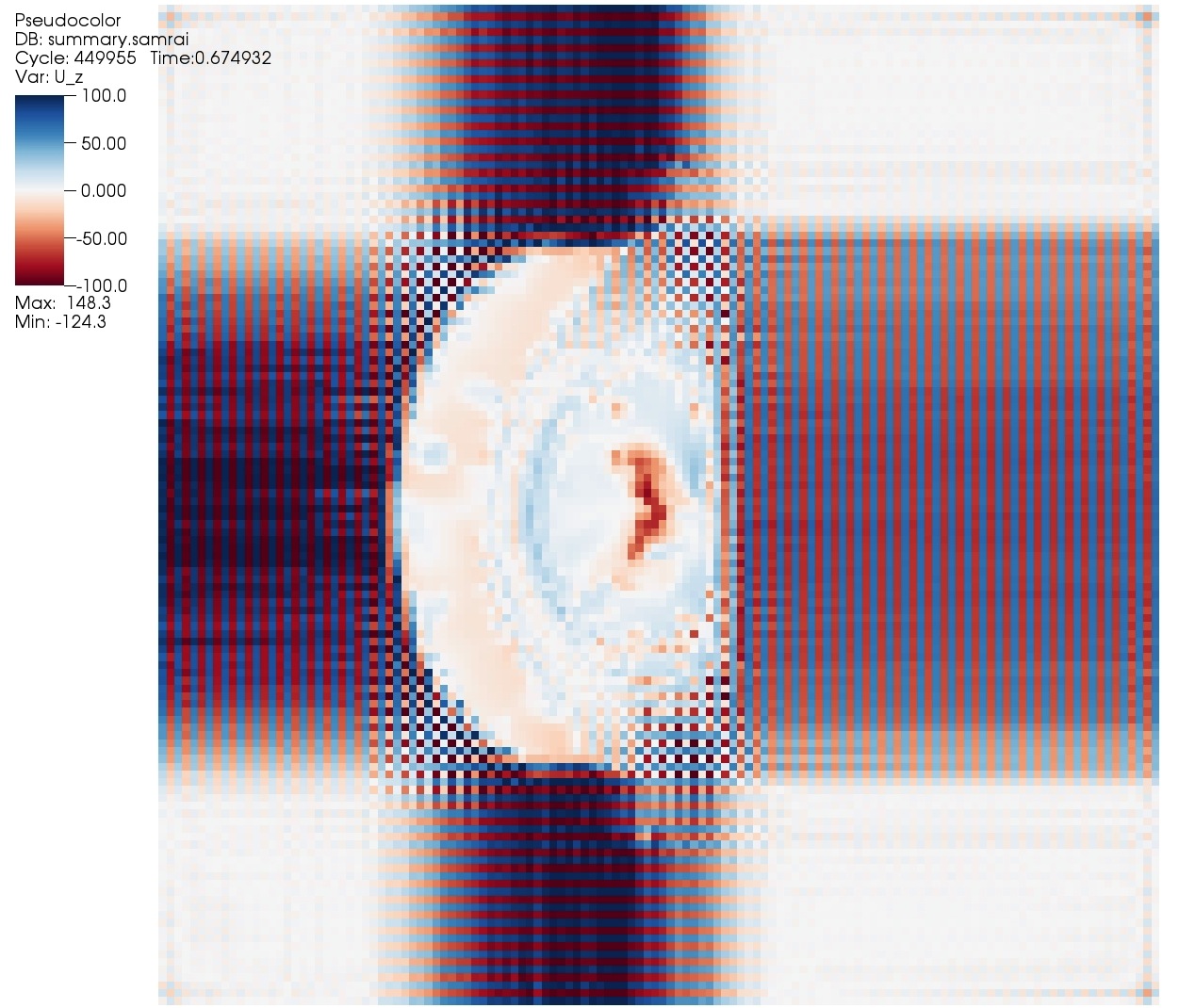}
\caption{Even/odd artifacts in $z$ component of velocity field. This slice view shows the $z = 0$ plane, so this is the velocity field at approximately the location of the valve partition. This simulation uses a delta function with the even/odd condition, but is otherwise identical to that of figure \ref{artifact_valve}, and shows the same time step.}
\label{checker_artifacts_top.jpeg}
\end{figure}

\begin{figure}[H]
\centering
\includegraphics[width=\textwidth]{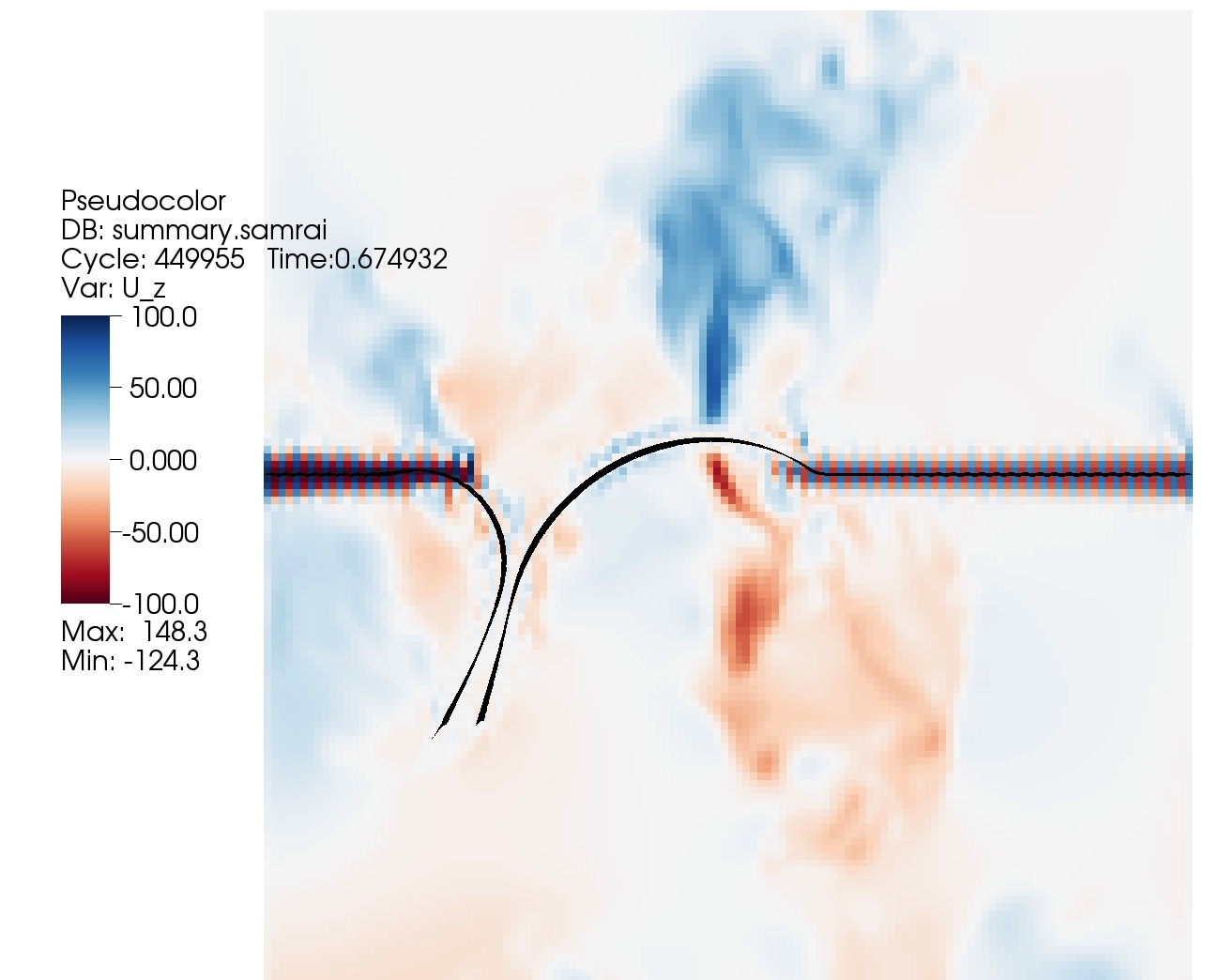} 
\caption{Even/odd artifacts in $z$ component of velocity field around the partition viewed from the side. Note the blue/red alternating pattern in a neighborhood of the partition, representing changing signs in the velocity field. This is the same simulation as shown in figure \ref{checker_artifacts_top.jpeg}, and is identical to that of figure \ref{artifact_valve} except for the delta function.}
\label{checker_artifacts_side.jpeg}
\end{figure}

Staggered grid methods do not suffer from these problems, and thus we have no need for an even/odd condition. 
Further, we discovered that the standard four point delta function \cite{ib_acta_numerica} can actually \emph{create} checkerboard type artifacts. 
When the valve partition, discussed in Section \ref{setup}, supports a pressure, sometimes there is a localized even/odd checkerboard mode. 
This is shown in figures \ref{checker_artifacts_top.jpeg} and \ref{checker_artifacts_side.jpeg}, which show dramatic even/odd artifacts. 

Consider the following one-dimensional example to explain this phenomenon. 
Suppose that the discrete velocity field is given 
\begin{align}
u_{j} = \alpha (-1)^{j}  . 
\label{even_odd_function}
\end{align}
Let $x_{j} = j h$, so $x_{j}/h = j$, where $h$ denotes the Eulerian grid spacing. 
Let $U( X)$ denote the interpolated field evaluated at arbitrary $X$. 
(In this example it is unimportant whether this is actually a Lagrangian point). 
Apply the one-dimensional discrete interpolation formula, shown in three dimensions in equation \ref{discrete_interpolate}. 
\begin{align}
U(X) 
&= \sum_{j } u_{j}  \phi( j - X/h) \\
&= \sum_{j  \text{ even}} \alpha \phi( j - X/h)  - \sum_{j  \text{ odd}} \alpha \phi( j - X/h)  \nonumber \\ 
&= 0 \nonumber
\end{align}
Thus, the Eulerian velocity field may be arbitrarily large in any norm, yet it interpolates to zero. 
(This example is due to Boyce Griffith.)

Now suppose that target points are used to hold a horizontal structure (such as the valve partition) in place, and that somehow a field with $z$ component of the form shown in equation \ref{even_odd_function} appears in the Eulerian velocity. 
The target point forces try to enforce that the Lagrangian velocity is zero, but have no direct relationship to the local values of the Eulerian field.
This explains why we observe target points moving very little, but large nonzero Eulerian velocities in a neighborhood of the target points. 
How the Eulerian velocity fields develop this pattern in the first place is an open question. 
Experience shows that if a large pressure is imposed across a immersed structure that is flat and grid aligned then such velocity fields are likely to appear. 
The valve partition is an example of such a structure and this is reproducible with a one-dimensional flat membrane in a two-dimensional flow.

This demands a delta function without an even/odd condition. 
One such function is the three point delta function, which is derived much like the four point in \cite{ib_acta_numerica} but lacks an even/odd condition. 
Tests run by our colleague Yuanxun Bao suggest that this function has the worst translation invariance properties of any delta functions we have used \cite{IB5_arxiv}. 
Additionally, there is a new version of the immersed boundary method that interpolates Eulerian velocity field a field defined everywhere that is divergence free in the continuous sense \cite{div_free_ib}. 
This method involves taking discrete convolutions with the derivatives of the approximate delta function, so we wish to have an approximate delta function that has at least two continuous derivatives. 
The three point delta function has only one continuous derivative, and is not twice differentiable.

\begin{figure}[H]
\centering
$ \begin{array}{cc}
\includegraphics[width=.475\textwidth]{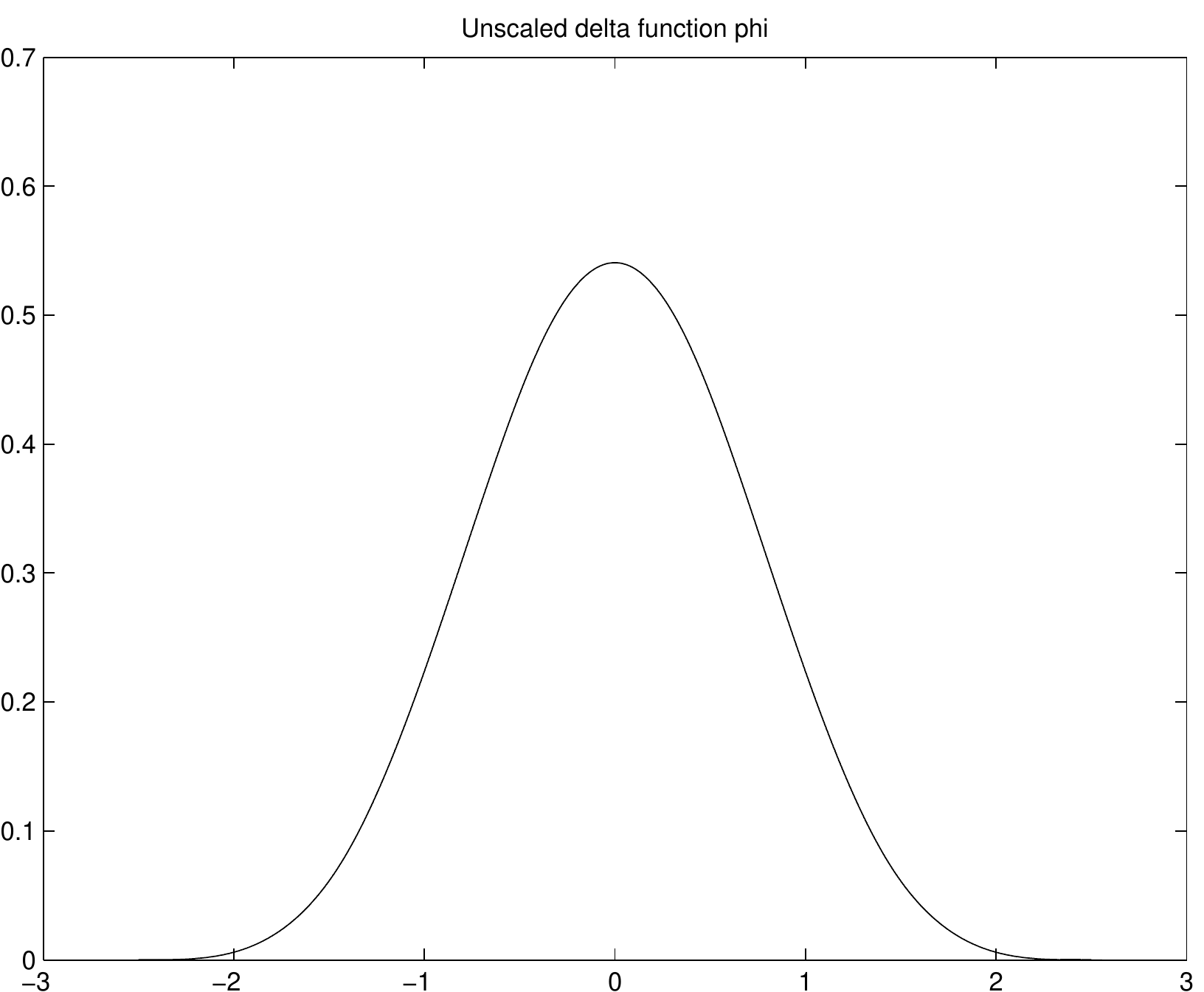}  & \includegraphics[width=.475\textwidth]{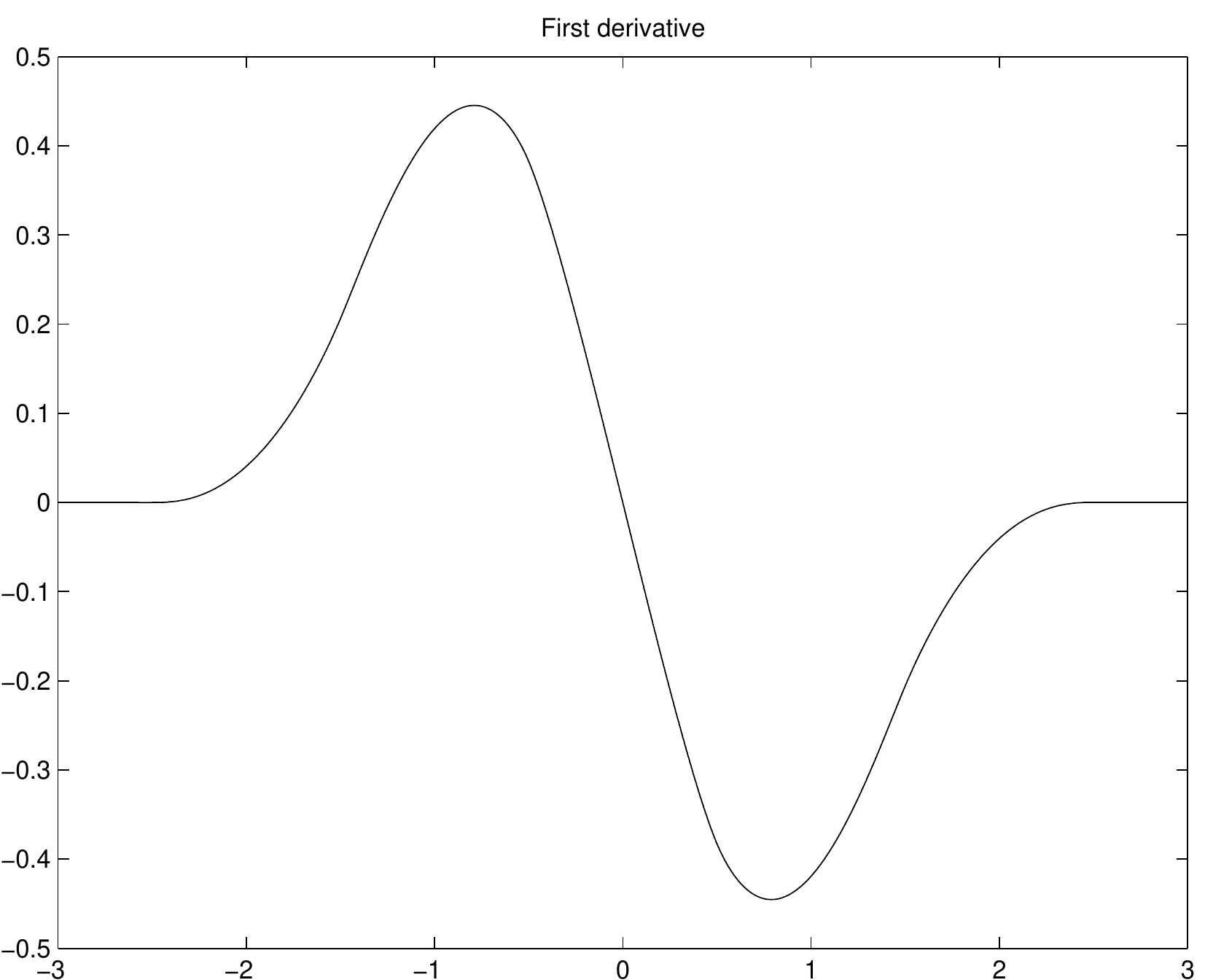}  \\ 
\includegraphics[width=.475\textwidth]{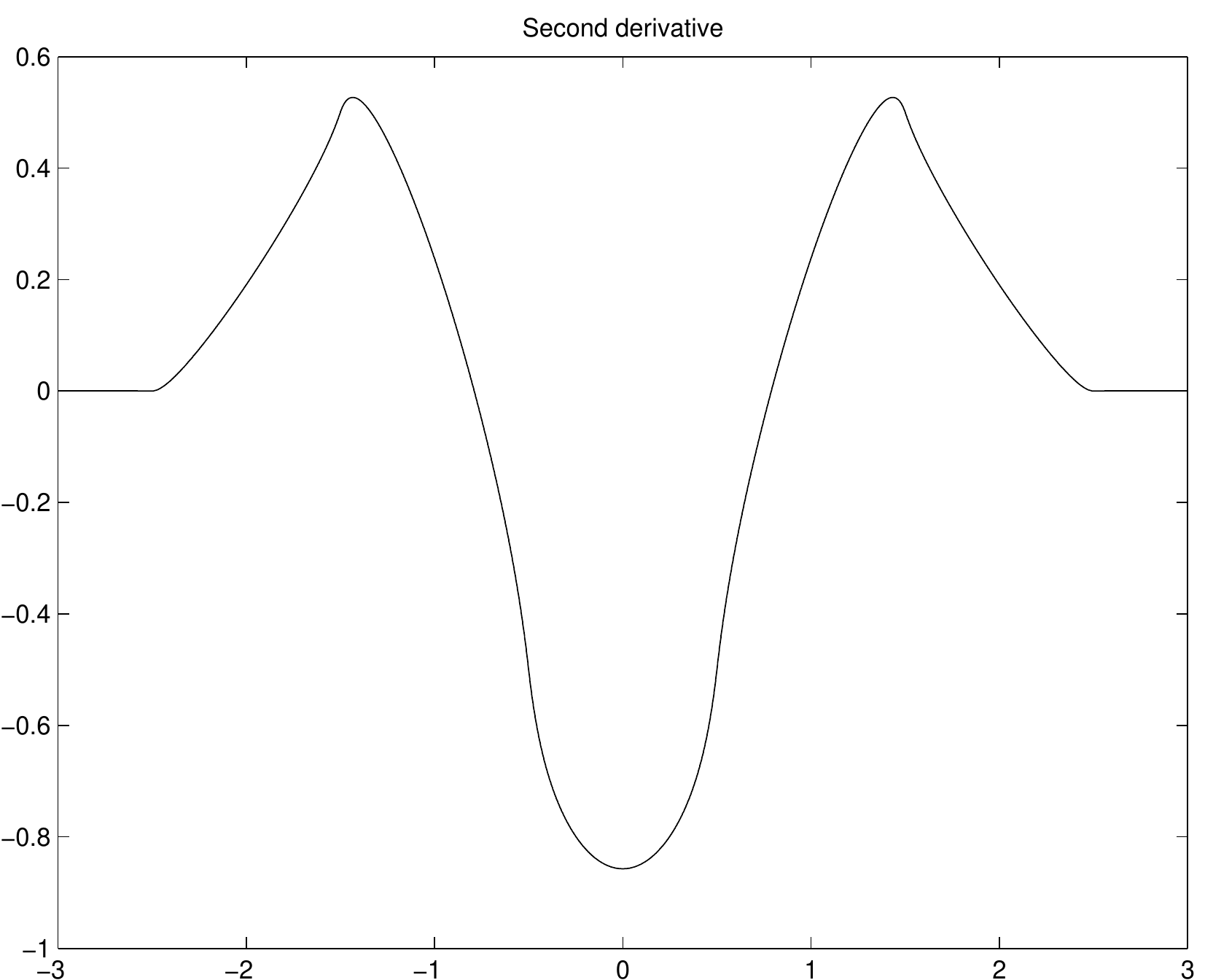}  & \includegraphics[width=.475\textwidth]{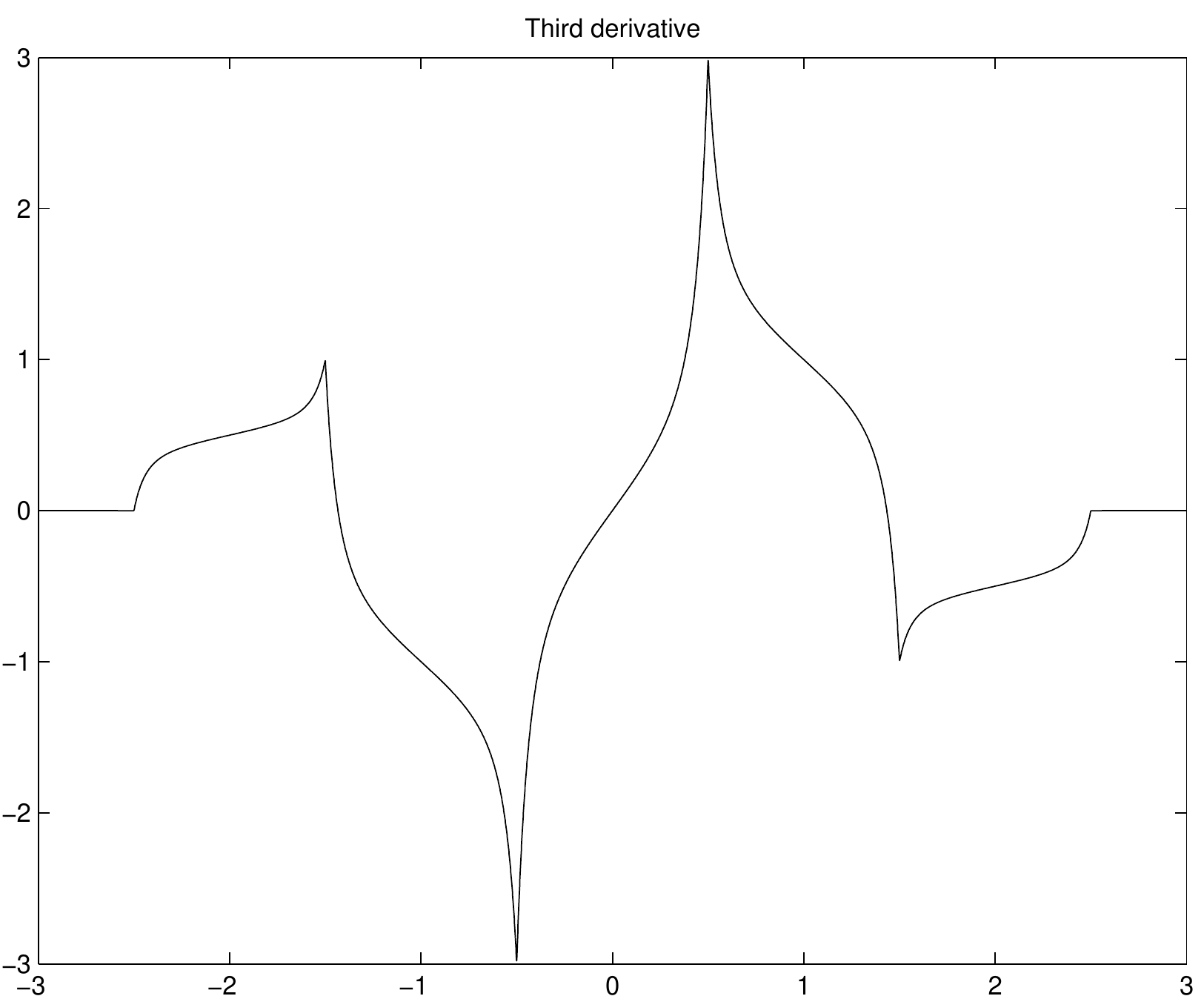}  \\ 
\end{array} $
\caption{The five point $\phi$ function and its derivatives.}
\label{delta5plots}
\end{figure}

Thus, we derive a new approximate delta function. 
The function $\phi$ is nonzero on at most five grid points, and has three continuous derivatives. 
We show plots of the function and its derivatives in figure \ref{delta5plots}. 
The derivation is also summarized in \cite{IB5_arxiv}, which was updated from the published version to include this delta function. 
Our colleague's test shows that the translation invariance of the new function drastically improves translation invariance over the conventional three and four point delta functions. 
Thus, we use it on simulations throughout.

The derivations is as follows. 
We assume that the support of $\phi$ is $|r| < 5/2$, so that there are a total of $5^{3}$ grid points at which $\delta_{h}(\bb x - \bb X)$ may be nonzero, for any shift $\bb X$. 
Although we only assume continuity of $\phi$ itself, it will turn out that $\phi$ has three continuous derivatives. 
The function $\phi$ has zeroth moment equal to 1, first moment equal to zero, second moment equal to a positive constant (independent of shift relative to the grid) to be determined, and third moment equal to 0. 
As in the case of many other IB delta functions, we also impose a sum-of-squares condition, equation \ref{sumsquares}, which closes the system and is indented to achieve approximate translation invariance for the IB method as a whole. 

The 5-point delta function described here and in \cite{IB5_arxiv} is closely related to a 6-point delta function, also described in \cite{IB5_arxiv}, that satisfies the even/odd condition as well as the postulates stated below. 
For the reasons discussed above, we consider it better \emph{not} to satisfy the even/odd condition, and it is certainly a practical advantage for $\delta_{h}$ to have a support containing $5^{3} = 125$ grid points instead of $6^{3} = 216$ grid points. 

The mathematical postulates that define the function $\phi(r)$ used in this thesis are 
\begin{align}
\phi(r) &\in C^{0}(\R) \\ 
\phi(r) &= 0 \quad \quad   (|r| \geq 2.5)       \\
\sum_{j} \phi(r - j) &= 1 \label{zeromoment5pt}\\ 
\sum_{j} (r-j) \phi(r - j) &= 0 \\ 
\sum_{j} (r-j)^{2} \phi(r - j) &= K \quad \quad (K \geq 0)\\ 
\sum_{j} (r-j)^{3} \phi(r - j) &= 0 \label{thirdmoment5pt} \\
\sum_{j} \phi(r - j)^{2} &= C(K) \label{sumsquares}
\end{align}
All sums are over $j \in \Z$, but the moment conditions and sum of squares condition, equations \ref{zeromoment5pt}-\ref{sumsquares}, must hold for all real values of $r$. 
The function $C(K)$ is to be determined, as is the value of $K$. 
We use the computer algebra system \emph{Mathematica} \cite{Mathematica} to symbolically solve linear systems and plot polynomials. 

As we shall see, the above postulates uniquely define $\phi$ if we regard the parameter $K$ as given and if $K$ is sufficiently small. 
There is, moreover, a particular value of $K$ at which $\phi$ has three continuous derivatives, and this is the value of $K$ that we shall use. 
This special value of $K$ can also be characterized as the smallest value of $K$ for which $\phi(r)$ is nonnegative for all $r$.

Suppose that $r \in [-1/2, 1/2]$.
Then there are at most five values of $j$ for which $\phi(r-j)$ is nonzero. 
They are given as 
\begin{align} 
\phi(r-2) , \phi(r-1) , \phi(r) , \phi(r+1) , \phi(r+2) . 
\end{align}
The moment conditions are four linear equations in these five unknowns, plus an unknown parameter $K$. 

Consider the special case that $r = 1/2$, so $\phi(r+2) = \phi(5/2) = 0$ and the number of unknowns is reduced to four. 
These equations can be solve symbolically in terms of the unknown parameter $K$. 
The solution is 
\begin{align}
\phi \left(-\frac{3}{2} \right) = \frac{1}{16} (4 K-1)   \\   \label{phivalues}
\phi \left(-\frac{1}{2} \right) = \frac{1}{16} (9-4 K)   \\
\phi \left(\frac{1}{2} \right) =  \frac{1}{16} (9-4 K)   \\
\phi \left(\frac{3}{2} \right) = \frac{1}{16} (4 K-1)
\end{align}
Substitute these values in to the sum of squares condition \ref{sumsquares} and simplify to find an equation for $C(K)$. 
\begin{align}
C(K) =\frac{1}{128} (9-4 K)^2+\frac{1}{128} (4 K-1)^2
\end{align}
Note that since $C(K)$ does not depend on $r$, this expression is valid in general, not just the special case $r = 1/2$. 

Now, return to the general case of $r \in [-1/2,1/2]$. 
Solve the linear system of equations given by the moment conditions symbolically for all other unknown functions in terms of $\phi(r)$. 
This gives 
\begin{align}
\phi(r-2)  &=   \frac{1}{12} \left(2 \phi(r)+3 K r+2 K+r^3+2 r^2-r-2\right)  \label{phi_m2} \\ 
\phi(r-1)  &=   \frac{1}{6} \left(-4 \phi(r)-3 K r-K-r^3-r^2+4r+4\right) \label{phi_m1} \\ 
\phi(r+1) &=   \frac{1}{6} \left(-4 \phi(r)+3 K r-K+r^3-r^2-4 r+4\right) \label{phi_p1}\\ 
\phi(r+2) &=   \frac{1}{12} \left(2 \phi(r)-3 K r+2 K-r^3+2r^2+r-2\right). \label{phi_p2} 
\end{align}
Substitute these values into the sum of squares condition \ref{sumsquares} to obtain a quadratic equation for $\phi(r)$. 
This has solutions 
\begin{align}
\phi(r)  =   \frac{1}{280}\left( -40 K-40 r^2+136  \pm \sqrt{ 2\beta(r) + 2\gamma(r)} \right), \label{phideffivepoint} 
\end{align}
where 
\begin{align}
\beta(r) &=  -12600 K^2 r^2+3600 K^2-8400 K r^4+25680 K r^2-6840 K+3123 \\ 
\gamma(r) &= -40 r^2 \left(35 r^4-202 r^2+311\right).
\end{align}
This implies that $\phi$ is an even function on $[-1/2,1/2]$, and combined with equations \ref{phi_m2}-\ref{phi_p2}, implies that $\phi$ is even on $\R$.

This expression must guarantee that the value of $C(K)$ is consistent with the previous result at $r = 1/2$, which said that $\phi(1/2) = \frac{1}{16} (9-4 K)$.  
Evaluating at $1/2$ and simplifying gives  
\begin{align}
\phi \left( \frac{1}{2} \right) = \frac{3}{560} \left(\pm \sqrt{(20 K-21)^2}+84\right)-\frac{K}{7} . 
\end{align}
The expression $\sqrt{(20 K-21)^2}$ is written in this form because it is machine generated, and Mathematica did not simplify it automatically because of branches.  
If $0 \leq K < 21/20$, then the negative root is 
\begin{align}
\frac{27}{80}-\frac{K}{28}
\end{align}
and the positive root is 
\begin{align}
\frac{1}{16} (9-4 K), 
\end{align}
so we select the positive root. 
Note that if $K > 21/20$ then the sign of the expression $20K-21$ changes and these expressions are swapped. 
We do not consider this further, since we wish to have the smallest possible $K$ that satisfies the postulates.

Finally, we select $K$ to achieve the desired regularity. 
To evaluate continuity at $1/2$, we compute the difference of $\phi(r)$ with $r = 1/2$ and $\phi(r + 1)$ with $r=-1/2$, and so forth for higher derivatives.  
For continuity, this gives 
\begin{align} 
\frac{1}{112} \left(20 K+\sqrt{(20 K-21)^2}-21\right)
\end{align} 
Since we have chosen the positive root, the expression $\sqrt{(20 K-21)^2}$ becomes $21 - 20K$ and the difference evaluates to zero. 
Note that this condition holds for any $K \in [0, 21/20]$.
For the first derivative, the difference is 
\begin{align} 
-\frac{(84 K-55) \left(20 K+\sqrt{(20 K-21)^2}-21\right)}{168 \sqrt{(21-20 K)^2}}
\end{align} 
Again, considering the branch this reduces to zero. 
This implies that for general $K \in [0,21/20)$ that $\phi$ is continuously differentiable at 1/2.
This does not hold at at the value $K = 21/20$, but this will not turn out to be relevant.

For the second derivative, the difference is 
\begin{align} 
\frac{5 \left(-20160 K^2+25716 K+9 \sqrt{(20 K-21)^2}-7889\right)}{126 \sqrt{(21-20 K)^2}}
\end{align} 
This is not trivially zero. 
We set the correct branch and simplify the numerator to obtain 
\begin{align} 
-28 \left(720 K^2-912 K+275\right) . 
\end{align} 
This quadratic polynomial has roots 
\begin{align} 
K =  \left\{ \frac{1}{60} \left(38-\sqrt{69}\right) , \quad  \frac{1}{60} \left(38+\sqrt{69} \right)  \right\} , 
\end{align}
which gives two possible values for the derivative at $r = 1/2$. 

For the third derivative, the difference is 
\begin{align} 
\Big[ &-1209600 K^3+2320560 K^2-180 \left(\sqrt{(20 K-21)^2}+8098\right) K \\ 
        &+189 \sqrt{(20 K-21)^2}+298531  
\Big] 
/ \Big[ 9 \sqrt{(21-20 K)^2} (20 K-21) \Big] 
\end{align} 
Again, manually select branches and simplify the numerator to obtain 
\begin{align} 
-20 \left(60480 K^3-116208 K^2+73260 K-15125\right) . 
\end{align} 
This has roots 
\begin{align} 
K = \left\{ \frac{55}{84}, \quad \frac{1}{60} \left(38-\sqrt{69} \right) , \quad \frac{1}{60} \left(38+\sqrt{69} \right)  \right\}
\end{align} 
Remarkably, this has two roots in common with the second derivative matching condition. 
Still more remarkable is that the equations for matching derivatives at $3/2$ and $5/2$ are scalar multiples of these conditions.

\begin{figure}[H]
\centering
$ \begin{array}{cc}
\includegraphics[width=.475\textwidth]{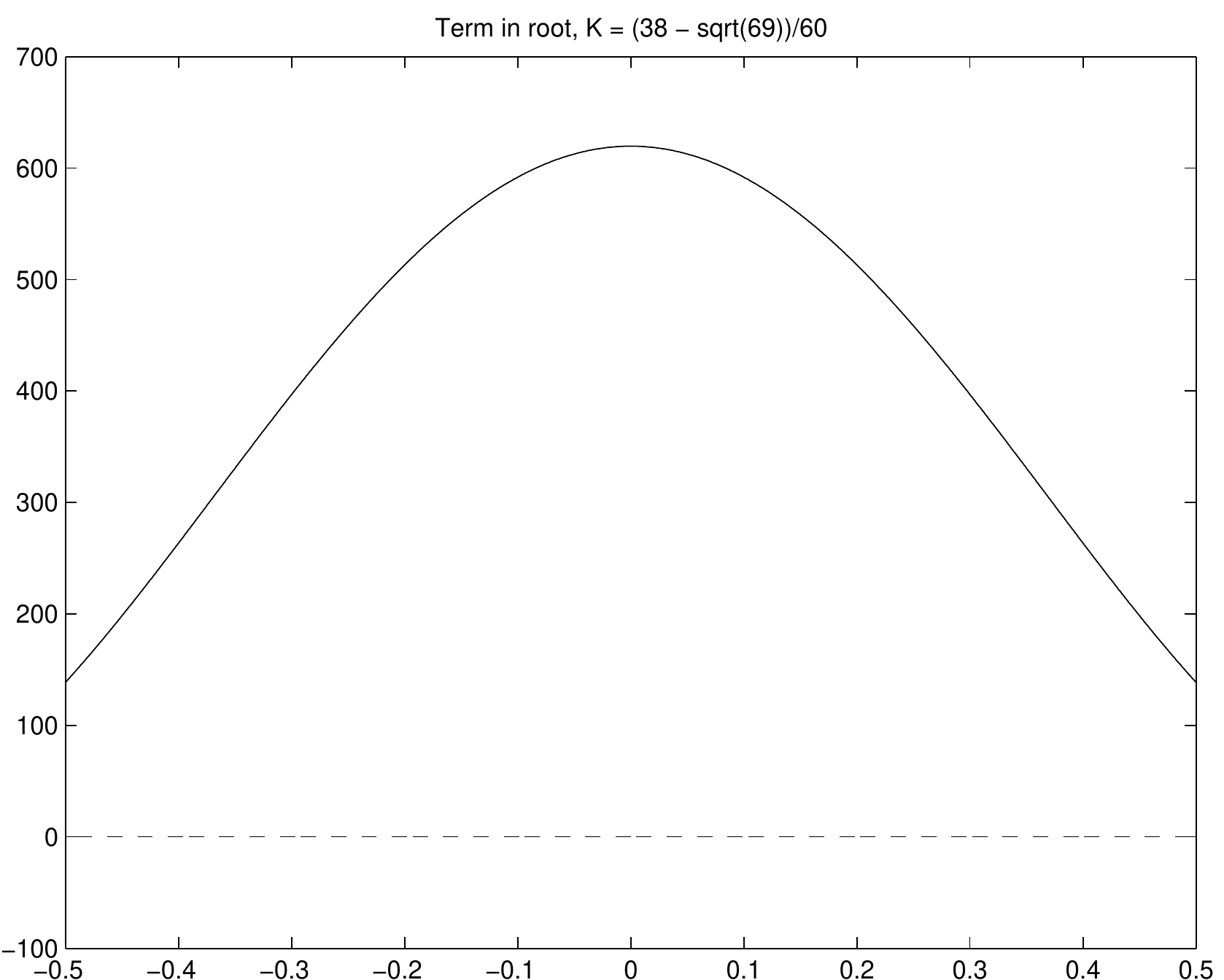} & \includegraphics[width=.475\textwidth]{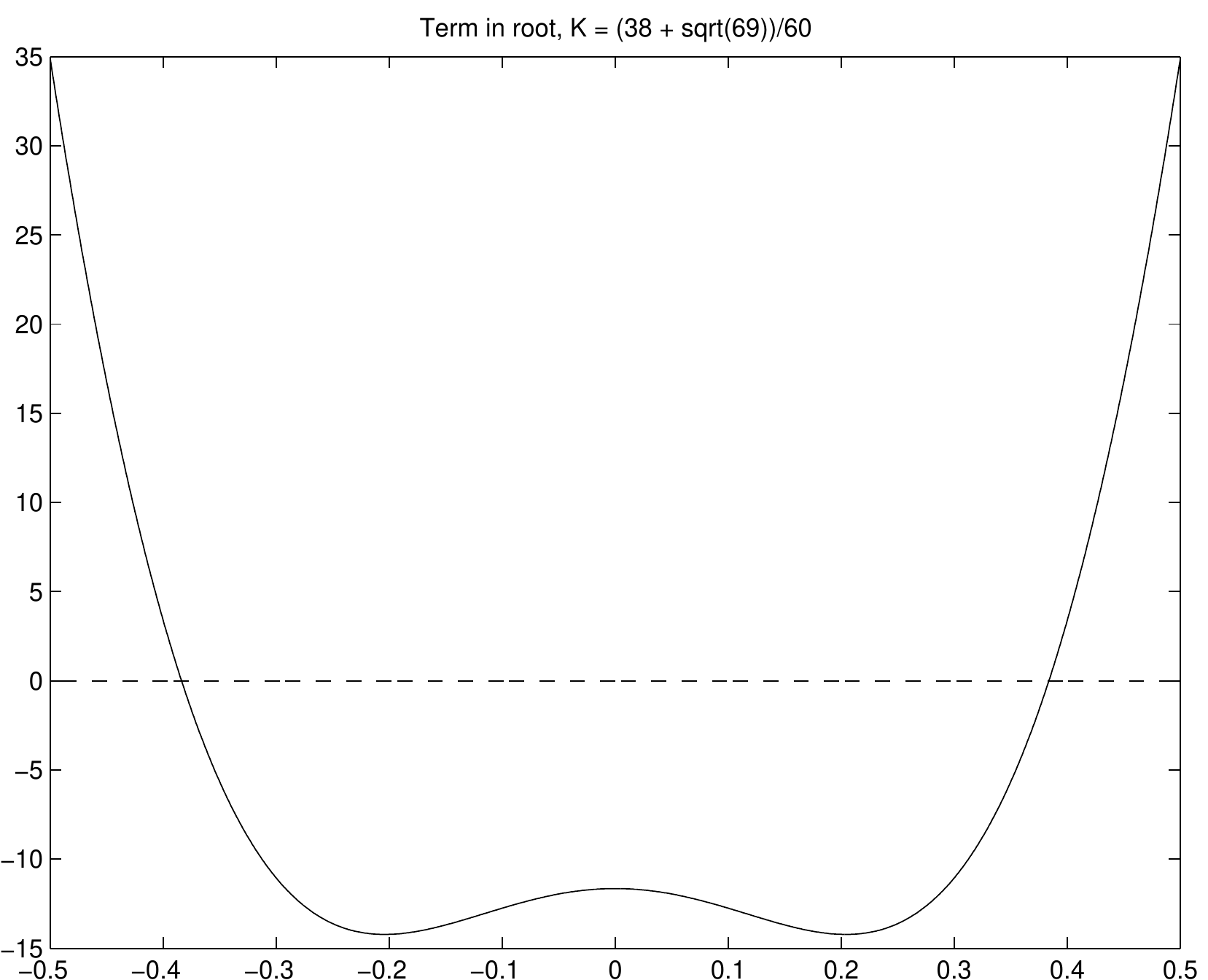}   
\end{array} $
\caption{Term in root of $\phi(r)$. 
Left, $K = \frac{1}{60} \left(38-\sqrt{69} \right)$, positive.  
Right,  $K = \frac{1}{60} \left(38+\sqrt{69} \right)$, negative for a portion of the interval.}
\label{K_root_term}
\end{figure}

The expression inside the square root in equation \ref{phideffivepoint} must be nonnegative to ensure real values of $\phi$. 
Plots of the expression, shown in figure \ref{K_root_term}, reveal that we must take $K =  \frac{1}{60} \left(38-\sqrt{69}\right)$. 

This value of $K$ is the smallest $K$ such that the function $\phi$ is nonnegative on $\R$. 
For general $K \in [0, 21/20)$, the left derivative at $5/2$ exists, and is given as 
\begin{align}
\phi'' \left(  r \right) \Big|_{r = 5/2^{-}} =  \frac{720 K^2-912 K+275}{-9 (21 - 20 K)}
\end{align}
Temporarily, suppose that $K \in \left[0, \frac{1}{60} \left(38-\sqrt{69} \right) \right)$. 
Since $\phi(5/2) = \phi'(5/2) = 0$, and the left second derivative is negative at $5/2$, this implies that $\phi$ is negative on an interval $r \in (5/2 - \eps, 5/2)$ for some $\eps > 0$.
However, at the special value $K = \frac{1}{60} \left(38-\sqrt{69} \right)$, the left second derivative is zero. 
The left third derivative is negative, and so $\phi$ is nonnegative in a neighborhood of $5/2$, and plotting reveals that $\phi$ is nonnegative on all of $\R$. 
Thus, the value for $K$ that makes $\phi \in \mathcal C^{3}$ is also is the smallest $K$ for which $\phi$ is nonnegative.

To summarize, for $r \in [-1/2,1/2]$ we have 
\begin{align}
K               &=  \frac{1}{60} \left(38-\sqrt{69} \right) \label{ib5final_first}
  \\ 
\beta(r)      &=  -12600 K^2 r^2+3600 K^2-8400 K r^4 \\
                  &\quad \; +25680 K r^2-6840 K+3123 \nonumber \\ 
\gamma(r) &= -40 r^2 \left(35 r^4-202 r^2+311\right) \\ 
\phi(r-2)     &=   \frac{1}{12} \left(2 \phi(r)+3 K r+2 K+r^3+2 r^2-r-2\right) \\
\phi(r-1)     &=   \frac{1}{6} \left(-4 \phi(r)-3 K r-K-r^3-r^2+4r+4\right) \\
\phi(r)        &=   \frac{1}{280}\left( -40 K-40 r^2+136  \pm \sqrt{ 2\beta(r) + 2\gamma(r)} \right) \\ 
\phi(r+1)    &=   \frac{1}{6} \left(-4 \phi(r)+3 K r-K+r^3-r^2-4 r+4\right) \\
\phi(r+2)    &=   \frac{1}{12} \left(2 \phi(r)-3 K r+2 K-r^3+2r^2+r-2\right) . \label{ib5final_last}
\end{align}
These formulas allow us to evaluate $\phi$ for any real value.

\section{Driving pressures}
\label{driving_pressures}
\begin{figure}[H]
\centering
\includegraphics[width=\textwidth]{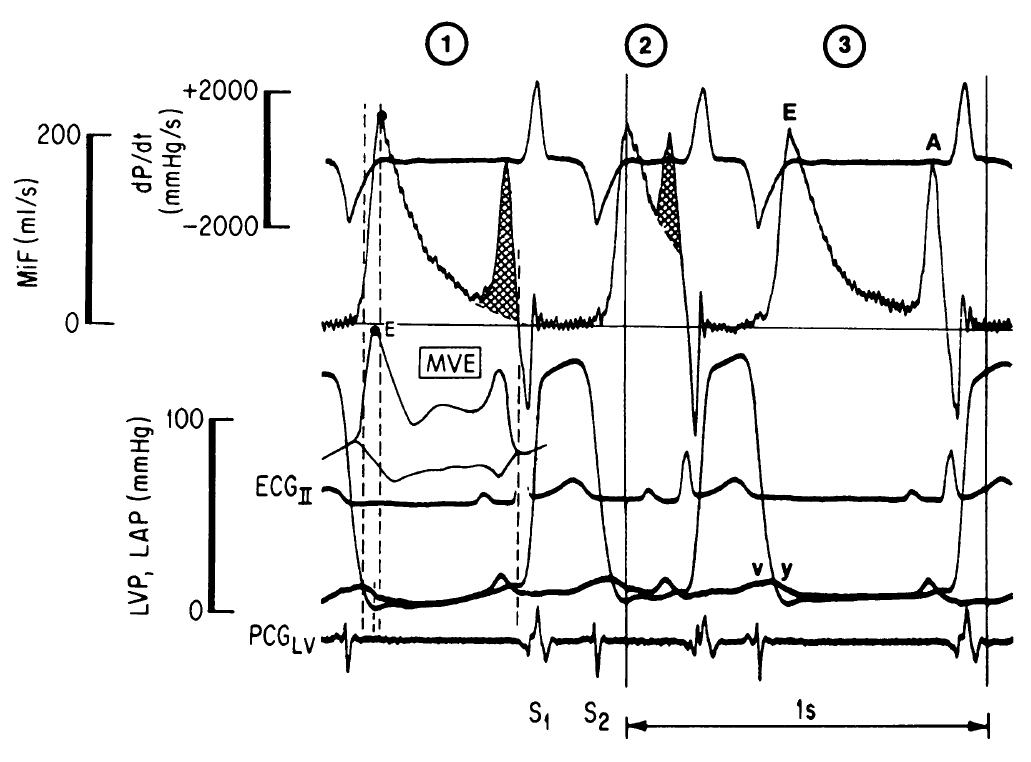}
\caption{Experimental records of mitral valve flow and surrounding pressures. 
Reproduced from \cite{yellin_book} with permission of Springer.}
\label{yellin_dog_flow_stats}
\end{figure}

To drive simulations, we use experimental records of atrial and ventricular pressure, taken from \cite{yellin_book} and shown in figure \ref{yellin_dog_flow_stats}. 
In these experiments, dogs were implanted with a number of probes to measure many aspects of left heart function, including flux through the mitral valve ring, and left atrial and left ventricular pressures. 
Their measurements are both qualitatively and quantitatively similar to those found in standard medical references on cardiovascular physiology \cite{mohrman2010cardiovascular}.

There are many notable features of the pressure and mitral flux curves as measured. 
Diastole, in which the left ventricle fills, starts at the leftmost dashed line, which marks the time at which atrial and ventricular pressures are equal.
Following this time, the ventricular pressure continues to fall rapidly. 
This creates a transient forward pressure difference, and causes mitral flux to increase rapidly. 
The atrial pressure quickly falls to near the ventricular pressure, and for an extended period there is only a small forward pressure. 
Mitral flow correspondingly decreases. 
Just before the third dashed line, the left atrium contracts. 
This is referred to as the atrial kick. 
This creates a bump in left atrial pressure and a second increase in mitral flux.

At the third dashed line, ventricular systole begins. 
The left ventricle contracts and the mitral valve closes. 
Ventricular pressure rapidly rises and a large negative pressure occurs across the mitral valve. 
There is a brief, large spike of reverse flux through the mitral ring. 
This is not really backflow or regurgitation; rather a small volume of blood has passed the mitral ring but is still in the space between the leaflets. 
Beginning with this spike, there is a large oscillation in flux that causes the ``S1'' heart sound, also denoted on the phonocardiogram.  
This oscillation decays rapidly, and mitral flux stabilizes near zero. 
(There is some additional vibration during closure on this record. We believe this to be instrumental noise.)
The pressure difference across the mitral valve during this time is very high. 
When the pressure begins to decrease, the valve unloads and there is a slow forward flow through the ring.

It is notable that the forward pressure for most of diastole is of order 1 mmHg, and the back pressure in systole is over 100 mmHg. 
Peak forward pressure is of order 10 mmHg. 
Thus the back pressure is about one order of magnitude stronger than the forward pressure at peak, and two orders of magnitude stronger in general. 
This says that the mitral valve must freely open and allow forward flow with tiny pressures, but resist huge pressures when closed. 
Because back pressures are so high, even the slightest failure to fully seal may cause significant regurgitation. 
These are demanding conditions.

From this record, we select the first beat to use in our simulations, as it has duration of approximately $0.8$ seconds, corresponding to a typical heart rate of 75 beats per minute. 
(The second beat in the experiment shown in figure \ref{yellin_dog_flow_stats} is much faster, and the third is not complete on the record.)
To represent each of the two pressure curves, one for the atrial and one for the ventricular pressure, we use a finite Fourier series with 600 frequencies. 
This ensures that each pressure is periodic in time and smooth.  
To do this, we first select relevant points by eye in this diagram. 
We construct a piecewise linear interpolant of these points.
To obtain a smooth function, we take the convolution of the piecewise linear interpolant with a smooth compactly supported function $\phi$. 
The function $\phi$ is supported on $|x| < \tau$ and is defined as 
\begin{align}
\phi(x) = 
\begin{cases}
\frac{1}{r} \cos^{2} \left( \frac{\pi}{2 \tau } x  \right)  & : \; |x| < \tau \\ 
0                         & : \; |x| \geq \tau . 
\end{cases}
\end{align}
Note that this function integrates to one for all $\tau$. 
We exclusively use $\tau = 0.05$ s here. 
The summed Fourier series are shown are figure \ref{pressure_and_flux}, upper panel.

\section{Results}
\label{control_results}

Finally, we have results. 
The simulation uses the valve model described throughout Chapter \ref{static} and finalized in Section \ref{general_model}. 
Every technique discussed in the current Chapter is used for the fluid-structure interaction simulations. 
We simulate flows using the immersed boundary method with the IBAMR software library. 
We place the valve in the test chamber described in Section \ref{setup}, hold the partition and papillary muscles in place with forces described in Section \ref{papillary_target}, and thicken the structure as describe in Section \ref{layers}. 
The delta function for all simulations is the 5-point delta function derived in Section \ref{delta_function_section} and fully expressed (without scaling) in equations \ref{ib5final_first}-\ref{ib5final_last}. 
The simulation is driven by pressures constructed from experiments as discussed in Section \ref{driving_pressures}.

\begin{figure}[H]
\centering 
\includegraphics[width=\textwidth]{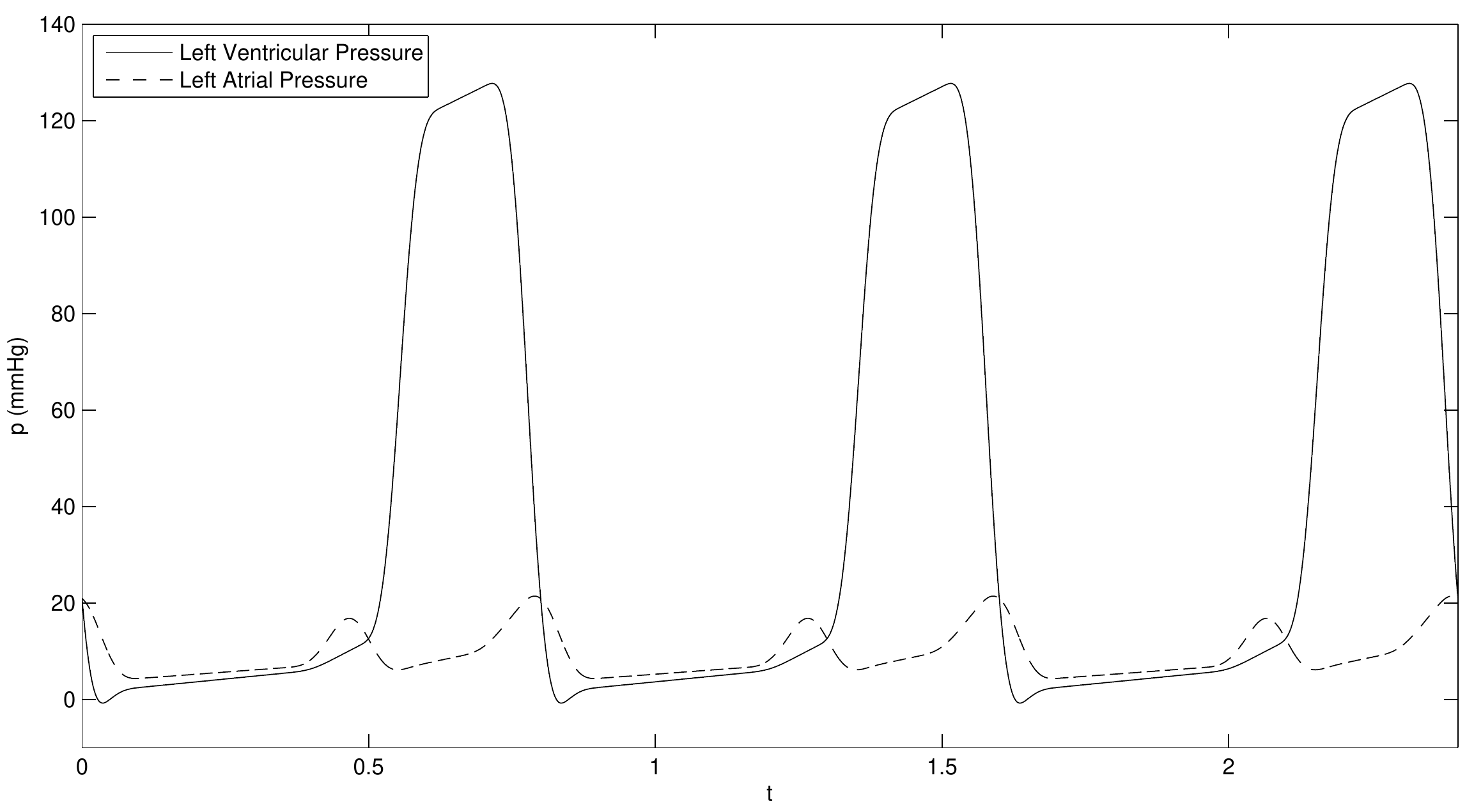}
\includegraphics[width=\textwidth]{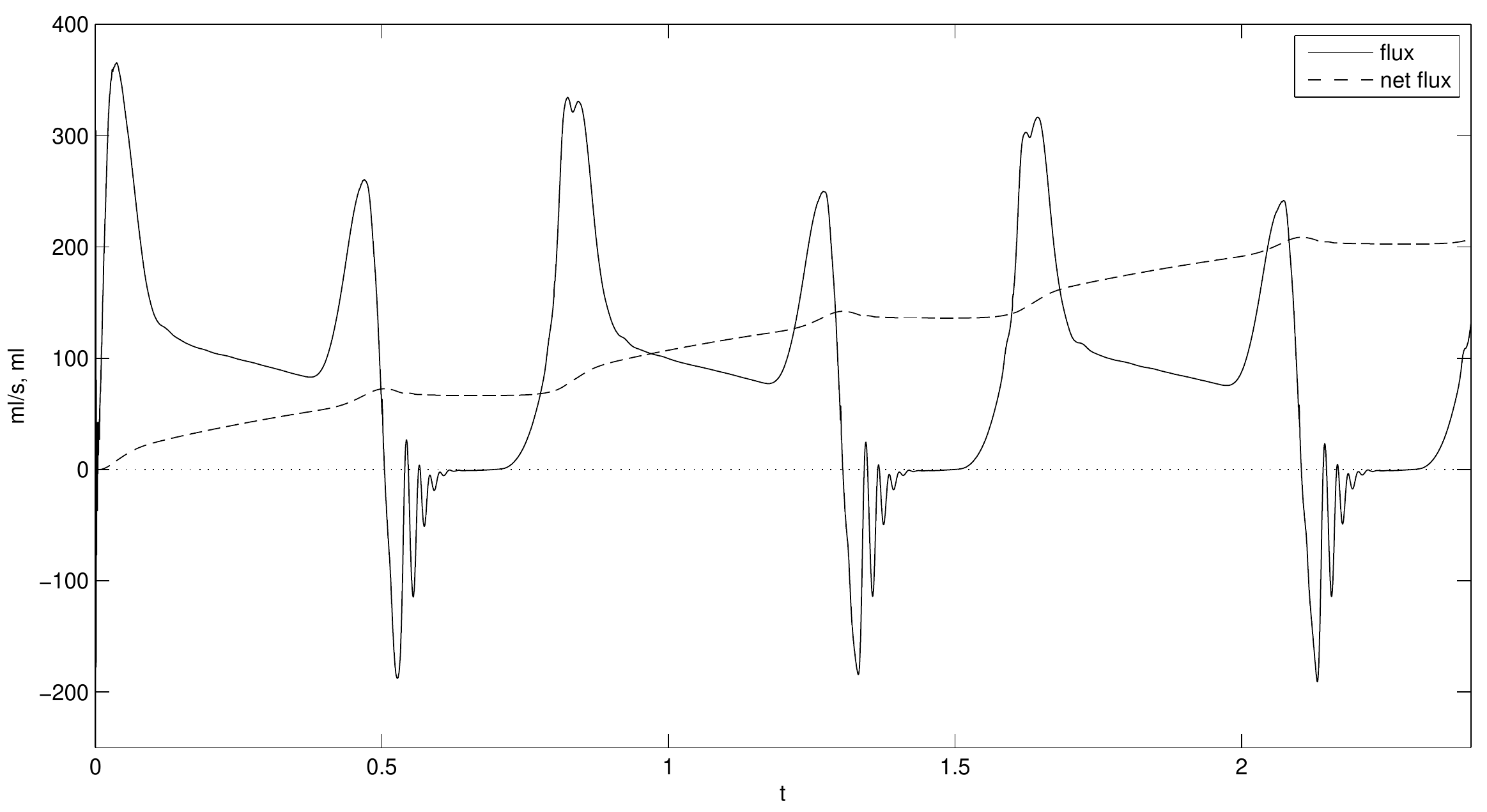}
\caption{Driving pressures (prescribed, top panel) and resulting flow through mitral ring (emergent from simulation, bottom panel).}
\label{pressure_and_flux}
\end{figure}

The simulation is driven by the pressures shown in figure \ref{pressure_and_flux} in the top panel. 
The bottom panel shows the emergent mitral flux, which is output from the simulation. 
The maximum systolic pressure difference is approximately 116 mmHg, which occurs near the end of systole. 

The emergent flux qualitatively matches the experimental flux; many features of the measured flux appear in the simulated flux. 
A rapid inflow occurs as ventricular pressure drops. 
Flux quickly decreases, then more slowly decreases in value until the atrial kick. 
Pressure caused by atrial contraction causes a brief increase in the flux, after which the valve begins to close. 
We observe that there is a single large spurt of apparent backflow. 
This is followed by a rapid oscillation, which is quickly damped out. 
This causes the ``S1'' heart sound. 
Further, we have synthesized the sounds from the flux, and they sound somewhat realistic (listen to the audio of the accompanying movies). 
Following this oscillation, the valve is fully pressurized and closed. 
Then, there is a slow rise in forward flow as the valve unloads. 
Note that some of backflow that occurred during closure is recovered here. 
To the extent that this is happening, the backflow is apparent rather than real, but we have not quantified this. 
Finally, the rise in flux becomes rapid and the cycle repeats. 
Opening on the first beat appears different from the others. 
Since the simulation starts at rest, and the initial position of the valve is not pressurized, this should not be a cause for concern.

We estimate the Reynolds number of this flow as follows. 
We see peak velocities of order $100$ cm/s during filling, and take the length scale as $L = 1$ cm, or slightly smaller than the valve ring radius. 
Density of blood is taken to be $1$ g/cm$^{3}$, and viscosity is $0.04$ g/(cm$\cdot$s). 
This implies that the Reynolds number is approximately 2500. 
The Reynolds number is much greater than one and the flow is inertially dominated.

\begin{figure}[H]
\centering 
\centerline{ \includegraphics[width=.88\textwidth]{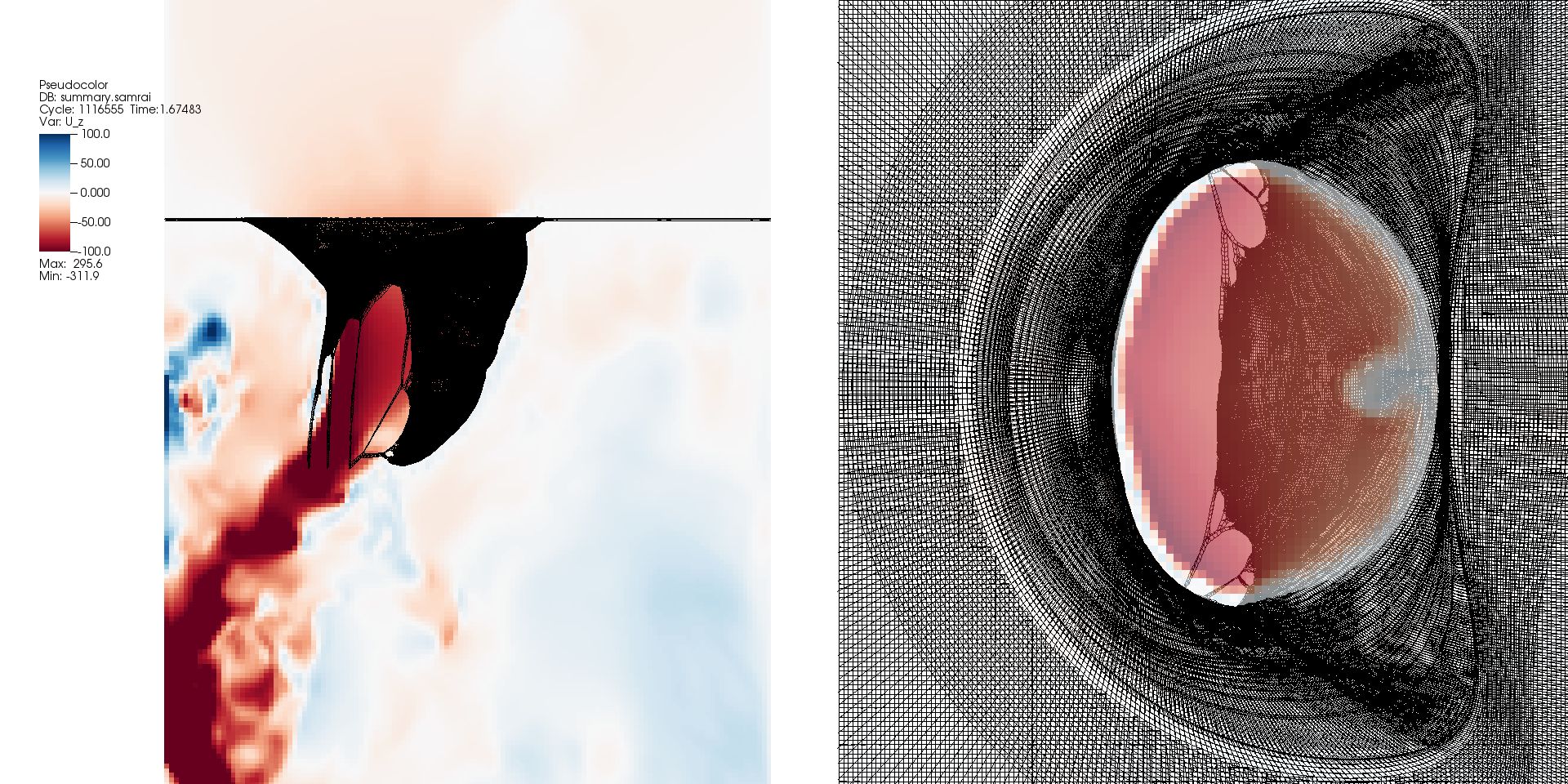} } 
\centerline{ \includegraphics[width=.88\textwidth]{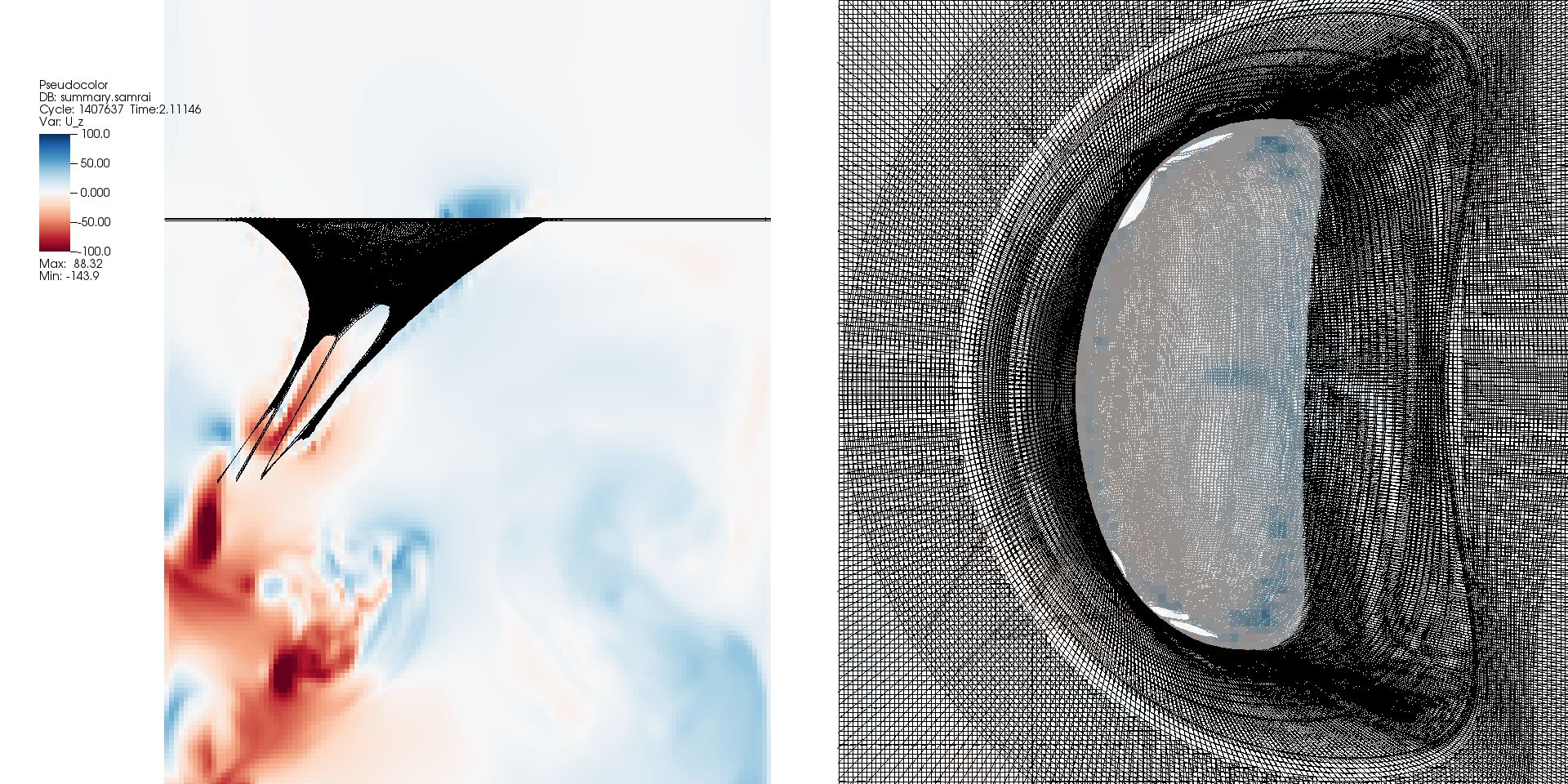} } 
\centerline{ \includegraphics[width=.88\textwidth]{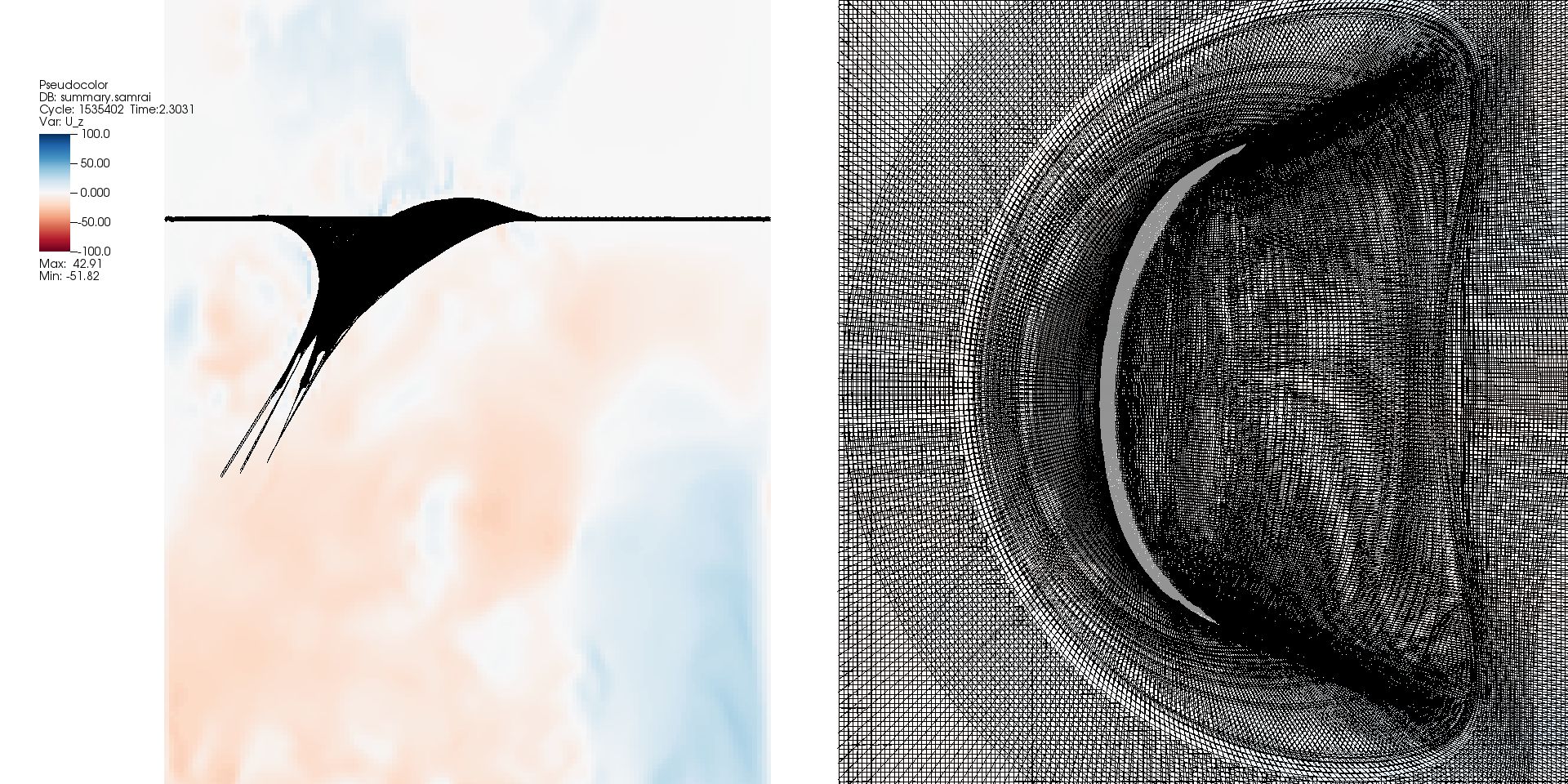} } 
\caption{Views of the mitral valve in simulation. The first shows filling, the second closing, and the third the fully closed position.}
\label{three_panels}
\end{figure}

Three time steps of the simulation are shown in figure \ref{three_panels}. 
These show the $z$-component of velocity in two slice views, along with the full Lagrangian structure. 
The first time shows diastole, near peak forward flow in the third beat of the simulation. 
The second shows the valve immediately during the start of closing. 
The jet of forward flow has begun to break up, and the leaflets have begun to come together. 
The third shows the valve in the fully closed position. 
The leaflets are pressurized and coapted, the chordae tendineae are tight.

\begin{figure}[H]
\centering
$ \begin{array}{cc}
\includegraphics[width=.475\textwidth]{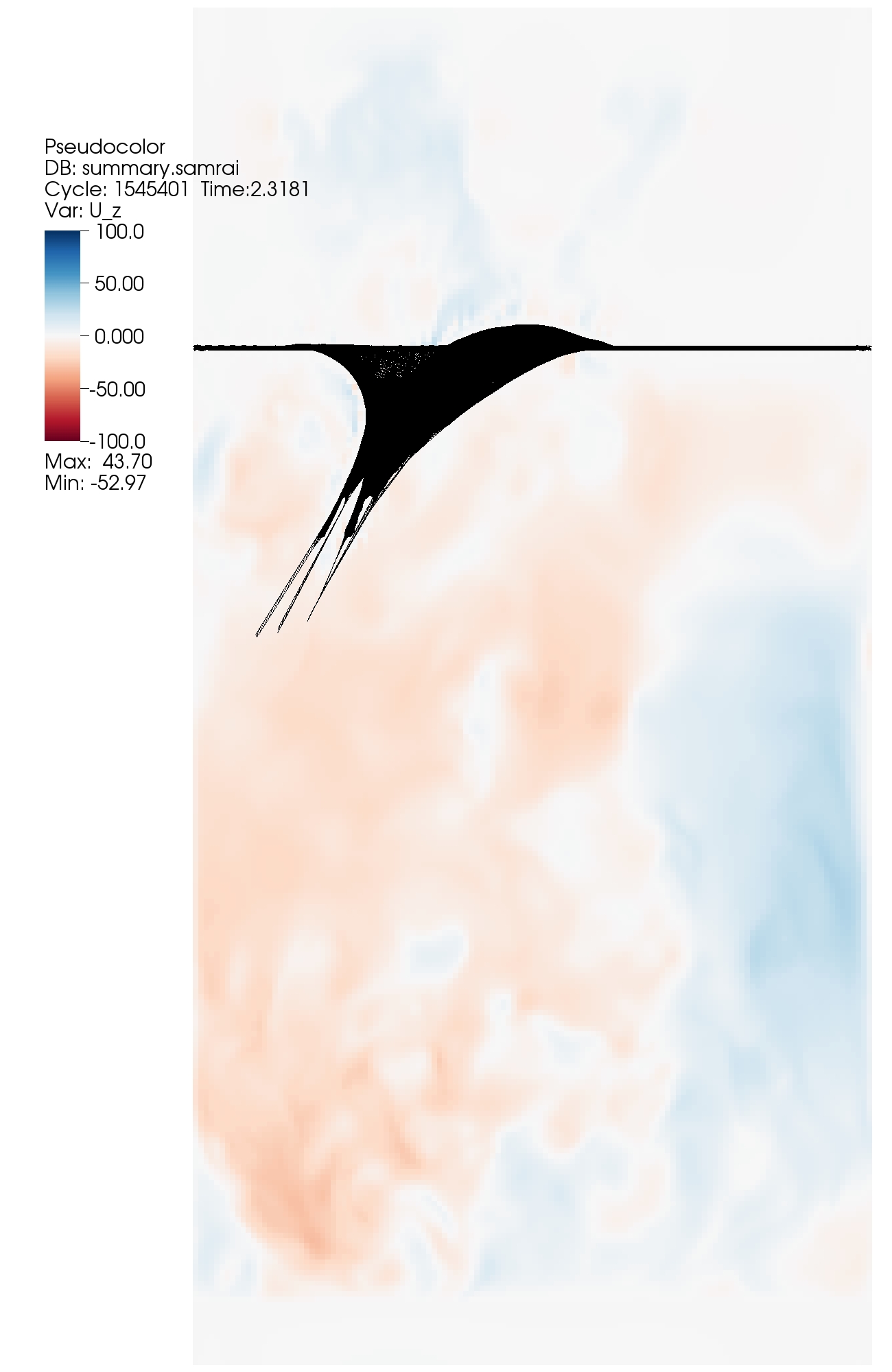}  & \includegraphics[width=.475\textwidth]{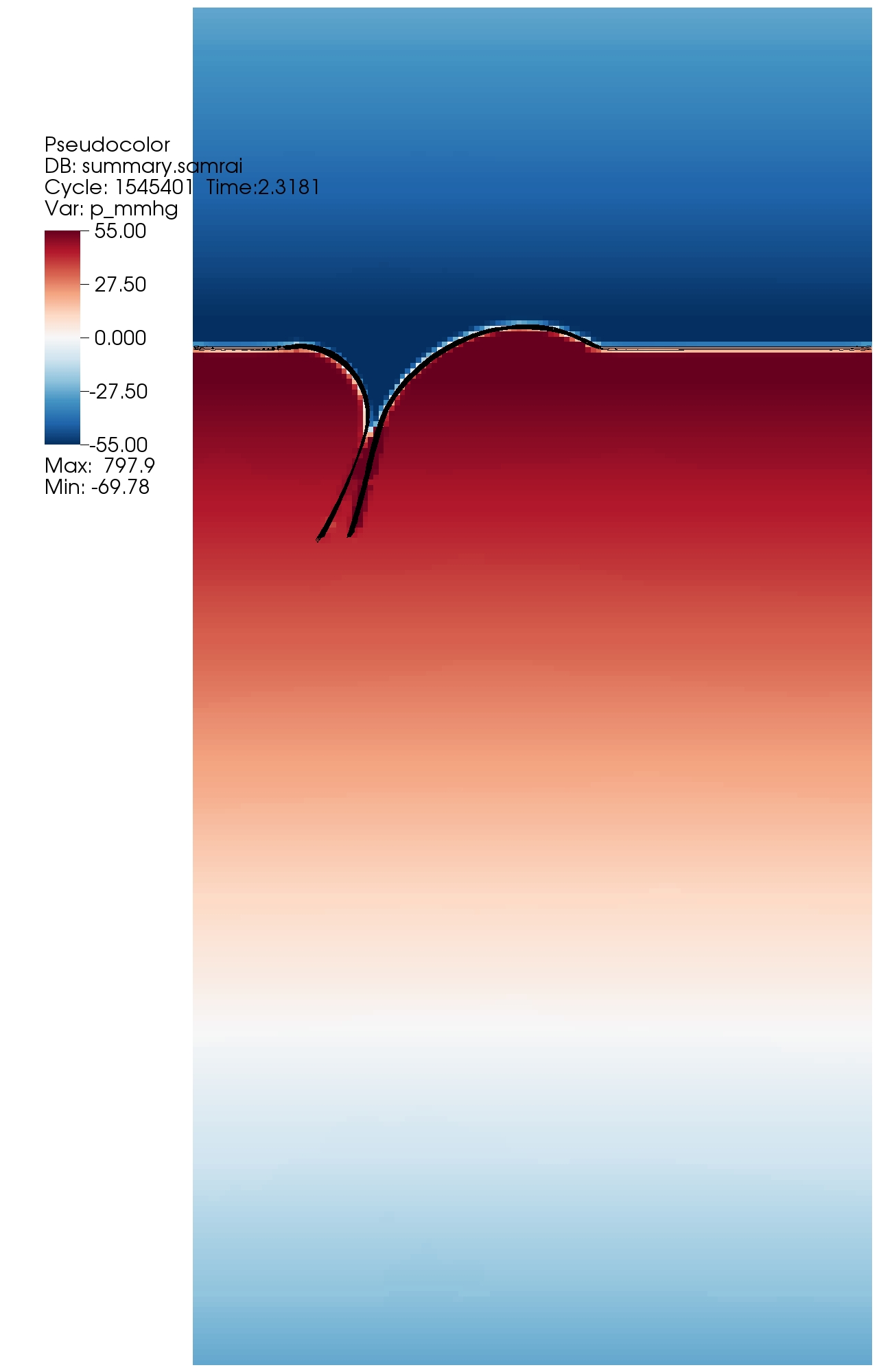}  
\end{array} $
\caption{Slice view of $z$ component of velocity (left) and modified pressure (right) during diastole. }
\label{pressure_systole}
\end{figure}

Two views of the valve during systole are shown in figure \ref{pressure_systole}. 
The valve is tightly closed at this time. 
The vertical component of velocity shows some motion, but is largely relaxed. 
The pressure shows a jump of approximately 110 mmHg across the model valve. 
Note that the modified pressure (see Section \ref{setup}) is approximately piecewise linear in $z$, with a jump at the valve, and the physical pressure is approximately piecewise constant.

When the valve closes, two mechanisms have been proposed that assist it to do so effectively. 
The first is called the ``breaking jet.''
During the atrial kick, there is a jet of forward flow between the leaflets. 
This rapid flow produces a low pressure due to a Bernoulli effect, resulting in a pressure differential across the leaflets around the free edge. 
This then sucks the leaflets together. 
The second is a large vortex. 
Vortices are shed from the leaflet, and a large vortex forms in the left ventricle. 
This vortex then comes back around and pushes the valve, especially the anterior leaflet closed. 
Since the geometry of our model valve tester is not shaped like the ventricle, shed vortices may not form a large vortex to push the valve back into place. 
This means that the vortex may not be as effective in helping the valve close in the rectangular box geometry. 
However, the presence of the vortex would still suggest that this mechanism is plausible. 

Here, we look at three views of the simulation, showing the velocity, pressure and vorticity in a plane. 
The pressure is shown on a 10mmHg scale so that fine details are visible. 
Moreover, these comparisons do not include pressures in systole that are an order of magnitude larger. 
The component of vorticity normal to the viewing planes is shown. 
Note that this is also normal to the approximate symmetry plane of the valve. 
These suggest that both the breaking jet and vortex are plausible, though the evidence here is not decisive enough to say that they are definitely responsible for assisting valve closure.

\begin{figure}[H]
\centering
$ \begin{array}{ccc}
\includegraphics[width=.3\textwidth]{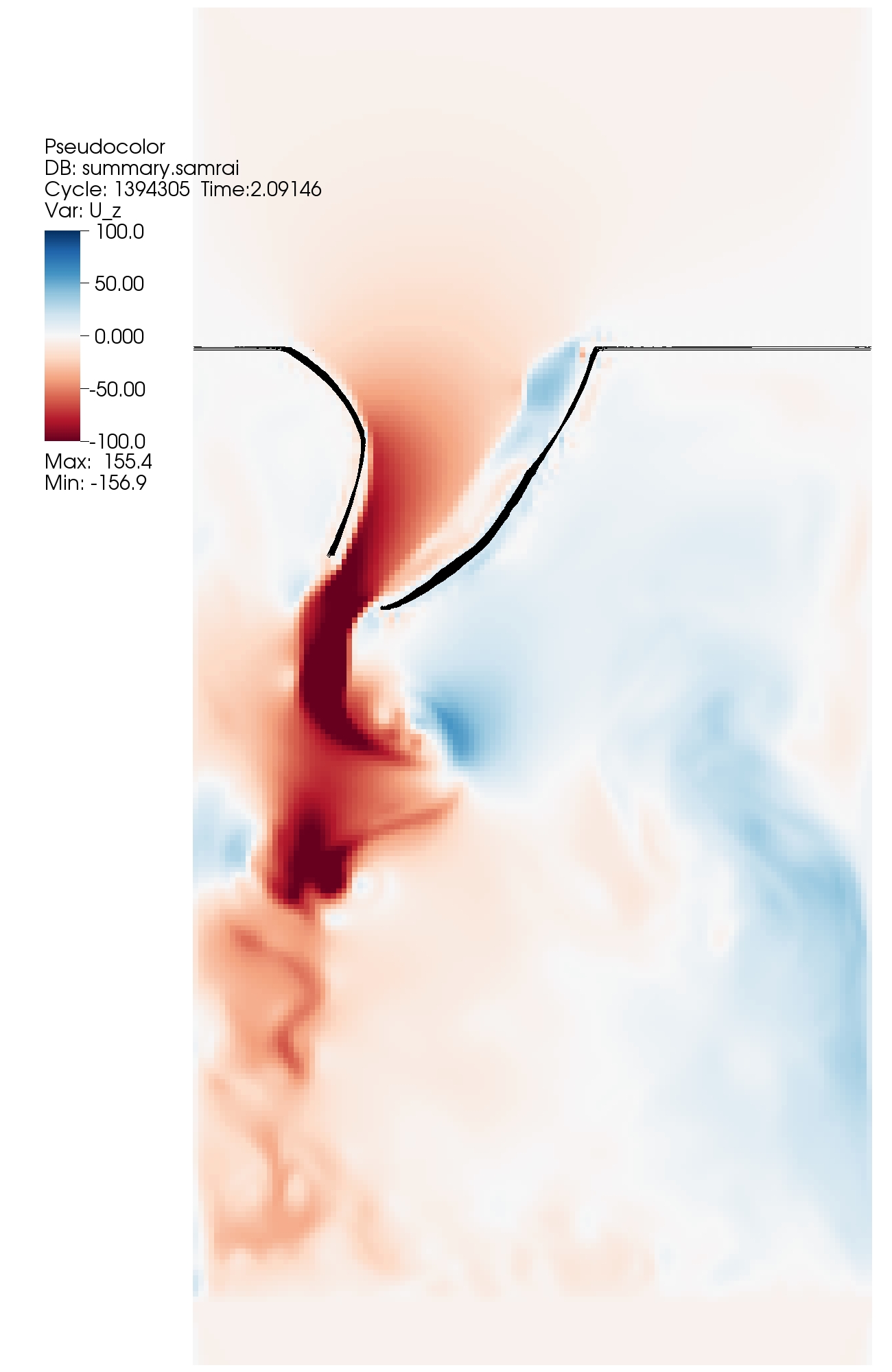}  
& \includegraphics[width=.3\textwidth]{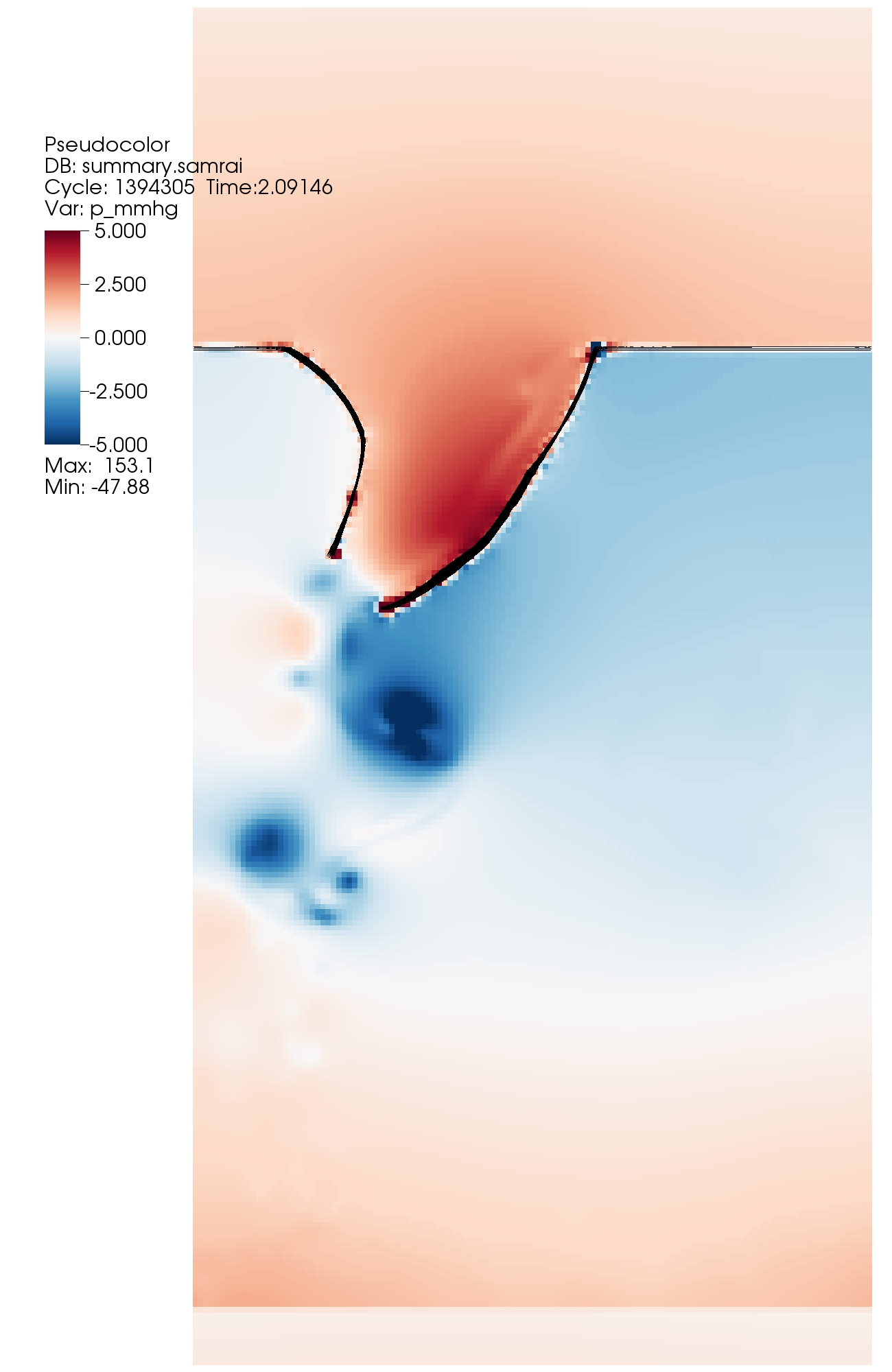}  
& \includegraphics[width=.3\textwidth]{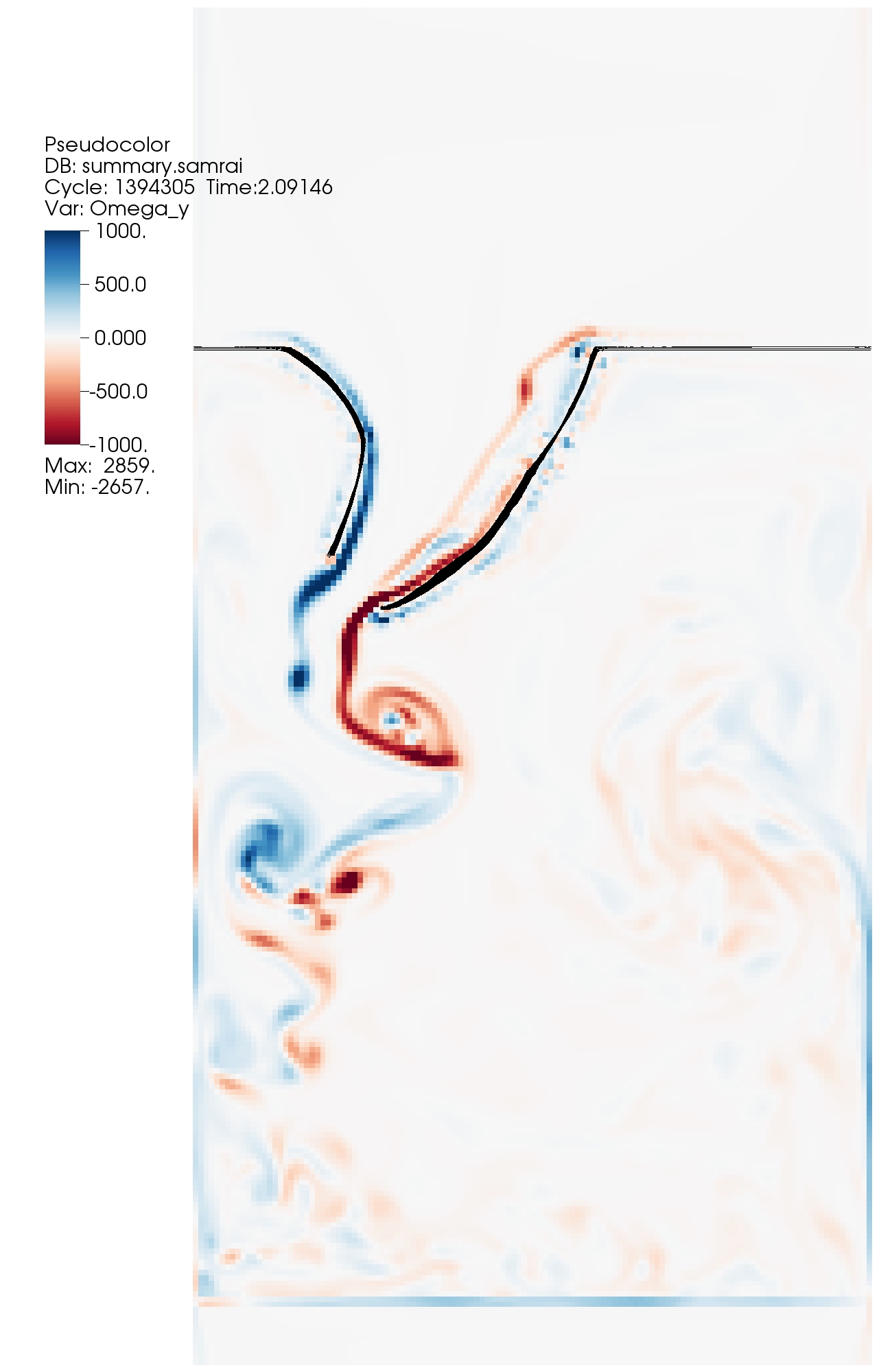}  
\end{array} $
\caption{Slice view of $z$ component of velocity (left) and modified pressure (center) and vorticity (right) early in the period when we expect the breaking jet.}
\label{three_velocities_breaking_one}
\end{figure}

The first collection displays the flow at $t = 2.09146$ s, near the end of the atrial kick during the third beat of the simulation, and is shown in figure \ref{three_velocities_breaking_one}. 
The velocity field contains an unbroken jet of forward flow between the leaflets. 
The pressure has a forward pressure drop across the valve. 
A number of tight vortices are visible below the leaflet in the vorticity field. 
These vortices are also visible in the pressure field, in which they appear as localized low pressures.

\begin{figure}[H]
\centering
$ \begin{array}{ccc}
\includegraphics[width=.3\textwidth]{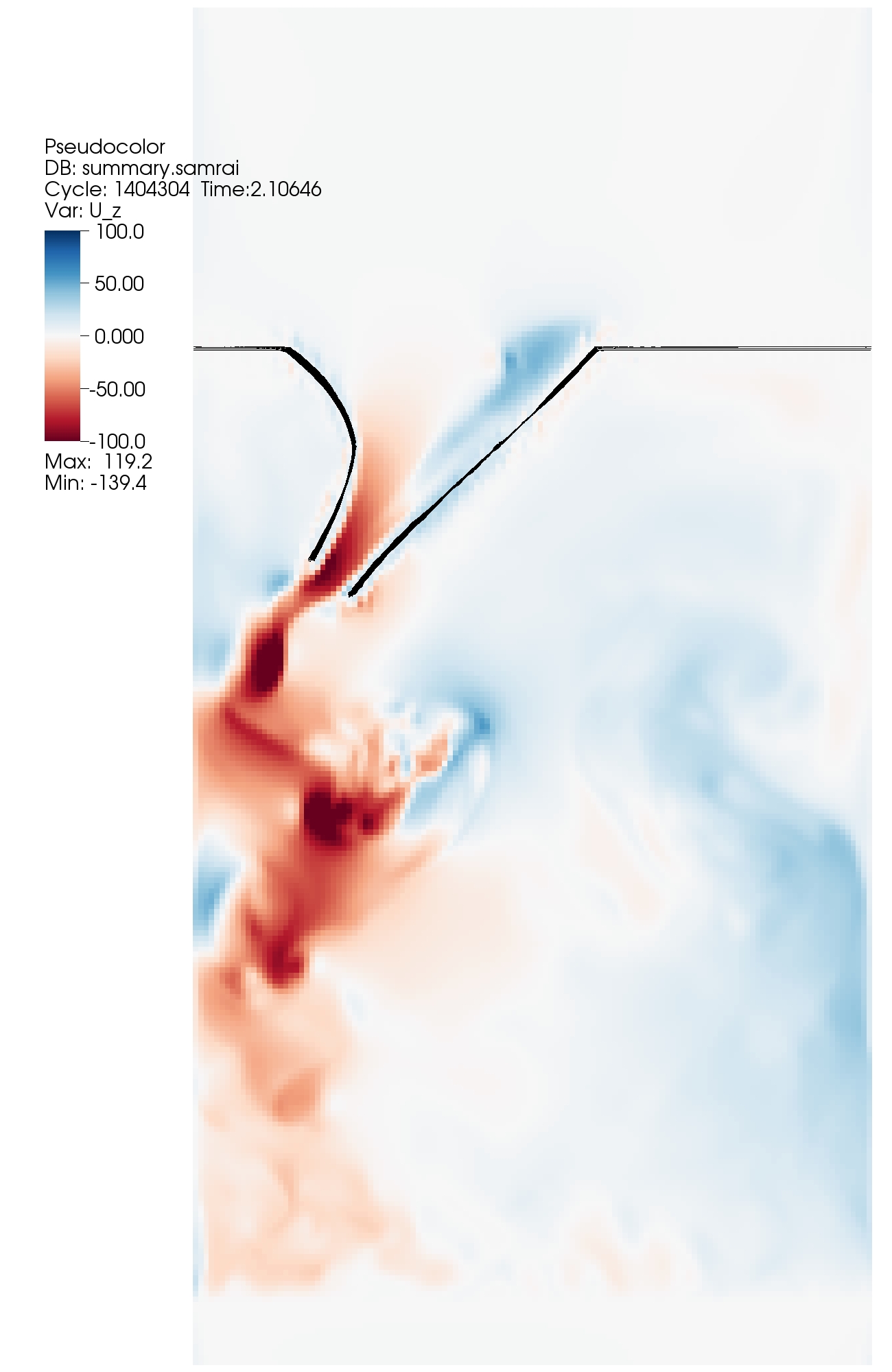}  
& \includegraphics[width=.3\textwidth]{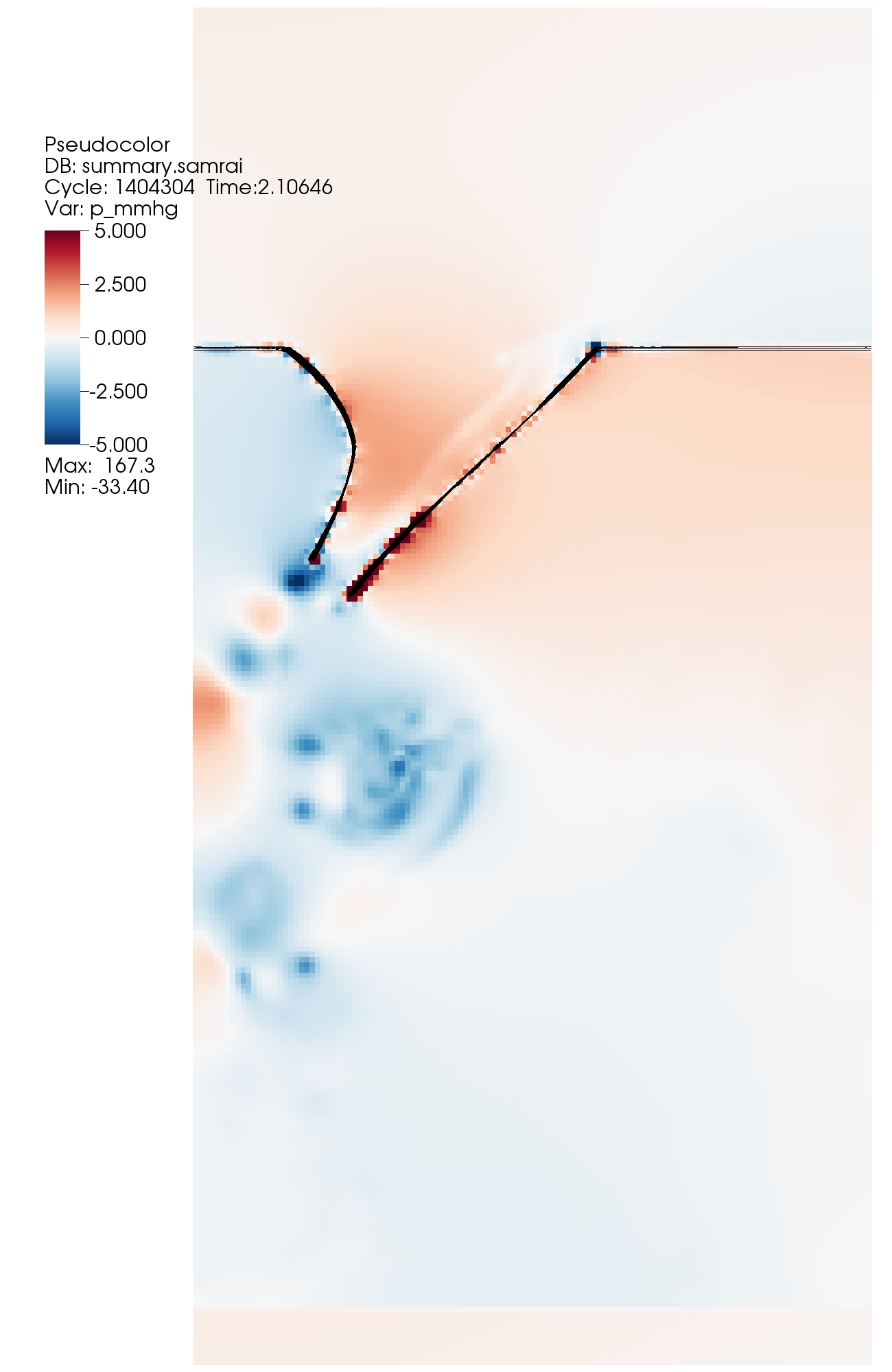}  
& \includegraphics[width=.3\textwidth]{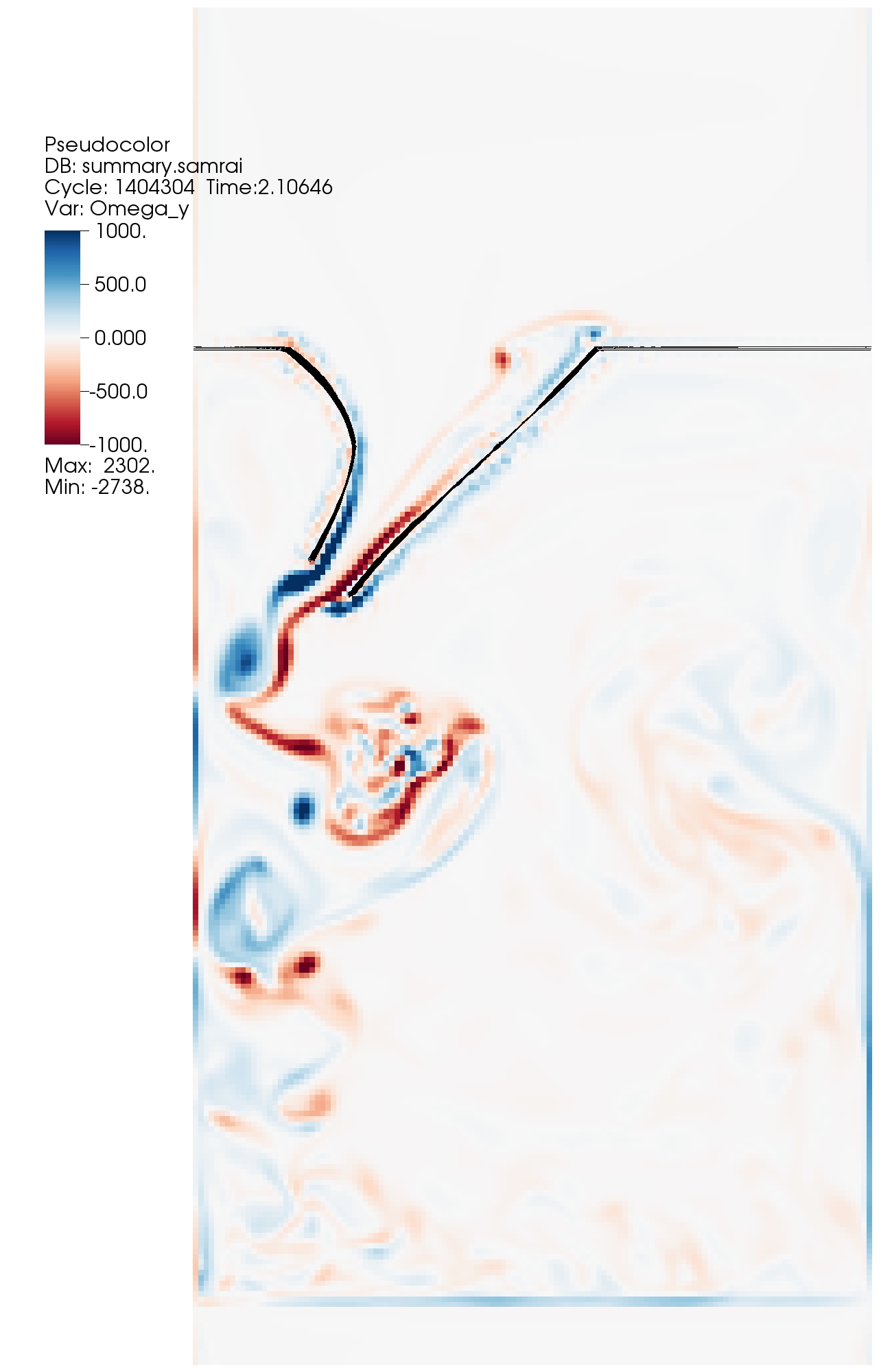}  
\end{array} $
\caption{Slice view of $z$ component of velocity (left) and modified pressure (center) and vorticity (right) in the center of the period when we expect the breaking jet.}
\label{three_velocities_breaking_two}
\end{figure}

Fifteen milliseconds later, at $t = 2.10646$, the jet appears to have begun to break, as shown in figure \ref{three_velocities_breaking_two}. 
Near the valve ring, there is an absence of forward flow. 
But in a small region between the free edges of the leaflets, there is a local high velocity and low pressure. 
We believe that this shows the jet in the process of breaking; that is, inertial effects carry the jet forward, even as the valve has otherwise begun to close. 
This creates a pressure differential across the anterior leaflet that may help the suck the leaflets together. 
There is a still lower pressure on the ventricular side of the posterior leaflet, which we suspect may be related to the geometry of the chamber. 

A larger but less organized set of vortices are present below the valve, connected to vorticity shed from the anterior leaflet. 
This structure cannot hit the ventricular walls and return to push the anterior leaflet closed, because there are no ventricular walls here.
However, were this field to occur in a heart, this structure that could plausibly create a vortex that would assist in valve closure. 
Additionally, a small but prominent vortex has just been shed from the posterior leaflet.

\begin{figure}[H]
\centering
$ \begin{array}{ccc}
\includegraphics[width=.3\textwidth]{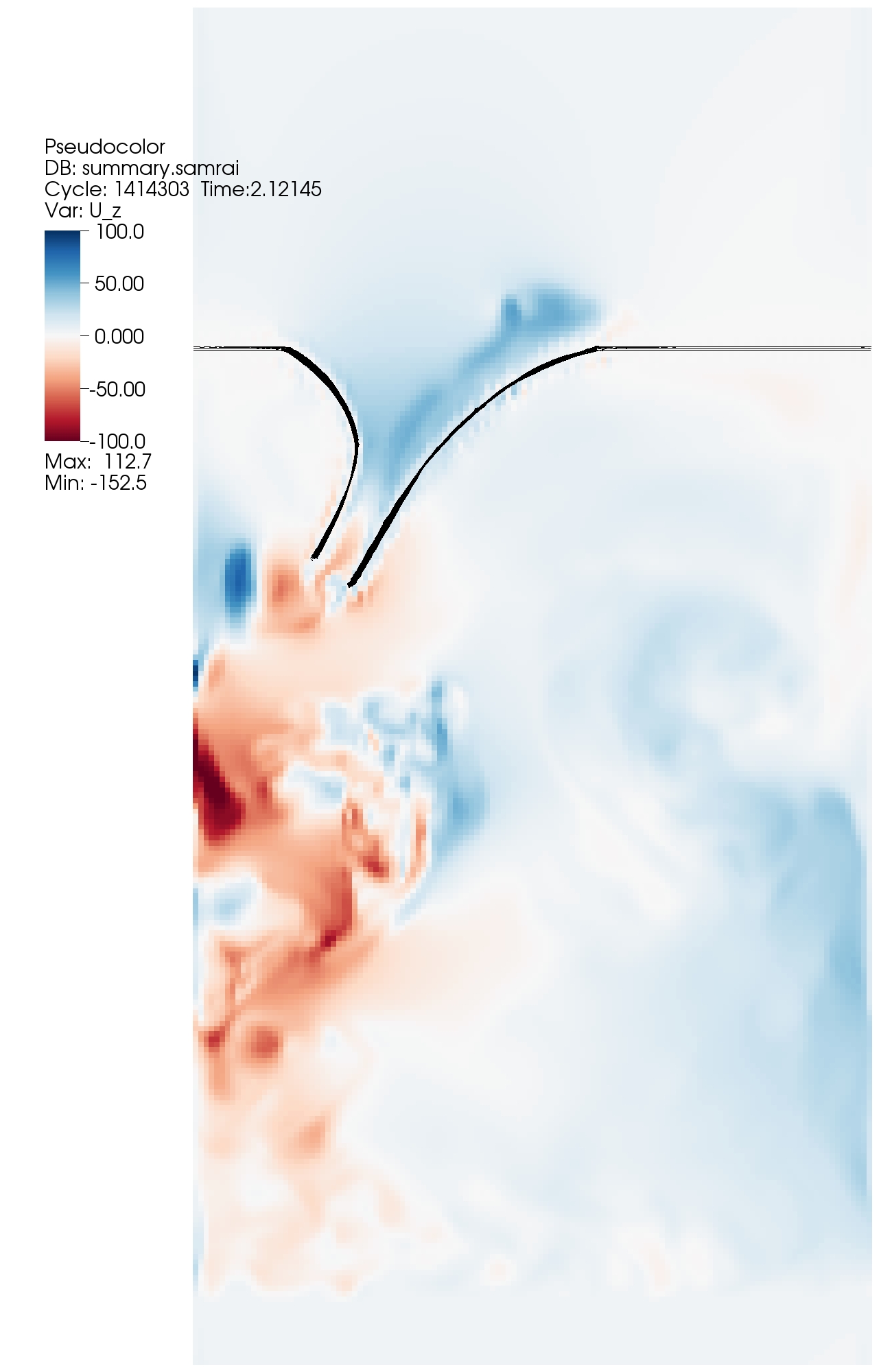}  
& \includegraphics[width=.3\textwidth]{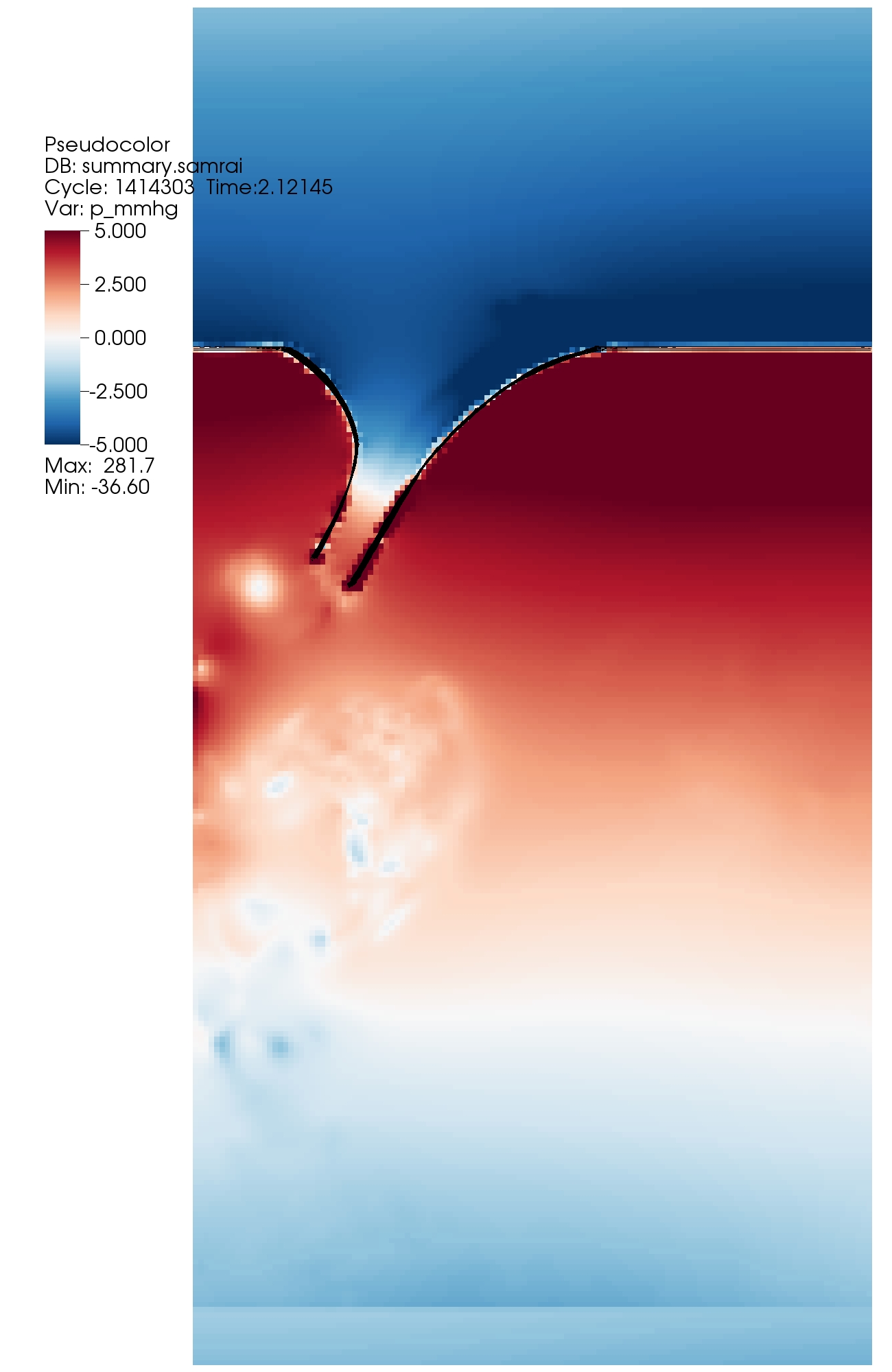}  
& \includegraphics[width=.3\textwidth]{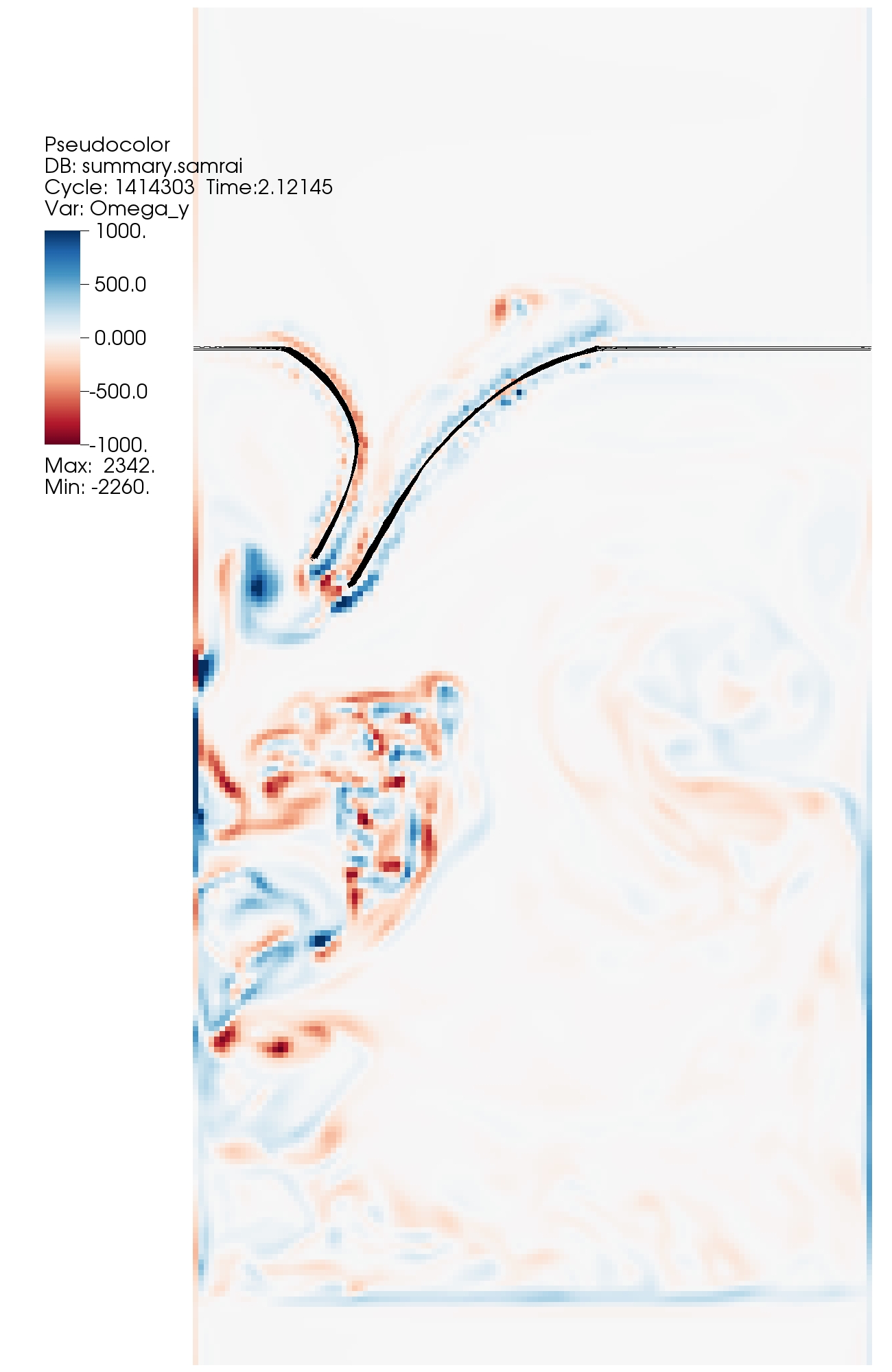}  
\end{array} $
\caption{Slice view of $z$ component of velocity (left) and modified pressure (center) and vorticity (right) late in the period when we expect the breaking jet.}
\label{three_velocities_breaking_three}
\end{figure}

Another fifteen milliseconds later, at $t = 2.12145$, a slight remnant of the jet remains between the free edges, as shown in figure \ref{three_velocities_breaking_three}. 
By this point, there is a back pressure differential across the entire valve and partition. 
There is a region between the free edges of the leaflets in which the pressure is lower than the pressure on the ventricular side of both leaflets. 
The large region of vorticity below the free edges remains, and appears less organized in this frame. 

We will now examine how this valve reacts to various driving conditions. 

\begin{figure}[H]
\centering 
\includegraphics[width=\textwidth]{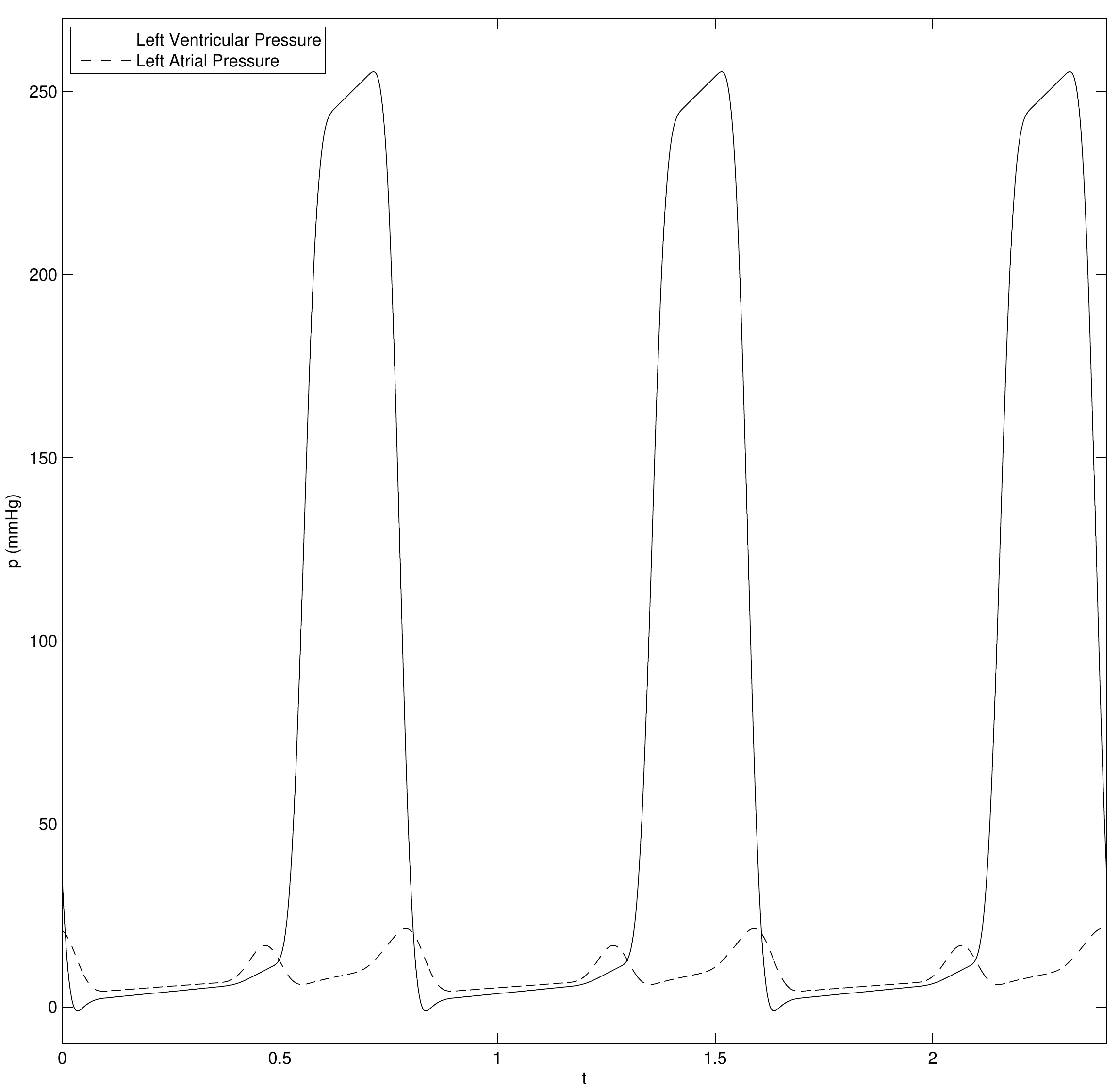}
\caption{Driving pressure with high systolic ventricular pressure.}
\label{pressure_high}
\end{figure}

We prescribe a much higher systolic pressure, and see if the valve still supports this pressure, or develops holes or leaks. 
High ventricular pressure during systole may occur in the body, for example in a patient with aortic stenosis, which restricts outflow from the ventricle. 
Here, we set the ventricular systolic pressure to approximately twice that of the standard simulation. 
The driving pressures are shown in figure \ref{pressure_high}. 

\begin{figure}[H]
\centering 
\includegraphics[width=\textwidth]{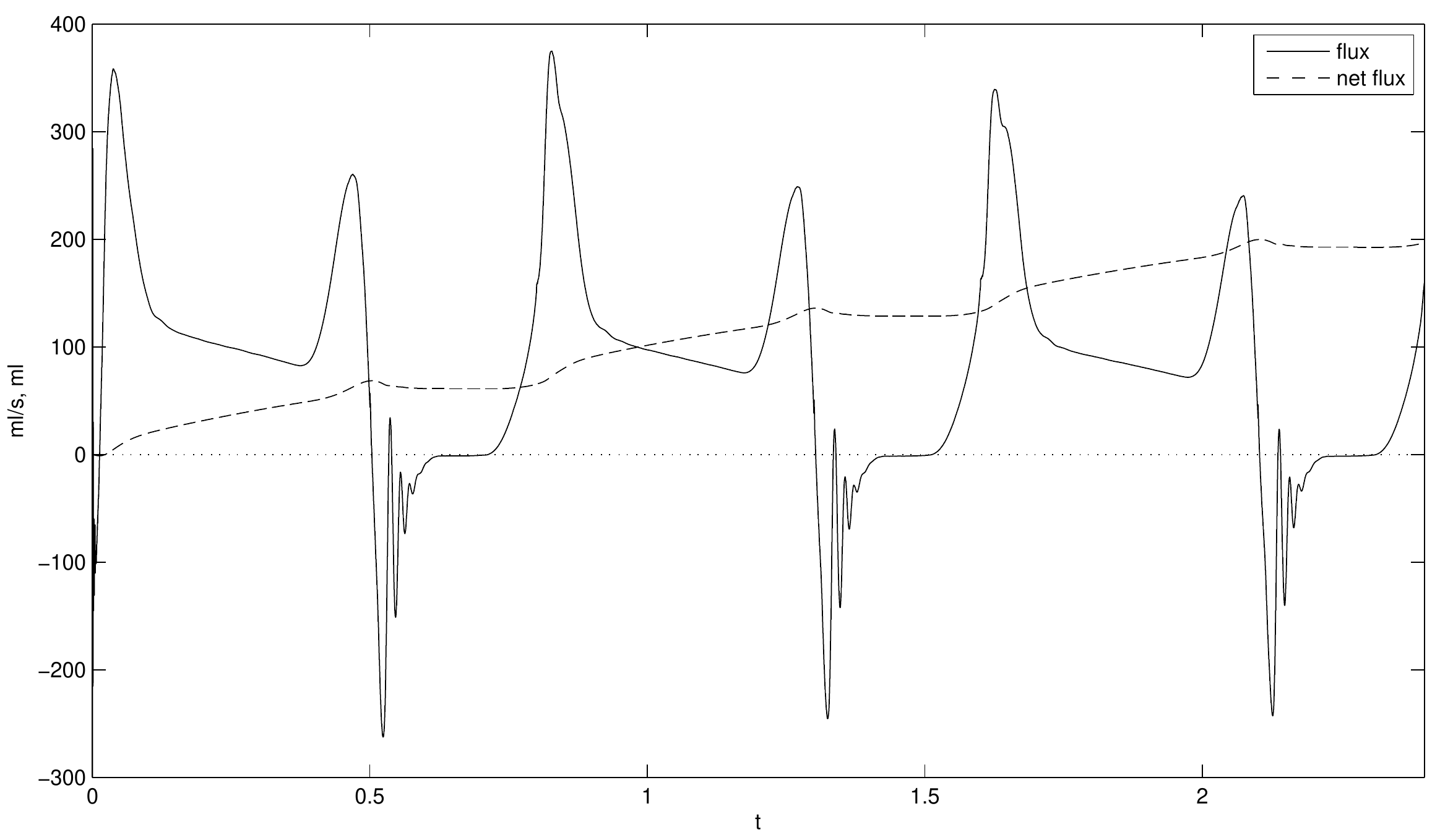}
\caption{Resulting flux through mitral ring with high systolic ventricular pressure.}
\label{flux_pressure_high}
\end{figure}

In the affine region of the constitutive law, shown in equation \ref{constitutive_summary}, the effective stiffness of every fiber in the valve model increases with increased strain. 
Thus, we expect the model to work effectively under higher pressures. 
The emergent flux is shown in figure \ref{flux_pressure_high}, which shows that this is the case. 
The most striking thing about this flux is that it is quantitatively similar to the flux with standard pressures shown in figure \ref{pressure_and_flux}. 
The initial spike of negative flux that occurs on closure is larger in magnitude, as expected. 
Some of the oscillations appear to stay below zero, perhaps because the valve is settling into a more loaded state.

\begin{figure}[H]
\centering 
\includegraphics[width=\textwidth]{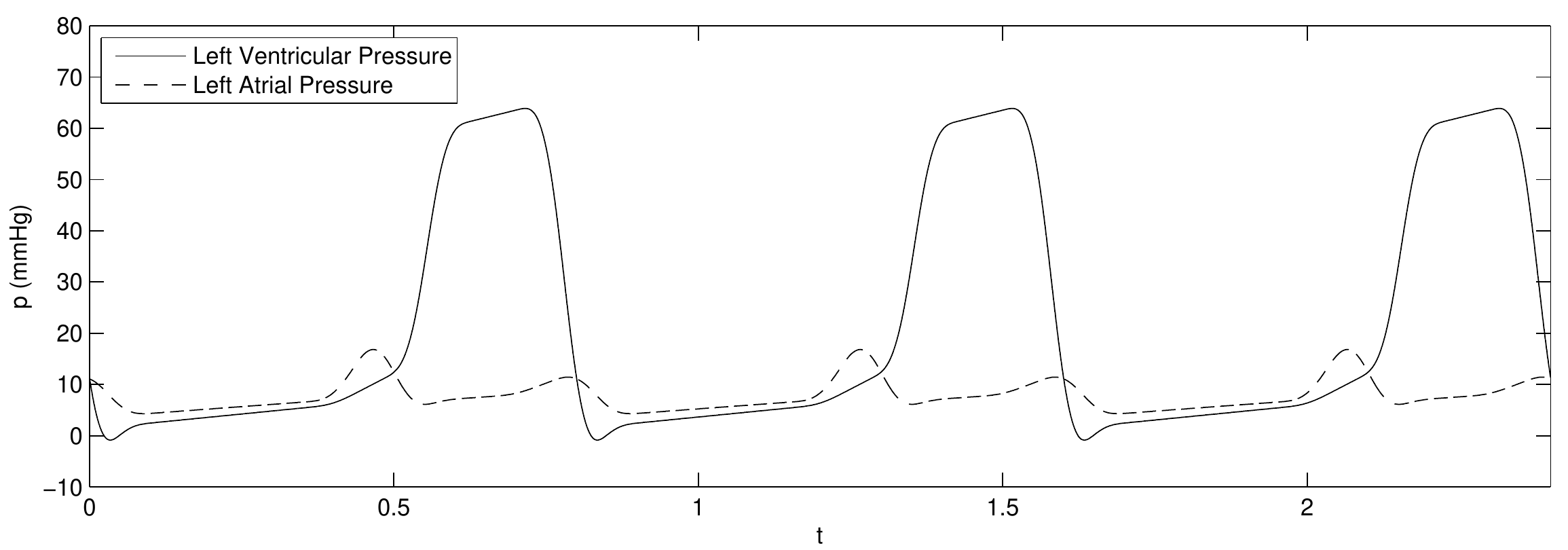}
\includegraphics[width=\textwidth]{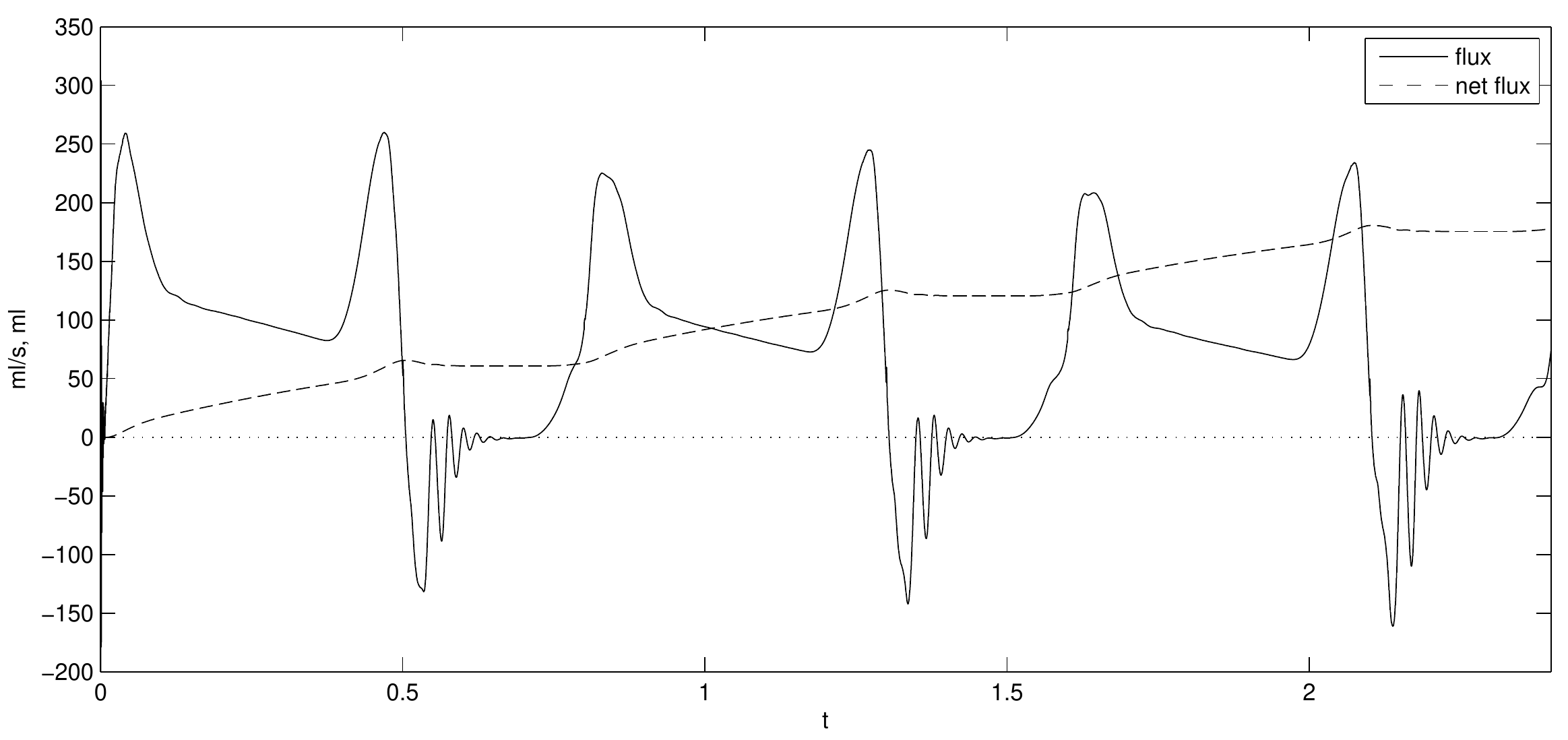}
\caption{Driving pressure and flux with low systolic ventricular pressure.}
\label{pressure_flux_low}
\end{figure}

Next, we prescribe a much lower ventricular systolic pressure, approximately half that of the standard.
This is shown in figure \ref{pressure_flux_low}, along with the resulting flux. 
The lower pressure creates a smaller load on the model valve, and thus lower strains throughout. 
It is possible at this pressure difference, which is much lower than the pressure difference for which it was tuned, the leaflets will not be strained enough to coapt properly. 
However, the flux shows this is not the case. 
The valve closes and seals effectively at these lower pressures as well. 

\begin{figure}[H]
\centering
\includegraphics[width=.55\textwidth]{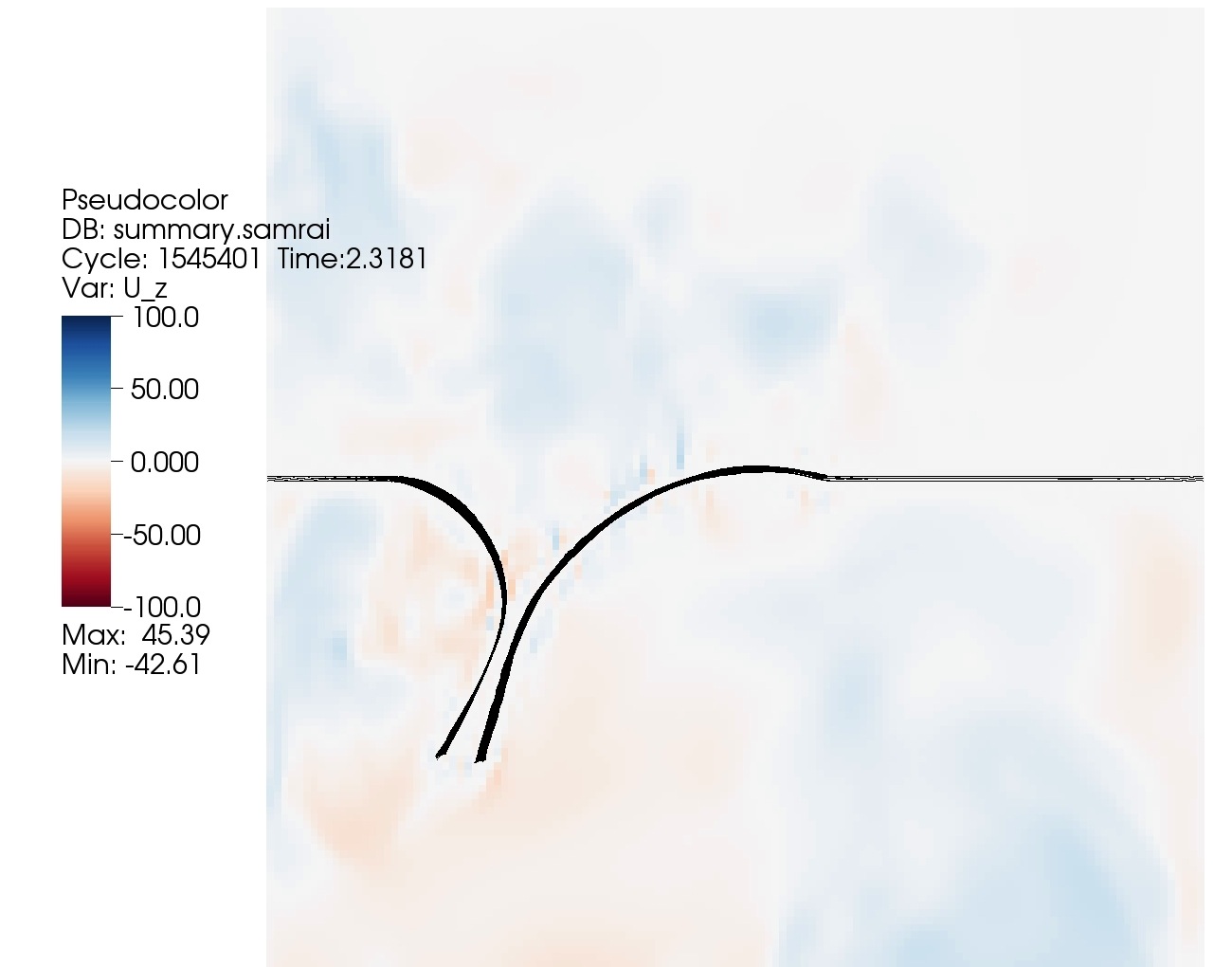}  \\ 
\includegraphics[width=.55\textwidth]{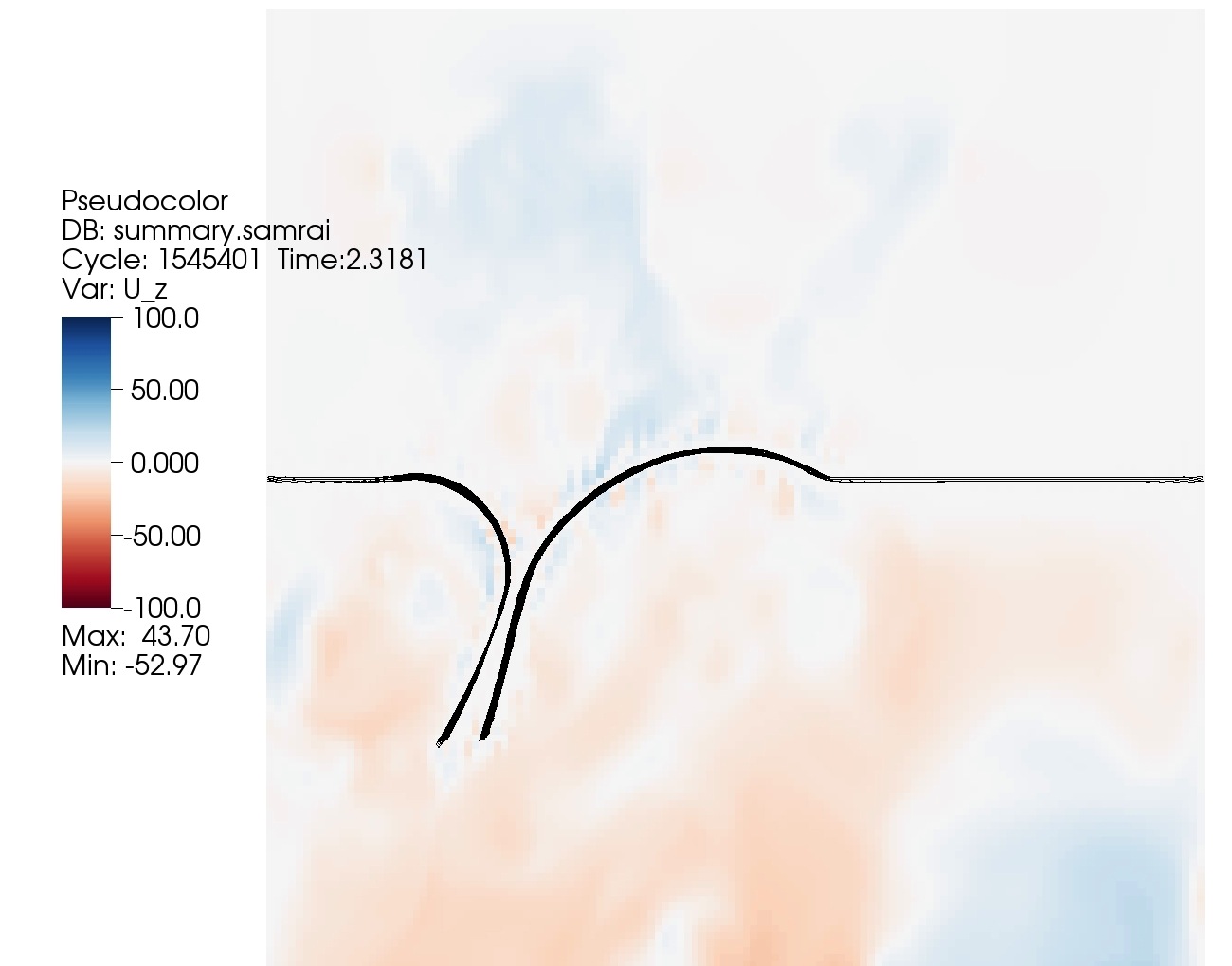}  \\ 
\includegraphics[width=.55\textwidth]{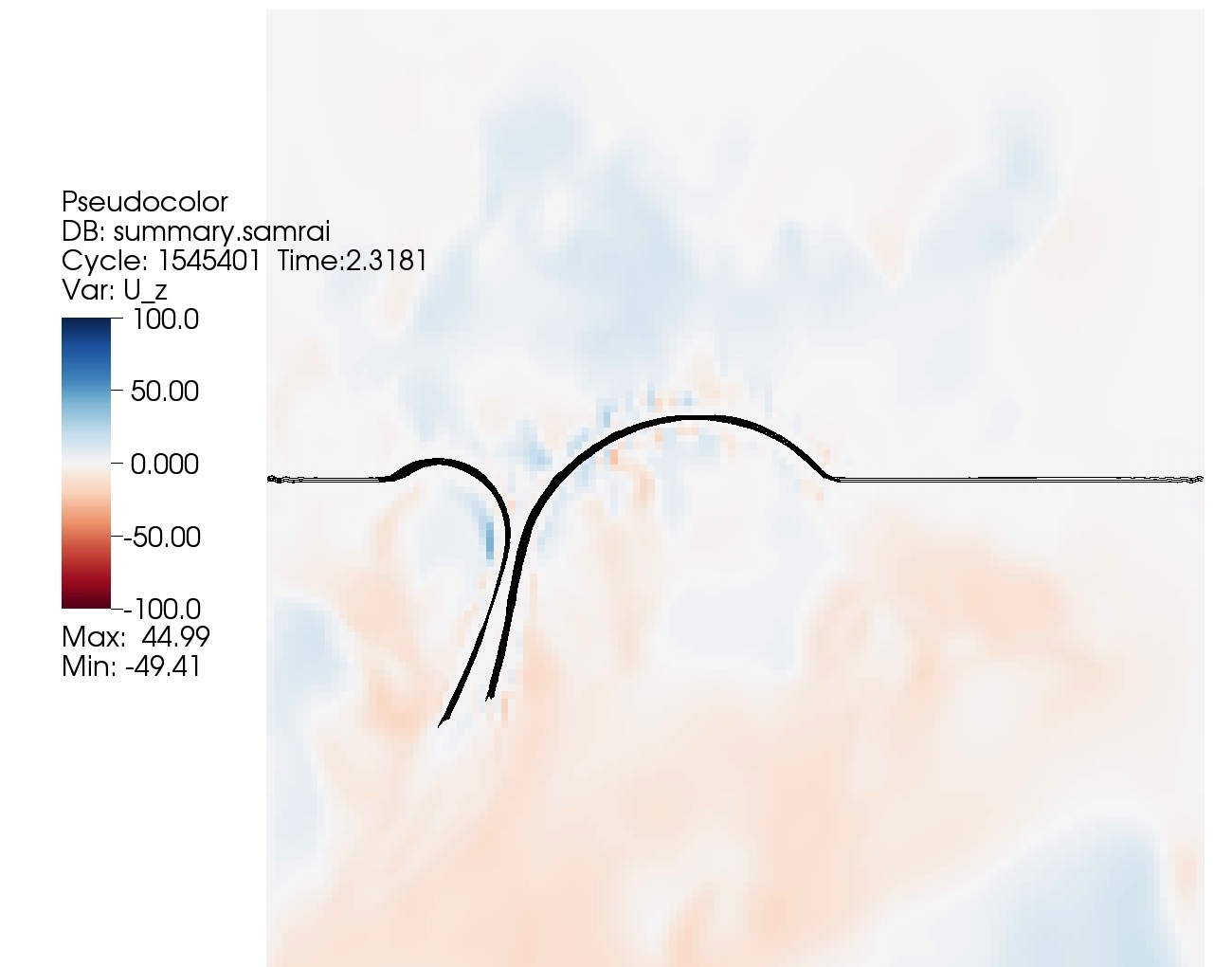}  
\caption{Slice view of $z$ component of velocity with low, standard and high ventricular systolic pressure.}
\label{three_pressure_slices}
\end{figure}

To compare, we view three slice views of the velocity field with the valve at $t = 2.32$, shown in figure \ref{three_pressure_slices}. 
This is approximately peak pressure difference during the third beat of the simulation. 
The prescribed pressure difference in the left frame is 55.0 mmHg, the center frame is 114.9 mmHg, and the right frame is 242.5 mmHg. 
The valve appears to close well in all three frames. 
The higher the pressure, the more loaded the valve appears and the smaller radius of curvature we see in the leaflets, as expected. 
Despite very different loading pressures, the velocity field in each of the three frames looks qualitatively similar. 

For the present work, this suggests that the model valve is robust to a number of driving conditions.  
A more detailed analysis of the strains the model undergoes would be instructive for future work.

\begin{figure}[H]
\centering 
\includegraphics[width=\textwidth]{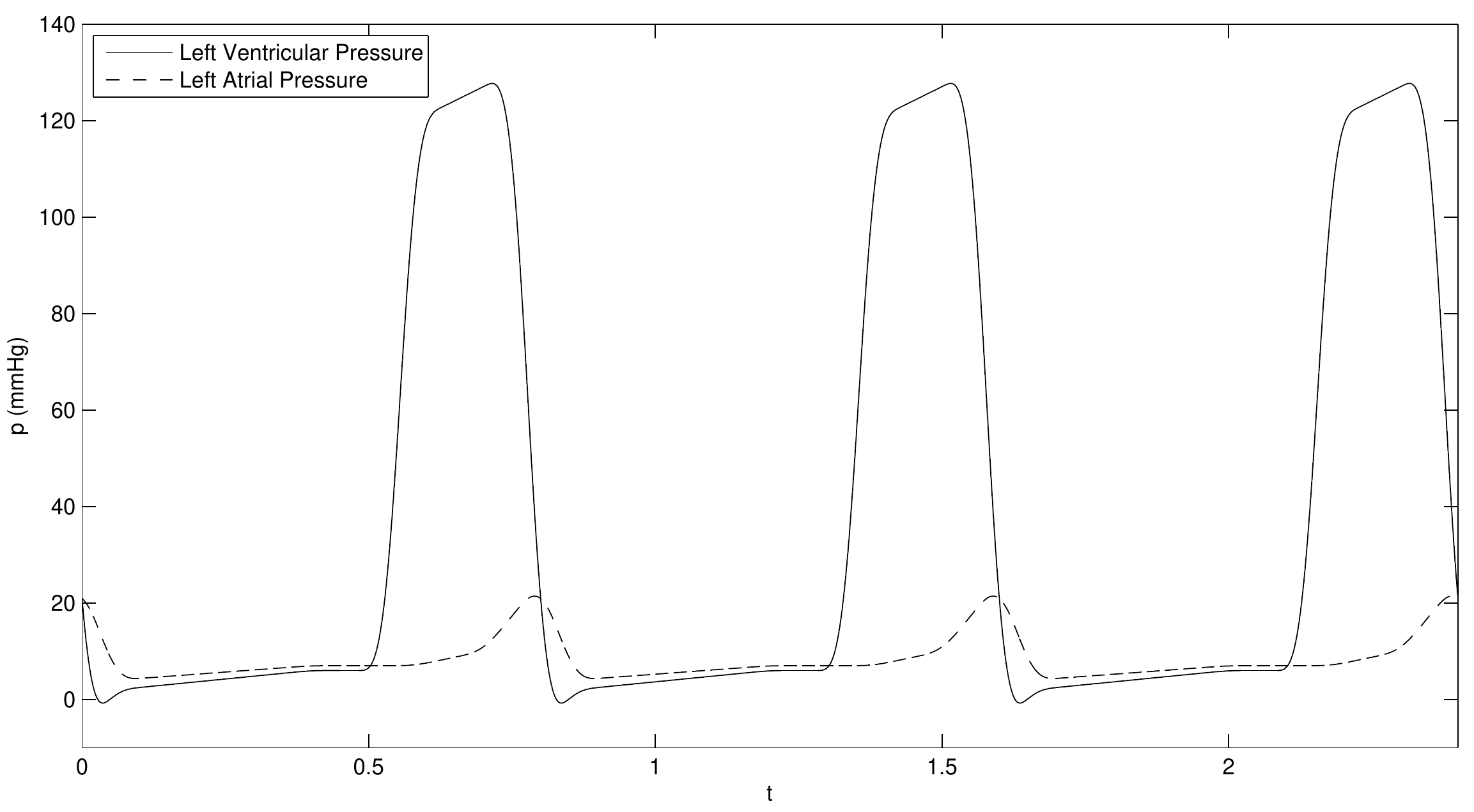}
\includegraphics[width=\textwidth]{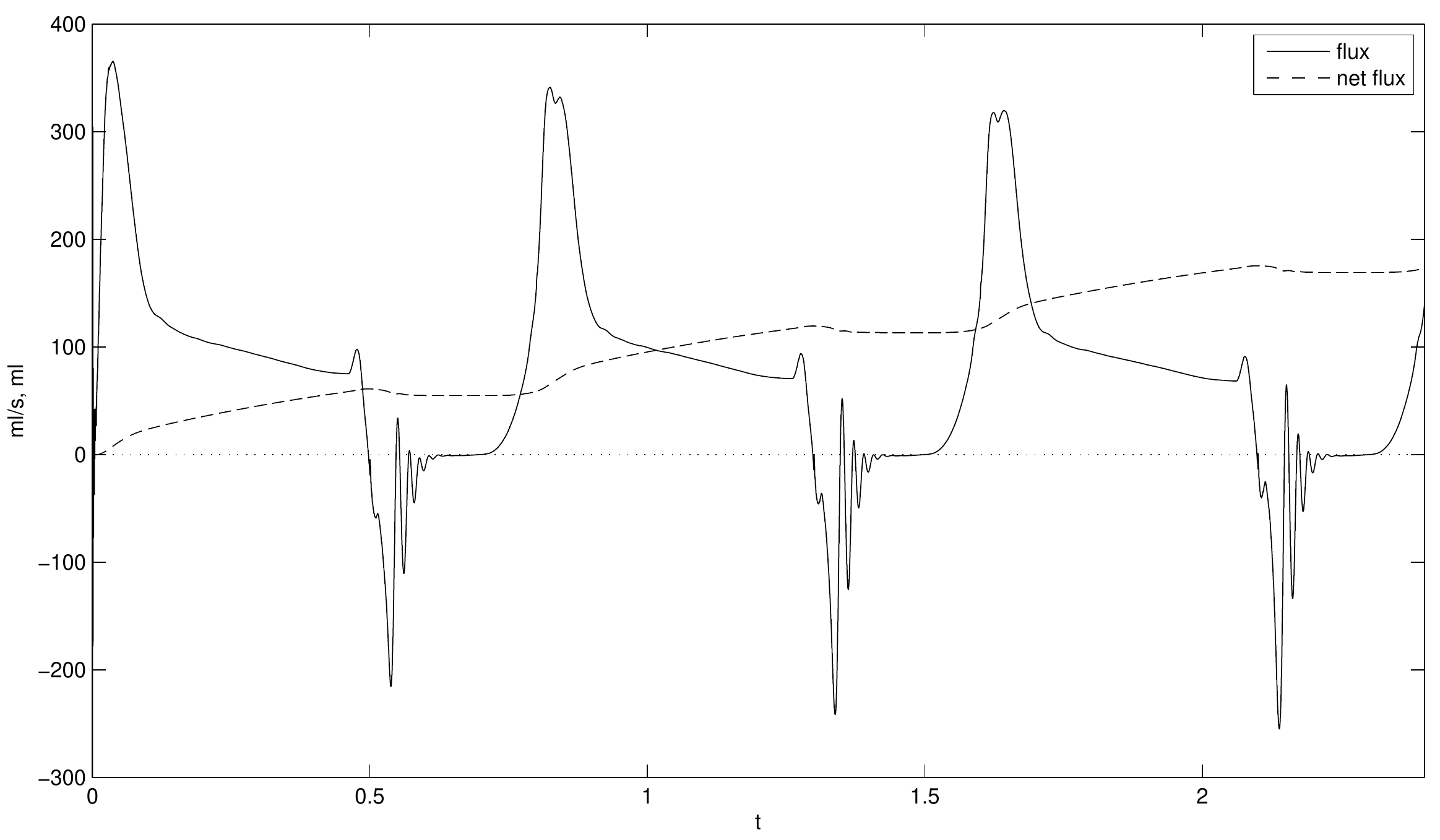}
\caption{Driving pressure and flux with no atrial kick.}
\label{pressure_flux_nokick}
\end{figure}

Finally, we drive the flow with no atrial kick present in the pressure. 
Since it has been suggested that the atrial kick affects closure of the valve, we wish to see if the model closes properly without it. 
The pressure and emergent flux are shown in figure \ref{pressure_flux_nokick}. 
The spike in forward flow during the time in which the atrial kick occurs is absent, as expected. 
The initial spike during closure has a slightly different pattern, but looks qualitatively similar to closure with the atrial kick. 

To see if we can identify the breaking jet and vortex, which are now possibly missing, we view this simulation at three time steps immediately before closure.
The same views are shown for the version with the atrial kick in figures \ref{three_velocities_breaking_one}-\ref{three_velocities_breaking_three}. 

\begin{figure}[H]
\centering
$ \begin{array}{ccc}
\includegraphics[width=.3\textwidth]{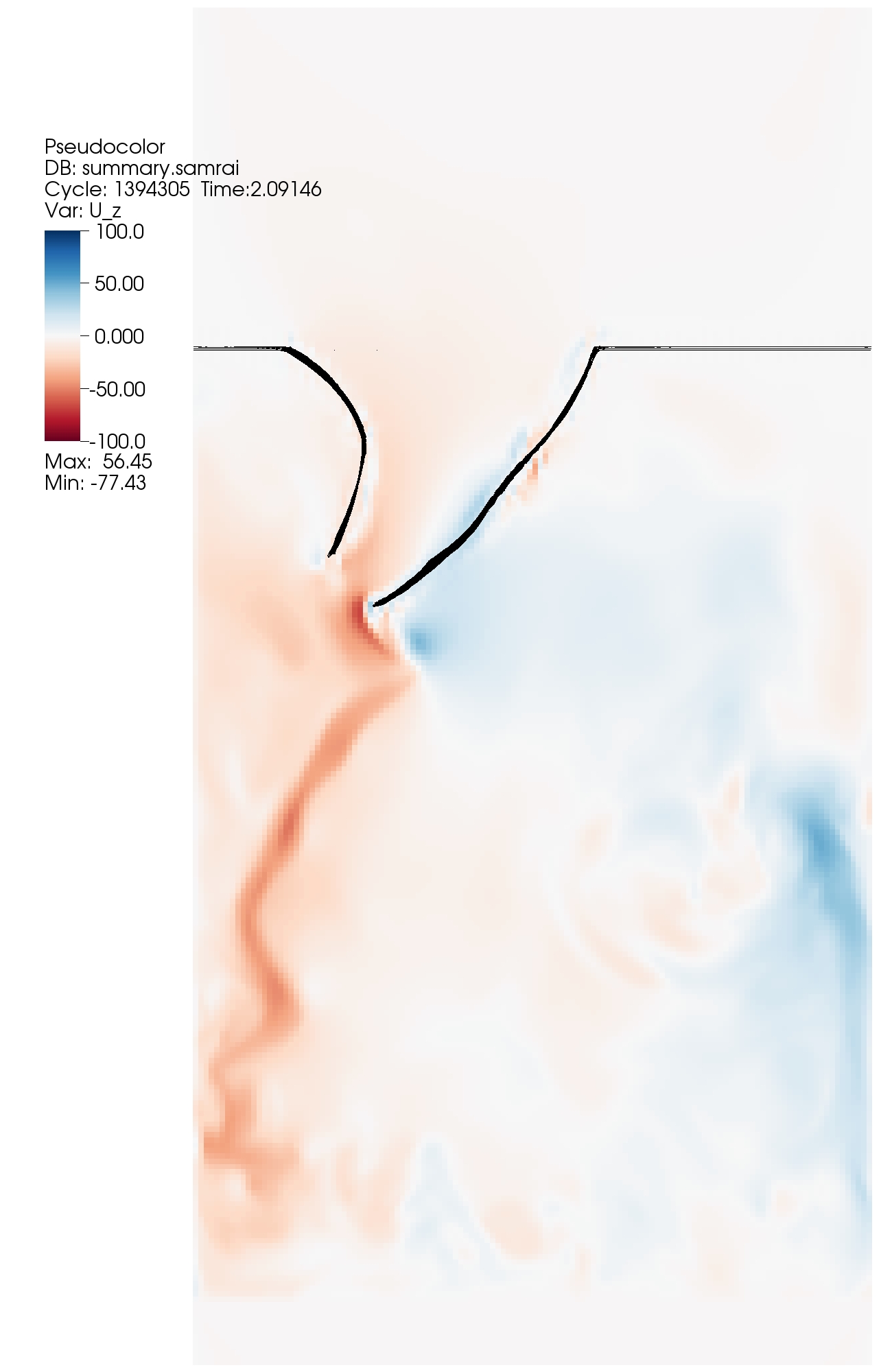}  
& \includegraphics[width=.3\textwidth]{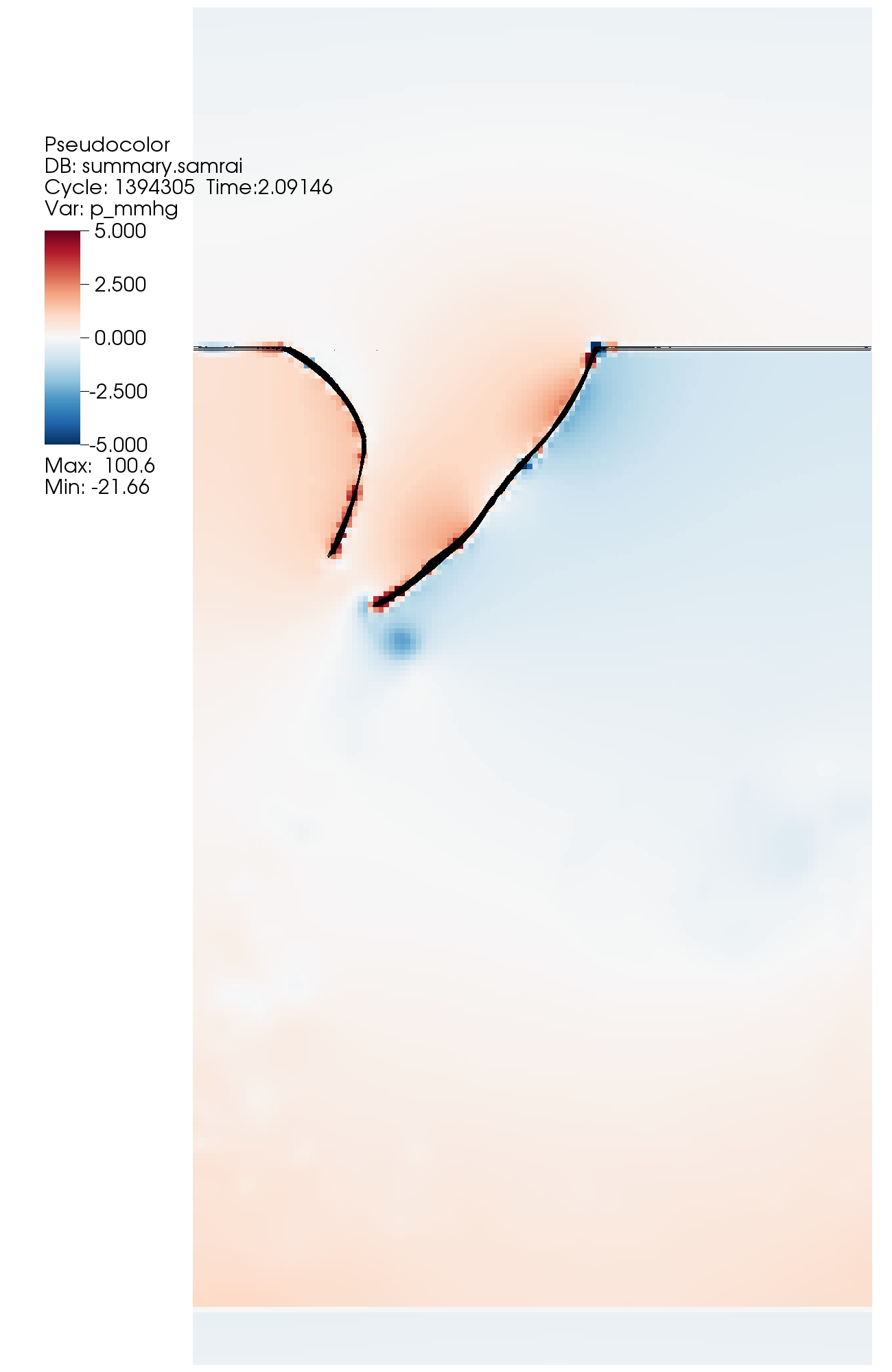}  
& \includegraphics[width=.3\textwidth]{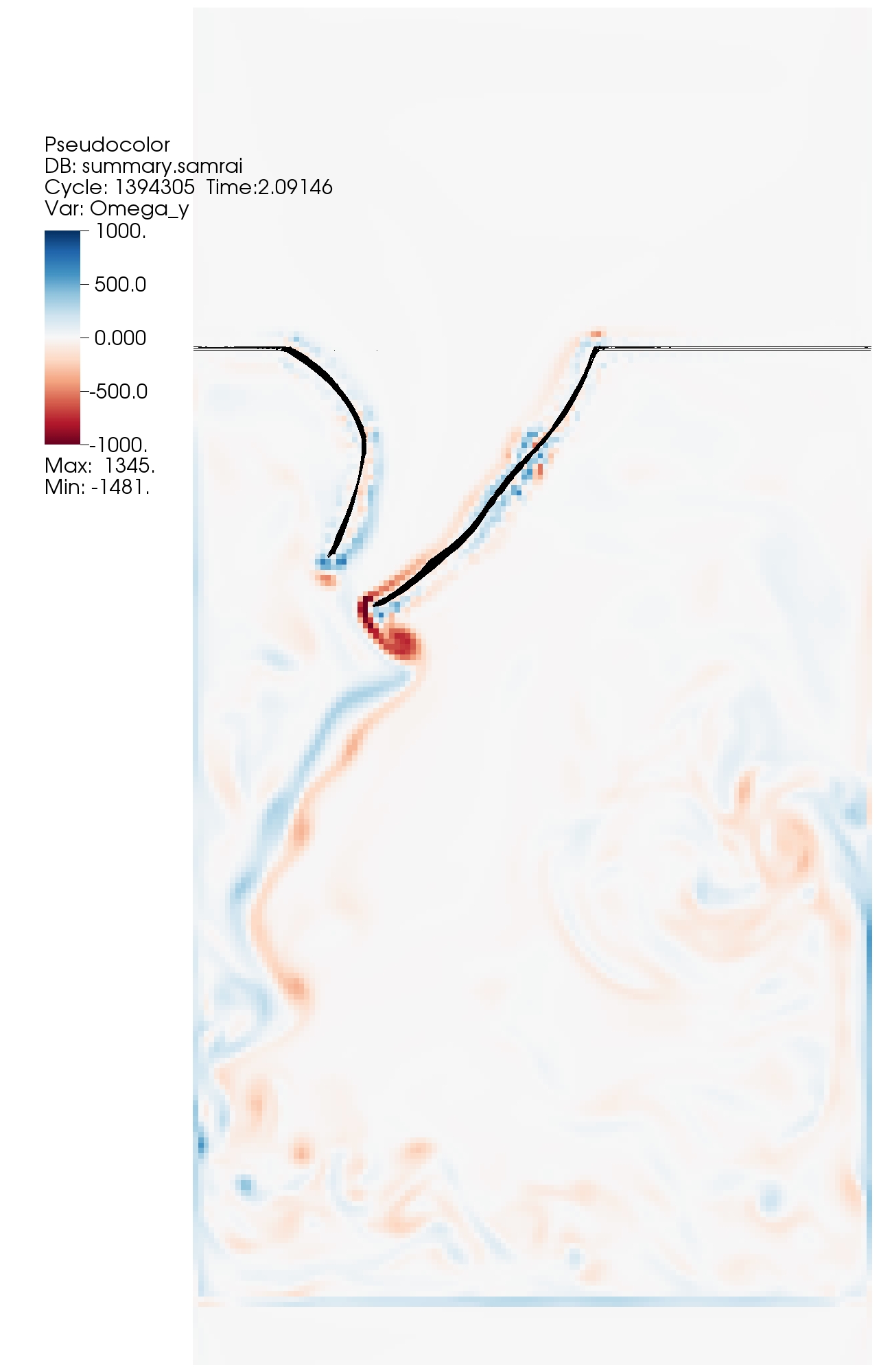}  
\end{array} $
\caption{Slice view of $z$ component of velocity (left) and modified pressure (center) and vorticity (right) early in the period when we might expect the breaking jet. Driving pressure has no atrial kick.}
\label{three_velocities_breaking_nokick_one}
\end{figure}

Figure \ref{three_velocities_breaking_nokick_one} shows the simulation without the atrial kick at $t = 2.09146$, during the time when the kick would normally occur. 
Forward flow is much reduced, there is little evidence of a large jet between the leaflets at this time. 
There is a small pressure drop over the anterior leaflet at this time. 
This may be because the papillary muscles are in motion, pulling the leaflet to the left in the frame. 
There is a small vortex being shed from the anterior leaflet, but no larger structures appearing below the valve. 

\begin{figure}[H]
\centering
$ \begin{array}{ccc}
\includegraphics[width=.3\textwidth]{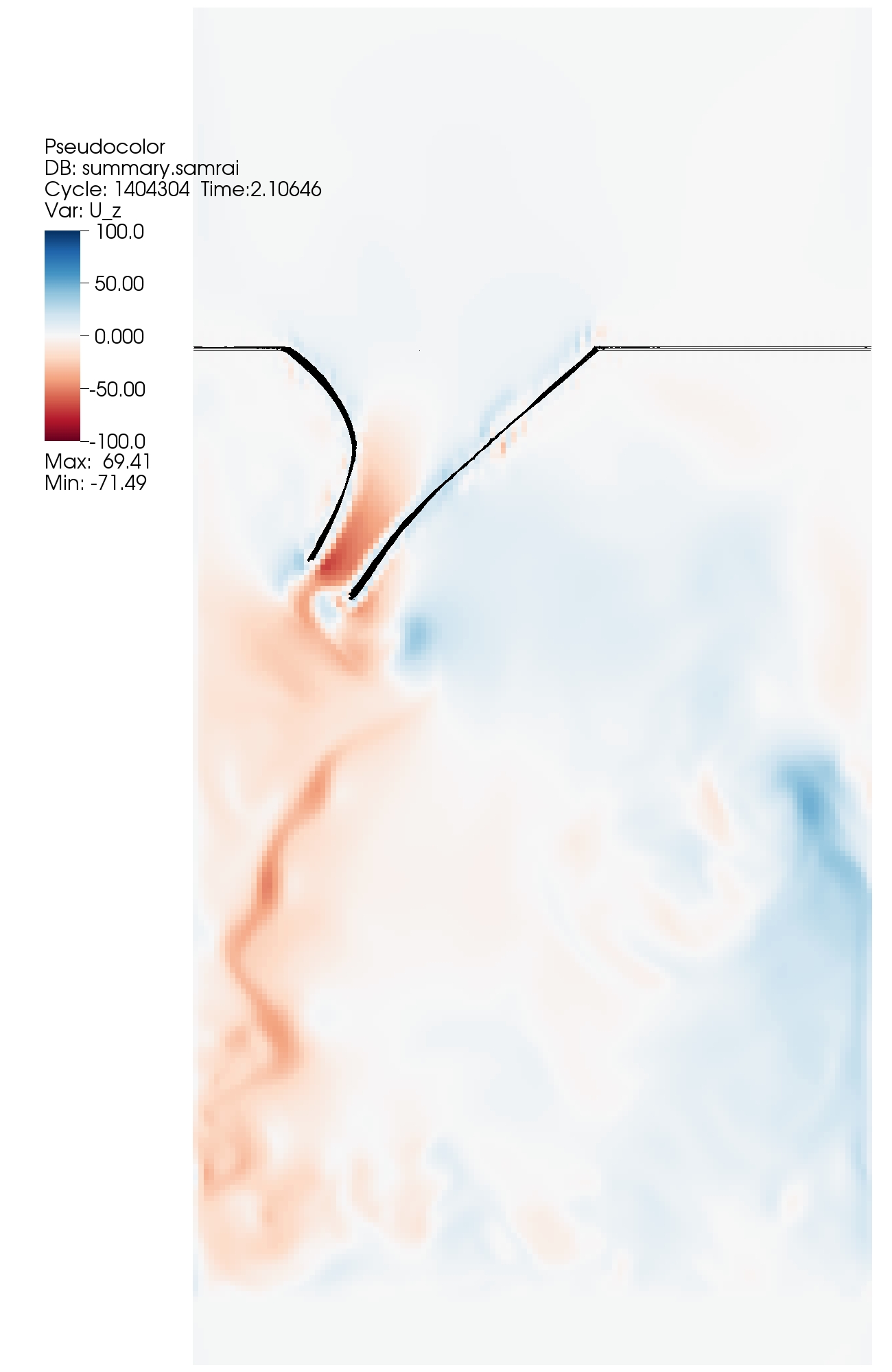}  
& \includegraphics[width=.3\textwidth]{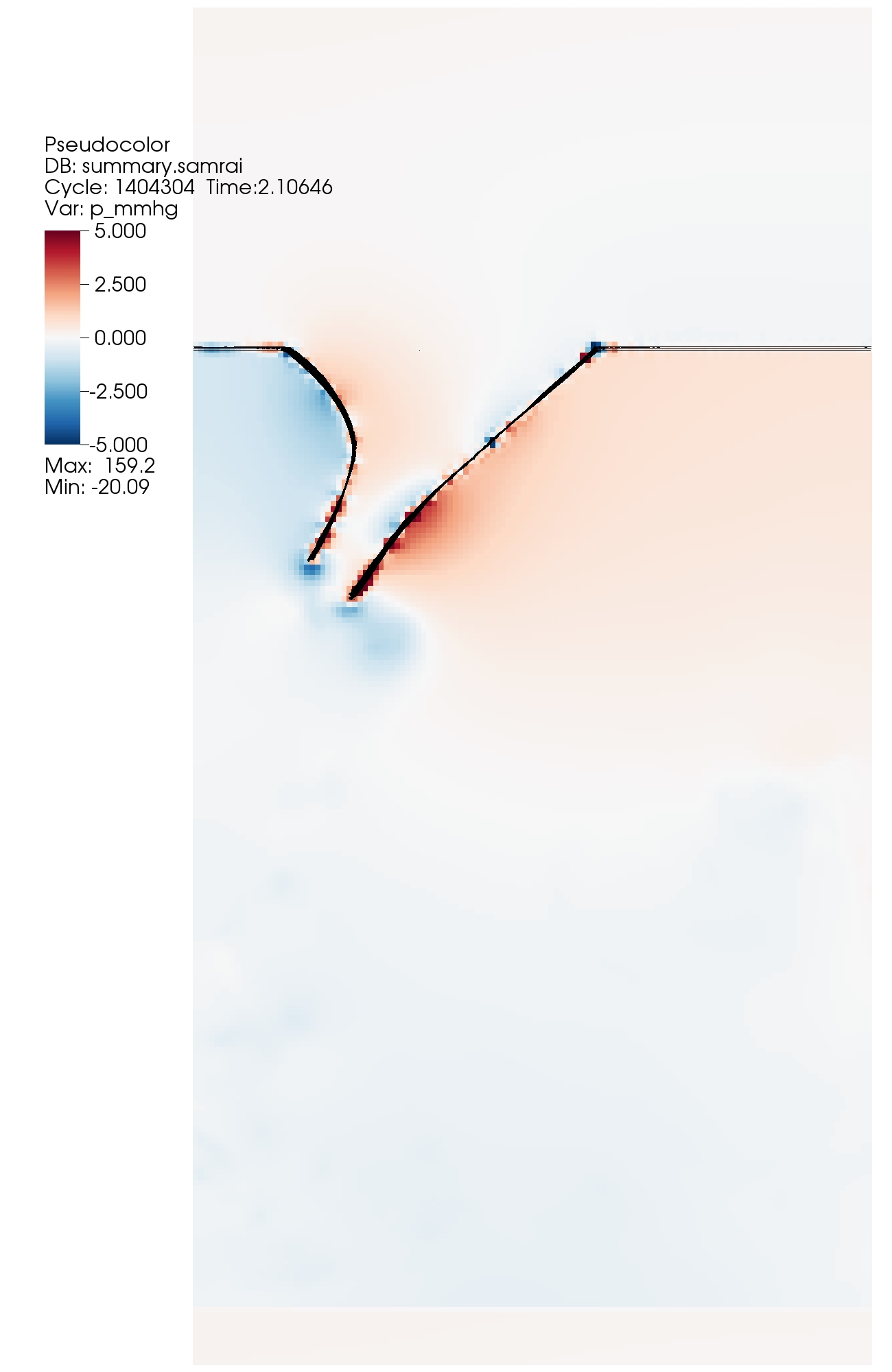}  
& \includegraphics[width=.3\textwidth]{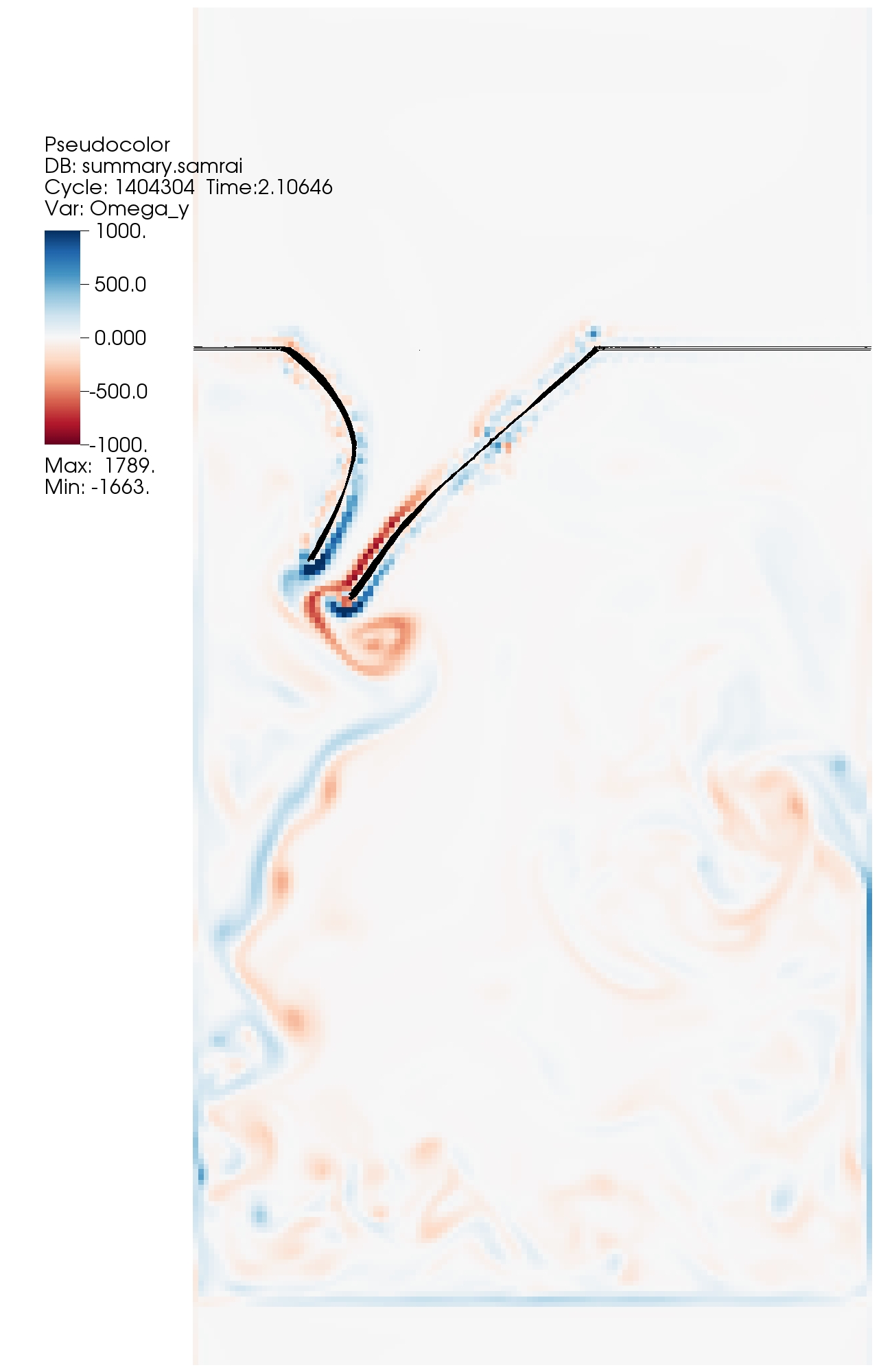}  
\end{array} $
\caption{Slice view of $z$ component of velocity (left) and modified pressure (center) and vorticity (right)  in the middle of the period when we might expect the breaking jet. Driving pressure has no atrial kick.}
\label{three_velocities_breaking_nokick_two}
\end{figure}

Figure \ref{three_velocities_breaking_nokick_two} shows the simulation at time $t = 2.10646$. 
There is forward flow remaining between the leaflets near the free edges, and a pressure difference across the anterior leaflet. 
This appears to be some amount of smaller breaking-jet-type phenomena than was observed when the atrial kick was present. 
The vortex shed from the anterior leaflet remains tight, and does not form a larger structure below the valve.

\begin{figure}[H]
\centering
$ \begin{array}{ccc}
\includegraphics[width=.3\textwidth]{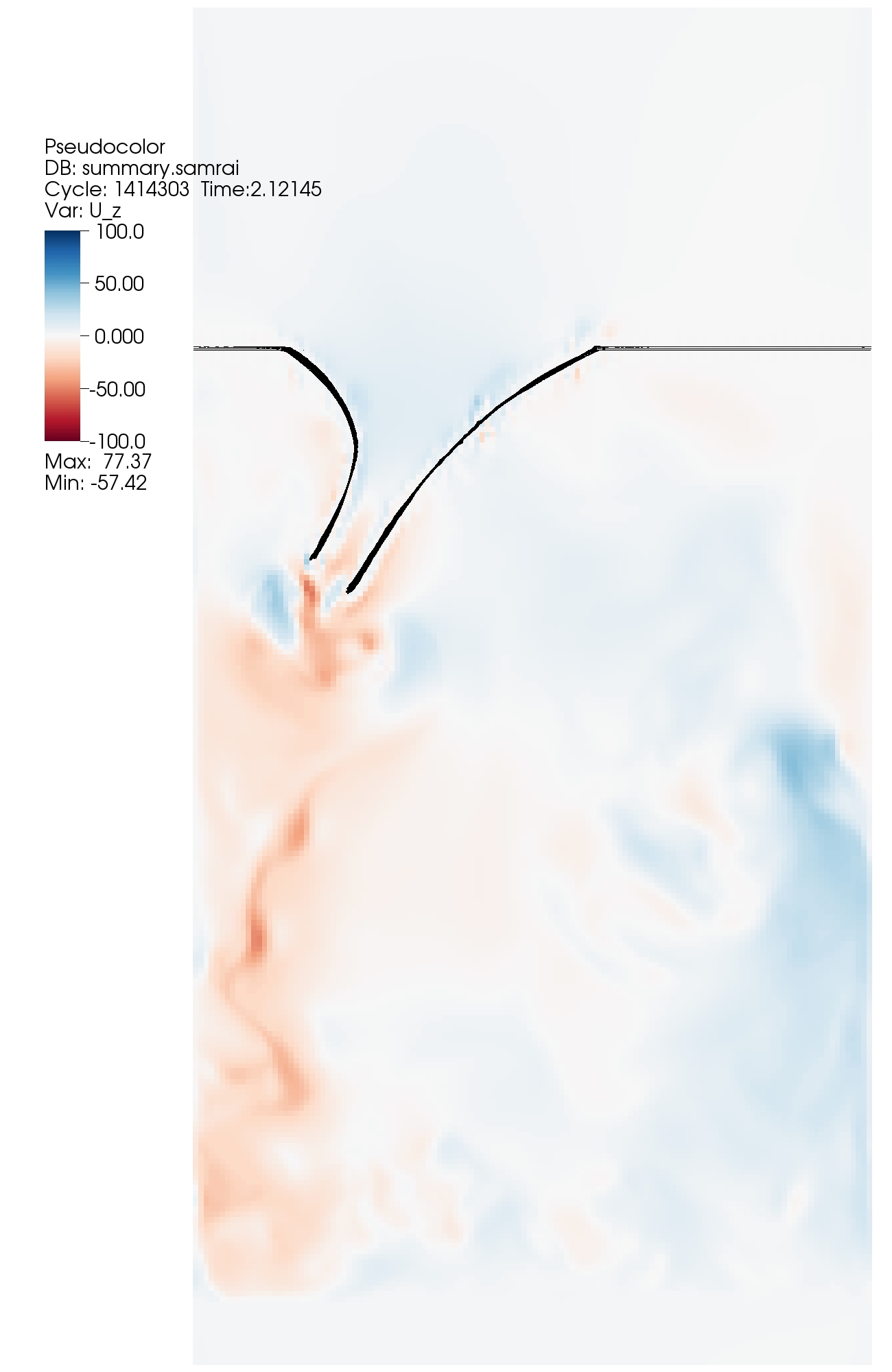}  
& \includegraphics[width=.3\textwidth]{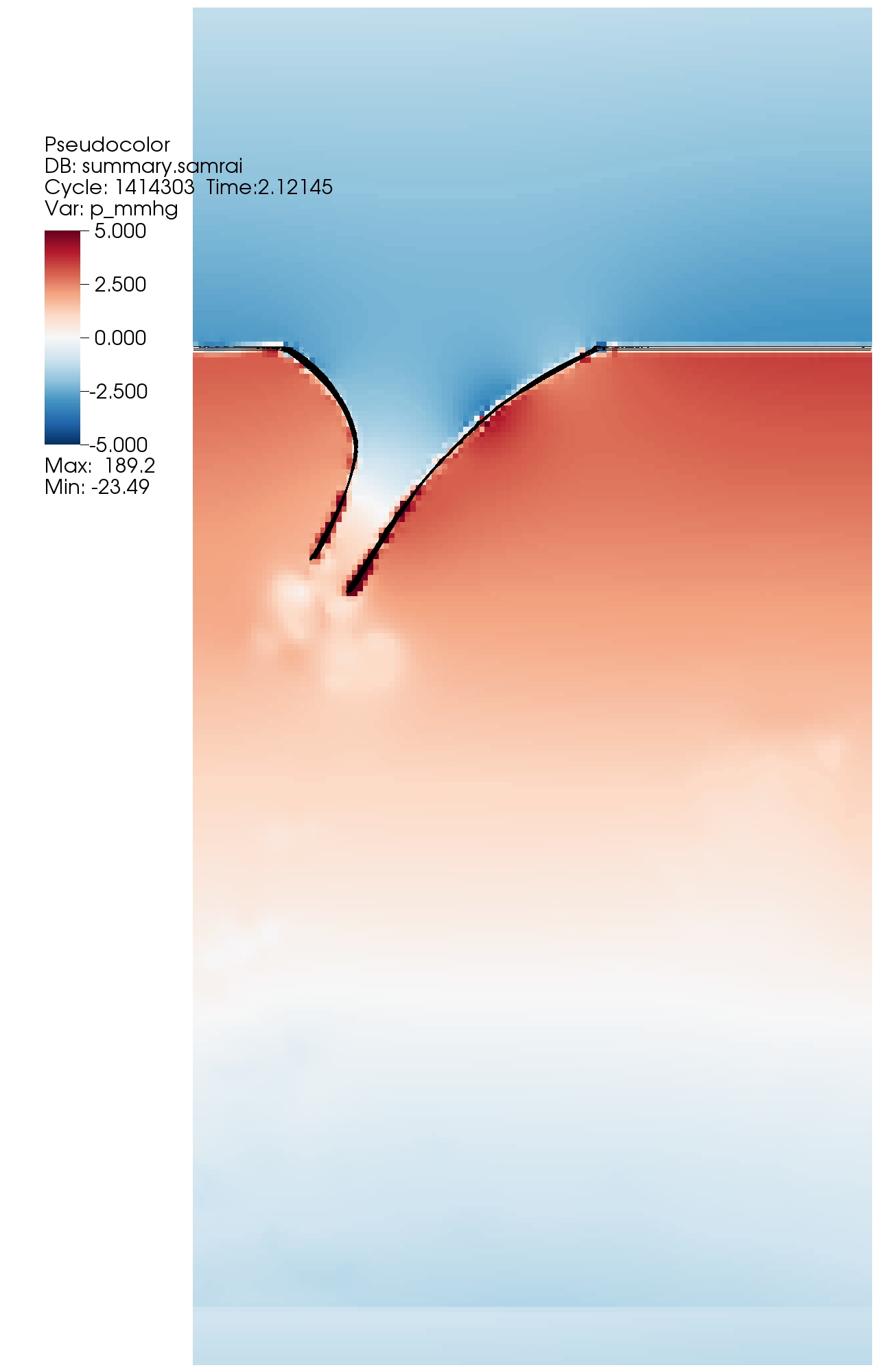}  
& \includegraphics[width=.3\textwidth]{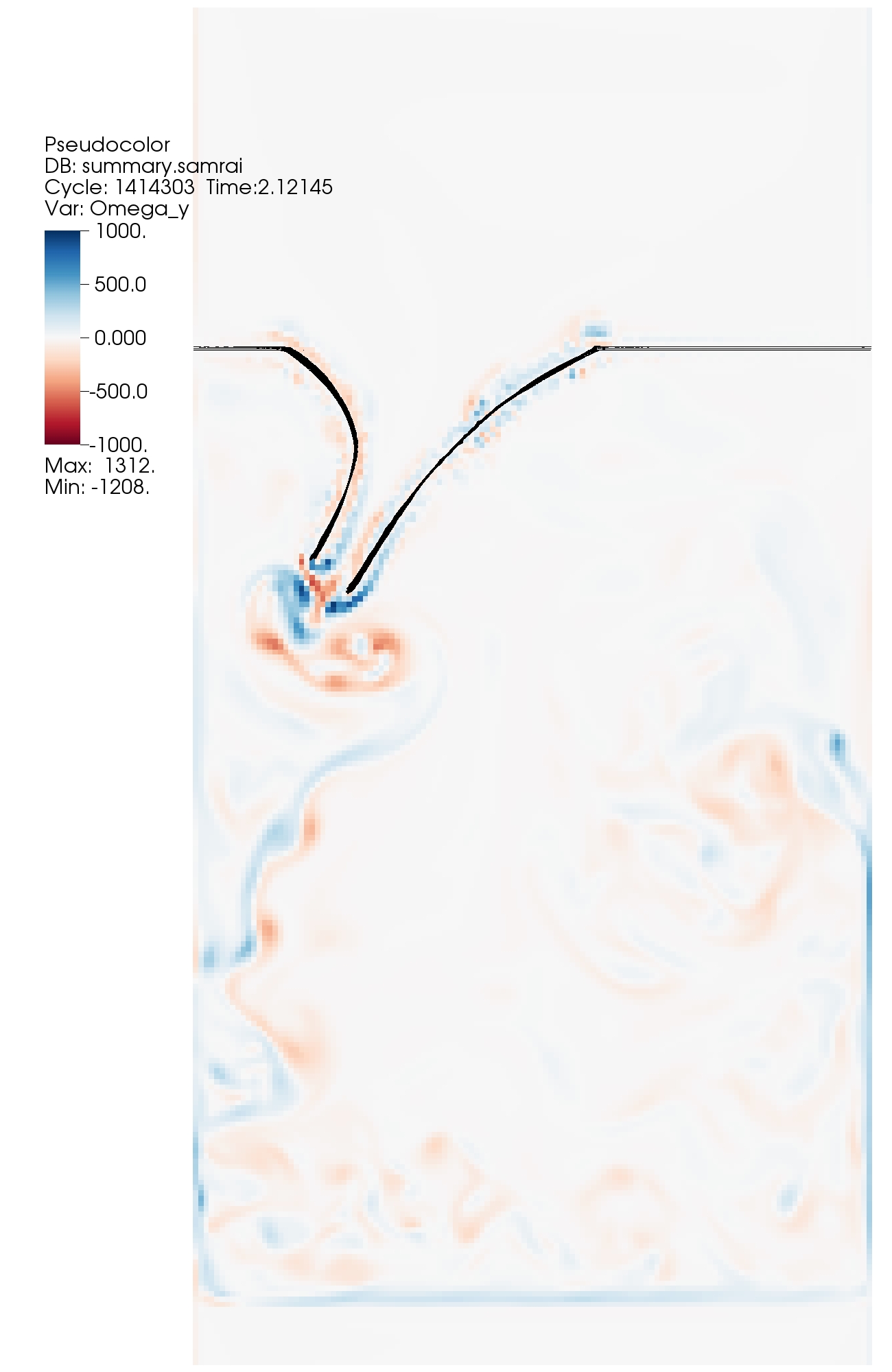}  
\end{array} $
\caption{Slice view of $z$ component of velocity (left) and modified pressure (center) and vorticity (right) at the end of the period when we might expect the breaking jet. Driving pressure has no atrial kick.}
\label{three_velocities_breaking_nokick_three}
\end{figure}

Figure \ref{three_velocities_breaking_nokick_three} shows the simulation at time $t = 2.12145$. 
As in the previous frame, there is still some forward flow between the leaflets when the flow at the ring is no longer forward.
The pressure may be lower between the leaflets, but this is not obvious. 
There is much less visible structure in the vorticity than was present with the atrial kick. 

Both the breaking jet and vortex phenomena are less present in this model without the atrial kick. 
This suggests that these phenomena are indeed helped by the atrial kick, and lends plausibility to the argument that these effects help the valve close. 
This supports, though does not prove, the assertion that the atrial kick assists in mitral valve closure.

\section{Model comparisons}
\label{model_comparison}

\begin{figure}[H]
\centering 
\includegraphics[width=\textwidth]{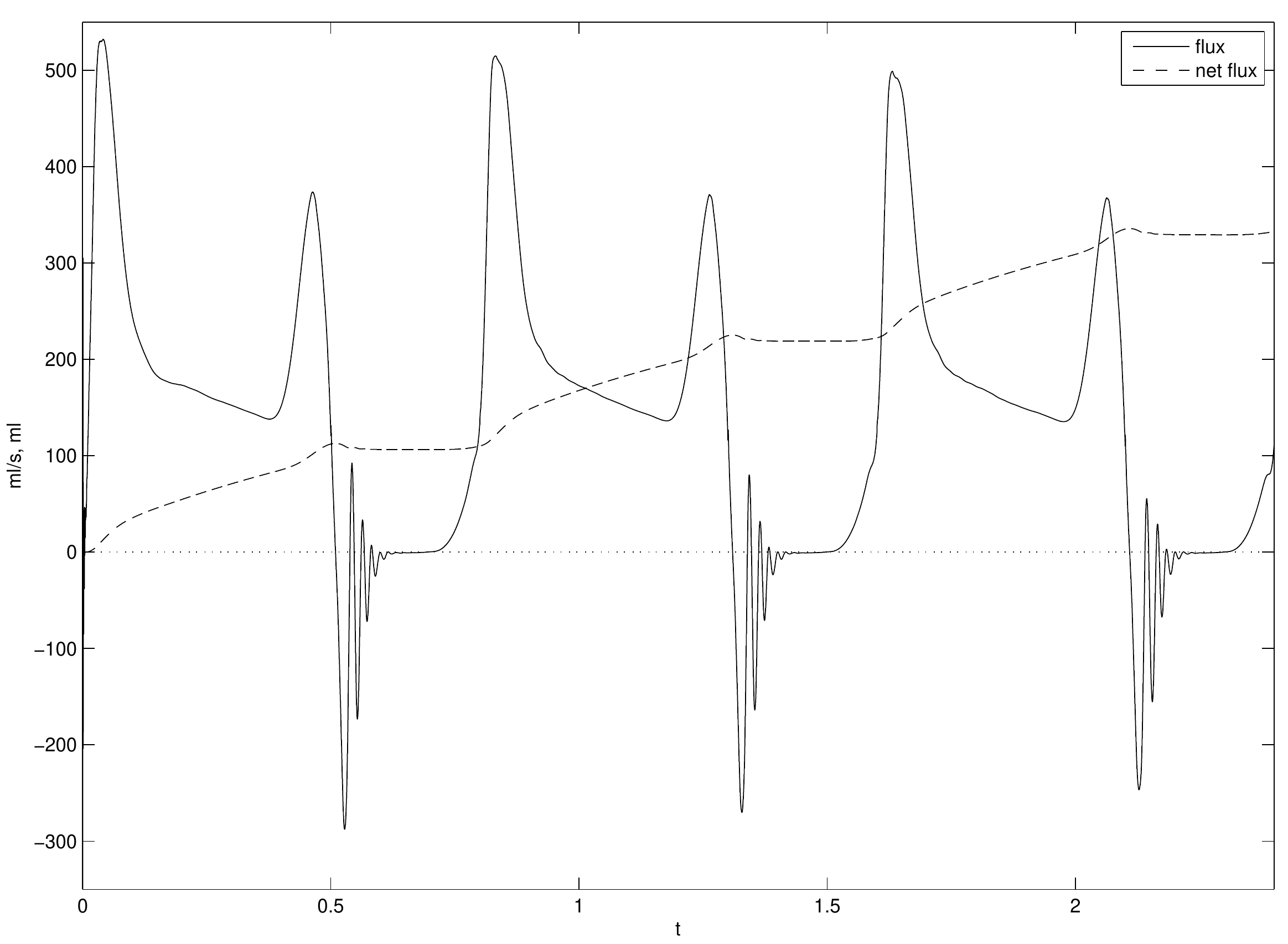}
\caption{Emergent flux for model with no edge connectors at commissures. Driving pressure identical to that of figure \ref{pressure_and_flux}.}
\label{flux_no_connector.pdf}
\end{figure}

The first concerns a model with no edge-connectors. 
In this model, there is no model tissue below the attachment of the chordae tendineae. 
Speaking anatomically, this has less tissue to connect the leaflets at the commissures. 
This model should open more freely, and create high diastolic flows. 
This is indeed the case. 
This model may be prone to leaks at the commissures, since there is less model tissue there to hold things together. 
To our surprise, this model does not leak, and appears fairly robust. 
This is contrary to many preliminary experiments, in which leaks at the commissures were a constant problem. 
However, with the model and setup tuned, this no longer appears to be the case. 
The simulation is driven with standard pressures shown in figure \ref{pressure_and_flux}. 
The emergent flux, which is approximately 50 percent higher during diastole, is shown in figure \ref{flux_no_connector.pdf}. 

\begin{figure}[H]
\centering 
\includegraphics[width=\textwidth]{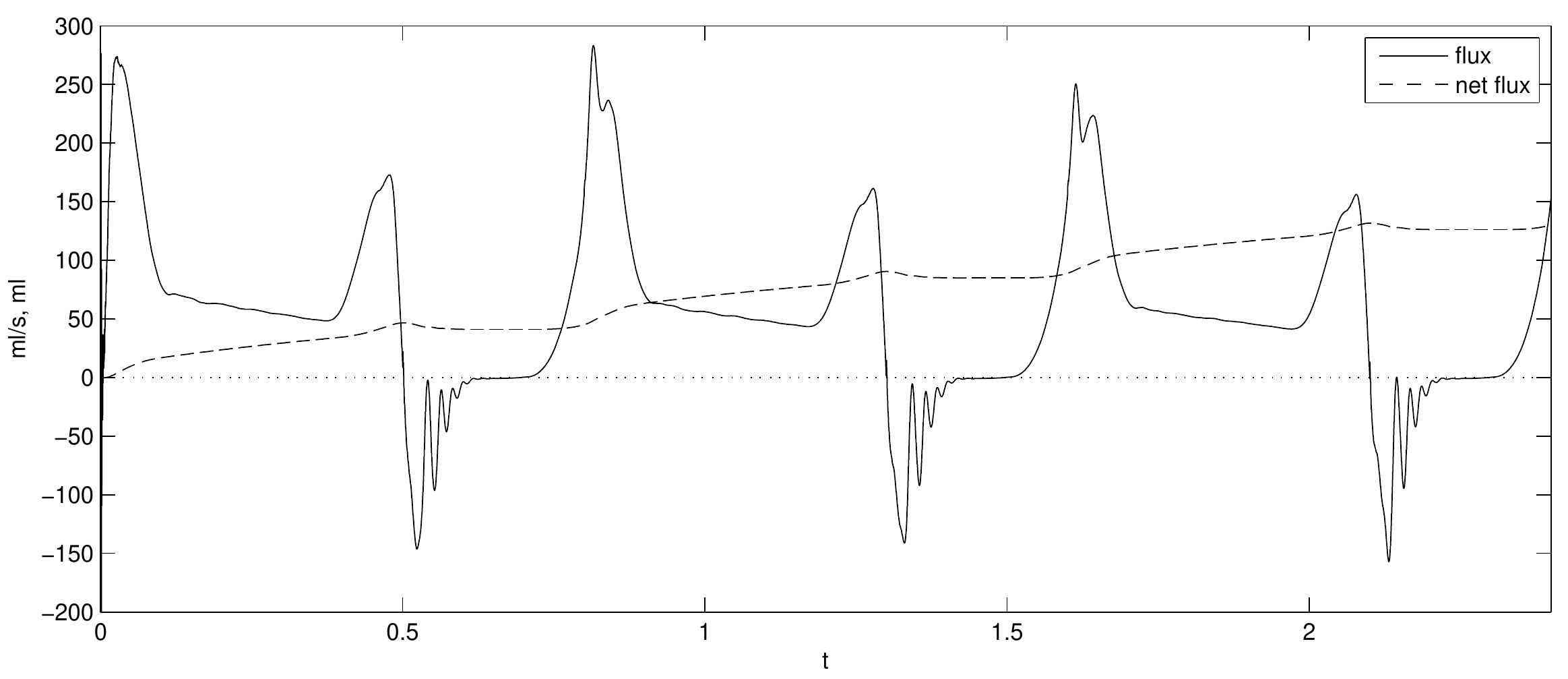}
\caption{Emergent flux for model with four times the edge connectors at commissures as in Section \ref{control_results}. Driving pressure identical to that of figure \ref{pressure_and_flux}.}
\label{flux_32_connector.pdf}
\end{figure}

The next comparison concerns a model with many edge connectors. 
In this model, there is approximately four times the connecting tissue at the commissures as that of Section \ref{control_results}. 
One might expect this to result in restrictive forward flow, and it does. 
The simulation is driven with standard pressures shown in figure \ref{pressure_and_flux}. 
The emergent flux, which is somewhat more restricted during forward flow, is shown in figure \ref{flux_32_connector.pdf}. 

\begin{figure}[H]
\centering
$ \begin{array}{ccc}
\includegraphics[width=.3\textwidth]{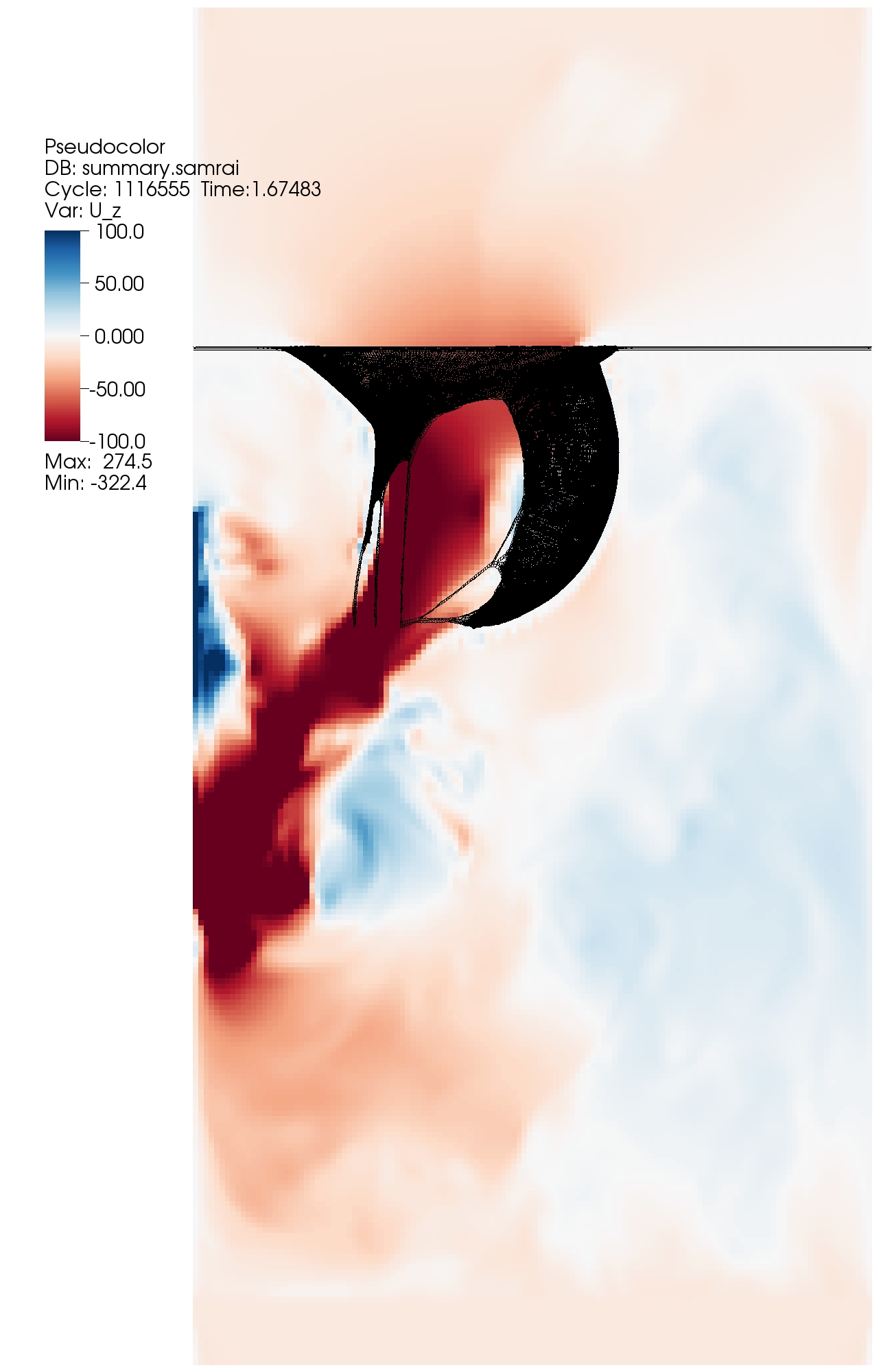}  
& \includegraphics[width=.3\textwidth]{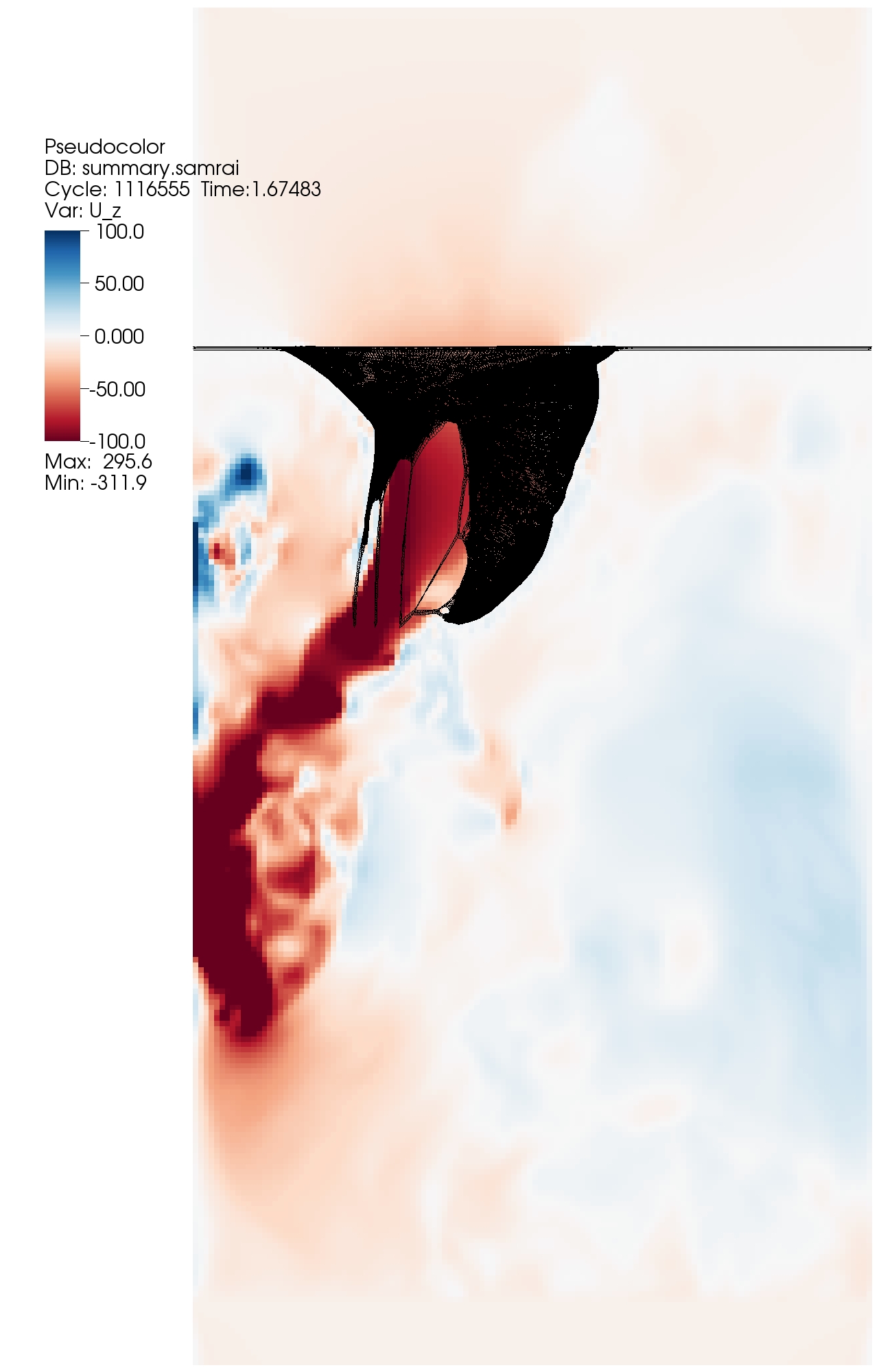}  
& \includegraphics[width=.3\textwidth]{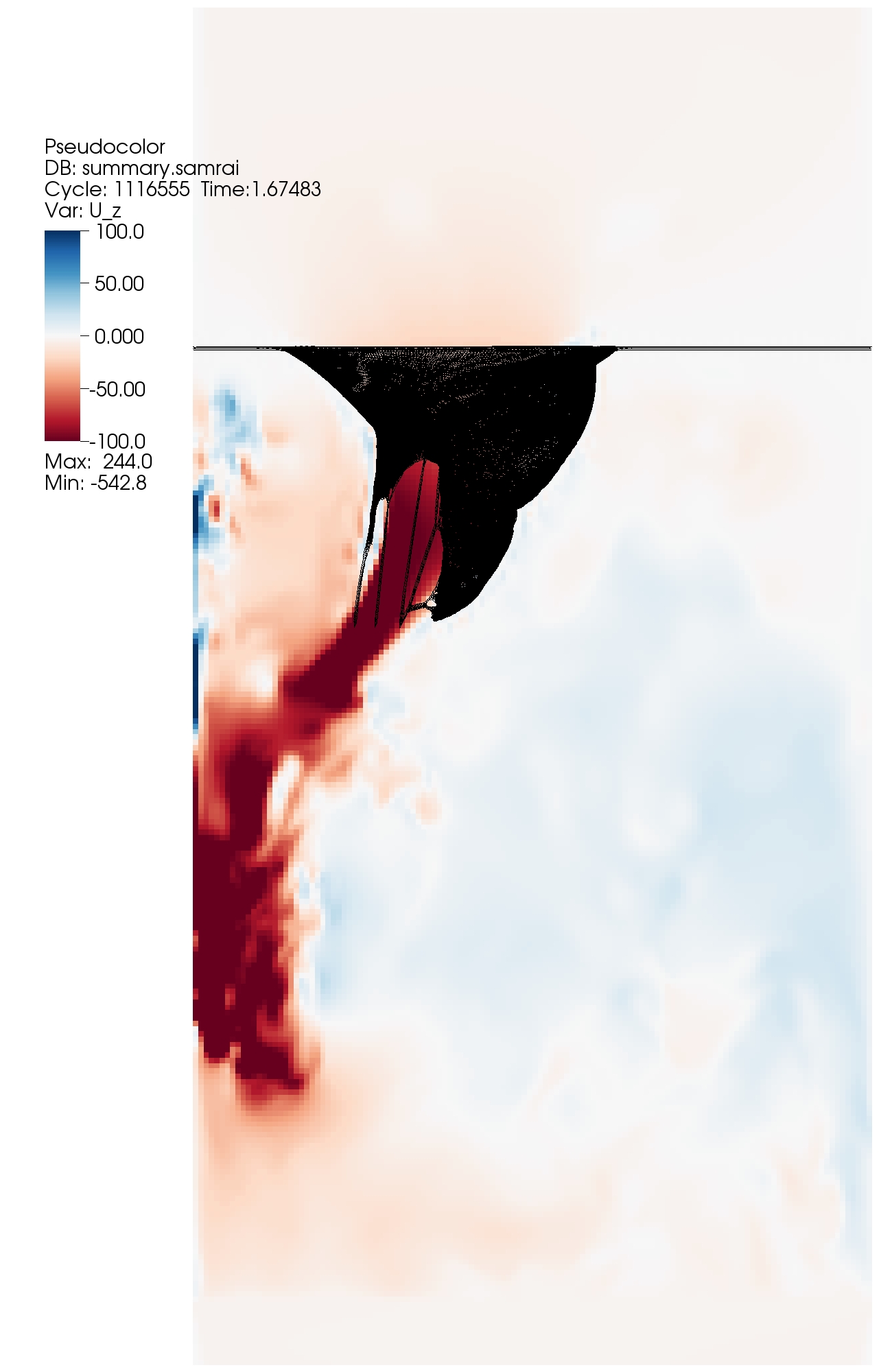}  
\end{array} $
\caption{Slice view of $z$ component of velocity during filling comparing models with no connectors (left), eight connectors (center) and 32 connectors (right).}
\label{connector_comparisons}
\end{figure}

Finally, figure \ref{connector_comparisons} shows a comparison of the three models during filling. 
Here, we show the whole valve, rather than the valve with clipping planes, to better illustrate the structure motion. 

\section{Setup comparisons}
\label{comparison}

Having shown basic results, we present results without some of the features in the simulation setup that we previously discussed. 
This serves to illustrate why we implemented certain features, and why we believe they were effective.

\begin{figure}[H]
\centering
$ \begin{array}{cc}
\includegraphics[width=.475\textwidth]{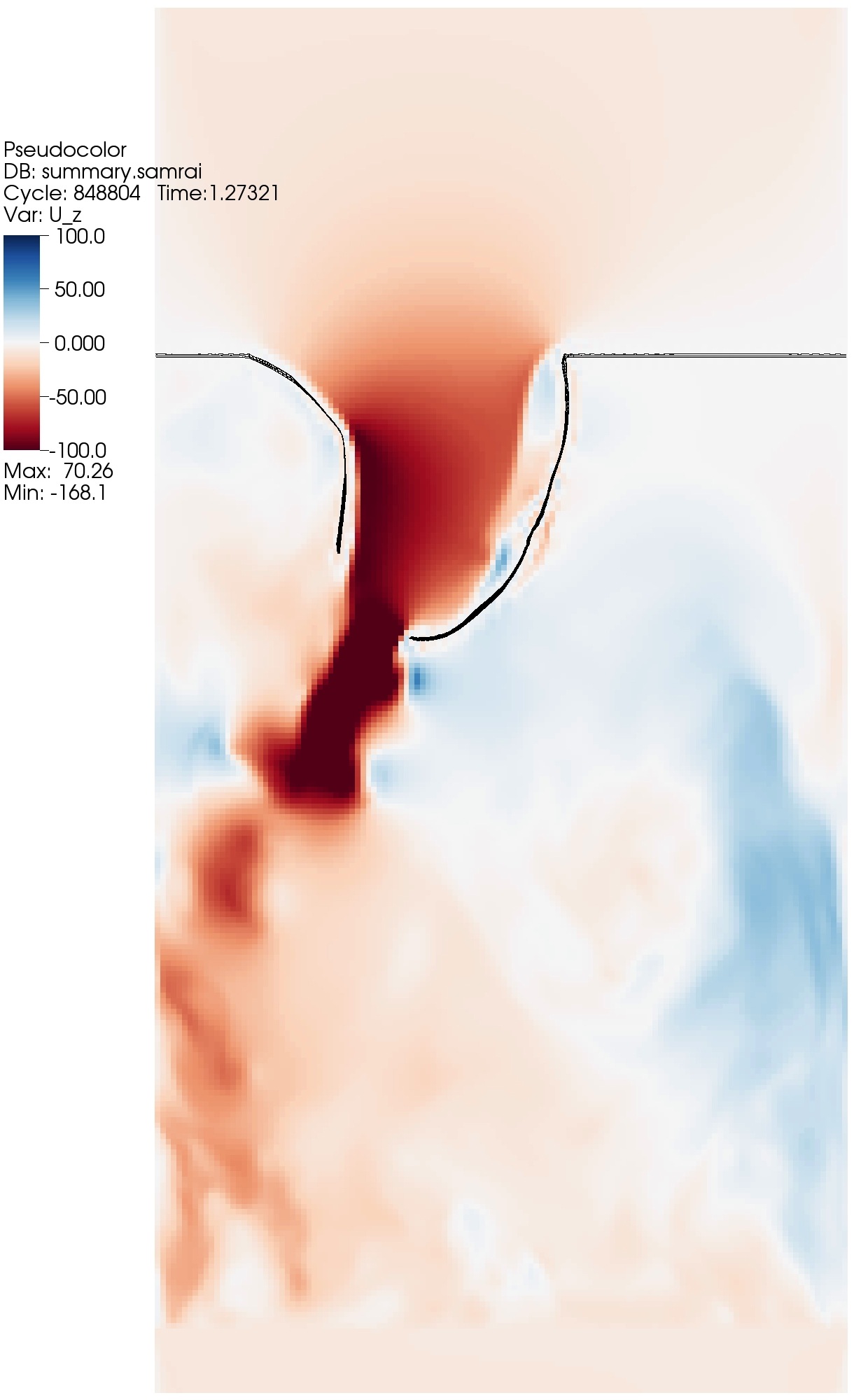}  & \includegraphics[width=.475\textwidth]{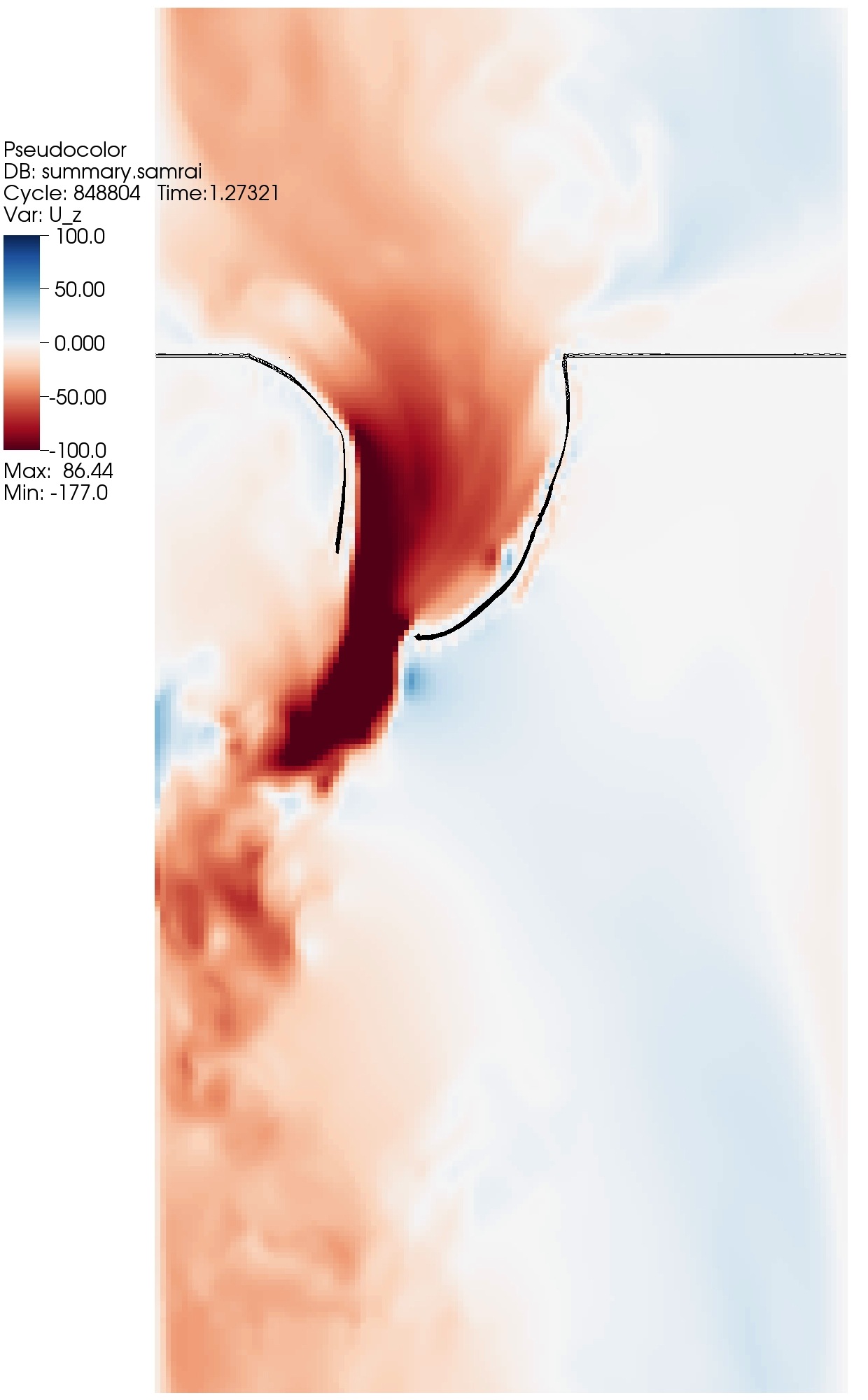}  
\end{array} $
\caption{Comparison of $z$-component of velocity with averaging force (left) and without averaging force (right) during diastole.}
\label{averaging_velocity_compare}
\end{figure}

\begin{figure}[H]
\centering 
\includegraphics[width=\textwidth]{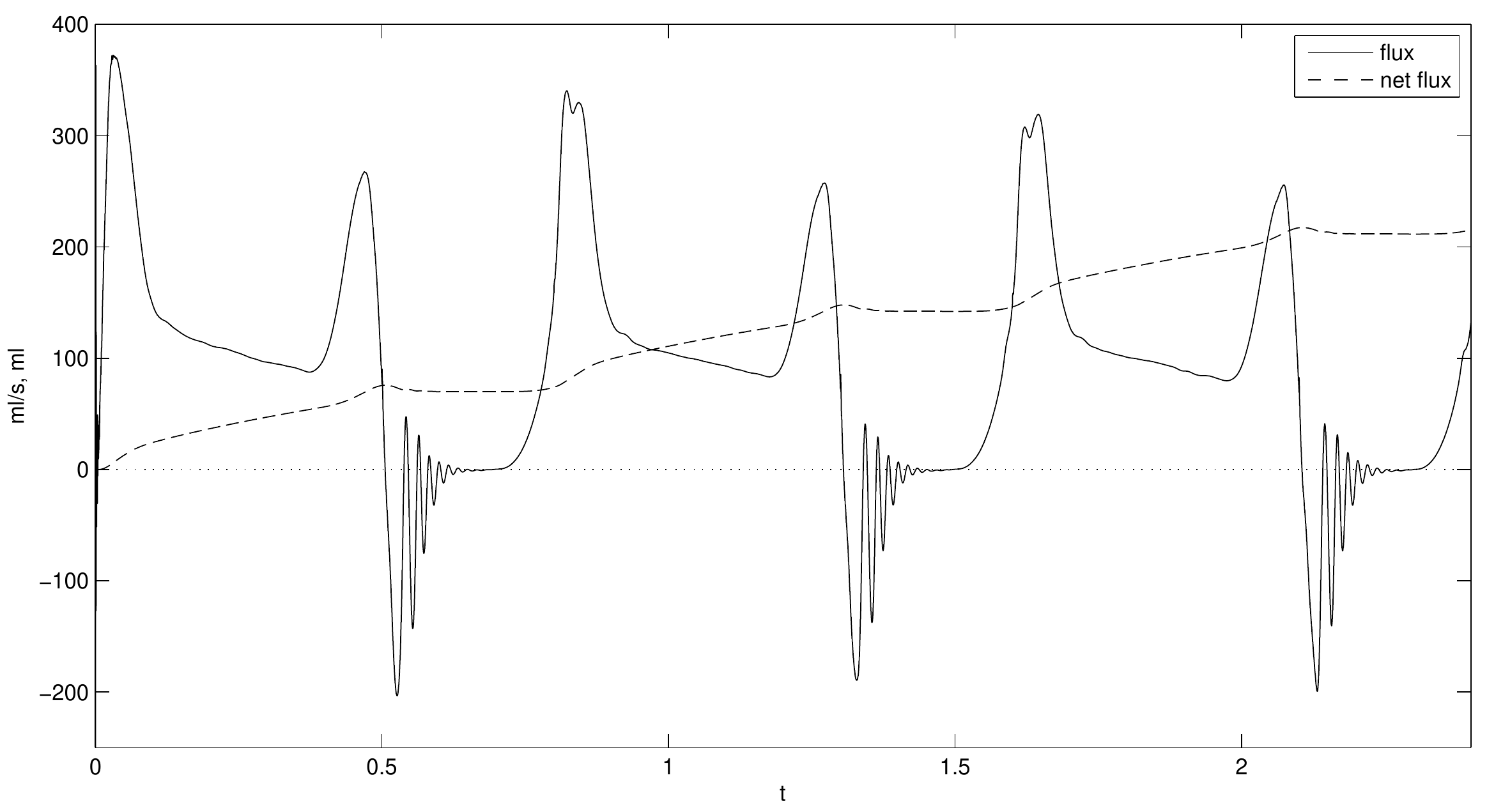} 
\caption{Flux without averaging force. }
\label{flux_no_average}
\end{figure}

Suppose that the averaging force is removed in equation \ref{straightener_force}, leaving only straightening forces in the $x,y$ directions. 
This force is given by 
\begin{align}
\bb s =  -\eta (u, v, 0) . 
\label{straightener_force_no_z}
\end{align}
This will result in a much less uniform inflow profile, and less damping in the overall system. 
A comparison of a slice view of the $z$-component of velocity is shown in figure \ref{averaging_velocity_compare}. 
Nonphysical effects due to periodicity are much more present without the averaging force.

Flux with no averaging force, given by \ref{straightener_force_no_z}, is shown in figure \ref{control_results}, which shows that there is indeed more damping in the system. 
Note that the valve meshes and driving pressures are identical to those of Section \ref{control_results}. 

\begin{figure}[H]
\centering 
\includegraphics[width=\textwidth]{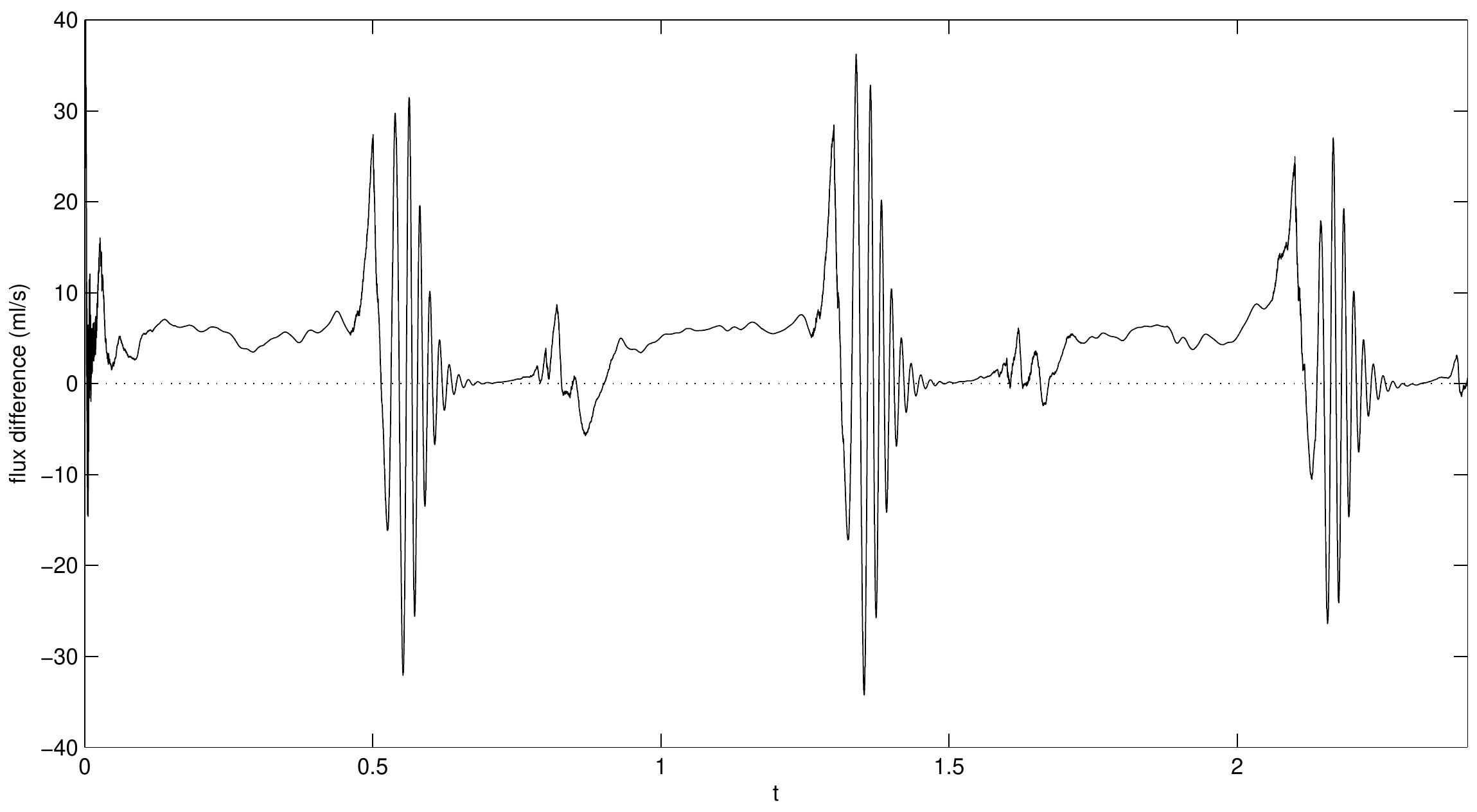} 
\caption{Difference of flux without and with averaging force.}
\label{flux_no_average_diff}
\end{figure}

Figure \ref{flux_no_average_diff} shows the flux without the averaging force in $z$ minus the flux with it. 
During diastole, the flux with the averaging force is slightly lower than that without. 
This is owing to the reduction of kinetic energy in from this force shown in equation \ref{ke_reduction}. 
The difference in damping causes the oscillations during closure to differ on the two simulations; this appears in this plot as still more oscillations during closure. 

\begin{figure}[H]
\centering 
\includegraphics[width=\textwidth]{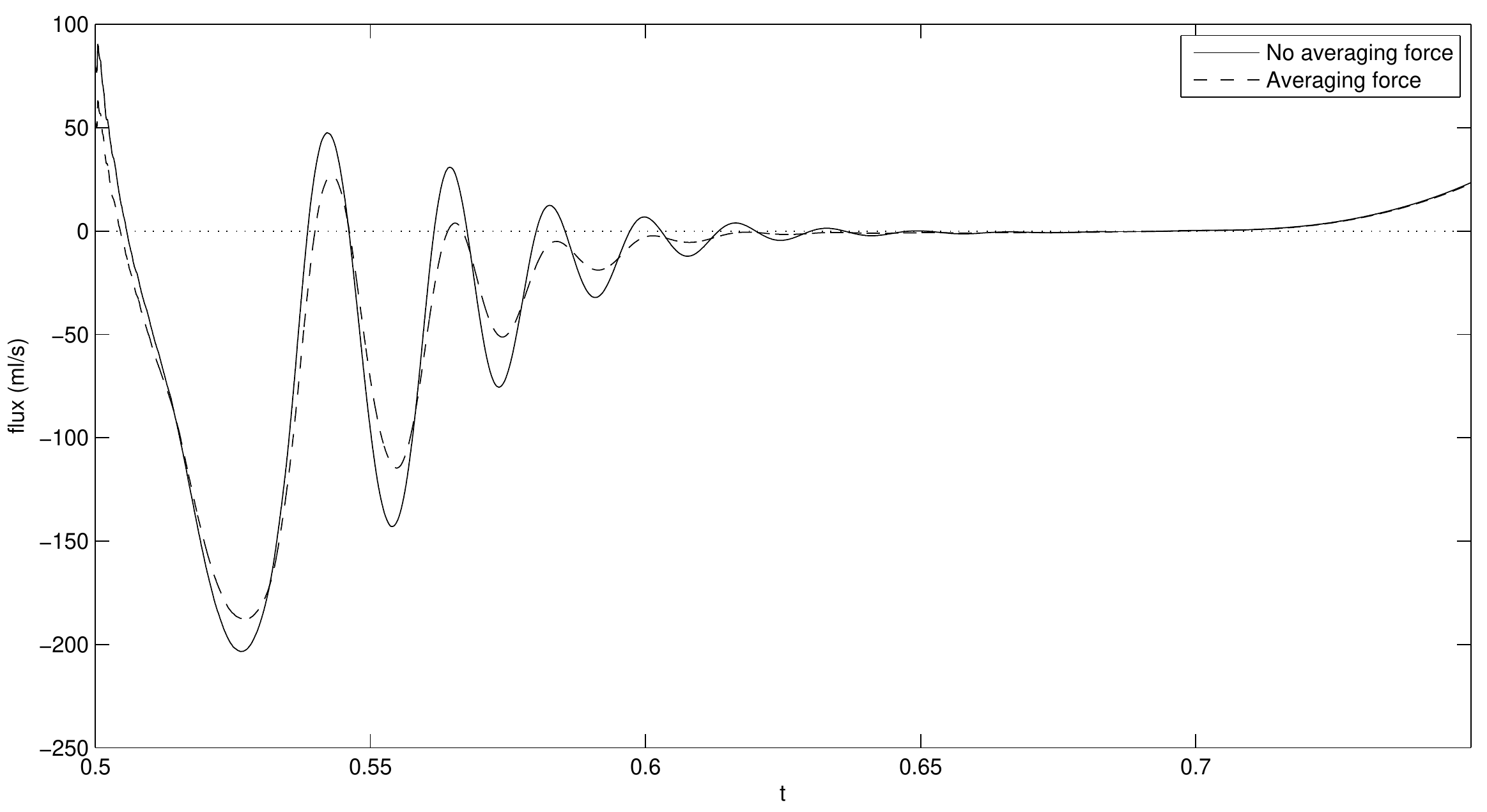} 
\caption{Comparison of flux with and without averaging force, showing a detail of oscillation during closure. }
\label{oscillation_comparison}
\end{figure}

Figure \ref{oscillation_comparison} shows a detail of the flux in systole with and without the averaging force. 
Note that the frequency of the oscillation remains roughly constant, but the version with the averaging force is more damped, as we expect. 

\begin{figure}[H]
\centering 
\includegraphics[width=\textwidth]{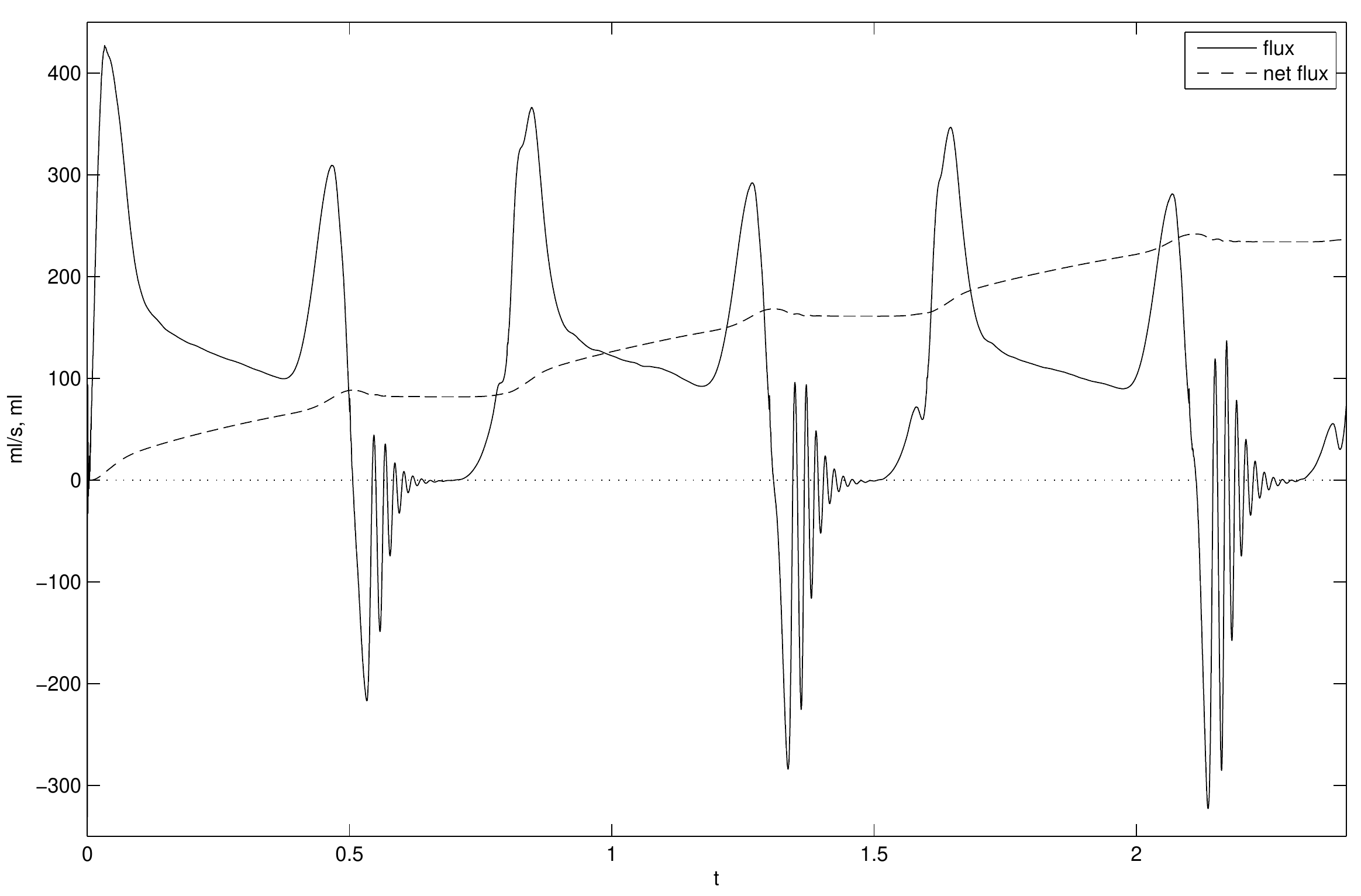} 
\caption{Flux without connection in layers. Note that this figure has the same scale as figure \ref{pressure_and_flux}, but that since the maximum fluxes are larger it is plotted on a larger range.}
\label{flux_no_layers}
\end{figure}

Consider a model with three layers, as discussed in Section \ref{layers}. 
Suppose temporarily that the layers are not attached; that is, there are three layers of the valve placed $\ds = \dx/2$ from each other. 
Since the layers move in a continuous velocity field, one might expect that they stay together. 
Indeed, tests without them attached behave reasonably. 
However, we observed that the oscillations in closing increase in amplitude as the simulation proceeds. 
This is shown in figure \ref{flux_no_layers}. 
This simulation also lacks the averaging force. 

\begin{figure}[H]
\centering 
\includegraphics[width=\textwidth]{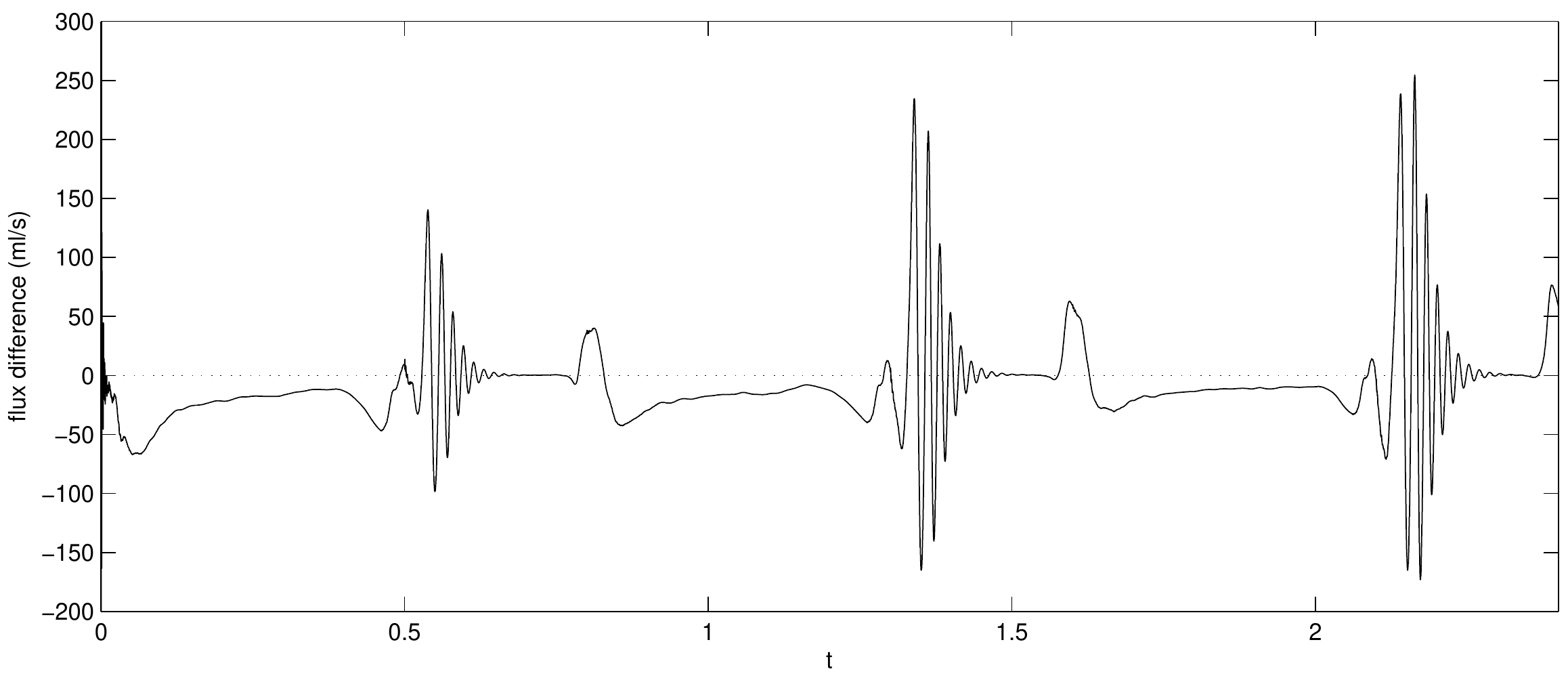} 
\caption{Difference of flux with layers not connected and layers connected. 
No averaging force is applied to either model. 
Note that this is the difference of the values shown in figure \ref{flux_no_average} and figure \ref{flux_no_layers}. }
\label{diff_no_straighten_and_no_layer_no_straighten}
\end{figure}

We believe that this is because the layers are not staying tightly together. 
Over the course of the simulation the mesh is distorted. 
This leads to a loss of valve performance as the simulation proceeds. 
Figure \ref{diff_no_straighten_and_no_layer_no_straighten} shows the difference between the version with and without connections between the layers. 
The version without connections has larger oscillations on each subsequent beat. 
This appears as increasing differences in the oscillations through each beat in this plot. 
The version with layers connected is has less forward flux through diastole, which is consistent with the fact that the layers add rigidity to the model. 

\chapter{A Mitral Valve Model with Commissural Leaflets}

In this chapter, we present a second model of the mitral valve that includes four distinct leaflets. 
In addition to the anterior and posterior leaflets, we add explicit commissural leaflets. 

The real valve possesses commissural leaflets, though whether they are fully distinct from the posterior leaflet is subject to debate and variable across individuals. 
For example, the valve shown in figure \ref{valve_basic} has a distinct commissural leaflet at the posteromedial commissure (left side of the photograph) but the anterolateral commissural leaflet (right side of the photograph) is not well-separated from the posterior leaflet.  
In other specimens, the commissural leaflets are very present and possess many chordae tendineae attachments,  as shown in figure \ref{pinned_ventricular_zoom_out}. 
This specimen has significant connecting tissue below the locations of the attachments, so these distinct regions are not well-separated from either the anterior or posterior leaflet. 
This is especially visible when viewed from the atrial side, which exposes the tube-like structure of the valve, as shown in figure \ref{pinned_atrial_zoom_out}. 
In all specimens we examined, this region typically has many chordae tendineae attachments, as shown throughout Chapter \ref{anatomy}.

We believe that model presented in this section improves on the realism of the anatomy, but the overall function of this model is not yet as robust as the two leaflet model. 
This model has more trees, more distinct regions of leaflet, and thus more parameters to tune. 
We present it as a work in progress, that will require more tuning and testing to be complete. 
The steps to build this model are identical to those presented in Chapters \ref{static} and \ref{Fluid-Structure_Interaction}.

The leaflet mesh contains four regions for leaflets. 
The anterior and posterior leaflets are trapezoids in parameter space.
A small region in the center of each free edge is flat. 
In the two leaflet models, the ``point'' of the triangle tended to spread out effectively. 
Here, we observed that without a flat section on the mesh, the free edge of the leaflets took on an un-anatomical point. 
The commissural leaflets are triangles in parameter space. 
Additionally, to mimic the tube-like structure we add many edge connectors below the points of attachment of the chordae. 
These regions are discretized as part of the leaflet and included in the static solve. 
We do not add any rings to the mesh below the mitral ring, because we view the edge-connector region as making a sufficiently tube-like structure. 
Also, we will attach some chordae at various heights and even directly to the mitral ring, as discussed below. 
The regions below which these chordae attach to the ring are viewed as edge connectors. 
Thus, we view these areas as having material properties depending on how they align with the chordae attachments, and do not wish to consider a region of rings with its own distinct coefficients.

Figure \ref{papillary_detail} shows the details of attachment of chordae to the papillary tips. 
The attachments are approximately evenly spaced on a portion of a roughly circular arc at the papillary tip. 
There appear to be about six major chordae that emerge from the papillary tip. 
(It may be seven, the chordae that attaches to the highest point in the frame could be considered as two.)
From this, we see two chordae that attach to the posterior leaflet at the top of the frame. 
The next has connections to the posterior leaflet, and to one side of the commisssural leaflet, which has a small, pointy free edge and is approximately centered in the frame. 
The next has connections to the commissural leaflet and anterior leaflet. 
The final two appear to connect exclusively to  the anterior leaflet. 
Figure \ref{pinned_atrial_zoom_out}, which shows the same valve, confirms this impression. 
Figure \ref{pinned_ventricular_zoom_out} shows more detail of how these chordae attach. 
The attachments in the commissural regions range from the free edge all the way to the mitral ring.

Following these observations, we include six trees of chordae per papillary muscle in this model, twelve trees in total. 
On each side, two attach to the posterior, one between the posterior and commissural leaflet, one between the commissural leaflet and the anterior leaflet, and two to the anterior leaflet. 
The trees that connect to both the commissural leaflet and the anterior or posterior leaflet attach directly to the ring in a small region. 
The number of leaves at which this occurs scales with the mesh parameter $N$ so there is a consistent total force on these trees.

\begin{figure}[H]
\centering 
\includegraphics[width=\textwidth]{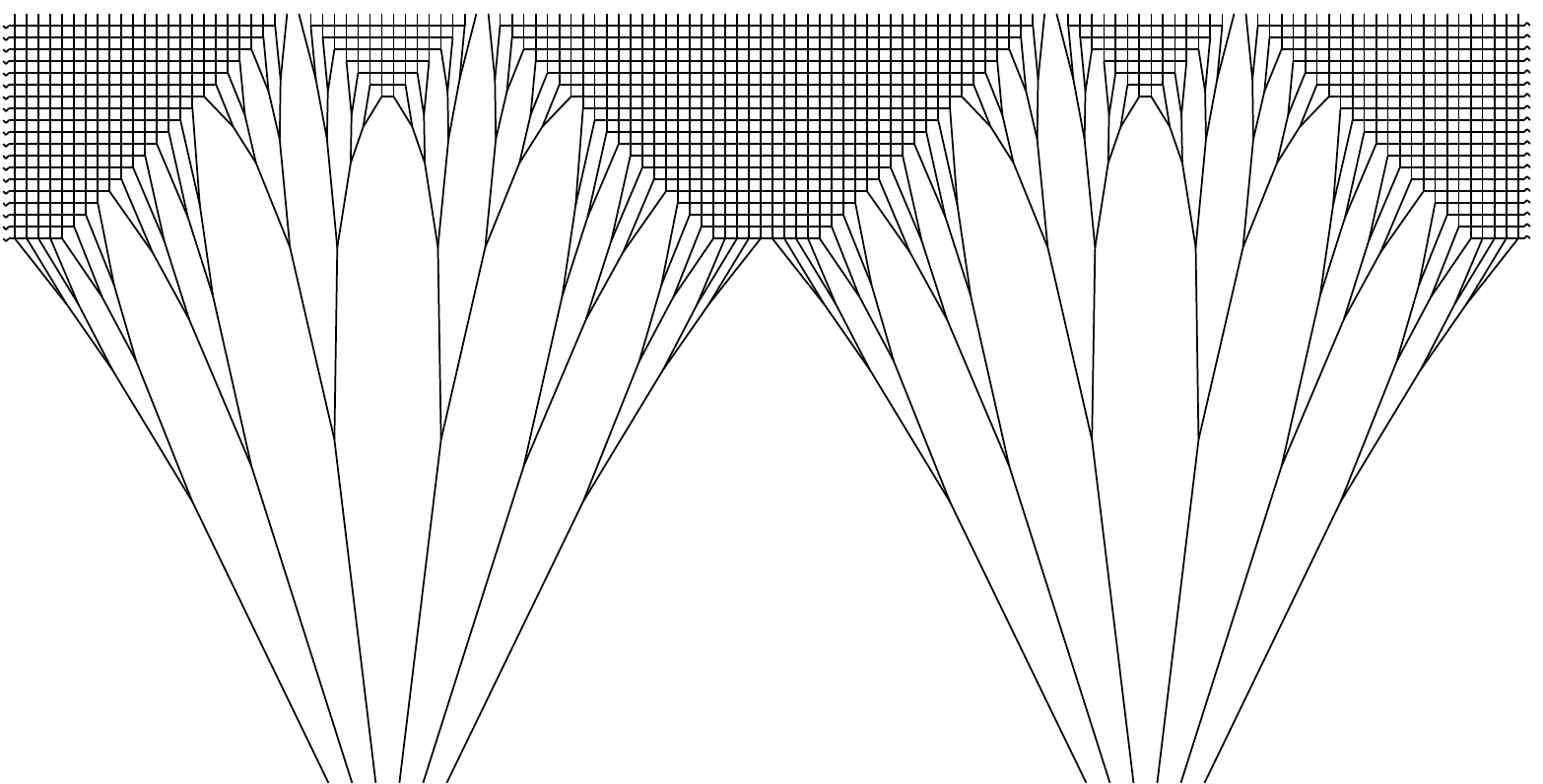} 
\caption{Mesh schematic with trees for commissural leaflet version, displaying only mesh at or above the chordae attachment. $N=128$.}
\label{mesh_schematic_commissures}
\end{figure}

\begin{figure}[H]
\centering 
\includegraphics[width=\textwidth]{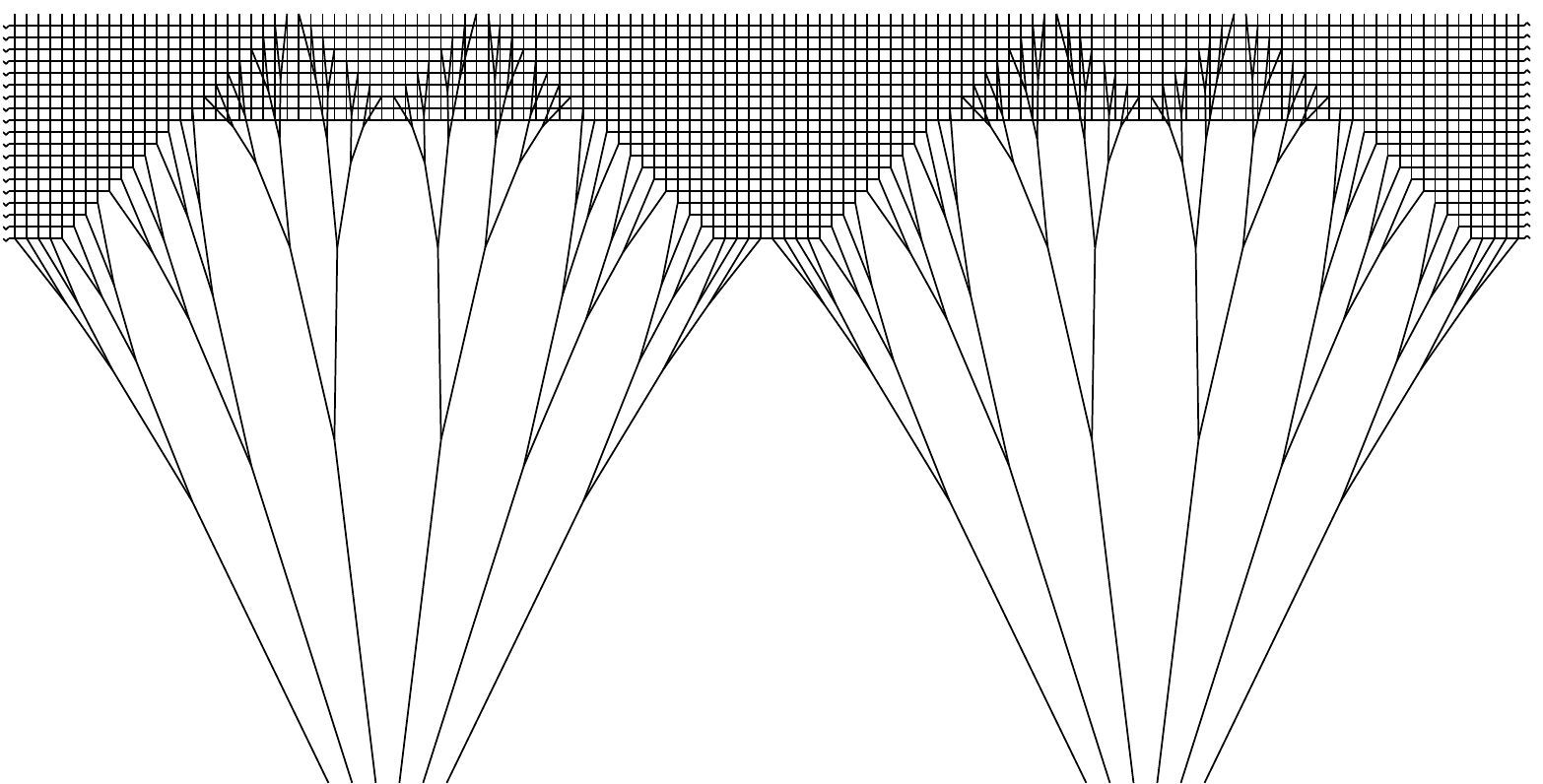} 
\caption{Mesh schematic with trees for commissural leaflet version, including model tissue below chordae attachment. $N=128$.}
\label{mesh_schematic_commissures_edge_connector}
\end{figure}

Figure \ref{mesh_schematic_commissures} shows a schematic of the mesh without any model tissue below the points at which the chordae tendineae attach. 
Figure \ref{mesh_schematic_commissures_edge_connector} shows a schematic including the model tissue below the chordae tendineae attachments at the commissures. 
These figures serve to illustrate the topology of the discretization only.
The locations of the trees of chordae are not used in solving the equations of equilibrium and have no physical interpretation. 
Thus no scale is included. 

\begin{figure}[H]
\centering
\includegraphics[scale=.74]{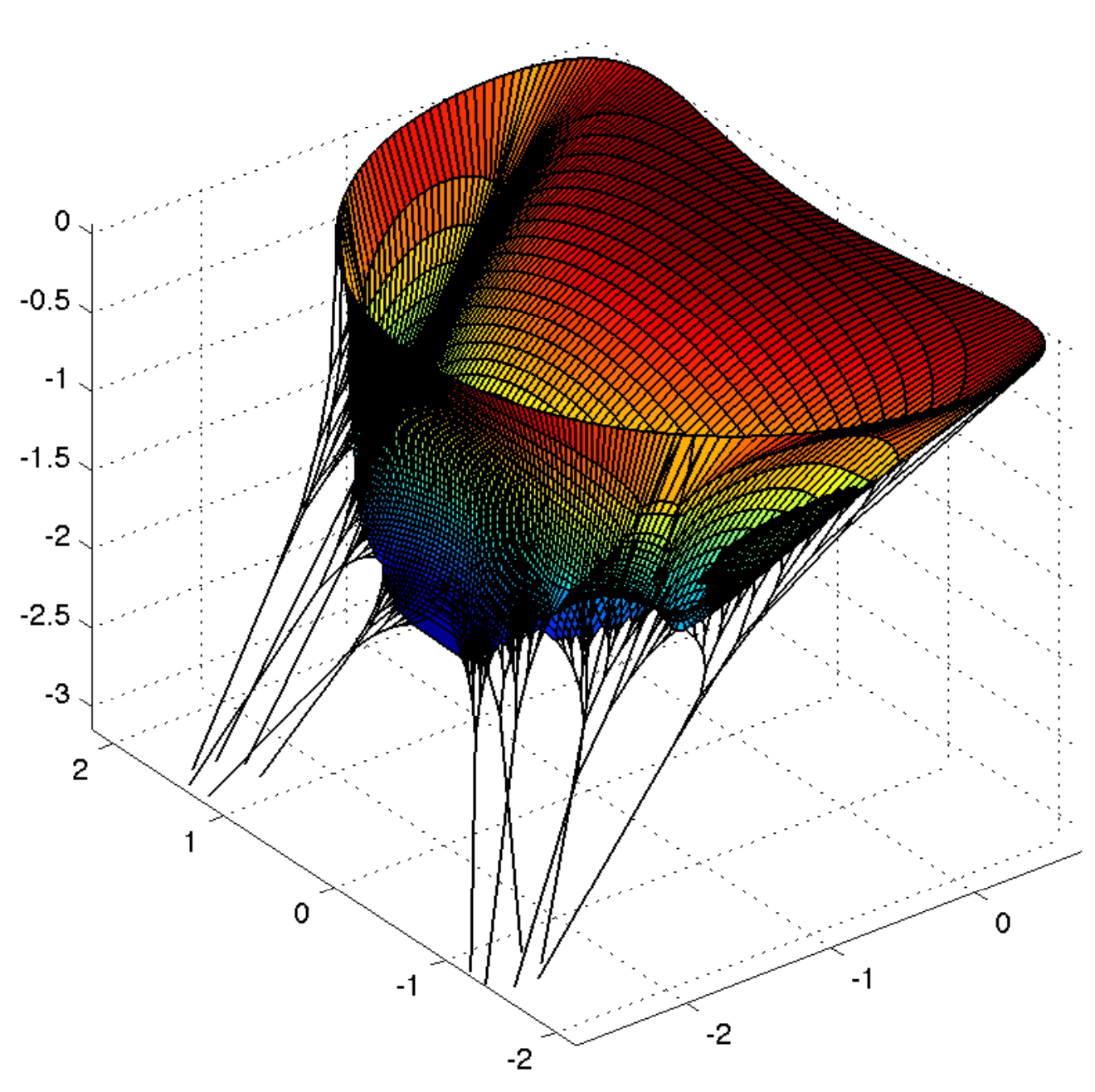} 
\includegraphics[scale=.74]{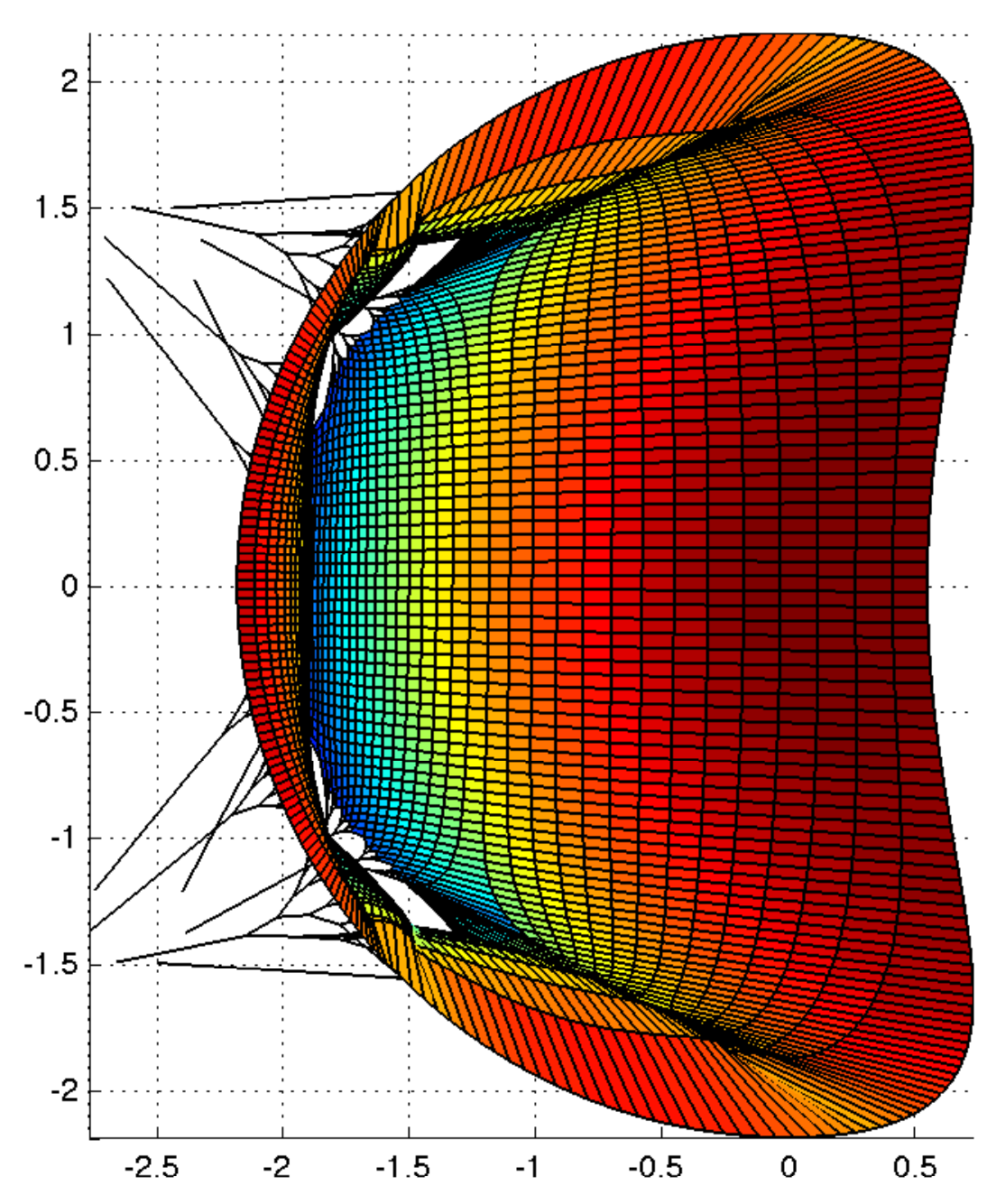}
\caption{Closed geometry of model valve with commissural leaflets. $N=256$. }
\label{static_valves_comm}
\end{figure}

We solve the equilibrium equations, \ref{equilbrium_eqn_discrete}, simultaneously with the equilibrium equations in the chordae as in Chapter \ref{static}. 
Two views of the pressurized model valve are shown in figure \ref{static_valves_comm}.

\begin{figure}[H]
\centering
\includegraphics[scale=.73]{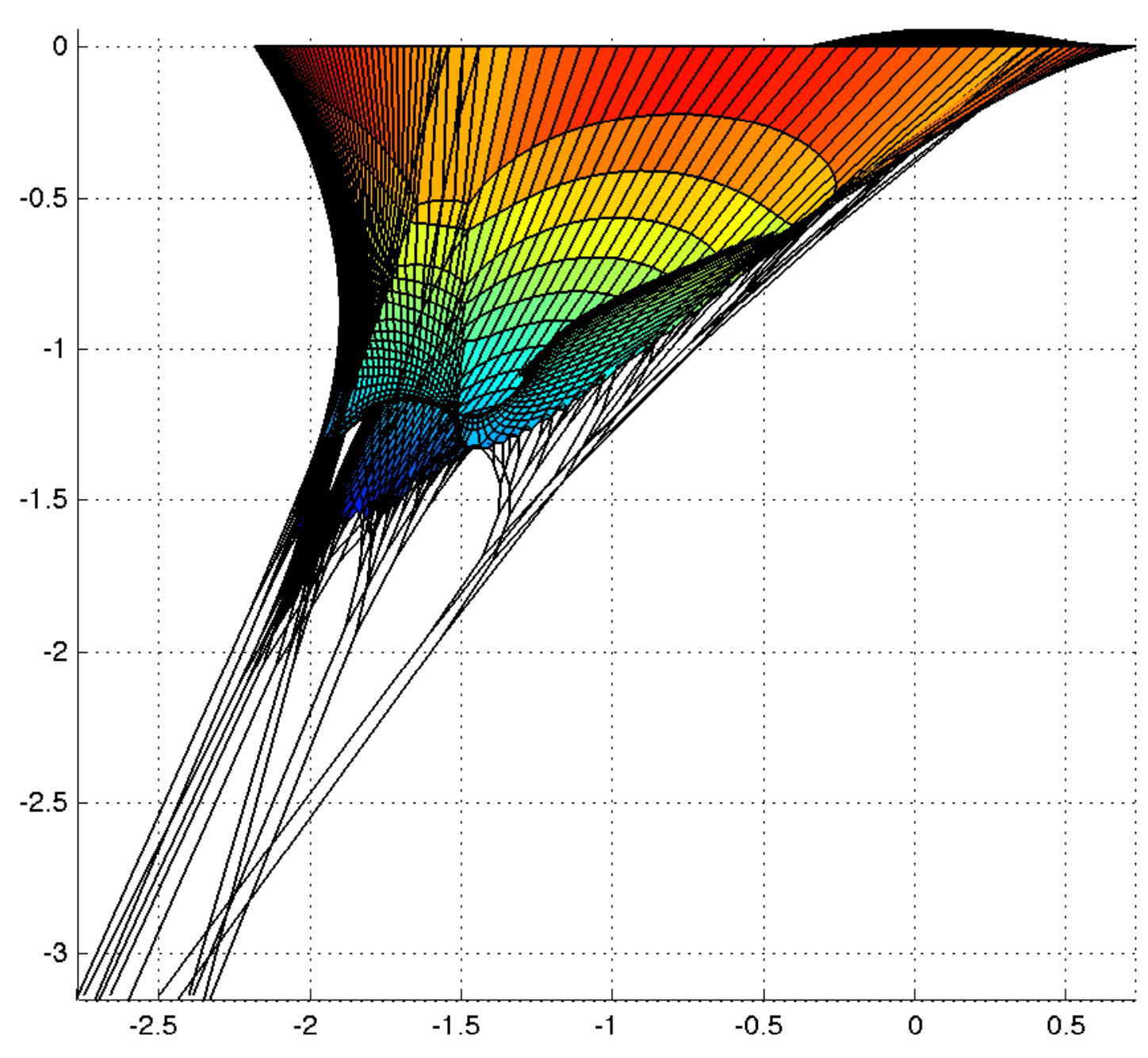} 
\includegraphics[scale=.73]{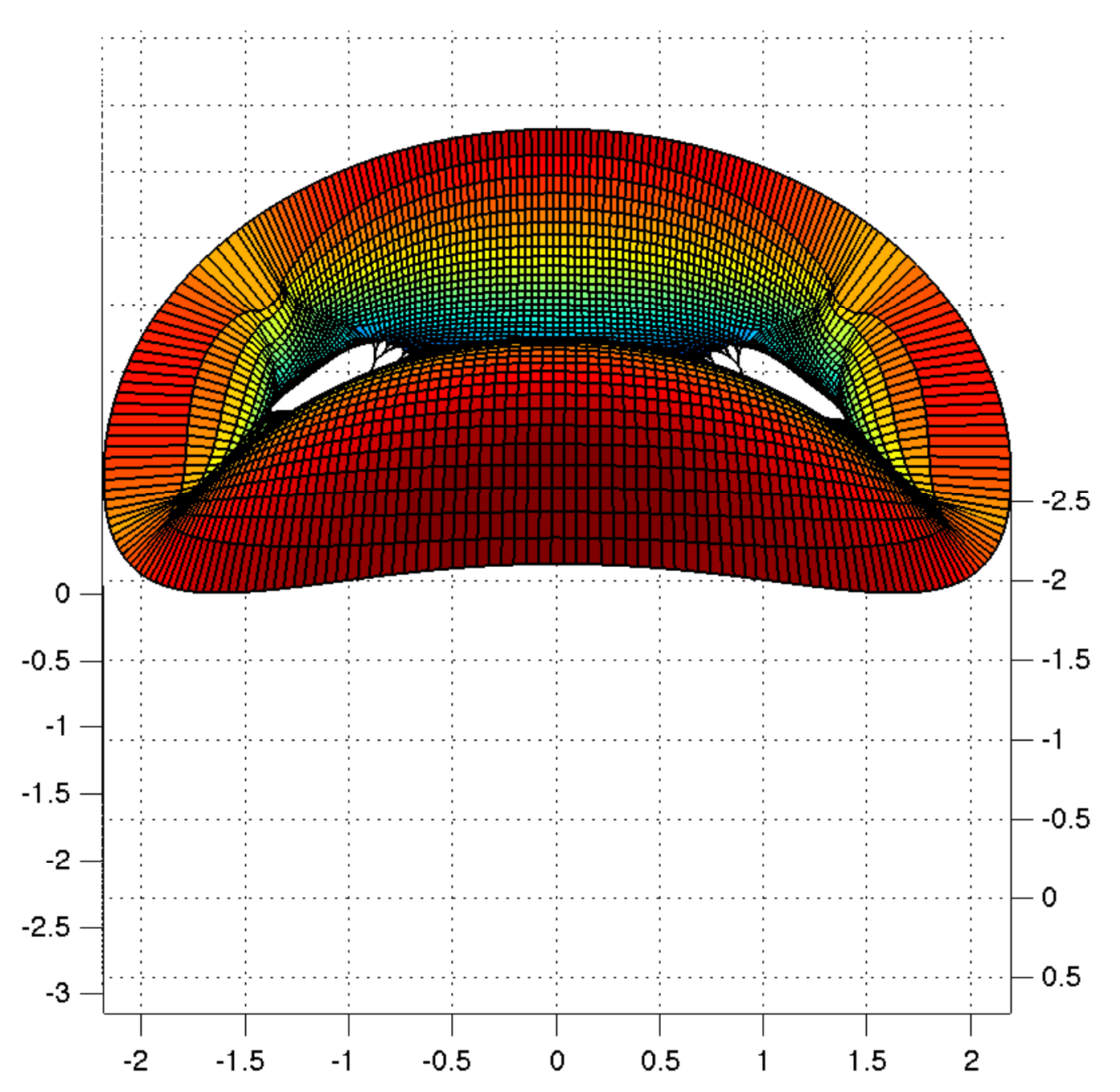}
\caption{Closed geometry of model valve with commissural leaflets, views to emphasize closure. $N=256$.}
\label{static_valves_closure_comm}
\end{figure}

This model has difficulty attaining full closure in static solves. 
It is prone to having small holes near the free edge around the commissures that are challenging to close. 
Two views to emphasize this are shown in figure \ref{static_valves_closure_comm}. 
Because of this, we set the pressure in equation \ref{equilbrium_eqn_discrete} to 80 mmHg, twenty percent lower than that of Chapter \ref{static}. 
This makes the valve structure slightly more compliant, and hopefully this compliance will allow the valve to fill these holes when placed in fluid.

\begin{figure}[H]
\centering
\includegraphics[scale=.74]{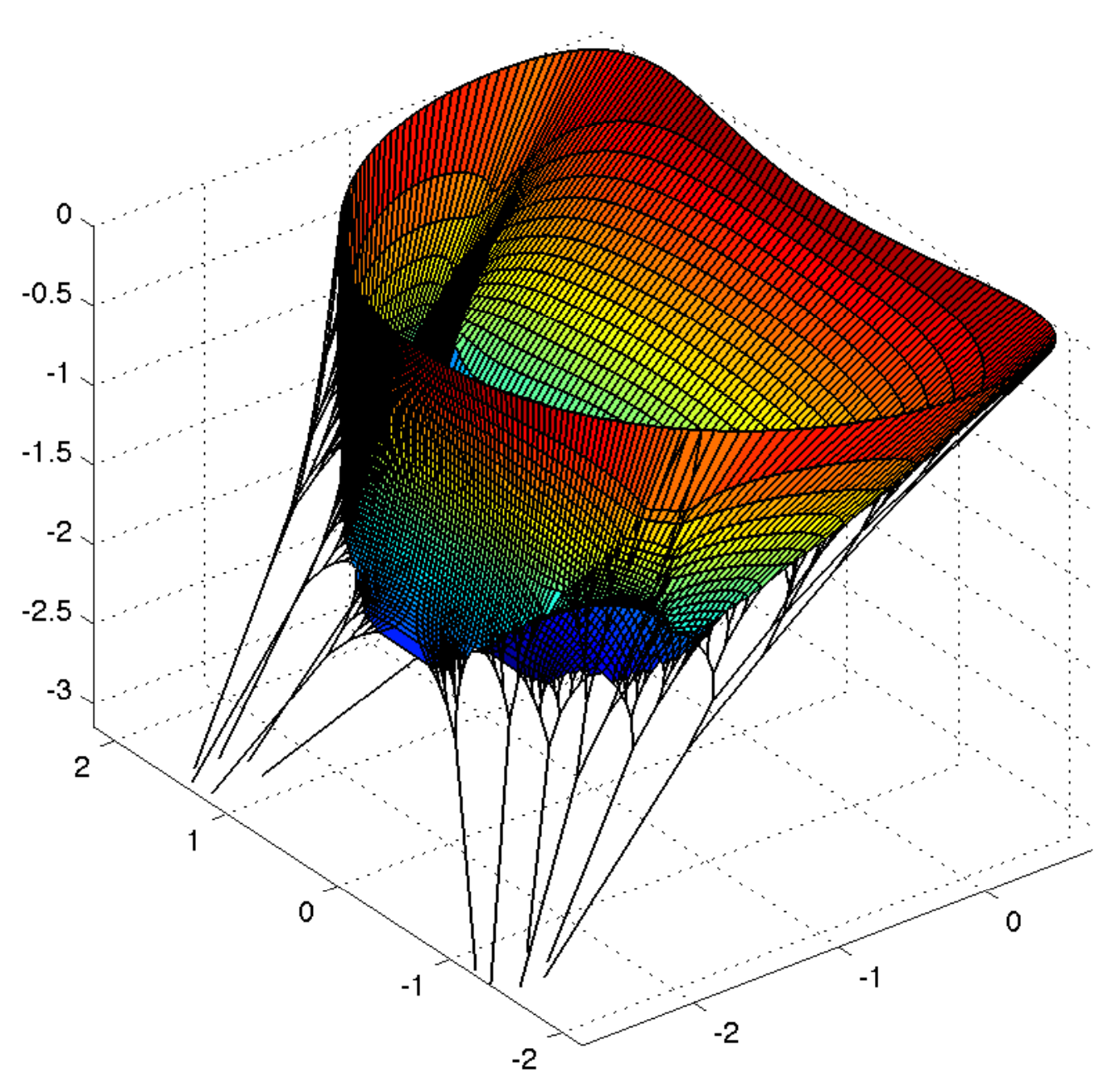} 
\includegraphics[scale=.74]{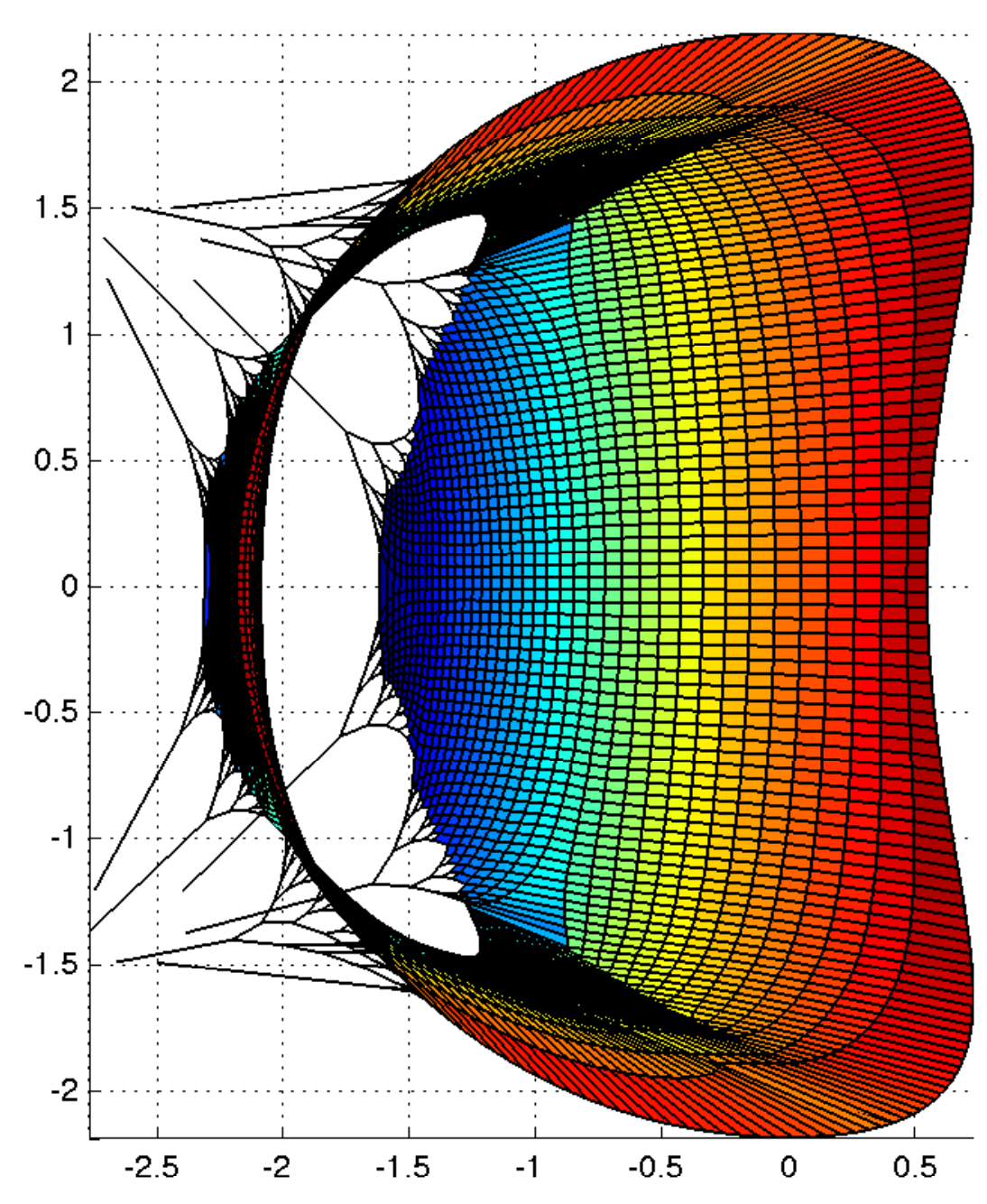}
\caption{Partially open geometry of model valve with commissural leaflets.}
\label{with_ref_comm}
\end{figure}

\begin{figure}[H]
\centering
\includegraphics[scale=.71]{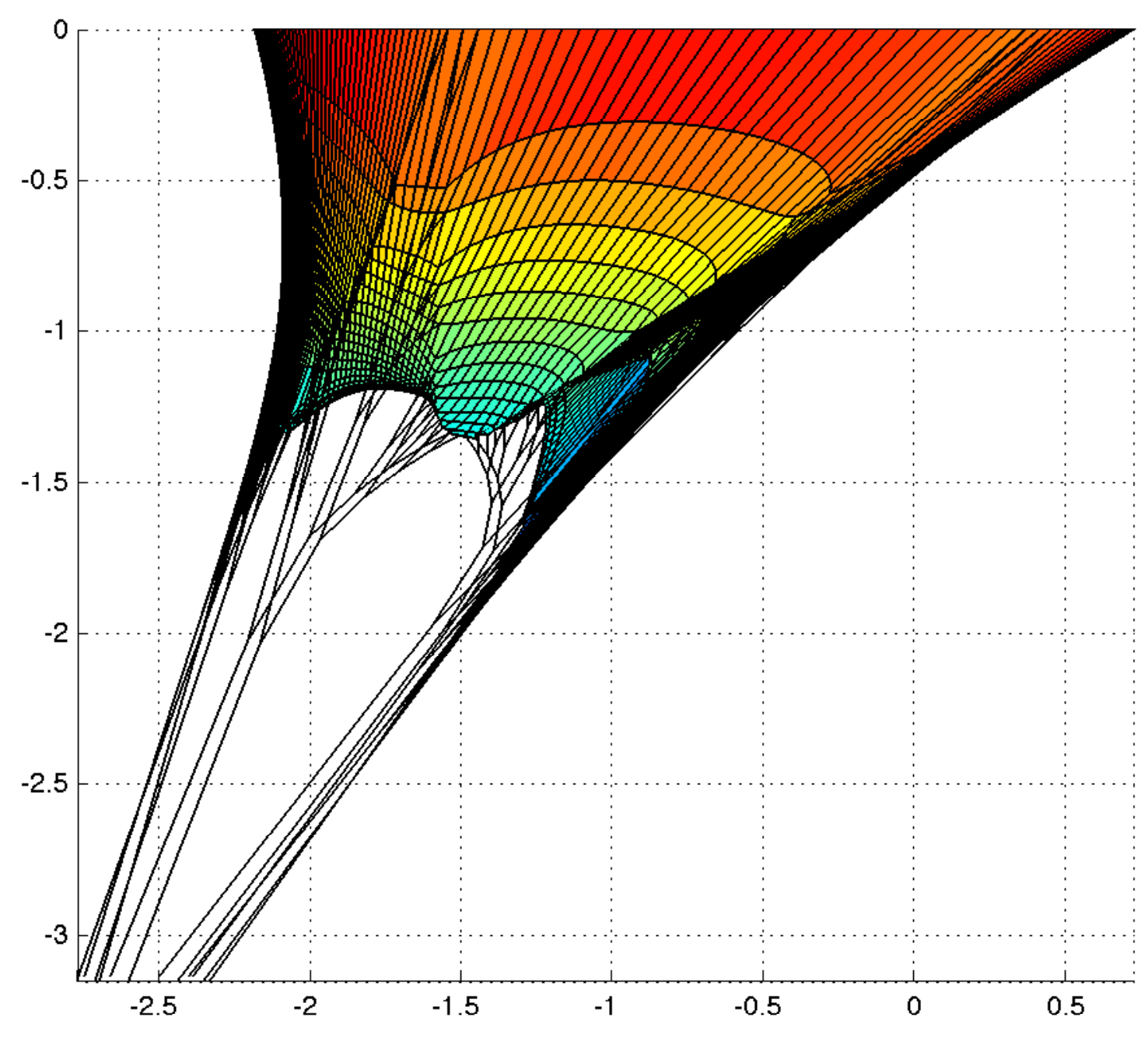} 
\includegraphics[scale=.71]{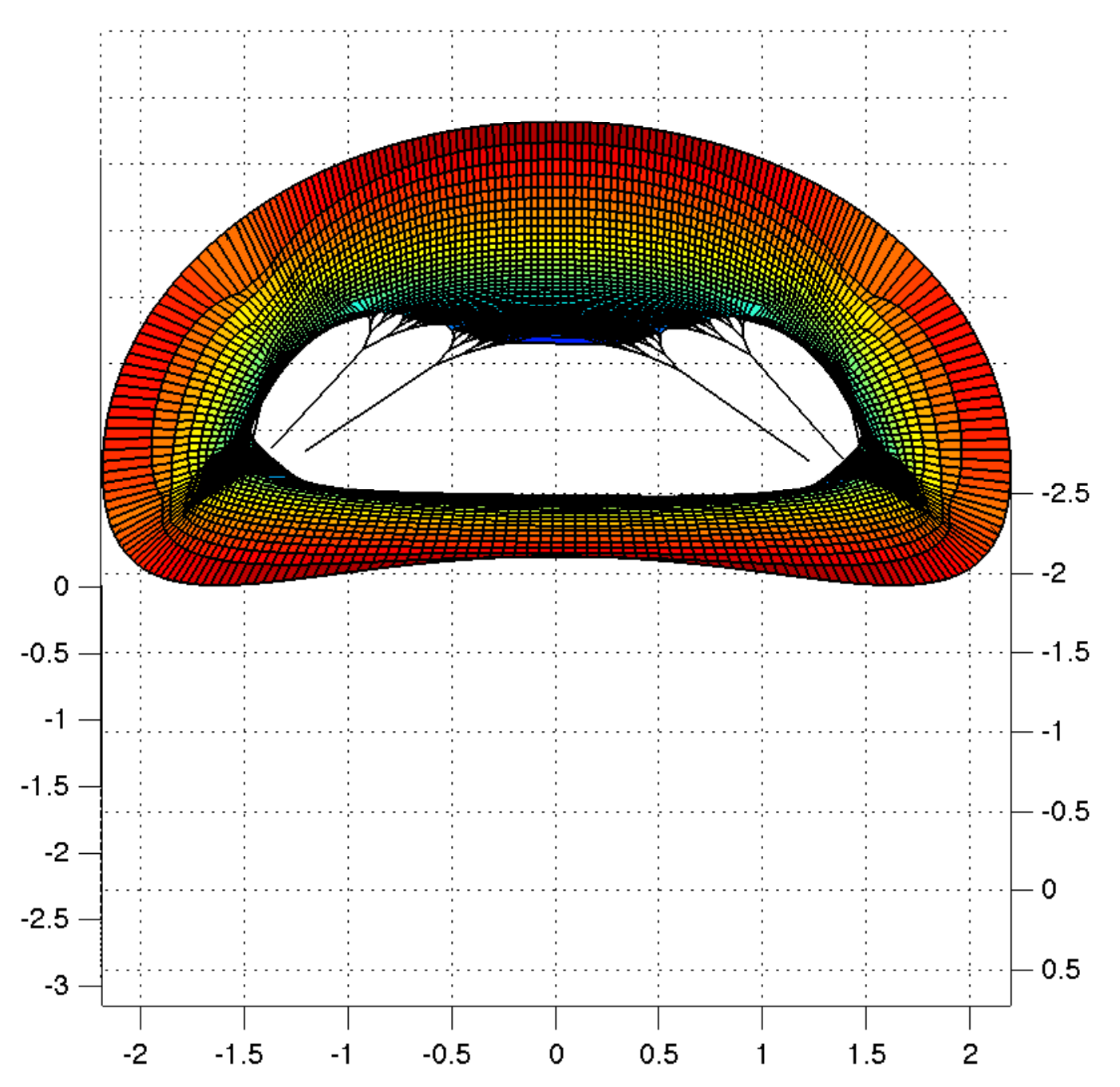}
\caption{Partially open views of model valve with commissural leaflets, additional views.}
\label{with_ref_2_comm}
\end{figure}

From this static solve, we build a general model for use in fluid following the steps in Section \ref{general_model}. 
This is shown in figures \ref{with_ref_comm} and \ref{with_ref_2_comm}.

\begin{figure}[H]
\centering
\includegraphics[width=\textwidth]{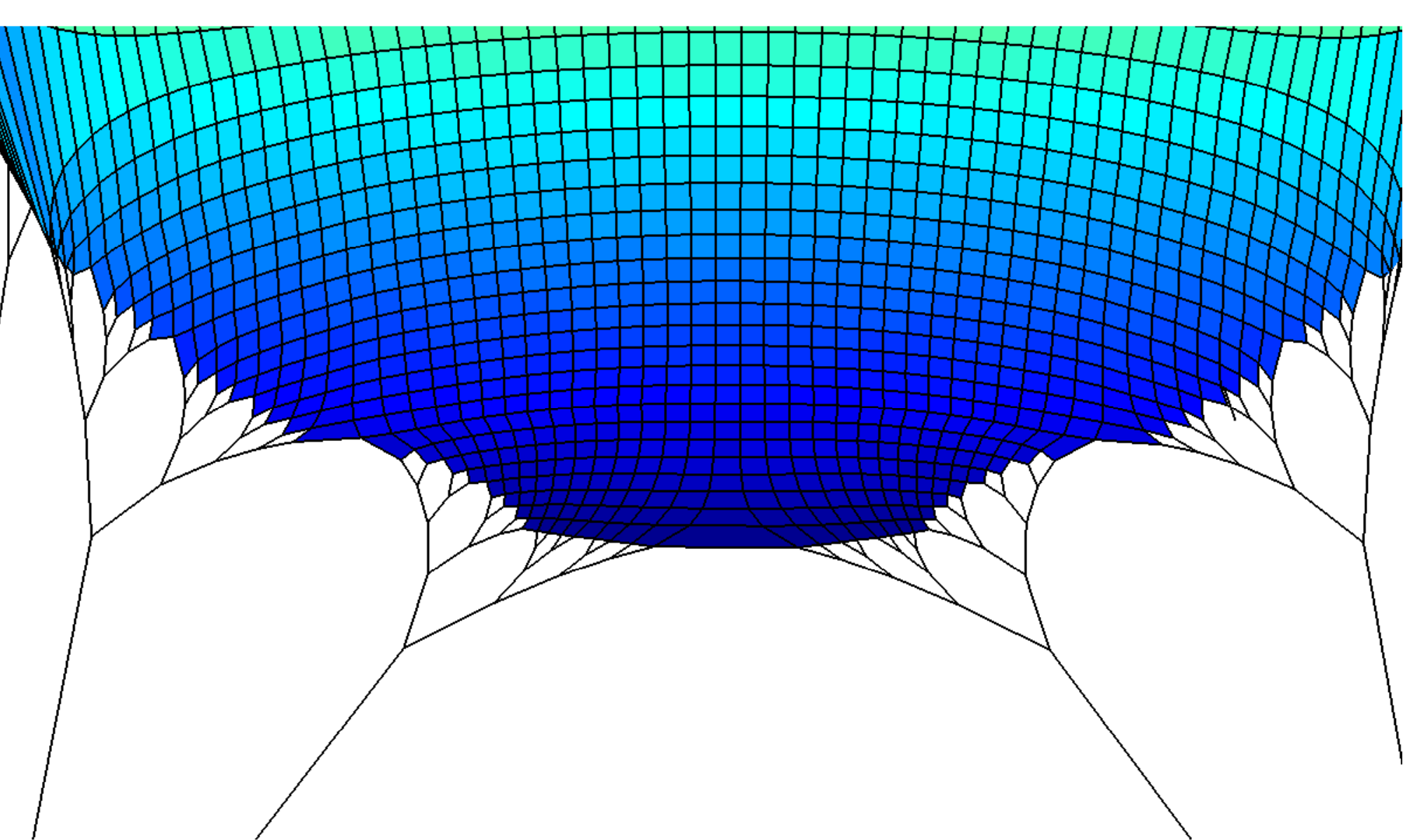} 
\caption{Detail of anterior free edge on model with commissural leaflets. }
\label{valve_with_ref_front_comm.pdf}
\end{figure}

Figure \ref{valve_with_ref_front_comm.pdf} shows a detail of the free edge of the anterior leaflet on this model. 
Note the arching circumferential fibers, which are reminiscent of the fibers seen in dissections of figure \ref{pinned_ventricular_zoom_out}. 

The simulation setup is nearly identical to that of Section \ref{setup}. 
We do not use the averaging force of equation \ref{straightener_force}, but rather the straightener force of equation \ref{straightener_force_no_z}, 
simply because we developed the averaging force after this model was investigated. 
The driving pressure is identical to that of figure \ref{pressure_and_flux}.

\begin{figure}[H]
\centering 
\includegraphics[width=\textwidth]{both_pressure_yellin_three_cycles.pdf}
\includegraphics[width=\textwidth]{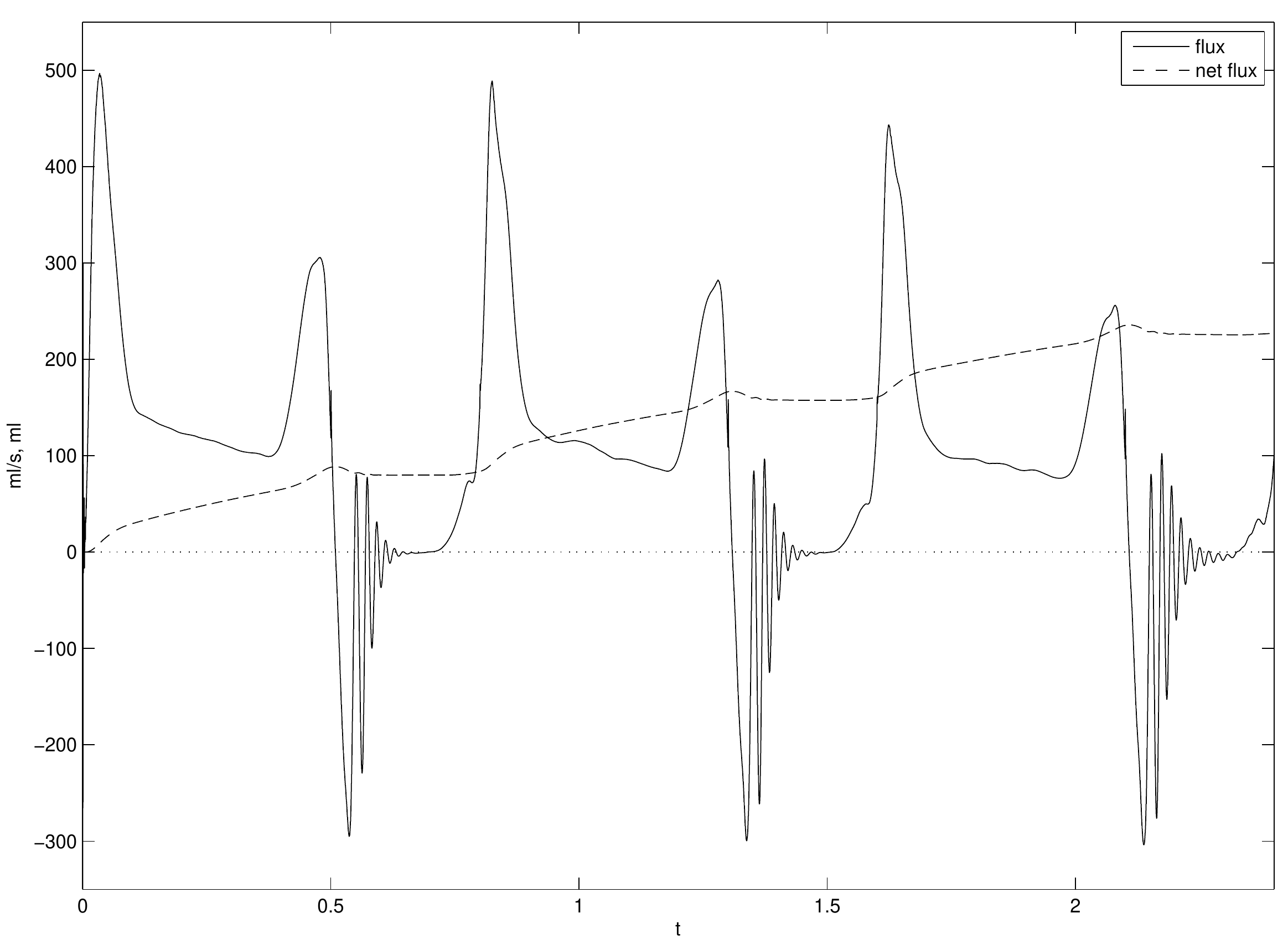}
\caption{Driving pressures and resulting flow through mitral ring with commissural leaflets.}
\label{pressure_and_flux_comm}
\end{figure}

The flux that emerges from the simulation is shown in figure \ref{pressure_and_flux_comm}, which includes the pressure again for completeness. 
The forward flux at its peaks during diastole is higher than the flux with the two leaflet model, and the vibration during systole is larger in magnitude. 
The first beat seals well.
However, this model develops slight regurgitation; it does not seal fully on the third beat. 
Additionally, there are changes in amplitude of the vibrations during systole. 
Since the first beat flux looks better than the second or third, it is possible that the layers of this model need to be held together more tightly.

\begin{figure}[H]
\centering 
\centerline{ \includegraphics[width=.86\textwidth]{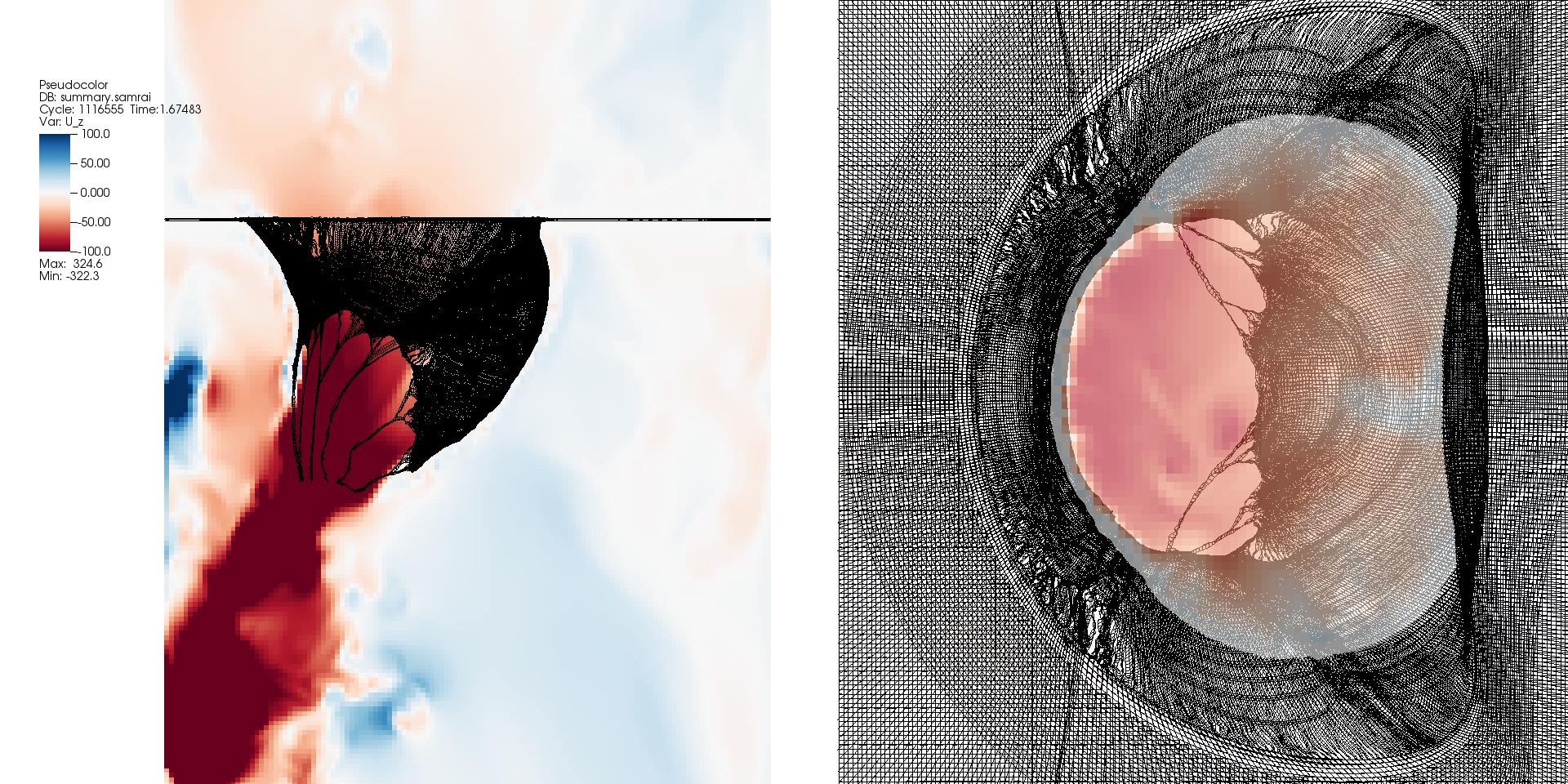} } 
\centerline{ \includegraphics[width=.86\textwidth]{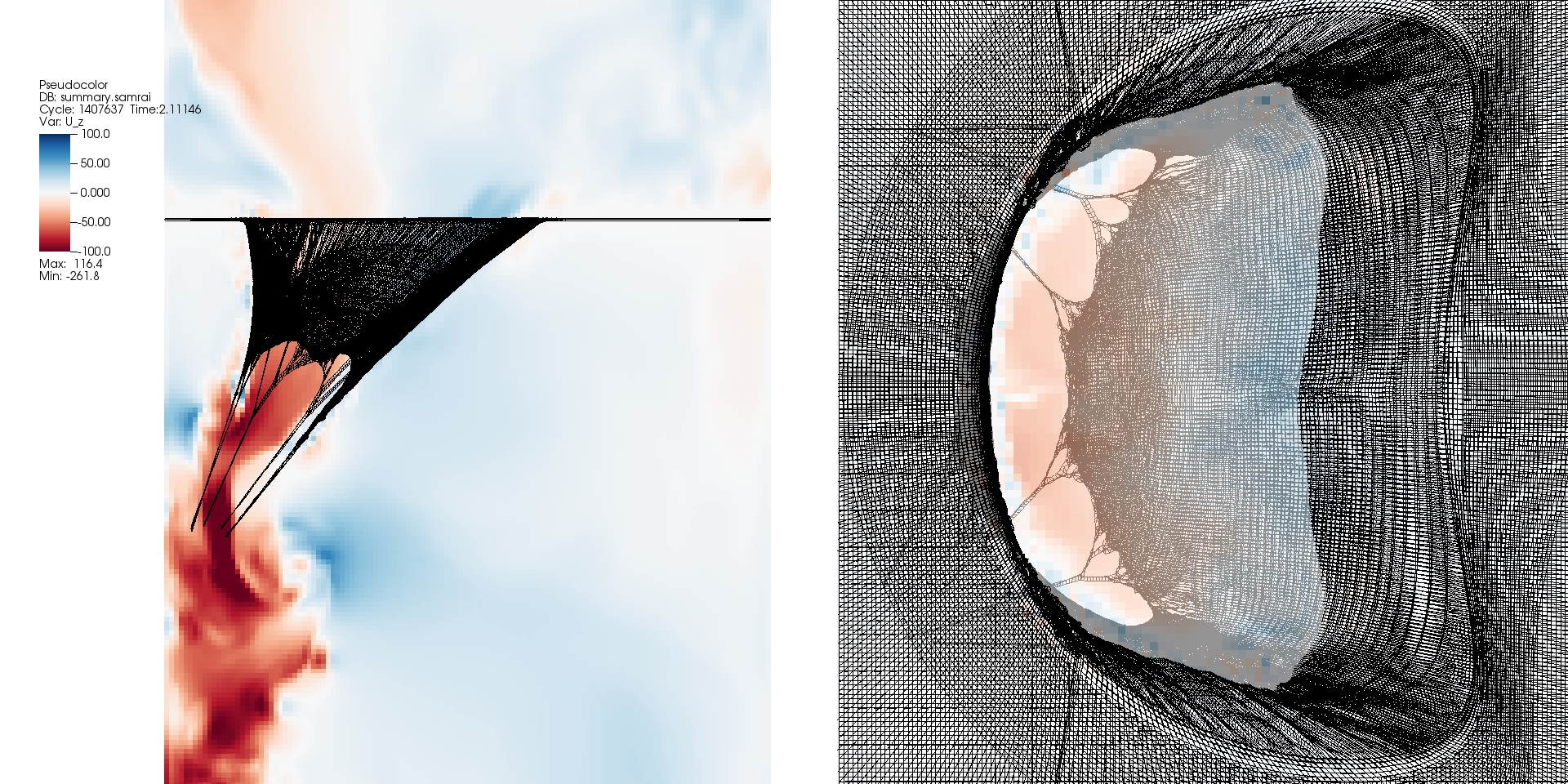} } 
\centerline{ \includegraphics[width=.86\textwidth]{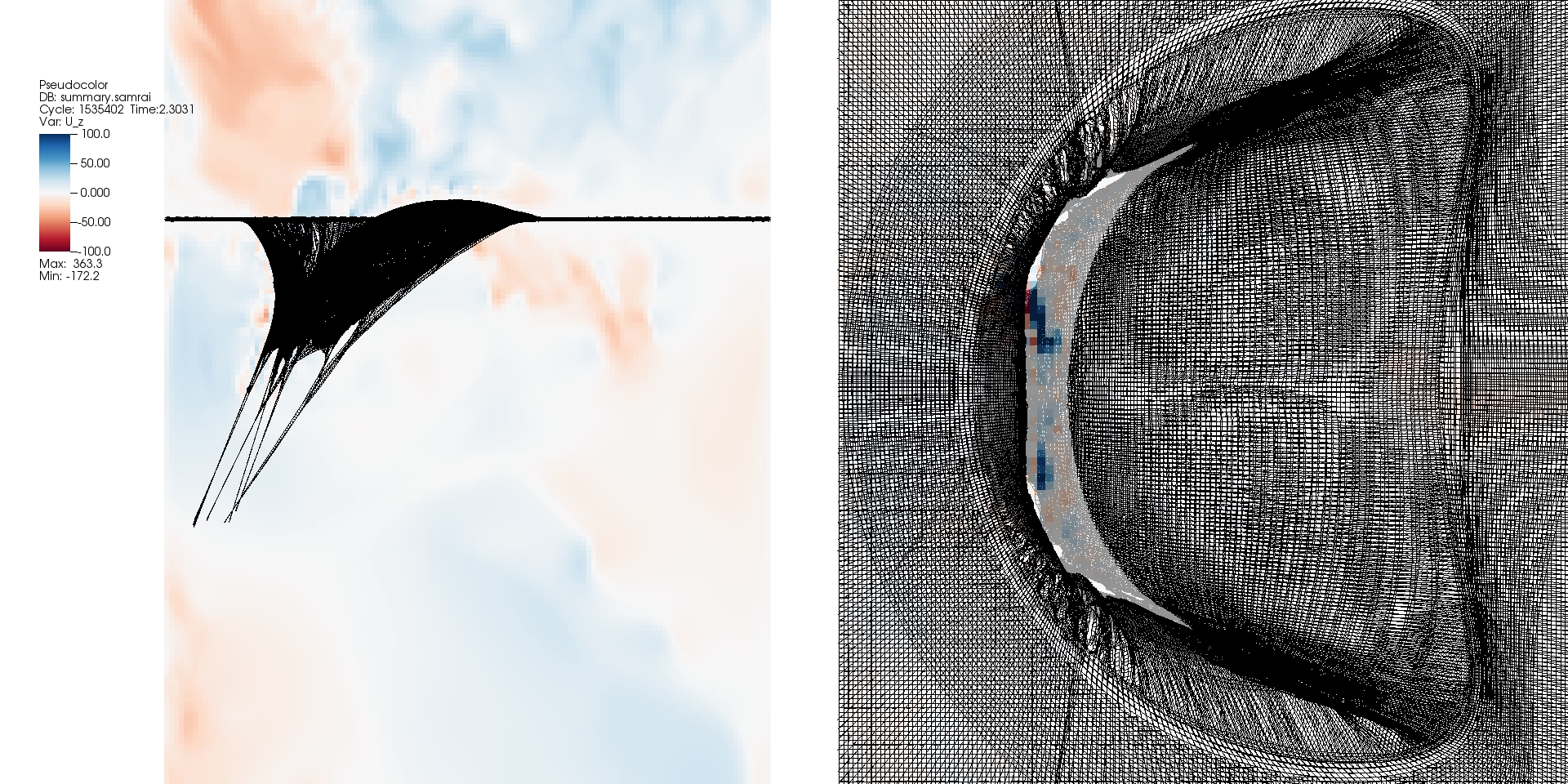} } 
\caption{Views of the mitral valve with commissural leaflets in simulation. The first shows filling, the second closing, and the third the nearly closed, slightly leaky position. }
\label{three_panels_comm}
\end{figure}

Three time steps of the flow are shown in figure \ref{three_panels_comm}. 
This shows the final beat, and the model valve is leaking and vibrating slightly in the third panel.

This model incorporates more features of the real anatomy, as it includes commissural leaflets. 
We believe that the trees and their attachments around the commissures are more anatomically correct, and that the additional tissue below the chordae attachment is realistic. 
However, this needs tuning and testing for reliability, which we will leave this for future work. 
We expect this model to become effective with improvements to chosen parameters.

\newpage 
\chapter{Conclusions and future work}

Using a design-based elasticity approach, we have built a model mitral valve. 
The model incorporates many anatomical details. 
The form of the constitutive law for this model is taken from experiments. 
The ever-important material constants emerge from the requirement that the model supports, through tensile forces, a physiological pressure, and the phrasing of this requirement as a system of partial differential equations.

This model works. 
When simulated under physiologically realistic driving pressures, it produces fluxes that match experimental records, both qualitatively and quantitatively. 
Features of the flow that are observed, such as the vibration leading to the ``S1'' heart sound, emerge. 
Features of the flow that are conjectured, such as the breaking jet and the vortices shed from the leaflets during filling, are supported. 
Since the simulations are driven with pressures, the valve really is allowing forward flow during diastole and supporting pressures during closure. 
Thus, we hit many of the metrics that the large review \cite{kheradvar2015emergingIV} discussed in Section \ref{literature} describes as difficult.

There are a number of future directions in which to take this work. 
We could use this information to build a prosthetic valve, as the design-based approach to elasticity suggests a design for prosthetics. 
Current prosthetic mitral valves generally do not include a full system of chordae tendineae \cite{kheradvar2015emerging}. 
This work suggests a way to build one that does.

The commissural leaflet version could be polished and improved. 
Perhaps there is something new to be revealed from the ways in which it is more anatomically correct.

In building the initial model, this framework does not allow fibers within a family to cross. 
We assume that we know where fibers start and end in the model, and that fibers within each family do not cross each other so the quadrilateral nature of the mesh is preserved. 
One could imagine a model in which the fiber topology was also unknown. 
How to approach this is not at all clear, and this would certainly be a significant technical challenge.

Further, we could investigate lumped parameter models to drive the simulations. 
This uses zero-dimensional models of other parts of the circulation to interact with the three-dimensional model described in this thesis. 
This would make it possible to study the interaction of mitral valve pathology with the rest of the circulation. 

The model of this thesis could easily be extended to model the tricuspid valve. 
More information about the histological layers of the valve could be included and added to the fiber models. 
We could experiment with various valve pathologies, changing material parameters or geometries accordingly to answer questions about mitral valve disease.

Finally, and perhaps most importantly, we could replace the model test chamber with a model left ventricle and atrium. 
This is the an essential next step to a more broad view on using these models to study heart disease.

\section{Support}

Alexander D. Kaiser was supported by the National Science Foundation Graduate Research Fellowship Program, grant DGE 1342536, and a Henry M. MacCracken Fellowship through New York University. 
Additional funding came from the Thomas Tyler Bringley Memorial Fellowship, which was used to fund laboratory expenses. 
I am pleased to acknowledge the support.

Research in Chapter \ref{anatomy} was performed in collaboration with Mark Alu and Cynthia Loomis of the Experimental Pathology Research Laboratory at the New York University Langone Medical Center. 
The Experimental Pathology Research Laboratory is partially supported by the Cancer Center Support Grant P30CA016087 at NYU Langone's Laura and Isaac Perlmutter Cancer Center.

\chapter*{Appendix: Polarized light microscopy}
\addcontentsline{toc}{chapter}{Appendix: Polarized light microscopy} 
\label{appendix}

\setcounter{figure}{0} 
\renewcommand{\thefigure}{A.\arabic{figure}}

In this appendix, we present some preliminary results using polarized light microscopy on a valve that has been cleared and stained for collagen. 
These results are preliminary; images and comments are presented as a work-in-progress.

The cleared valve discussed in Section \ref{clearing} was stained with Picrosirius Red. 
When viewed with polarized light, collagen stained with this protocol produces a phenomena called birefringence, which allows for striking pictures of collagen fibers. 
A basic description of this technique is available in \cite{laszlo1991organization}. 
A polarizing filter called the \emph{analyzer} is placed above the specimen, between the specimen and the camera. 
An additional filter is placed below the below the specimen, between the specimen and the original light source. 
When these filters are orthogonal, light that is not altered by the specimen is blocked.
Refractive properties of the material vary with fiber direction. 
Incident light is split in two by the material, subsequently some light becomes non-aligned with the analyzer filter. 
This light can pass through the analyzer and be photographed. 
When they are not orthogonal, the optics are more complicated. 
We show some such pictures but recommend caution to the reader when viewing them, as they are difficult to interpret. 
Further description of polarized light microscopy, with examples and methods of interpretation are shown in \cite{CHANGOOR2011126}. 
An example of use of Picrosirius Red with polarized light microscopy to examine collagen is shown in \cite{arun2015study}. 
More sophisticated and quantitive techniques are described in \cite{rich2005collagen}.

As with the previous stain, the protocol was modified from the laboratory's internal procedure for a whole mount stain. 
The standard nuclear stain was omitted, because we are looking for collagen which is part of the extracellular matrix. 
The valve was placed in phosphomolybdic acid hydrate for 2 minutes, followed by the primary stain of trinitrophenol (sometimes referred to as picric acid) and RED-3BA (proprietary, from Polysciences) for one hour. 
Subsequently we performed differentiation in 0.1 molar hydrochloric acid for two minutes.

This valve was viewed on a Leica DMLM optical microscope at 10x magnification.  
A Nikon SF Fi2 camera is mounted to the microscope. 
All photographs have the ``analyzer'' arranged to be vertical, as it is fixed on the microscope. 
Angles of polarization describe the angle between the analyzer (vertical) and the filter between the light source and valve. 
Except where specifically noted, the valve is oriented such that the ring is at the top of the frame, so circumferential orientation is horizontal and radial is vertical. 
Scale bars are 100 $\mu$m, or 0.1 mm.

Images in this section confirm hypotheses and findings in literature, especially of circumferential fiber orientation near the free edge of the anterior leaflet. 
Additional work would be required to map the fiber structure of the whole valve using these techniques, especially through the distinct histological layers. 
We view this as an important and challenging goal.

\begin{figure}[H]
\centering
\includegraphics[width=.8\textwidth]{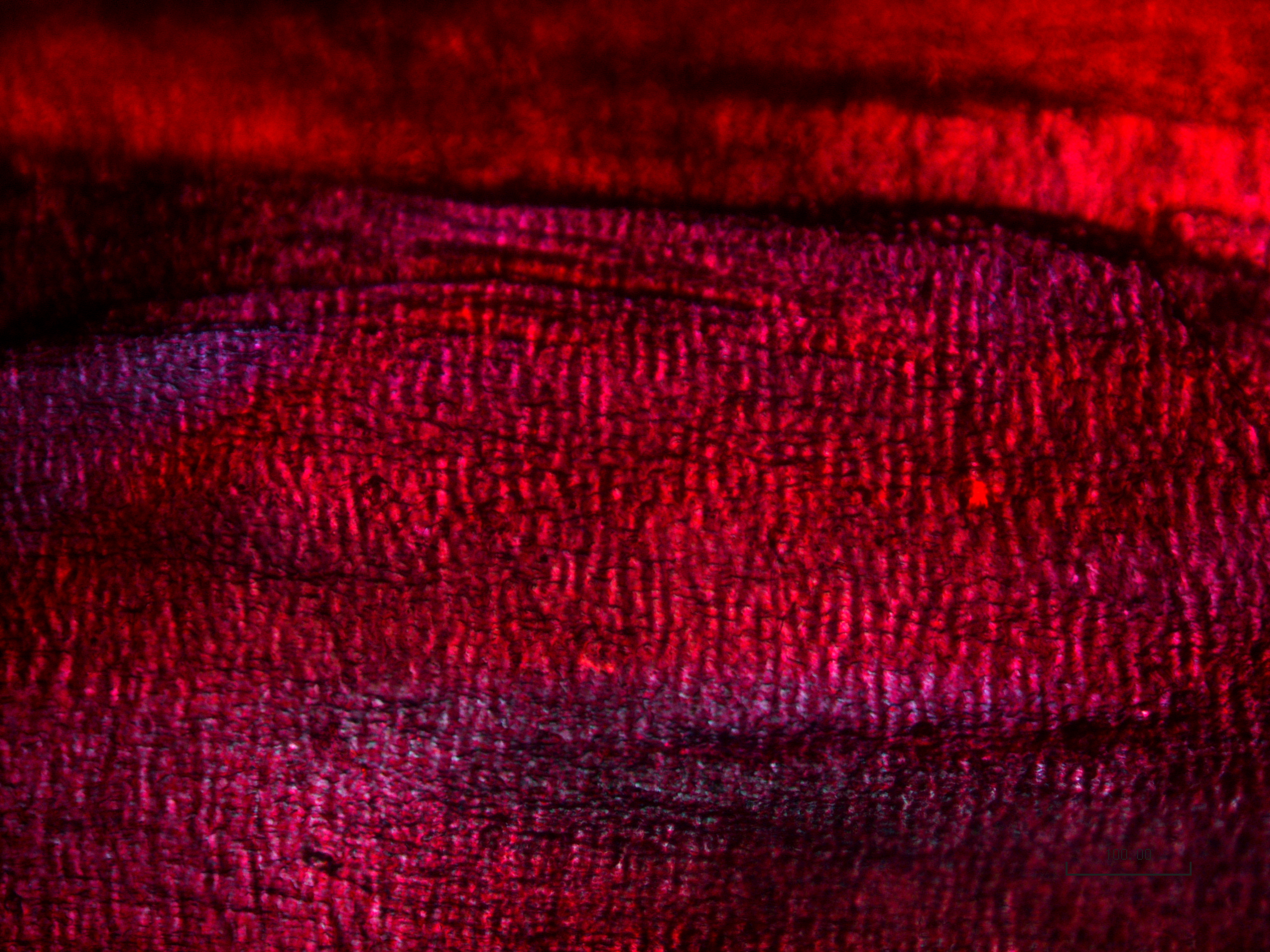} 
\caption{
Collagen fibers in anterior leaflet near free edge. 
Horizontal corresponds to circumferential orientation in the valve.  
Polarization angle, 90$^{\circ}$; microscope: Leica; camera, Nikon.}
\label{free_edge_scope}
\end{figure}

Figure \ref{free_edge_scope} shows the ventricular surface of the valve just above the center of the free edge. 
This image was taken on a region that has heavy circumferential orientation to the naked eye.
Note that this figure has some appearance of vertical black/bright alternating bands. 
We hypothesize that this is the appearance of collagen fibers oscillating normal to the plane of the frame. 
The fiber structure appears to run horizontally here, which suggests that the leaflet has largely circumferential orientation near the free edge. 
This is expected, since at the center of the the free edge there is nothing to support radial fibers. 
(In our mathematical models, the radial fibers like to ``pull away'' from this region, either spreading out away from the center of the leaflet or taking on an orientation that locally approaches circumferential. See figures \ref{valve_with_ref_front.pdf} and \ref{valve_with_ref_front_comm.pdf})

\begin{figure}[H]
\centering
$ \begin{array}{ccc}
\includegraphics[width=.3\textwidth]{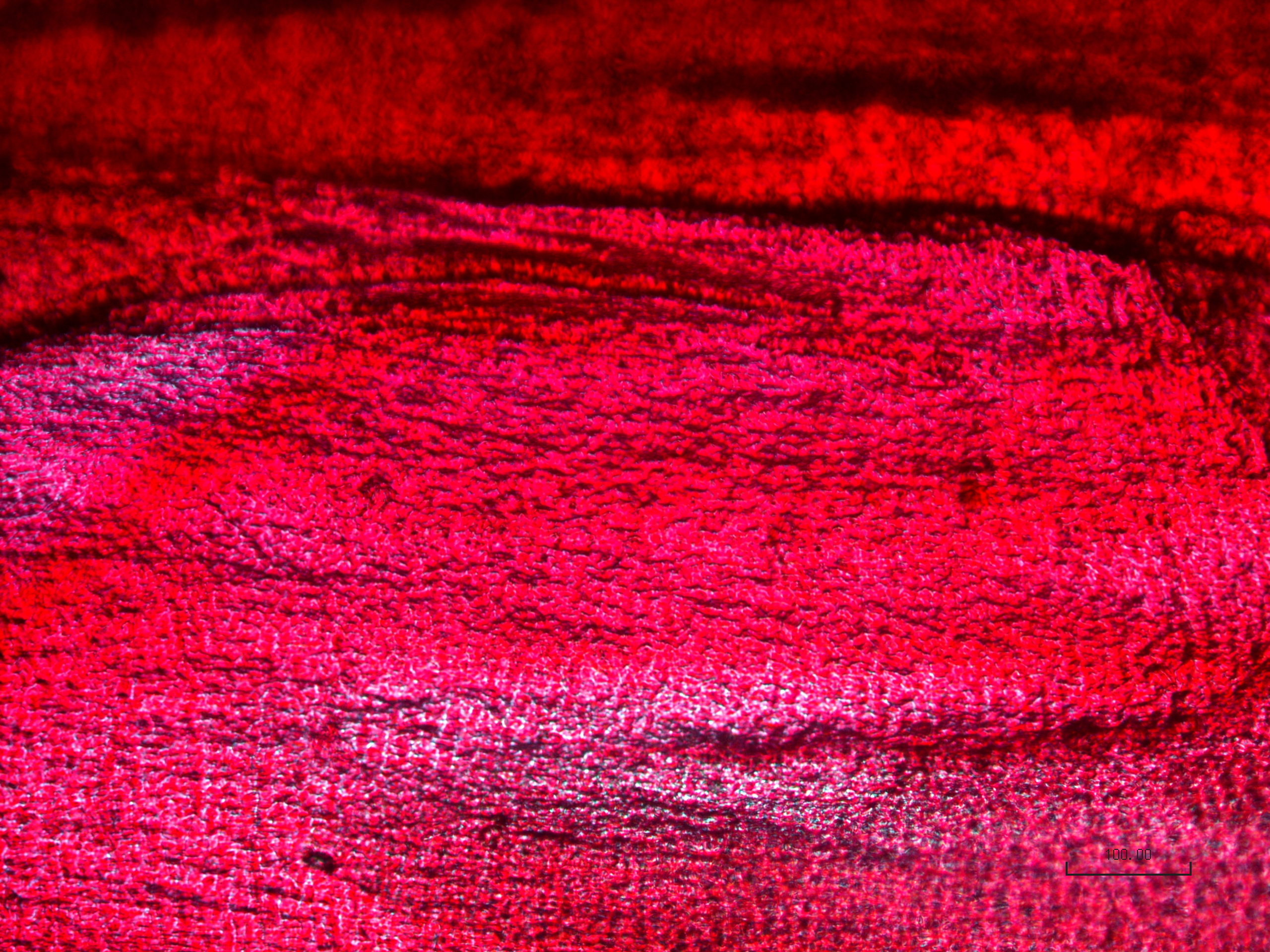} & 
\includegraphics[width=.3\textwidth]{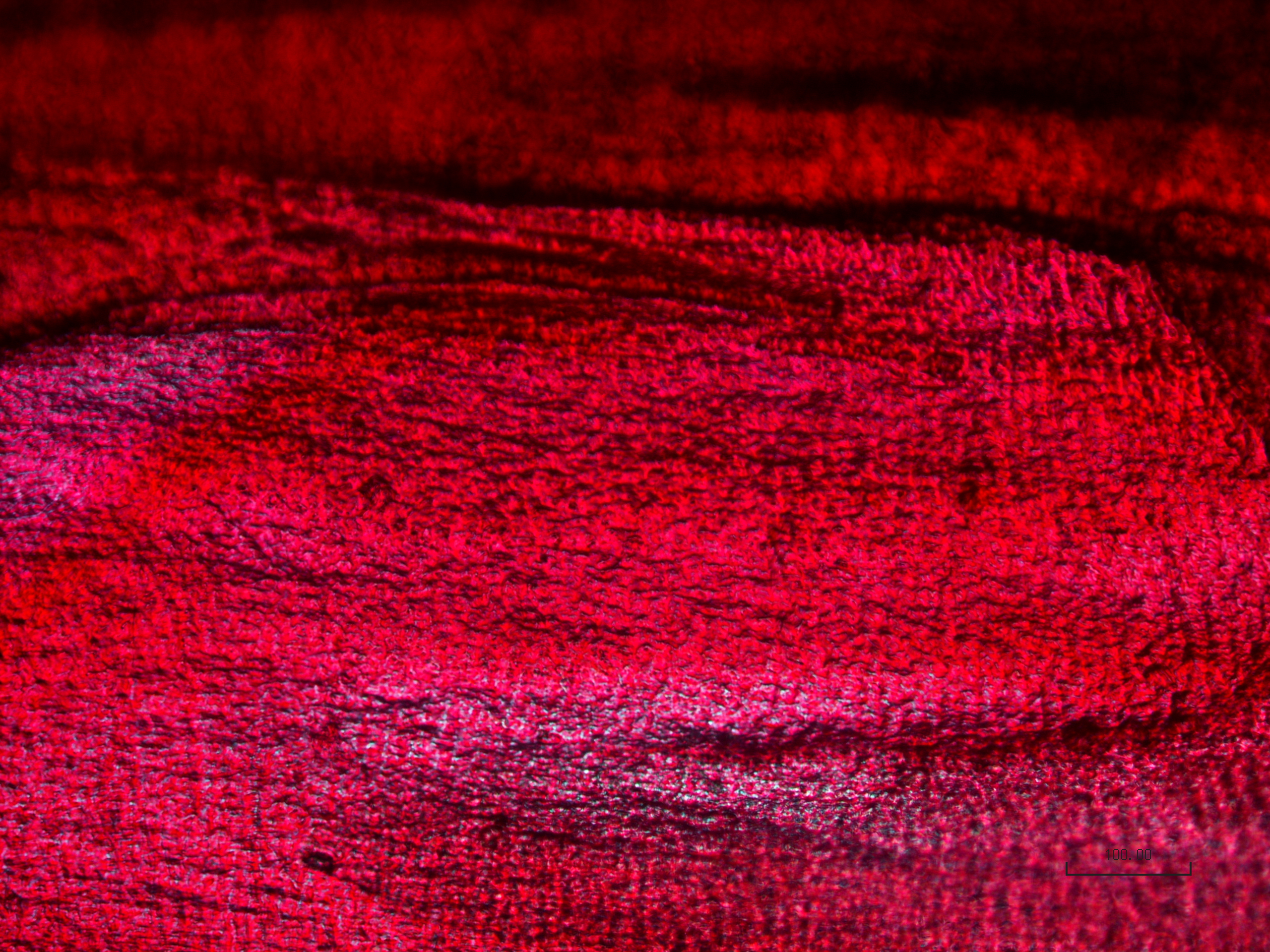}  &  
\includegraphics[width=.3\textwidth]{5c01003612.JPG} 
\end{array} $
\caption{
Similar to figure \ref{free_edge_scope} but showing three angles of polarization. 
Polarization angles, 0, 45, 90$^{\circ}$; microscope, Leica; camera, Nikon.}
\label{free_edge_scope_three}
\end{figure}

Figure \ref{free_edge_scope_three} shows the same image as figure \ref{free_edge_scope}, but also includes two additional polarization angles for comparisons. 
It is unclear at the time of writing how to interpret their differences, but we include them for the interested reader to draw their own conclusions. 

\begin{figure}[H]
\centering
\includegraphics[width=.8\textwidth]{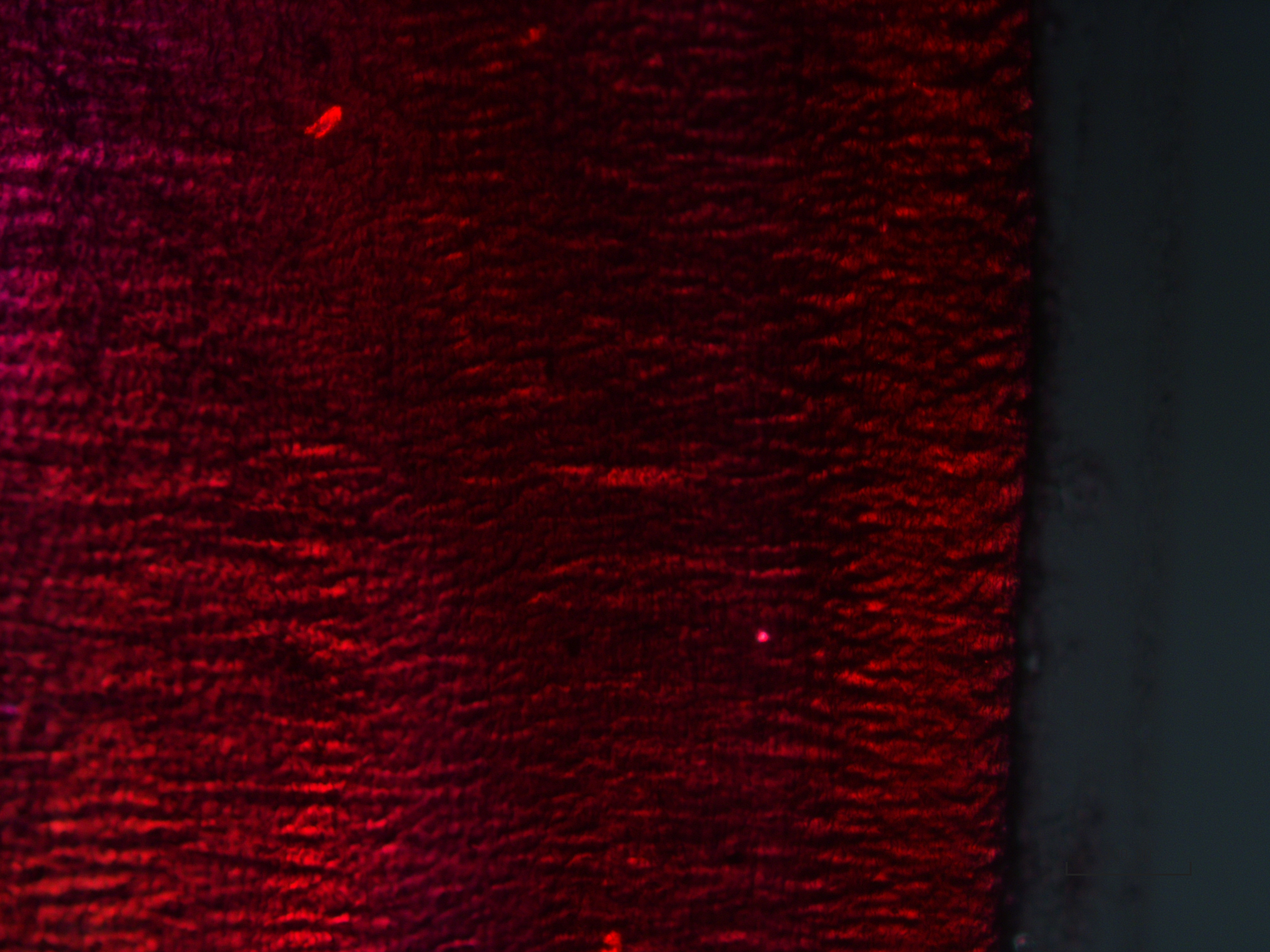} 
\caption{
Collagen fibers in anterior leaflet near free edge. 
The specimen is rotated such that the free edge runs vertically in the frame. 
That is, vertical orientation in the photograph corresponds to circumferential orientation in the valve. 
Compare to figure \ref{free_edge_scope}, which shows the valve placed 90$^{\circ}$ off to this frame. 
Polarization angles: 0, 45, 90$^{\circ}$; microscope: Leica; camera, Nikon.}
\label{free_edge_90}
\end{figure}

Figure \ref{free_edge_90} shows approximately the same views of fibers near the free edge, but rotates the specimen such that circumferential fibers are aligned with the analyzer. 
We also see fibers that appear to have circumferential orientation.

\begin{figure}[H]
\centering
$ \begin{array}{ccc}
\includegraphics[width=.3\textwidth]{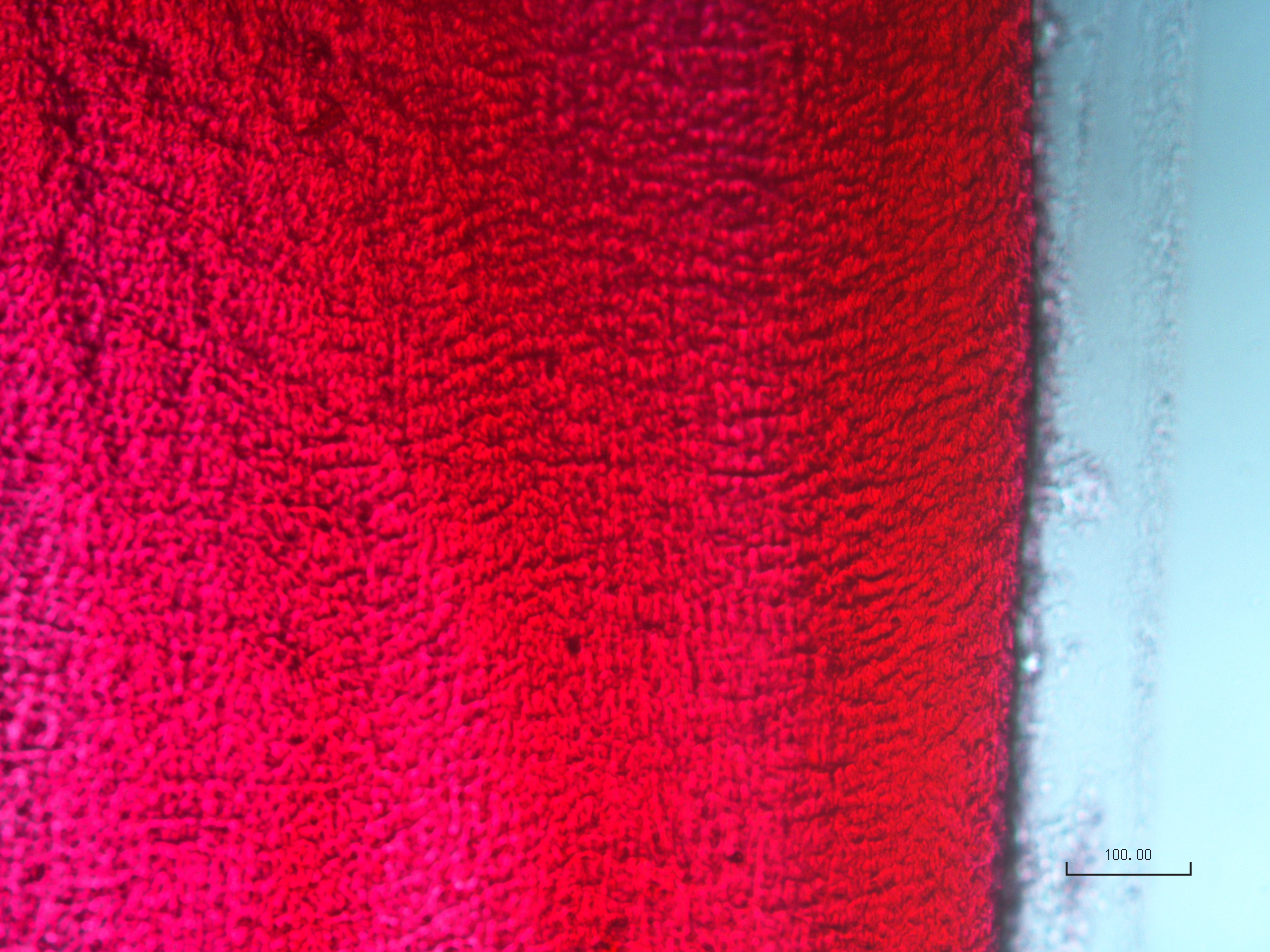} & 
\includegraphics[width=.3\textwidth]{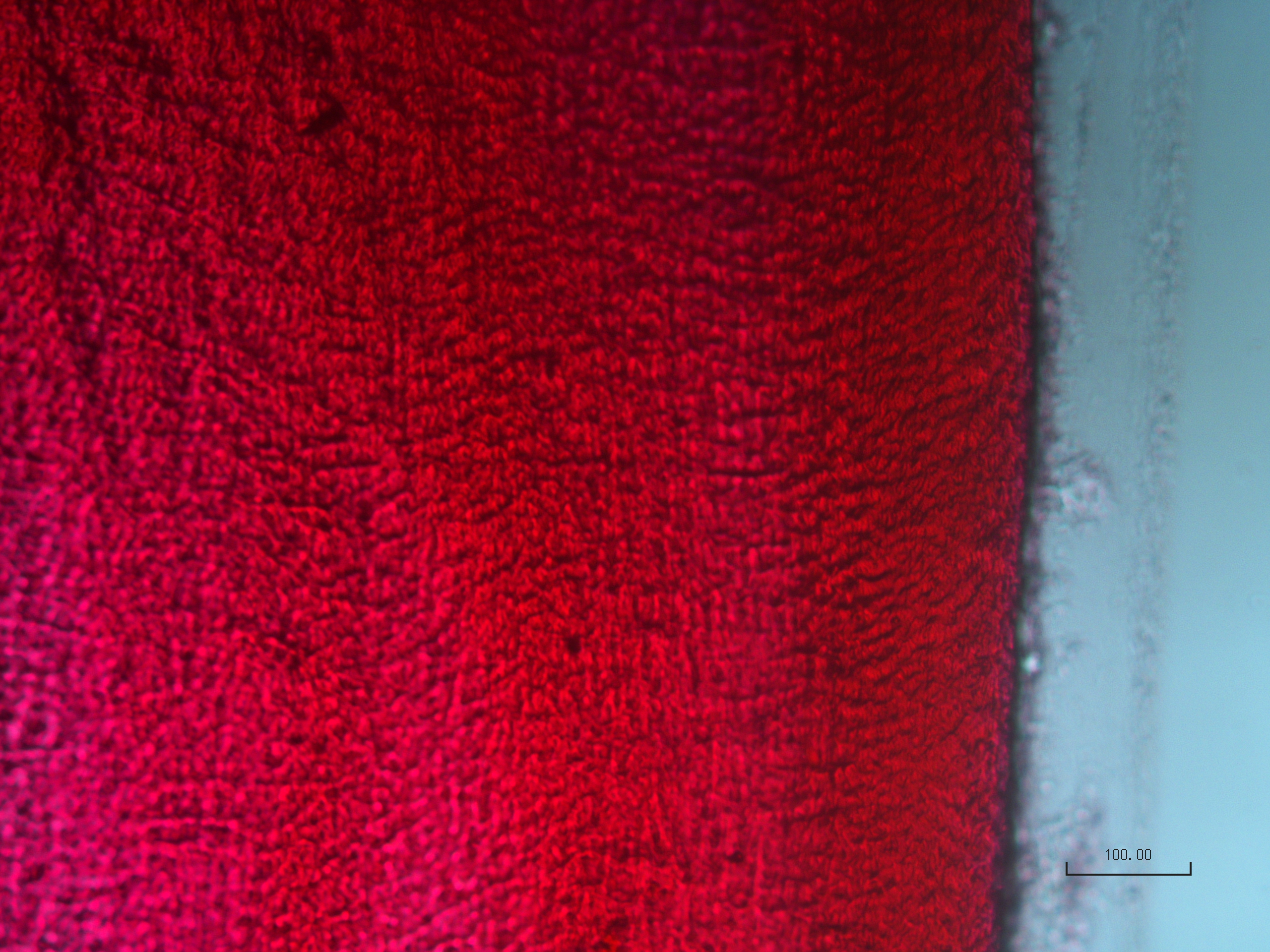}  &  
\includegraphics[width=.3\textwidth]{01000914.JPG} 
\end{array} $
\caption{
Similar to figure \ref{free_edge_90} but showing three angles of polarization. 
Polarization angles, 0, 45, 90$^{\circ}$; microscope, Leica; camera, Nikon.}
\label{free_edge_90_three}
\end{figure}

Figure \ref{free_edge_90_three} repeats the image shown in figure \ref{free_edge_90}, but includes three polarization of angles. 

\begin{figure}[H]
\centering 
\includegraphics[width=\textwidth]{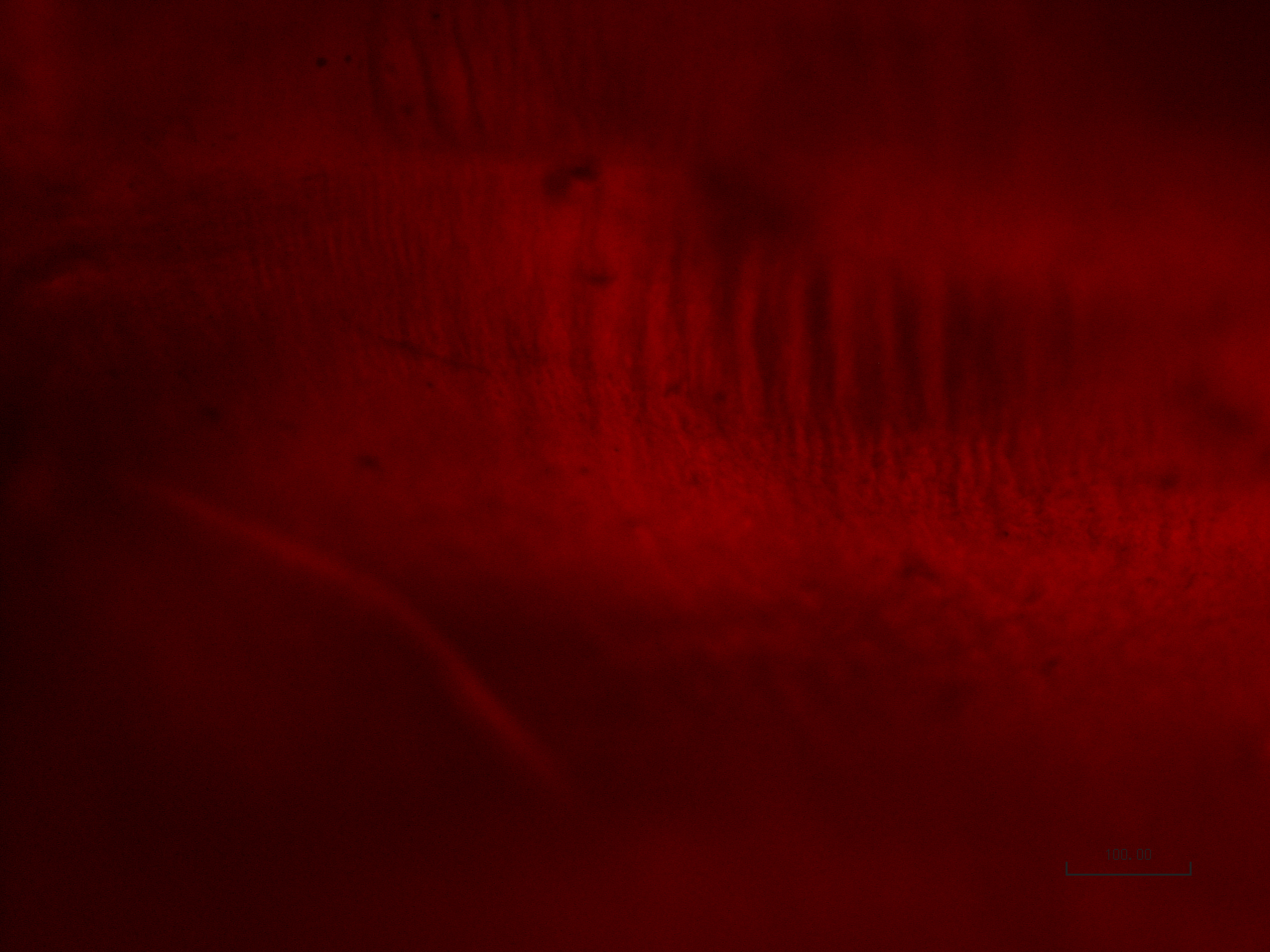}
\caption{Atrial side of the anterior leaflet away from the free edge. 
The thickness and opacity of the leaflet makes imaging this region difficult. 
Polarization angle, 90$^{\circ}$; microscope, Leica; camera, Nikon.}
\label{atrial_side_bad}
\end{figure}

We attempted to image locations further from the free edge, but the thickness of that area makes the tissue too opaque. 
Additionally, the roughness of the ventricular side makes the atrial side difficult to image, because the valve material physically resists sitting flat on the microscope. 
An example of this is shown in figure \ref{atrial_side_bad}. 
It is challenging to draw conclusions from this image; it is presented to illustrate the difficulty in imaging most of the mitral valve due to its thickness. 
Further study of this would be necessary to resolve the fiber structure of the entire leaflets. 
Perhaps a more sophisticated sectioning technique would allow the histological layers to be separated from each other and be imaged using polarized light microscopy. 

\begin{figure}[H]
\centering 
\includegraphics[width=\textwidth]{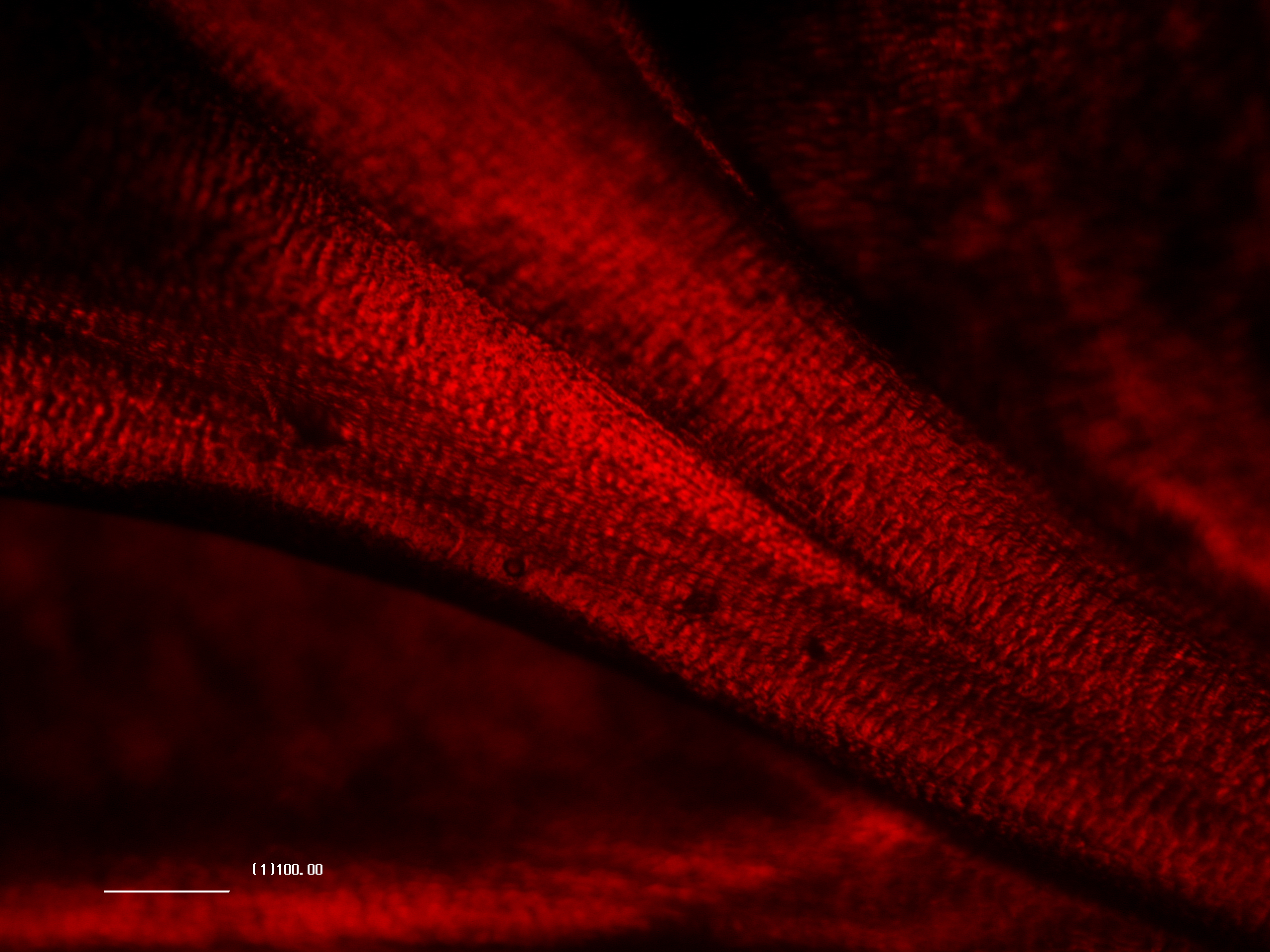}
\caption{Collagen fibers in chordae tendineae near insertion on anterior leaflet. 
Polarization angle, 90$^{\circ}$; microscope, Leica; camera, Nikon.}
\label{chordae_scope}
\end{figure}

Finally, for aesthetic value, we show one view of the chordae tendineae insertion on the ventricular surface in figure \ref{chordae_scope}. 

\newpage 
\addcontentsline{toc}{chapter}{Bibliography} 
\bibliographystyle{acm}
\bibliography{mv_refs}

\end{document}